\documentstyle[epsfig]{mn}

\newcommand{\etaflow}{\mu|{acc}}
\def\modmod#1{{#1\@}} 
\def\modC{\modmod{A}} \def\modA{\modmod{B}} \def\modB{\modmod{C}}
\def\modE{\modmod{D}} \def\modK{\modmod{F}} \def\modG{\modmod{E}}
\def\modD{\modmod{H}} \def\modF{\modmod{G}}

\makeatletter 
\def\@cite#1#2{(\if@tempswa #2 \fi #1)}
\makeatother
\newcommand{\aj}{AJ}
\newcommand{\ana}{A\&A}

\newcommand{\apj}{ApJ}
\newcommand{\apjl}{ApJL}
\newcommand{\apjs}{ApJS}

\newcommand{\araa}{ARAA}
\newcommand{\mn}{MNRAS}

\newcommand{\nat}{Nat\@.}

\newcommand{\bv}[1]{\mbox{\boldmath $#1$}}
\mathcode`\|="8000		
{\catcode`\|=\active \gdef|#1{_{\rm #1}}}
\def\pdif#1#2{\mathchoice{\partial{#1}\over\partial{#2}}
{\partial{#1}/\partial{#2}}{\partial{#1}/\partial{#2}}
{\partial{#1}/\partial{#2}}}
\def\refeq#1{{(\ref{#1})}}
\font\ninebmi=cmmib10 at 9pt \skewchar\ninebmi ='177
\font\sixbmi=cmmib10 at 6pt \skewchar\sixbmi ='177
\font\fivebmi=cmmib10 at 5pt \skewchar\fivebmi ='177
\newfam\bmifam
\textfont\bmifam=\ninebmi
\scriptfont\bmifam=\sixbmi
\scriptscriptfont\bmifam=\fivebmi
\def\bmi{\fam\bmifam\ninebmi}
\newcommand{\bld}[1] {{\bmi #1}}
\def\rd{{\rm d}}                                      
\def\rdif#1#2{\mathchoice{\rd{#1}\over\rd{#2}}{\rd{#1}/\rd{#2}}
{\rd{#1}/\rd{#2}}{\rd{#1}/\rd{#2}}}                            
\def\id{{\rm\,d}}                                
\def\Zsun{{\rm Z}_\odot}
\def\Lbol{L_{\rm Bol}}
\def\Mstar{M_\star}

\def\unit#1{\,{\rm {#1}}}
\def\umult#1#2{\ifx#2\unit\def\tmpa{\umulti#1#2}\else\toks0=\expandafter{#2}%
\edef\tmpa{\noexpand\umulti\noexpand#1\the\toks0}\fi\tmpa} 
\def\umulti#1#2#3{\ifx#2\unit\unit{#1#3}\else%
\message{\noexpand\umult error: must precede unit or \noexpand\unit}%
\unit{#1}#2#3\fi} 

\def\mega{\umult{M}}
\def\kilo{\umult{k}}

\def\micro{\umult{\mu}}

\def\metre{\unit{m}}

\def\kms{\unit{km\,s^{-1}}}

\def\secnd{\unit{s}}
\def\yr{\unit{yr}}
\def\Myr{\mega\yr}

\def\micron{\micro\metre}
\def\cm{\unit{cm}}

\def\parsec{\unit{pc}}

\def\erg{\unit{erg}}
\def\ergs{\unit{erg\,s^{-1}}}

\def\eV{\unit{eV}}

\def\Msun{\unit{M_\odot}}
\def\Lsun{\unit{L_\odot}}

\def\Zsun{\unit{Z_\odot}}

\def\Kelv{\unit{K}}

\def\refeq#1{{(\ref{#1})}}
\def\ee#1{\ifmmode\times10^{#1}\else$\times10^{#1}$\fi}
\def\etal{{et~al.}}
\def\eg{{e.g.}}
\def\ie{{i.e.}}
\def\etc{{etc.}}

\def\cf{{cf.}}
\renewcommand{\thefootnote}{\fnsymbol{footnote}}
\title[Hydrodynamics of the ISM in Symbiotic AGN]
{Symbiotic starburst-black hole AGN -- I.\\
Isothermal hydrodynamics of the mass-loaded ISM}
\author[R.J.R. Williams, A.C. Baker \& J.J. Perry]
{R.J.R. Williams$^{1}$\footnotemark[1]
A.C. Baker$^{2,3,4}$\footnotemark[1] 
\& Judith J. Perry$^{2}$\footnotemark[1]\\
$^1$Department of Physics and Astronomy, The University, Leeds, England
LS2 9JT\\
$^2$Institute of Astronomy, Madingley Road, Cambridge, England CB3 0HA\\
$^3$Service d'Astrophysique, C. E. A. Saclay, Orme des Merisiers
B\^{a}t. 709, F91191 Gif-sur-Yvette CEDEX, France\\
$^4$Department of Physics and Astronomy, Cardiff University, PO Box 913, Wales CF2 3YB}

\date{Received **INSERT**; in original form **INSERT**}
\pagerange{\pageref{firstpage}--\pageref{lastpage}}

\begin{document}
\label{firstpage}
\maketitle

\begin{abstract}

Compelling evidence associates the nuclei of active galaxies and
massive starbursts.  The symbiosis between a compact nuclear starburst
stellar cluster and a massive black hole can self-consistently explain
the properties of active nuclei.  The young stellar cluster has a
profound effect on the most important observable properties of active
galaxies through its gravity, and by mass injection through stellar
winds, supernovae and stellar collisions.  This mass-loss, injected
throughout the nucleus, creates a hot nuclear interstellar medium
(nISM). The cluster both acts as an optically-thin fuel reservoir and
enriches the nISM with the products of nucleosynthesis.  The nISM flows
under gravitational and radiative forces until it leaves the nucleus or
is accreted onto the black hole or accretion disc.

The radiative force exerted by the black hole--accretion disc
radiation field is not spherically symmetric. This results in complex
flows in which regions of inflow can coexist with high Mach number
outflowing winds and hydrodynamic jets.  We present two-dimensional
hydrodynamic models of such nISM flows, which are highly complex and
time variable.  Shocked shells, jets and explosive bubbles are
produced, with bipolar winds driving out from the nucleus.  Our results
graphically illustrate why broad emission line studies have
consistently failed to identify any simple, global flow geometry.  The
real structure of the flows is {\it inevitably}\/ yet more complex.

The structure of these nISM flows is principally determined by two
dimensionless quantities.  The first is the magnitude of the stellar
cluster velocity dispersion relative to the sound speed in the nISM.
These speeds measure the gravitational and thermal energies in the
nISM respectively, and, therefore, whether the gas is initially bound,
or escapes in a thermal wind.  The second parameter is the Mach number
of the ill-collimated nISM flow which is driven away from the central
black hole.  We discuss a two-parameter classification based on this
observation which, in future papers, we will relate to
empirical classifications.

The interplay between the nucleus and the wider galaxy depends
critically on the exchange of radiative {\it and mechanical}\/ energy.
The outbound mechanical energy transfer is governed by the nuclear
stellar cluster.  Active galactic nuclei will only be understood once
the symbiotic relationships between the black hole, the stellar
cluster, and the galaxy are considered.  It is impossible to treat
correctly any isolated component. Our conceptually simple and
self-consistent {\it Symbiotic} Model explains the observed complexity
of active galaxies without {\it ad hoc} measures.

\end{abstract}

\footnotetext[1]{\raggedright E-mail: rjrw@ast.leeds.ac.uk (RJRW),
a.baker@astro.cf.ac.uk (ACB), jjp@ast.cam.ac.uk (JJP)}
\renewcommand{\thefootnote}{\fnsymbol{footnote}}

\section{Introduction}

\label{s:intro}

There is strong evidence for black holes with a wide range in mass in
the centres of galaxies, and unambiguous evidence for dense stellar
clusters in galactic cores.  Activity in the nuclei of galaxies appears
to be intimately linked not only to these supermassive black holes (BH)
and nuclear starbursts, but also to interactions between galaxies.  The
evidence for this link -- both direct and circumstantial -- has been
reviewed in detail by Perry~\cite[1993a, 1999; see
also]{filippenko92}.  The exact nature of the link remains uncertain:
interactions, starbursts and activity certainly can not be equated, and
not all systems with these components are active.  In particular,
galactic interactions do not always result in activity
\cite{Sanders96}. However, the evidence does unequivocally demonstrate
that the active nucleus is {\em not} a `parasite' upon a `host'
galaxy, but rather that activity is a galaxy-wide phenomenon which is
most clearly identified by the dramatic events in the galaxy nucleus.
In this paper, we consider the hydrodynamics of this basic nuclear
system.

The most dramatic of the nuclear phenomena is the highly luminous
($10^{10}$ to $10^{14} \Lsun$), time-variable, broad-band (radio to
$\gamma$-ray) radiation emitted from regions as small as a few
light-seconds.  This radiation is accompanied in many AGN by broad
emission lines from heavy elements in cold ($\approx 10^4 \Kelv$)
dense ($\approx 10^9$ to $10^{12} \cm^{-3}$) gas.  In the unified
model of active nuclei, the broad line region is a constituent of all
active nuclei, although it is not observed in some as a result of
continuum beaming or intervening obscuration \cite{kb86,barthel}.
Line widths, which reflect local gas velocities, are typically several
$10^3 \kms$. Average equivalent widths are remarkably consistent
across orders of magnitude differences in total bolometric luminosity
(although with a marked scatter between individual objects at a given
luminosity), indicating a tight relationship between the production of
cold, condensed radiating gas structures and the production of the
central luminosity itself.


Although accretion onto black holes seems clearly to dominate the
physics of the innermost regions of AGN (within $\sim 100$
Schwarzschild radii), and to provide a satisfying explanation of the
radiation processes which create the luminous broad-band continuum,
pure black hole accretion models have always suffered from an
inability to explain the structure of the line emitting regions and
had great difficulty in explaining the fuelling, before the importance
of stars in the wider parsec sized core was recognised.

The stars which play the dominant role in the energetics and mass
budget of galactic nuclei are young, mass-losing stars: these are most
abundant and important in just those nuclear stellar clusters --
starbursts -- which are thought to be involved in activity.  Current
theoretical models linking activity and starbursts come in two
variants. In this paper, we explore the structure of the model
\cite{taiwan} based on the early work of Perry \& Dyson \cite[1985,
hereafter{}]{PD} which unites starbursts and black holes as compact
symbiotic systems in the nuclei of active galaxies
\cite[Collin-Souffrin \etal\ 1988, hereafter CDMP;]{NS,taiwan,WP94}.
In contrast, the pure starburst model equates the two phenomena,
considering AGN to be but extreme examples of nuclear starbursts
\cite{terl90,cid97}.  Interestingly, radiative shocks around
supernovae play a central r\^ole in both theories.  We shall refer to
the starburst-black hole paradigm for AGN as the Symbiotic Model.

Since powerful winds and supernovae dominate the energetics of
starbursts, understanding the dynamic link between black holes and
starbursts requires careful analysis of the stellar- and hydro-dynamics
of the combined system.  In this paper we present the first
comprehensive treatment of the hydrodynamics of such a symbiotic
system.

In the rest of this introduction, we outline the physical context and
assumptions of our model.  These assumptions will be discussed in more
detail in subsequent sections.

\subsection{Symbiotic models}

In the Symbiotic Model of AGN, the nuclear starburst stellar cluster
feeds the black hole through the nuclear interstellar medium (nISM)
created by stellar mass-loss.  The formation of shocks within the nISM
is an {\it inescapable} consequence of the interaction between
individual (mass-loading) stars and the global nISM.  It is important
to emphasise that the presence of shocks in the nISM is a model
independent phenomenon and is a direct consequence of basic
thermodynamics, stellar dynamics and hydrodynamics in galactic nuclei.
In QSOs, the shocks around individual supernovae are sufficiently
large that the shocked nISM gas has time to cool radiatively into
thermal equilibrium with the central radiation field at $\approx
10^4\Kelv$, so forming broad line emitting structures (PD). The
thermal pressure in such shock confined structures equals the
stagnation pressure of the flow, which agrees well with the pressures
deduced from observations of the emission lines.  PD showed that using
the observed equivalent widths of the high-ionization lines (HILs) to
deduce the average properties of the nISM, the implied flow rates in
the nISM are sufficient to fuel the black hole accretion.

This satisfyingly self-consistent picture did not initially explain
the two-(or more)-component nature of the broad emission line region
(BELR): the low-ionization lines (LILs) are observed to be both
red-shifted with respect to the HILs, and to require a harder ionizing
continuum.  However, CDMP then showed that the nISM flow can
back-scatter X-rays which provide `deep heating' of the accretion disc
sufficient to produce the LILs.  This two-component model resolved
several outstanding questions concerning the BELR.  The transient
nature of the BELR clouds solved the `cloud confinement problem'.  The
existence of two distinct regions of broad line emission also gave a
simple explanation for the observed systematic velocity shifts and
multiple component structure of the broad line profiles
\cite{cf94,broth94,arav98}.  Our results, in this paper, add another
dimension to this understanding of the structure of the BELR: we find
that the nISM itself forms distinct multi-component regions of
differing characteristic excitation levels and velocities.

The structure of the emission line region which emerges is that of
distinct local clumps of line-emitting gas confined by transient
shocks.  Recently, Arav \etal{} used the new generation of
observations to update the limits which smooth line profiles put on
the number of such centres of emission \cite[\eg]{abc}.  Their
analysis was based on the fact that discrete emitting units give a
smooth line profile only when the number of emitting units is very
large.  Their method is sensitive to structures with velocity widths
between 50 and $200\kms$; within this range they find essentially no
perturbations in the wings of the Balmer lines, to very tight limits.
This rules out many of the simplest scenarios for the formation of the
lines by discrete clumps.  Arav \etal{} were able to rule out emission
from less than $3\ee6$ smooth shell sources with expansion velocity
$100\kms$. However, such sources, with small velocity width and
perfect symmetry, are not representative of shock-confined emission
line clouds in galactic nuclei.  Structures which are intrinsically
smoother in emission than those they specifically considered, such as
accretion discs (which are thought on other grounds to be important
contributors to the Balmer emission, CDMP), or strong wind sources and
bowshocks ($v
\ga 200\kms$) are not ruled out by these observations.  The sonic
velocity in the nISM is at least $400 \kms$; this is the minimum
expected shock velocity in the region.  In fact, characteristic
velocities of the shocks which arise in symbiotic nuclei are $ \times
10^3 \kms$ [the combination of stellar ejection velocities (often $\ga
3 \times 10^3 \kms$) and the stellar orbital velocities ($\ga 600
\kms$ within the central parsec)].  Furthermore, the shocks are
non-uniform: they are either open bow shocks with streaming tails, or
closed shells distorted by their interaction with the wind.  The
density, radial velocity and surface emissivity -- for any shock --
all vary significantly around the shock-front, generating broad,
complex lines with wings which are much less steep than those Arav et
al. assumed.  As Arav et al.  pointed out, any flow which produces
smoothly varying structures on velocity scales of 500 -- $1000\kms$ is
allowed by the observations. We therefore conclude that their elegant
analysis provides support for the PD model, and underlines the
importance of hydrodynamic models of the BELR.

\subsection{Physical Ingredients}

The physical parameters characterising the centre of a galaxy are the
mass of the black hole, and the mass, stellar density profile, IMF and
age of the stellar cluster.  The IMF and age govern the mass-loss rate
from stars due to stellar evolution; collisions and tidal disruption
are negligible except in the central regions of the densest clusters.
The mass lost from the stars inevitably creates a nuclear interstellar
medium (nISM)\@.  If there is a black hole accretion flow, either from
this nISM or from some other reservoir (\eg{} flow into the nucleus
from a galactic bar), then radiation processes near the black hole can
generate a powerful broad-band continuum (although not if the flow
into the black hole is hot or optically thick \cite{ny94,bb99}). For
the present, we take this as an additional input to our simulations.
The system which we consider in this paper is one in which such
accretion processes are effective.

The stellar cluster surrounding the black hole helps to determine the
structure of the nISM flow on the scale of the BELR, through gravity
and distributed stellar mass loss.  The strength of this mass-loading
is principally determined by the age and initial mass function (IMF)
of the stellar cluster.  We shall assume that the cluster is young and
dominated by high mass stars, and take our basic parameters from the
stellar synthesis models of Williams and Perry \shortcite{WP94}.

Mass input from the stellar cluster is discrete and local.  Because we
are interested in the generic properties of the global flows, we will
concentrate on the luminous QSOs. In such models, the mass processing
rates are high enough that we can average over the numerous small
scale structures resulting from the interaction between individual
stars and supernovae with the global flow.  Future papers will
investigate the detailed structure of individual mass input sources,
and the effects of the discrete mass input on the global flow.

We have focussed on mass injection from stars within the cluster
rather than from the external ISM, primarily for two reasons.
Importantly, the mass flux from the external ISM is unlikely to be able
to provide a significant fraction of the fuelling requirements of the
AGN as a continuous source \shortcite[Shlosman \etal{}]{sbf90}.  The
wind from the nucleus is in some cases strong enough to drive the
external ISM away.  Also, this is the situation described in PD, where
the shocks around the mass-loading centres also generate the cool gas
which emits the observed emission lines.  Therefore, rather than
include the additional free parameters specifying the distribution of
the ambient ISM at the outer boundary of our models in this first
paper, we have chosen to concentrate on the case where the phase of
accretion from tens or hundreds of parsecs to within a parsec of the
central black hole has already ceased.

We have assumed that any relativistic jet driven from close to the
central black hole is absent, or at least narrow, transferring little
momentum to the surrounding global flow.  If this assumption is
reasonable, our models apply equally to radio-loud and radio-quiet
objects.  Radio structures on parsec scales are seen in radio-quiet
objects as well as radio-loud ones: the explicit inclusion of jets
in our models would be an interesting extension.

\subsection{Formation scenarios}

The starburst-black hole system is in a symbiotic relationship with
the wider galaxy. We study the structure of the nuclear system in its
active phase in this paper. We leave the vexed questions of the
formation of the nuclear system and the details of its symbiosis with
the wider galaxy to later work. This follows the tradition well
established in stellar physics: first understand structure, then
evolution, lastly formation.  Nonetheless, there are clear
indications, both from observations and from theory, of possible
formation scenarios \cite{p97}.  Interactions between galaxies disturb
the equilibrium of the ISM of the galaxies, and gas can flow towards
the nucleus where it can form a nuclear starburst.  The size scale of
the starburst depends on whether the gas can shed its angular
momentum, and on whether it hangs up at a Lindblad resonance
\cite{lamb97}.

Direct optical imaging of galaxies with bright active nuclei is
difficult, but HST and near-IR imaging are now revealing the
morphological types
\cite{Dunlop93,McLeod94a,McLeod94b,taylor96,bahcall97,boyce97,mcleod97}.
Radio and quasar galaxies appear to be large and luminous, and to show
signs of past merger events. Galaxies with radio-loud active nuclei
are essentially all spheroidal, whereas those with radio-quiet cores
are approximately equally divided between spheroidal and
disc-dominated structures \cite{kukula97a}. This may indicate
\eg different interaction histories for radio-loud and radio-quiet
systems.

There is much circumstantial evidence that similar environmental
factors influence the frequency of both active nuclei and starburst
galaxies \cite{nefh92}.  Cooling flows around massive elliptical
galaxies may provide a source of fuel for both phenomena
\cite{bt95,ciotti97,bremea97,friac98}.  Ultraluminous IR galaxies may
represent a transition phase as dense shrouds of dust lift to reveal
bright optical quasar nuclei, since the environments, molecular gas
content, near-IR colours and optical emission line diagnostics of the
two classes are similar \cite{Sanders88}.  Processed material with
near-solar metallicities in the nISM is seen in AGN even at the
highest redshifts. At redshifts $ 2 < z < 3 $, absorption line studies
reveal that most of the gas in galaxies typically has $Z/\Zsun \approx
0.01 \mbox{--} 0.1$ \cite{pettea94,pettea97}.  To reach the high metal
abundances observed in QSOs after short cosmic times, massive star
formation, probably {\it in situ}, must be intimately linked to the
evolution of active galaxies \cite{taiwan,ferea96,korea96}. Populations of
star-forming galaxies at $z \sim 3$ have now been discovered
\cite{steidel96}.  This redshift range is hence a critical period in
the evolution of both normal and quasar galaxy populations
\cite{mcleod97}.

Whatever the likely formation scenario, the nISM in our models reaches
a structure characteristic of its long term behaviour -- equilibrium,
oscillations or explosions -- on dynamical timescales typically a few
thousand years. These timescales are much shorter than any plausible
formation or evolution times \cite{WP94}, and so we treat the
structure of the system independently from its formation and
evolution.

\subsection{Hydrodynamics of the nISM}

Our aim in this paper is to explore the hydrodynamics of those nuclear
black-hole/stellar cluster systems which {\em are}\/ active.  Williams
\shortcite{will97} summarises previous hydrodynamic models of the flow
structure of active nuclei.  Our study is unique in that it treats the
distributed mass input through the nucleus and the gravitational field
of the stellar cluster.

We treat the global hydrodynamics of the nISM, as it interacts with
the central black hole, the stellar cluster and the accretion
disc. These physical components are essential to the starburst-black
hole model of AGN -- {\it and}\/ to most other dynamical models of
active nuclei.  We have chosen to simplify the properties of this
system as far as possible.  This approach allows us focus on the most
important properties of the global hydrodynamic interactions between
stellar cluster, black hole and accretion disc.

Independent of the detailed model, it is possible to estimate
characteristic values of the flow density and velocity required for
the accreting nISM flow to be able to fuel the AGN.  PD used such
simple scaling arguments to discuss the structure of the nISM flow,
and to investigate the consequences for the fuelling of AGN and for
the creation of the BELR.  If the local nISM is too dense, it obscures
the system; if it is too diffuse, it fails to provide the necessary
fuel for the AGN. Those flows consistent with these constraints can be
shown to have ram pressures similar to the pressures deduced from
observations of the broad line clouds.  The ram pressure of the nISM
will therefore have an important dynamical effect on the BELR clouds,
whatever their confinement mechanism.  Although these simple arguments
provide interesting overall constraints, many important properties of
the AGN depend on the structure of the flow, which PD did not address.
Detailed hydrodynamic modelling is necessary to determine these flow
structures.  More broadly, it is clearly necessary to consider the
hydrodynamics of the nISM in order to develop fully {\em any}\/ model
of AGN\@.

In the absence of a black hole, in a dense stellar cluster with
negligible external pressure, gas accretes onto a hydrostatic core,
and a weak wind forms (\cf{} Section~\ref{s:spherical}).  The core
grows until it reaches the sonic radius, and the wind from the cluster
increases to an equilibrium in which it transports the mass injection
from the stellar cluster out into the wider galaxy.  Once a black hole
is added to the system, with an accretion-generated radiation field,
many more complex behaviours can occur, some of which we investigate
in this paper.  Broadly, however, the black hole can accrete matter at
the centre of the flow, weakening the eventual wind, and the radiation
field may drive it away.  In the absence of radiative driving, the
sonic (or Parker) radius is $GM/2c|s^2 = 1.4\parsec$ for a $10^8\Msun$
BH mass and gas at a temperature of $10^7\Kelv$, which is at least as
large as the broad emission line region even in QSOs. The velocities
in pure thermally driven winds only reach $\sim360\kms$ (the sound
speed in the nISM) at such radii.  Such low velocities are in conflict
with the strong evidence for outflowing gas in and around AGN at far
higher velocities, most clearly in broad absorption line QSOs
\cite{turn88,ost91}.  If radiative driving occurs at all angles, then
the hole may well be starved of fuel. This will reduce the luminosity,
giving rise to a limiting feedback process.  However, there is strong
evidence of beaming in all AGN. This asymmetry in the radiation field
means that the radiation force can be greater than the gravitational
attraction of the black hole on the axis of the accretion disc, but
not in its plane. In such cases, mass can fall inwards towards the
plane and be driven away on axis.  Such flows are typical of our
simulations.

The structures and the global kinematics of these nISM flows are
likely to be generic features.  However, we will see that small-scale
features such as gas cooling may play an important role in determining
the {\it observable} properties of the flow.  Where cooling is
significant, the average opacity of the gas will be increased, which
will re-scale the importance of the central radiation field.  The
central broad-band (radio to X- and $\gamma$-rays) radiation field is
powered by mass accretion through a disc and down into the black
hole. We model this radiation field as the sum of an unresolved point
source and an accretion disc component which follows a simple cosine
law.  We concentrate on models appropriate to QSOs, since there is
wide agreement that they accrete at or near the Eddington limit.  In
contrast, the lower luminosity Seyferts appear to span a much wider
range in mass-to-light ratio. We discuss the consequences of scaling
our models to Seyfert properties in Section~\ref{s:seyfert}.

Because we do not model the accretion disc explicitly, the effective
Eddington ratio, $f|{Edd}$, is an arbitrary input parameter. A
goal of future research is to determine the relationship between
$f|{Edd}$ and the properties of the nISM, since this must be
self-consistent (\ie, the accretion rate onto the central black hole
must be sufficient to power the luminosity of the AGN).  The
characteristics of the disc should be determined by the rest of the
nuclear system.  Although $f|{Edd}$ changes on timescales much shorter
than the evolution of the overall system, these are much longer than
the dynamic timescales of the nISM\@. It is not currently feasible to
determine $f|{Edd}$ by modelling the disc and nISM simultaneously. We
have chosen a simplified parameterisation so we can, to first order,
rescale $f|{Edd}$, (see Section
\ref{s:outline}).  The rescaled solutions will have the same global
properties as the real flows.  Subsequent papers will fill in the
important details in this picture.

The flows in individual simulations are complex and varied, whilst
exhibiting a generic bipolar circulation pattern.  These results
suggest why over 30 years of careful and detailed BELR observations
and simulations have failed to identify a simple, global flow
geometry.  If anything, real flows will be more complex, not simpler,
than the flows we find here.

These generic bipolar flow patterns are characterised by both an
accretion inflow and a wind outflow in the nISM\@.  The gas which
accretes can generate the luminosity assumed by the flow models.
Bipolar outflows -- and even hydrodynamical jets -- are generated, so
long as there is sufficient outward force along the accretion disc
axis.  These structures will be found independent of whether the
nuclei are radio-loud or radio-quiet: they are in general less
collimated and lower velocity than the most dramatic radio structures
observed, although when a plume of gas escapes the nucleus it may be
observable as a radio structure.

The strength and structure of these flows depends primarily on the
ratio of the gravitational and thermal energies in the nISM.  Some of
the inflowing gas is driven away into the wider galaxy, while some may
accrete onto the black hole.  The wind expelled from the nucleus can
carry both significant mechanical energy and heavy elements into the
wider galaxy.  Thus, the nISM flows will affect significantly the
structure of the ISM of the galaxy. At early times, the flows will
enrich the galaxy ISM and may drive shocks and shells of gas out to
distances of kiloparsecs
\cite{BP,DFP,FPD,sp98}.

These nISM flows can only be observed indirectly.  The nISM itself is
hot -- typically $\sim 10^7 \Kelv$, the Compton temperature
characteristic of the central radiation field -- and does not give
rise to observable line radiation, although limits on its column
density are set by Thomson optical depths.  The BELR is formed within
the nISM, and so the broad line profiles will reflect the nISM
velocities as well as the stellar orbital velocities.  If the BELR is
formed behind shocks driven by the stars as they move through the nISM
(as in the models of PD) then the characteristic velocities of the
line emitting material will be determined by stellar orbital, stellar
mass-ejection and nISM velocities.  The stellar velocities will
broaden the nISM contribution to the line widths, but the fundamental
shape and asymmetries of the lines will reflect the nISM flows.
Empirical models of such profiles \cite{ap99} show most of the
characteristics of observed line profiles and their diversity.  In
this paper, we study the flow structures which underly the
structures determined by such empirical studies.

We find that we can characterise the structure of the nuclear
interstellar flows on a two-dimensional plot.  This is our first step
in developing a theoretical diagnostic diagram for AGN flows,
analogous to the Hertzsprung-Russell diagram which provides a
systematic understanding of stellar structure.  Stellar spectral type,
luminosity, colour and temperature -- characteristics which are
determined by the basic physical parameters of age, metallicity and
mass -- are used as the axes of the HR diagram.  Similarly, the axes
of our plot can be expressed in a number of ways. Although it would be
desirable to use the fundamental physical parameters listed above as
the axes, we find that the most natural variables are derived
parameters which are closely related to the underlying physics (see
Section~\ref{s:parms}).

\subsection{Scope and layout}

We present numerical models of cylindrically symmetric, isothermal
flows in black hole-starburst AGN\@.  We have used global assumptions
which are physically justifiable for QSOs. Various local deviations
from those global approximations will be examined in detail in a
future paper.

In Section~\ref{s:physical}, we specify the parameters of the physical
model of a supermassive black hole and a starburst stellar cluster.
In Section~\ref{s:equations}, we give the hydrodynamical equations and
discuss their scaling behaviour and their relationships to AGN
observation.  We also introduce the reduced set of dimensionless
parameters with which we describe the gross features of the flows, to
serve as a guide to the numerical models which follow.  The numerical
methods used to implement the hydrodynamic representation of our
physical model are outlined in Section~\ref{s:code}.  We then describe
the results of our detailed modelling.  First, in
Section~\ref{s:spherical}, we describe results for purely spherical
flows.  We verify that our methods reproduce the analytic results, and
summarise the general properties of such flows, and those described in
the literature.  Next, in Section~\ref{s:results}, we present the
results from our aspherical models, calculated for parameters
appropriate for luminous QSOs.  In Section~\ref{s:seyfert}, we discuss
models where the Eddington ratio is very small, as is suggested to
occur in some Seyfert galaxies.  In Section~\ref{s:discuss}, we
discuss these results and their physical implications in the context
of the current observational and theoretical understanding of activity
in galactic nuclei.  We compare our results with the models
of the flow structure in active nuclei of PD and Perry (1993a).  We
sketch the global constraints on the models.  Readers who are
most interested in the observational consequences of our work, and
ways in which our model can be tested, could now turn straight to the
figures in Section~\ref{s:results} and then read
Section~\ref{s:discuss}, where we show that the symbiosis between a
nuclear starburst stellar cluster and a supermassive accreting black
hole, via the nISM, provides a self-consistent model for nuclear
activity in galaxies.  In Section~\ref{s:concl}, we draw together our
conclusions.

Illustrative animations are available at

http://ast.leeds.ac.uk/\verb!~!rjrw/agn.html.

\section{Physical Structure of the Active Region}
\label{s:physical}

\begin{table*}
\caption{Glossary of non-standard symbols and abbreviations widely used in the
text.}
\label{t:glossary}
\begin{tabular}{ll}
\hline
Symbol		& Description \\\hline 
BELR		& Broad emission line region\\
$c|s$		& Sound speed in the nISM\\
$f|{disc}$	& Fraction of luminosity in accretion disc (anisotropic)
component\\
$f|{Edd}(\theta)$ & Ratio of outward radiative force (from Thomson 
scattering) to inward gravity of the black hole\\
$\langle f|{Edd}\rangle$ & Angle average of $f|{Edd}$, \cf{}
eq.~\protect\refeq{e:fangle} \\
$\bar{f}(\theta)$ & Form function for $f|{Edd}$, so $f|{Edd}(\theta)
= \langle f|{Edd} \rangle \bar{f}(\theta)$\\
$g$, $\bld{g}$	& Body-force acceleration (gravity and radiation)\\
$h$		& Index of stellar distribution in stellar cluster halo,
\cf\ eq.~\refeq{e:cluster}\\
$H$		& Thickness of accretion disc\\
$L|{Bol},L_{46}$	& Accretion-generated bolometric luminosity of the BH,
$L|{Bol}/10^{46}\ergs$\\
$L|{disc}$	& Accretion disc (anisotropic) component of $L|{Bol}$\\
$L|{Edd}$	& Eddington luminosity of the black hole\\
$\cal M$	& Flow Mach number\\
$\dot{M}|a$	& Rate of accretion of mass onto black hole\\
$M|c$		& Mass of stellar cluster within $r|c$ (core mass)\\
$M|{cl}$,$M|{cl,8}$
& Total mass of the stellar cluster, $M|{cl}/10^8 \Msun$\\
$M|h$,$M|{h,8}$	& Mass of the black hole, $M|h/10^8 \Msun$\\
$m|l$		& Low mass cut-off for stellar cluster IMF\\
$\Mstar$	& Mass of an individual star\\
nISM		& Nuclear interstellar medium\\
$n_{\star}(r)$	& Radial star number density of stellar cluster\\
$n|c$		& Star number density at edge of stellar cluster core\\
$n$		& Total particle density\\
$Q, Q_{-8}$	& Rate of mass-loading from stellar cluster as a fraction of 
stellar density, $Q/10^{-8}\yr^{-1}$\\
$\dot{q}(r)$    & Mass loading rate\\
$\dot{q}|c$, $\tilde{\dot{q}}(r/r|c)$ 
	& Mass loading rate in the core, variation of mass loading
scaled to this value\\
$r|c$		& Radius of stellar cluster core\\
$r|s$		& Sonic radius (for spherical models)\\
$R|{\star}$	& Radius of an individual star\\
$R$		& Radius of accretion disc\\
$s$		& Index of stellar distribution in stellar cluster core\\
$t|C$		& Compton timescale at constant volume\\
$T_4$		& Temperature $T/10^4\Kelv$\\
$T|C,T|{C,7}$	& Compton temperature, $T|C/10^7\Kelv$\\
$v|K$		& Keplerian velocity (of circular stellar orbit) at cluster edge\\
$v_\star$	& Escape speed from individual star\\
$\eta|{acc}$	& Accretion efficiency factor\\
$\lambda$     & Dimensionless length scaling parameter\\
$\etaflow$	& Fraction of stellar mass loss accreted by BH\\
$\mu|\star(r/r|c)$ & Stellar ass enclosed with $r$ as a fraction of total cluster mass \\
$\rho_{\star}(r)$, $\bar{\rho}|\star(r/r|c)$  	& Radial mass density of stellar cluster\\
$\rho|{\star,c}$	& Mass density of stellar cluster core\\
$\sigma$	& Velocity dispersion of the stars in the stellar cluster\\
$\tau$          & Dimensionless time scaling parameter\\
$\tau|{f}$      & nISM characteristic flow timescale, $\sim\ 10^{10}\mbox{--}10^{12}\secnd$\\
$\tau|{coll}$	& Stellar collision time\\
$\tau|{dyn}$	& Cluster dynamical timescale, \ie{} $r|c/v|K$\\
$\tau_\star$	& Characteristic stellar lifetime\\
$\Xi$		& Ionization parameter\\
$\phi$                & Dimensionless mass-input scaling parameter\\\hline
\end{tabular}
\end{table*}
We model the nuclear system of an active galaxy as an accreting
supermassive black hole, a dense starburst stellar cluster, and the
resulting hot, flowing interstellar medium.  In principle, mass,
momentum and angular momentum will be exchanged across the interface
between the active nuclear region (radius $\sim 1\parsec$) which we
explicitly model, and the wider galaxy. In Section~\ref{s:timesc}, we
discuss the theoretical and observational timescales which
characterise our model.  These determine the physical effects which
dominate the structure of the hot phase flows.  We define the
parameters of our model in Section~\ref{s:outline}, and describe our
treatment of the interfaces between the flow and the accretion disc, a
possible nuclear jet and the surrounding galaxy in
Section~\ref{s:interf}.  We provide a more detailed physical and
observational background to our assumptions in
Appendix~\ref{a:detail}.

Table~\ref{t:glossary} is a glossary of non-standard symbols and
abbreviations widely used in the text.

\subsection{Timescales}
\label{s:timesc}

\begin{table*}
\caption{Approximate physical timescales.}
\label{t:times}
\begin{minipage}{\textwidth}
\footnotetext[1]{Variation of timescale with $(\lambda, \tau, \phi)$,
assuming all dimensionless parameters (\eg\ nuclear spectrum,
Eddington ratios) and `microphysical' values (\eg\ cool phase
temperature) are constant.}  \addtocounter{mpfootnote}{1} \centering
\begin{tabular}{lcll}
\hline
Property			& Timescale ($10^{10}\secnd$) 
& \multicolumn{2}{c}{Scaling$^a$}\\
&		& 
Physical & $(\lambda, \tau, \phi)$ \\ \hline
\\
Stellar collisions\footnote{\eg{} Perry \& Williams \protect\shortcite{pw93},
assuming the stellar cluster dominates the gravitating mass.}
& $3\ee{7}$ & $r|{pc}^{7/2} M|{cl,8}^{-3/2}$
& $\lambda^{-1}\tau^3$\\
Stellar relaxation in nucleus	& $1.6\ee{7}$ & $r|{pc}^{3/2} M|{cl,8}^{1/2} 
M_{\star,\odot}^{-1}$
& $\lambda^3\tau^{-1}$ \\
Galaxy interactions		& $3\ee{5}$
& \dots 	& \dots \\
Stellar evolution		& $3\ee{5}$ & $Q_{-8}^{-1}$ 
& $\phi^{-1}$ \\
Disc viscosity\footnote{$\alpha \la 1$ is the 
Shakura-Sunyaev \protect\shortcite{ss73} 
viscosity parameter, $T|{d,4}$ is the
temperature of the disc gas in $10^4\Kelv$ \cite[see also]{sbf90}.}
& $2\ee{4}$ & 
$\alpha^{-1} M|{h,8}^{1/2} r|{pc}^{1/2} T|{d,4}^{-1}$
& $\lambda^2\tau^{-1}$ \\
Orbital timescale
& $30$ & $r|{pc}^{3/2} M|{cl,8}^{-1/2}$
& $\tau$ \\
Global sound crossing time	& $10$ & $r|{pc}T_7^{-1/2}$
& $\tau$ \\
Flow time			& $1$ & $r|{pc} (v/0.01c)^{-1}$ 
& $\tau$ \\
Supernova evolution		& $0.2$ & $n_{4}^{-1/3} (v|{SN}/0.01c)^{-1}$ 
& $\tau^{1/3}\phi^{-1/3}$\\
Post-shock cooling\footnote{Applies if post-shock cooling is dominated by bremsstrahlung 
cooling from $10^8\Kelv$ -- for shock temperatures smaller than this, there
is an extra factor $T^{1/2}\equiv \lambda/\tau$.}
& $0.2$ & $p|{s,14}^{-1}$ & 
$\lambda^{-2}\tau^3\phi^{-1}$\\
Compton cooling			& $0.1$ & $r|{pc}^2 L_{46}^{-1}$ 
& $\lambda^{-1}\tau^2$ \\
Supernova frequency		& $0.03$ & $Q_{-8}^{-1}M|{cl,8}^{-1}$
& $\lambda^{-3}\tau^2\phi^{-1}$ \\
Cloud sound crossing		& $10^{-3}$ & $(d/10^{13}\cm)T_4^{-1/2}$ 
& \dots \\
Hot-phase ion collisions	& $10^{-4}$ & $T_7^{3/2}n_4^{-1}$
& $\lambda^3\tau^{-2}\phi^{-1}$ \\
Cool-phase recombination	& $3\ee{-7}$ & $n_9^{-1}$ 
& $\lambda^{-2}\tau^{3}\phi^{-1}$ \\ \hline
\end{tabular}
\end{minipage}
\end{table*}

We shall concentrate on global nuclear interstellar medium (nISM)
flows, which have characteristic timescales $\tau|{f} \sim\
10^{10}\mbox{--}10^{12}\secnd$, and on their response to global
gravitational and radiative forces.  In this regime, cooling
timescales are short enough that the gas can be taken as in
equilibrium.  By comparison, relative timescales $\ga 10^4 $ times
larger are inferred for large-scale radio structures. In
Table~\ref{t:times}, we compare various physical timescales relevant
to flows in AGN\@.

Dynamical evolution timescales are long enough that the hydrodynamics
is insensitive to secular changes in the gravitating mass
distribution, although the details of the mass loading do depend on
the secular evolution of the cluster \cite{mcd,mp}. Stellar
collisions, dynamical relaxation, galaxy interactions, stellar
evolution and disc viscosity all have characteristic timescales far
longer than the flow time through the nucleus.  These processes are
all important in determining the secular evolution of the nucleus.
Integrated over the nucleus, mass loss due to passive stellar
evolution dominates that due to collisions for all but the densest
($\ga 10^9 \Msun \parsec^{-3}$) and most massive clusters.  Mass loss
due to relaxation of stars into the loss cone of plunging orbits are
always insignificant.  If the cluster IMF extends well below $1
\Msun$, collisions can dominate mass input very close to
the black hole \cite{mp}.  Hence, if low luminosity AGN are
characterised by small, centrally condensed starburst stellar clusters
mixed in a massive old cluster, stellar collisions may dominate
stellar evolution as the mechanism for mass loss in such objects
\cite[\cf{}]{mcd,WP94,mp}.

The similarity of the timescales for galaxy interactions and stellar
evolution suggest that star formation will be important through much
of the life of the active nucleus.  However, threshold processes
operating at intermediate scales within the galaxy may serve to reduce
substantially the timescales which characterise the periods of mass
input to the nucleus.  The accretion disc viscous timescale is far
longer than the dynamical timescale of the hot phase flow, even in the
very central regions, $\sim 10^{-3}\parsec$, where temperatures reach
$3\ee4\Kelv$ (eqs~\ref{e:remit}, \ref{e:trad}).  We expect that the
hot phase gas will add to the disc at radii significantly outside
$10^{-3}\parsec$, so the feedback between flow and central luminosity
will have a very long delay (roughly $10^3$ times the flow time,
Table~\ref{t:times}).  This justifies our approach of calculating the
flow structure for an applied central luminosity rather than
attempting to find fully self-consistent solutions.

Small timescale variability is most often measured in relatively
low-luminosity Seyfert galaxies.  Typical values are $10^{-6} \tau|{f}$ for
continuum variability in X-rays, $\ga 10^{-6} \tau|{f}$ for the UV continuum
and $\ga 10^{-4} \tau|{f}$ for the optical continuum in Seyferts, which are
both believed to be driven by the high energy source
\cite[\eg{}]{agnw4}.  Variations of 10 per cent in QSOs are found over
timescales of $10^{-2}\tau|{f} $
\cite{hookea94}.  In Seyfert galaxies, optical-UV emission lines also
vary on a $\ga 10^{-4} \tau|{f}$ timescale, with evidence for evolution of the
distribution of line-emitting gas over a $10^{-2} \tau|{f}$ timescale
\cite{pvgw94,wanp96}.  The flow will be
perturbed by sound waves driven by short timescale variations in
radiation forces.  The steady luminosity we assume here may be taken
as a time-average, which we assume drives a `averaged' flow.

The thermal evolution of the nISM may affect the flow at the 10 per
cent level. This may be qualitatively important if instabilities
result.  The timescale for evolution of supernovae is comparable to
the flow timescales, and so may be significant -- individual supernova
events can significantly perturb the flow, particularly in
lower-luminosity nuclei \cite{pwd98}.  The short timescale for hot
phase collisions is a necessary (if not sufficient) condition for our
use of single-temperature hydrodynamics; the timescales for sound to
cross clouds and for recombination suggest that the hot phase may be
taken as a steady background in cloud structure problems.

As we shall see (Section~\ref{s:results}), some flows seem to reach
no equilibrium, undergoing a chaotic sequence of explosions or large
magnitude oscillations; others seem to be steady on a dynamical
timescale, but can take far longer than this to reach equilibrium.
Flows with the latter behaviour can be modelled analytically in
spherical symmetry (see Appendix~\ref{a:apspherical}).  However, even
those flows which do not reach equilibrium will adjust to changes in
the structure of the nucleus, such as an increase in the central
luminosity, within a flow crossing timescale.

\subsection{Components of Active Nuclei}
\label{s:outline}

\begin{figure*}
\epsfysize = 6.5cm
\epsfclipon
\begin{centering}
\begin{tabular}{ll}
a) & b) \\
\epsfbox{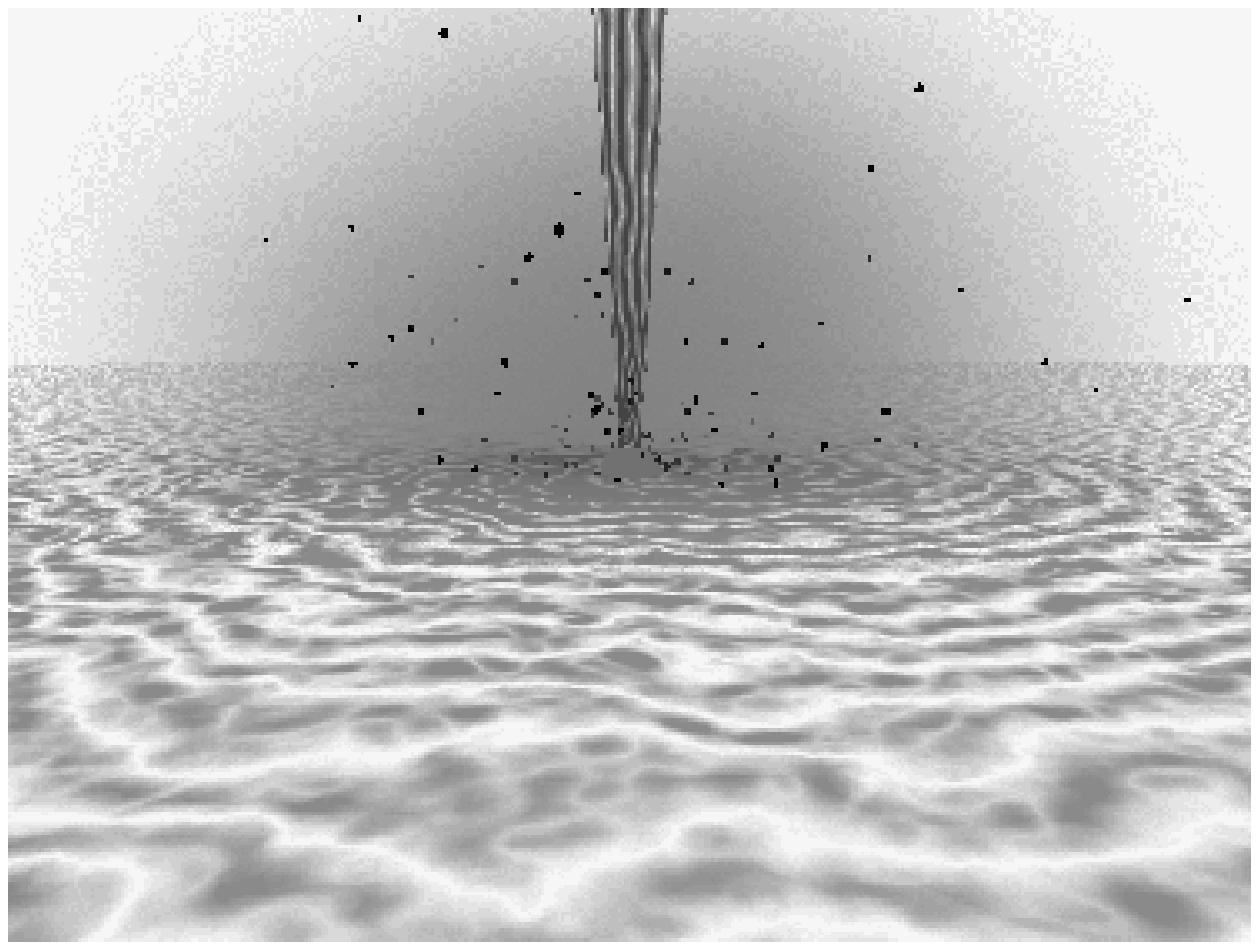} &
\epsfbox{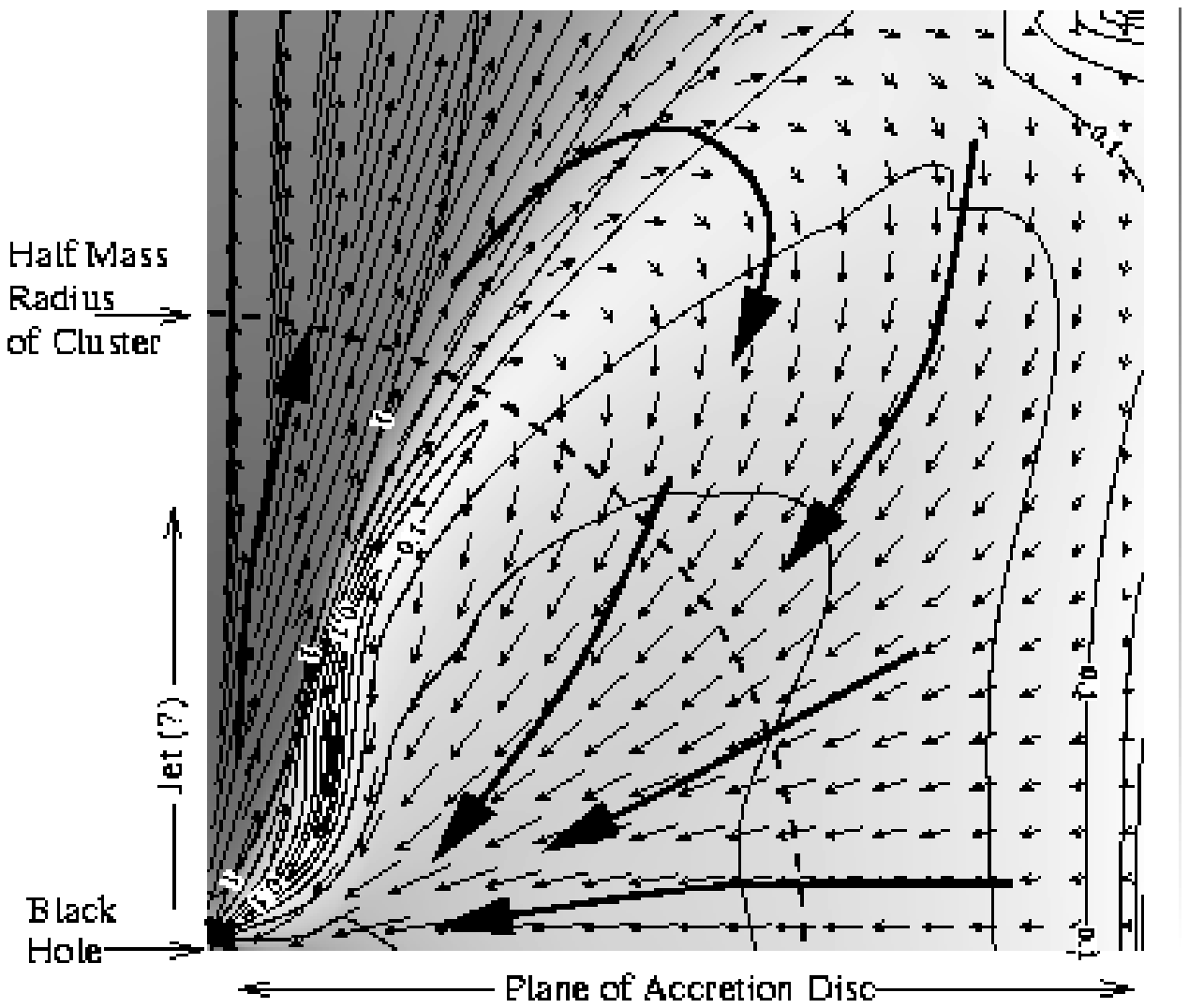}
\end{tabular}
\end{centering}
\caption{Illustrations of the assumed structure of the nucleus.  In
a), a schematic view of the nucleus shows the extensive disc which
surrounds the black hole.  Along the axis of the disc, a radio jet is
driven outwards in about 10 per cent of active nuclei.  Around the
black hole, the stellar density is high enough to be treated as a
distinct stellar cluster -- a small fraction of these stars are
indicated by dots in this figure.  The nebulosity which fills the
region within the cluster illustrates the location of the nISM which
we simulate.  In b), we illustrate how these components relate to the
structure of the simulations shown in the rest of this paper.  The
black hole is at the origin of the grid, the accretion disc in the
horizontal plane, and any jet would travel along the vertical axis.
The half-mass radius of the cluster reaches out a substantial way
through the grid: mass loss from the cluster falls inwards in the
plane of the accretion disc, and is driven outwards along its axis.}
\label{f:cartoon}
\end{figure*} 
This paper concentrates on the structure of the hot global ISM which
fills much of the volume of an active nucleus.  In this section, we
describe the other basic physical components of the nucleus which
dominate its mass content and energetics, and which give rise to and
act on the global flow to determine its structure.

\subsubsection{The Black Hole and Stellar Cluster}

We assume that an AGN contains a supermassive black hole, of mass $M|h
= 10^8 M|{h,8} \Msun$, surrounded by a starburst stellar cluster, of
total mass $M|{cl} = 10^8 M|{cl,8}\Msun$.  For the fiducial model of a
luminous QSO, the cluster has a mass of $M|{cl,8}\simeq 10$.
Dynamical models \cite{mcd} show that the stellar cluster dominates
the gravitational potential of the nucleus, and that the black hole
typically represents only 20 -- 50 per cent of the mass of the system.
Observational support for this is discussed in
appendices~\ref{a:black} and \ref{a:cluster}.  Mass is liberated from
the stars within the cluster and distributed through the volume of the
nucleus.

The stellar cluster is chosen to be steady and spherically symmetric
with a stellar density given by
\begin{equation}
\rho|\star = \rho|{\star,c}\bar{\rho}|\star(r/r|c),
\end{equation}
where $\bar{\rho}|\star$ is the dimensionless spatial profile
function normalised at the core radius, $r|c$.  In this paper, we take
a broken power-law form for $\bar{\rho}|\star$ (cf{}
Appendix~\ref{a:cluster}):
\begin{equation}
\label{e:cluster}
\bar\rho_\star = \left\{
\begin{array}{ll}
(r/r|c)^{-s} & r<r|c\\
(r/r|c)^{-h} & r>r|c,
\end{array}
\right.
\end{equation}
The power-law indices, $s$ and $h$, are required to be $s<3$ and $h>3$
for the cluster mass to be finite.  In fact, we choose $s=0$ and $h=5$
in most of the models presented, similar to the Plummer model initial
conditions of MCD\@.  Experiments with different stellar distributions
did not greatly alter the results that we discuss.  Where we take
$h=5$, the core radius is often chosen to be ${2/3}\parsec$, in order
that the average radius of mass injection is $1\parsec$.  With these
parameters, the central stellar density within $r|c$ for our canonical
model is $\rho|{\star,c} = 3.2\ee8 \Msun \parsec^{-3}$.  This value is
considerably larger than that inferred from observations of nearby
galactic nuclei \cite{Watson97}, but the lifetimes of the clusters we
consider are short and their presence may be masked by the luminosity
of the QSO they feed (for a more detailed discussion, see
Appendix~\ref{a:cluster}).  If the gravity and radiation forces are
negligible in such a model, the outflow becomes transonic at
$0.89\parsec$.

\subsubsection{Stellar Mass Loss}

There are various sources of mass for the flow in the nISM\@.  Stars
in the stellar cluster lose mass through the normal processes of
stellar evolution, and also as a result of collisions within the
stellar cluster and tidal disruption as they pass close to the black
hole. The mass-loading rate, $\dot{q}$, is chosen to be spherically
symmetric, and time steady over the period of our calculations.
Clearly the mass-loss rate evolves over the lifetime of the cluster
and is a function of the age and IMF of the cluster.  However, the
timescales of this evolution are much longer than the typical
dynamical timescales over which our calculations run.  We write
\begin{equation}
\dot{q}(r) = \dot{q}|c \tilde{\dot{q}}(r/r|c).
\end{equation}
In the present paper, we restrict ourselves to the case that the
mass-loading rate to be proportional to the stellar density, at a rate
\begin{equation}
\label{e:qdot}
\dot{q}(r) = Q\rho|\star(r),
\end{equation}
so that $\dot{q}|c = Q \rho|{\star,c}$.  $Q \simeq 10^{-8}\yr^{-1}
Q_{-8}$, and $Q_{-8}\sim1$ for a young cluster of massive stars
\cite[\cf{} Appendix~\ref{a:massloss};]{WP94,mp}.

\subsubsection{Accretion Disc}

Accretion from the global flow, directly and through the accretion
disc, powers emission from the vicinity of the central black hole
\cite{zcc95}.  The galaxy ISM may also feed an accretion disc {\em
if}\/ it does not hang up at an intermediate radius, and {\em if}\/
the extended ($\sim 1\parsec$ scale) regions of the disc are stable
\cite{hcbp94}.

The accretion disc which forms around the black hole can also act as both
a source and a sink of gas.  In the hot central regions, a wind may be
driven from the disc surface.  Further out, gas may condense onto its
surface, while self-gravitating instabilities could generate clumps of
cool gas which would then be ablated by the nISM flow as it streams
past.  Eventually, the mass lost from the stellar cluster must either
reach the black hole, to power the luminosity of the AGN, or be blown
out of the nucleus in a wind.

\subsubsection{Luminosity}
\label{a:lum}

The total integrated luminosity is
\begin{equation}
\Lbol \equiv 10^{46} L_{46} \ergs. \label{e:Lbol}
\end{equation}
We model the angular distribution of the radiation field,
$L|{Bol}(\theta)$, by two components, one isotropic -- representing the
radiation directly associated with the black hole -- and the other
anisotropic, representing the radiation of a thin accretion disc.  This
accretion disc luminosity is the only component of our model which is
not spherically symmetric (Appendix~\ref{a:rad}).  As we discuss in
Appendix~\ref{a:nISM}, we assume for the moment that any variation in
the continuum spectrum with angle is unimportant.  As we assume that
the dominant opacity is electron scattering, we can describe the
luminosities by Eddington ratios (\ie, ratios of outward radiative
force to the inward gravity of the black hole) for a given black hole
mass, where the radiation flux at $(r,\theta)$ is
\begin{equation}
F(r,\theta) = f|{Edd}(\theta) {L|{Edd}\over 4\pi r^2},
\label{e:frtheta}
\end{equation}
and where the Eddington limit luminosity (at which the radiation driving
balances the gravitational acceleration) is given by
\begin{equation}
L|{Edd} \simeq 1.3\ee{46} M|{h,8} \ergs.
\end{equation}
The angularly dependent Eddington ratio, $f|{Edd}(\theta)$, is
specified by its average over solid angle
\begin{equation}
\langle f|{Edd}\rangle = {1\over 4\pi}\int f|{Edd}(\theta)\id\Omega,
\end{equation}
and an angular form function $\bar{f}(\theta)$ which depends on the
fraction of the luminosity in the anisotropic component, $f|{disc}$
and its angular dependence (Appendix~\ref{a:rad}), so $f|{Edd}(\theta)
= \langle f|{Edd} \rangle \bar{f}(\theta)$.  Studies of AGN have
indicated that typical values of $f|{Edd}$ vary from $10^{-3}$ to
values close to unity \cite{laor90,cols93}, with the higher ratios
found for the brighter nuclei.  However, even amongst Seyfert galaxies
with very similar source luminosities the Eddington ratio appears to
vary between $10^{-4.5}$ \cite[NGC~4258,]{lasea96} and 0.5
\cite[NGC~1068,]{green96}.

\subsubsection{Thermal Equilibrium}

Consistent with our choice of electron scattering as the dominant
opacity source, we assume that the gas is maintained everywhere in
isothermal equilibrium, at the mean Compton temperature of the central
luminosity source, $T|C = T_7\times10^7\Kelv$.  This temperature is
determined by the spectrum of the radiation, with values $0.6\la
T_7\la15$ predicted for in the majority of AGN
\cite{kmt,fab86,matfer87}.  We discuss the physical limits on the
validity of this assumption in Appendix~\ref{a:nISM}.  The sound speed
in fully ionized gas with solar abundance is $c|s=360T_7^{1/2}\kms$:
we have used $c|s = 10^{-3} T_7^{1/2}c$ for simplicity.

The assumption of isothermality allows us to scale our models to a wide
range of situations; however, there are some limits on the
applicability of this assumption.  The presence of strong shocks in our
simulations means that including the energy equation explicitly would
allow gas to cool: the production of such cool gas is important in
deriving observational diagnostics for the flow, in calculating the
rate at which gas can be added to the accretion disc (though limits can
be put on the amount of gas which can possibly be added to the disc in
this manner), and in increasing the radiative force per unit mass,
which may have important dynamical effects.  However, in the context of
global models, the short time- and length-scales which will
characterise this cool gas mean that it is difficult to accurately
model both small and large-scale flows.  Modelling the details of
mass-injection from individual stellar sources presents similar
difficulties.  In consequence, we have decided to focus first on the
global properties of flows which remain closely isothermal, and treat
the finer-scale structures in future papers.

\subsubsection{Forces}

\label{s:smooth}

As well as hydrodynamic effects, body forces act on the nISM as a
result of gravity, radiation pressure, and friction with the mass
injected into the flow.  The gravity is dominated by the black hole
and the stellar cluster (Appendix~\ref{a:gravity}).  We assume that
the mass loss from the cluster is, on average, at rest with respect to
the black hole and stellar cluster and must therefore be accelerated
by the flow.

The radiative driving acceleration due to Thomson opacity is given by
\begin{equation}
g|{rad} = {\sigma|T F(r,\theta)\over m|H c} =
f|{Edd}(\theta){GM|h\over r^2},
\end{equation}
where $\sigma|T$ is the electron scattering cross section.  The
gravitational force on the nISM at $r$ is due to both the mass of the
black hole and the mass of the stellar cluster which lies within $r$;
the latter can be written as
\begin{equation}
\label{e:mustar}
M|{cl}({\mbox{$<$}}r) = M|{cl} \mu|\star(r/r|c),
\end{equation}
where $\mu|\star(r/r|c)$ is the fraction of the total cluster mass
within $r$.  The form of $\mu|\star$ for our broken power-law is given
in Section~\ref{a:gravity}.

The inverse-square laws for the central radiative and gravitational
forces hold at all radii outside, respectively, $\sim 10^{-3}\parsec$
(where the finite spatial extent of the radiating accretion disc becomes
important) and $\sim 10^{-4}\parsec$ (where relativistic corrections
need to be made), in QSOs.  In addition to these limits to the
inverse-square law, the effects of the angular momentum of the hot
phase, pressure from a thick central disc, and extra heating processes
(such as stimulated Compton scattering) may effectively work counter
to the gravitational force.  We have chosen to mimic these effects by
smoothing the forces in a region close to the central black hole.
Instead of a pure inverse-square law, we take the forces to vary as
\begin{equation} 
g \propto {r\over\left(r^2+\epsilon^2\right)^{3/2}},
\end{equation} 
where in general we take the smoothing radius, $\epsilon$, to be far
smaller than the cluster radius, but significantly larger than the
grid resolution where possible.  In this way, the grid resolves the
flow structures throughout the nucleus.  

We shall see that the link between the gravitational velocities at the
radius at which the outflow is driven and the broad line kinematics
leads to one of the fundamental distinctions between the results of the
current paper and previous models of global flows in active nuclei.
The low-ionization lines may be formed in exactly the regions of the
accretion disc which dominate the anisotropic continuum emission
\cite{CDMP}, which is one physical origin for such a smoothing term.
This provides a link between LILs and the driving of the global wind in
which the high-ionization lines are formed since these HIL are formed
behind shocks within the global flow. The shock velocities, reflected
in the line widths, represent the superposition of the nISM velocities
upon the stellar kinematics \cite{p97}.  Both the low- and
high-ionization emitting regions have velocities determined by the
gravitational potential. The high-ionization emitting gas has an
additional component of velocity due to the nISM flow. Hence the
similarity of high- and low-ionization line-widths \cite{corb91} arises
naturally in these models, allowing for the centroid shift between the
lines \cite{espey89,tytf92,sulea95}.

\subsection{Interfaces: Accretion discs, jets, the galaxy}
\label{s:interf}

We now discuss the interfaces of the nuclear flow at the edges of our
computational grid.  In the plane $z=0$, the flow interacts with the
accretion disc.  Along the grid axis, $r=0$, there may be a
relativistic jet.  At large radii, the flow in the nucleus will
interact with the galaxy ISM.

\subsubsection{Disc interface}

The nuclear starburst stellar cluster inevitably creates an nISM\@.
No available treatment of accretion disc physics has considered the
myriad physical processes which may arise in such an environment.  The
nISM flow near the disc plane can clearly have a wide range of
detailed structures.  At one extreme, the nISM and accretion flows are
completely separated by a contact discontinuity, and may be modelled
by a mirror boundary condition.  At the other extreme, if the pressure
is sufficient that cooling can occur, all the material raining onto
the disc will be lost to the global flow and incorporated into the
disc.  We model the latter case by specifying a region of the
accretion disc over which we treat the disc plane as a mass sink.

The scale height, $H$, of a thin accretion disc is given by
\begin{equation}
\frac{H}{r} \simeq 0.015 \left(\frac{T_4r|{pc}}{M|{h,8}}\right)^{1/2},
\end{equation}
where $T_4 \simeq 1$ is suggested by many models of the inner
disc. This will be reduced still further by the gravity of the stellar
cluster at radii beyond the cluster core.  We can, at best, marginally
resolve the cool disc in a (uniformly gridded) global model of the
flow in AGN\@.  The thermal equilibrium time of the disc gas is of
order the recombination timescale, roughly $3\ee4\secnd$, even at a
density as low as $10^{8}\cm^{-3}$.  To accurately model the structure
of the disc would require a cell size $\sim10^{-8}$ times the size of
the global grid.

\subsubsection{Jet interface}

The other inner boundary condition on the models is along the axis of
the disc.  This will form the interface with a nuclear jet, if one is
present.  While radio-loud AGN, with their strong jets, form only 10 per cent
of the population, there is evidence for radio emission on BELR scales
in many more nuclei \cite{Dunlop93,kukula97b}.

In the present paper, we treat the axis as a mirror boundary, in
effect assuming there is no jet or that its width is $\ll1\parsec$.
In reality, a boundary layer will form at the interface between global
flow and jet.  We leave the detailed treatment of such boundary layers
for the present.  In some cases we do find that well-collimated
non-relativistic jets form spontaneously in our model, driven by the
wide opening angle outflow from the central black hole.

\subsubsection{Galaxy interface} 

In real active nuclei, mass, momentum and angular momentum will be
exchanged between the active nuclear region and the wider galaxy.  The
rate of this exchange will depend upon both the properties of the
nuclear region and of the galaxy.  The timescales for stellar
evolution and mass inflow from galaxy interactions are comparable, but
far longer than those over which the structure of the nuclear ISM is
determined.

The rate of mass input by the galaxy to the nucleus has been the
subject of much discussion.  Shlosman \etal~\cite[1990, see also, \eg,
the proceedings edited by]{sh94} suggest that the rate may well be
controlled by galactic interactions and subsequent instabilities.
While some inflow of the hot phase from the ISM of a normal galaxy may
occur, it is difficult for this gas to lose its angular momentum and
so the accretion rates will be a very small component of the mass
budget of the nucleus (and may be further suppressed by any nuclear
wind).  Where a galaxy interaction has occurred, however, the angular
momentum constraint is removed for gas distributed on kiloparsec
scales.  The similarity between interaction timescales and the stellar
formation and evolution timescales within the nucleus might then
suggest that most active nuclei would be observed during the actively
star forming phase.  However, when the details of the fuelling process
are considered, it is found that the infalling gas tends to remain
well outside the nucleus until a sufficient mass has collected that it
gas go through a self-gravitating instability and can then fall
inwards once more (the so-called `bars within bars' picture).  The
fuelling thus depends on this threshold process operating on scales of
$\sim100\parsec$, where dynamical timescales are far shorter than
those of galactic interactions.  It is thus reasonable to concentrate
upon the case in which mass input from the wider galaxy is negligible
in this first paper.

Making only the reasonable assumption that the galaxy succeeds in
creating a nuclear stellar cluster (\cf{} Watson 1997, and references
therein) we conclude that {\it in situ} feeding of the black hole by
the stellar cluster is highly probable.

\section{The theoretical model: towards an HR diagram for AGN}
\label{s:equations}

We next discuss the basic equations, moving on to the scaling of the
solutions to different sets of parameters (Section~\ref{s:scaling}),
the scalings which may characterise observed AGN
(Section~\ref{s:nature}), and the parameters which dominate the flow
structures (Section~\ref{s:parms}, which also introduces our
diagnostic plot).

The interstellar medium is evolved using the equations of
hydrodynamics, with a distributed mass-loading source function and
gravitational and radiative forces.  These are conservation of mass:
\begin{equation}
\label{e:masscons}
\frac{\partial \rho}{\partial t} + \nabla . \left( \rho \bld{v} \right)
= \dot{q},
\end{equation}
the equations of motion,
\begin{equation}
\label{e:motion}
\rho\frac{D \bld{v} }{ Dt } = - \nabla p - \dot{q}\bld{v}
- \rho\bld{g},
\end{equation}
and the energy equation,
\begin{eqnarray}
\pdif{}{t}\left({1\over2}\rho v^2+ {p\over\gamma-1}\right)
        & =&  -\nabla.\left[\bv{v}\left({1\over2}\rho v^2+
        {\gamma p \over \gamma-1}\right)\right]\\
&&\nonumber
	- \rho \bld{g}.\bld{v} + \rho(\Gamma-n\Lambda)
	 + {\dot{q}c|q^2\over\gamma-1},
\end{eqnarray}
together with the equation of state $p = \rho k|BT/\mu$, where $\mu$
is the mean mass per particle, $\gamma = 5/3$ is the adiabatic
constant, $\Gamma$ and $\Lambda$ are the usual heating and cooling
coefficients, and $c|q$ is the isothermal sound speed which
characterises the internal energy of the mass injected.

These equations include both the effects of applied gravitational and
radiative forces through the combined effective acceleration term,
\begin{equation}
\label{e:accel}
\bld{g} = {G[(1-f|{Edd}(\theta))M|h + M|{cl}(\mbox{$<$}r)]\over
(r^2+\epsilon^2)^{3/2}}
\bld{r},
\end{equation}
where $M|{cl}(\mbox{$<$}r)$ is given by equation~\refeq{e:mint}.  Mass
loading from the stellar cluster, at a rate $\dot{q}$ per unit volume
per unit time, is included both in the mass conservation equation and
as a frictional term in the equations of motion.  The only radiative
driving which we have included is that due to Thomson scattering.

\subsection{Isothermal Flows}
\label{s:isothermal}

In the present paper we assume that the nISM is maintained in radiative
equilibrium at the Compton temperature and hence do not solve the
energy equation; instead we use the isothermal equation of state $p =
\rho c|s^2$ where the isothermal sound speed $c|s$ is determined by
Compton equilibrium (we discuss the validity of this assumption in
Appendix~\ref{a:nISM}). The input parameters to the equations
governing the system are therefore the mass of the black hole $M|h$;
the mass of the stellar cluster $M|{cl}$, the radial distribution of
the mean stellar density $\bar{\rho}|\star(r/r|c)$, the cluster core
radius $r|c$; the mass loading rate density in the core $\dot{q}|c$,
and its radial variation $\tilde{\dot{q}}(r/r|c)$; the bolometric
luminosity $L|{Bol}$, the sound speed in the nISM $c|s$ and the
distribution of the effective Eddington ratio $f|{Edd}(\theta)$.

Spherically symmetric, electron scattering driven, point source
isothermal flows ($f|{Edd}(\theta) = {\it const}$ [less than unity],
$\rho|\star = \dot{q} = 0)$ have the well-known Parker solutions, with
a characteristic length given by the sonic point radius, $r|s$,
\cite[$r|s = (1-f|{Edd})M|h/c|s^2$, Kippenhahn \etal{} 1975,
hereafter]{KMP} and a characteristic velocity by the sound speed,
$c|s$.  In such flows, the radiation driving reduces the effective
mass of the central gravitating object, $M|h$, to $ (1-f|{Edd})M|h$.
As is well known, the solution (wind or accretion flow) which is
realised in any particular physical situation is determined by the
boundary conditions.

This universal solution topology is removed by the radial variation of
the gravitational force or the distributed mass loading (both of which
may introduce internal shocks and additional sonic point to the flow;
mass loading also may introduce stagnation points), or, most
dramatically, by the symmetry breaking introduced by the angular
dependence of the radiation.  Each set of $f|{Edd}(\theta)$,
$\rho|\star(r/r|c)$ and $\dot{q}(r/r|c)$ gives rise to a different
family of solutions, further differentiated by the values of four
dimensionless ratios between the other physical input parameters, as
discussed below.  Although this may, at first sight, appear to render
any general conclusions impossible, the physical expectation that most
AGN share similar spatial forms for $f|{Edd}$, $\rho|\star$ and
$\dot{q}$ even when they differ in their total $M|h$, $M|{cl}$ and
$L|{Bol}$ means that we can classify families of solution which apply
over a broad range of objects, as we now discuss.  In order to do so
we write the equations in scaled, dimensionless form.

\subsection{Dimensionless Equations}
\label{s:dimensionless}
\label{s:scaling}

We can cast equations~\refeq{e:masscons}, \refeq{e:motion} and
\refeq{e:accel} into dimensionless form by scaling all distances to
the cluster core radius, $r|c$ (\ie\ $\bv{r} = r|c\tilde{\bv{r}}$,
$\nabla \to (1/r|c) \tilde\nabla$ and $\epsilon = r|c
\tilde{\epsilon}$) and all velocities to the isothermal sound speed,
$c|s$, (\ie\ ${\bv{v}}({\bv{r}},\theta,t) = c|s
\tilde{\bv{v}}(\tilde{\bv{r}},\theta,\tilde{t})$), from which it
follows that all times scale as $t = r|c/c|s\tilde{t}$.  Furthermore
$\rho(\bv{r},\theta, t) = (\dot{q}_cr|c/c|s)
\tilde{\rho}(\tilde{\bv{r}},\theta,\tilde{t})$.  We then have
\begin{equation} 
\label{e:mcdimless} 
\frac{\partial
\tilde\rho}{\partial \tilde t} +
	\tilde\nabla . \left( \tilde\rho \tilde{\bld{v}} \right) =
	\tilde{\dot{q}}, 
\end{equation} 
and 
\begin{equation}
\label{e:mmdimless} 
\tilde\rho\frac{D \tilde{\bld{v}} }{ D\tilde t } =
       - \tilde\nabla \tilde\rho -
	\tilde{\dot{q}}\tilde{\bld{v}} -\tilde\rho\tilde{\bld{g}},
\end{equation} 
where 
\begin{equation} 
\label{e:acceldimless}
\tilde{\bld{g}} = {GM|{cl} \over r|c c|s^2}
	{[(1-\langle f|{Edd}\rangle \bar{f}(\theta))(M|h/M|{cl}) +
	\mu|\star(\tilde{r})] \over
	(\tilde{r}^2+\tilde\epsilon^2)^{3/2}} \tilde{\bld{r}},
\end{equation} 
Hence, from equations~\refeq{e:mcdimless}, \refeq{e:mmdimless} and
\refeq{e:acceldimless} we see that the dimensionless solution ($
\tilde{\bv{v}}, \tilde{\rho}$) depends on

\begin{itemize}
\item[({\sc i})] $\bar{f}(\theta)$, $\bar\rho|\star(\tilde{r})$ and $
\tilde{\dot{q}}(\tilde{r})$ -- three dimensionless spatial form
functions: these specify the angular variation of the radiation, the
spatial variation of the stellar density and the mass loading,
respectively,
\end{itemize}

\noindent
and the six physical parameters, $M|h$, $M|{cl}$, $r|c$, $L|{Bol}$,
$c|s$ (or equivalently $T|{nISM}$), and $\epsilon$, through the four
dimensionless ratios:

\begin{itemize}
\item[({\sc ii})]
$GM|{cl} / r|c c|s^2$ -- essentially the ratio of the stellar
cluster velocity dispersion to the sound speed,

\item[({\sc iii})]
the ratio of the black hole to total cluster mass,
$M|h/M|{cl}$,

\item[({\sc iv})]
the Eddington ratio, $\langle f|{Edd}\rangle$, and

\item[({\sc v})]
the dimensionless smoothing length, $\tilde\epsilon =
\epsilon/r|c$.
\end{itemize}

Note that no value dependent on $\dot{q}|c$ appears in the list, since
the isothermal equations are homogeneous in $\rho$ and $\dot{q}$.
Whereas $\dot{q}|c$ sets the physical scale of $\rho$, it can be
ignored for isothermal flows as far as the dynamics is concerned.
This is a result of our assumption that $f|{Edd}(\theta)$ [or,
equivalently, $L|{Bol}(\theta)$] is a free parameter.  If we had
included the structure and dynamics of the accretion disc in the
hydrodynamics then the rate of mass-loss, $\propto \dot{q}$, would
determine $L|{Bol}(\theta)$ and $f|{Edd}(\theta)$ directly.

By varying ({\sc i}) through ({\sc v}), each of the numerical
solutions we present can be rescaled to a wide variety of AGN\@.  In
fact, as we shall discuss, two particular dimensionless ratios
dominate the differences between solutions.  The similarity of form of
the solutions for a wide variety of AGN allows us to predict how, for
example, the dynamics will scale with \eg\ changing luminosity or
Eddington ratio and spectral shape (or temperature) or luminosity and
cluster core radius.

For the solutions to scale, there are also constraints on the boundary
conditions: for most of our simulations, these are observed as, for
instance, free-flow boundary conditions which have no dimensional
dependence; where accretion is allowed over some fraction of the disc,
however, the size of this region must be specified as a fraction of
the cluster core radius for the models to scale.  In reality, several
of these dimensionless parameters will be less important than terms
that break the scaling, such as heating and cooling of the nISM gas.

As we shall see, within broadly similar bipolar nISM structures common
to all symbiotic AGN, there are systematic differences in topology (see
Fig.~\ref{f:xxx}).  AGN of widely different bolometric luminosities can
have topologically identical flows if they all share a common stellar
density profile (\ie\ core halo structure $\tilde{\rho}(\tilde{r})$),
have stellar mass loss rates which are directly proportional to
$\rho|\star$, have the same radiation pattern $\bar{f}(\theta)$,
ratios ({\sc ii}) through ({\sc v}), and scaled boundary conditions.
I.e. if their underlying physical structure is similar, their nISM will
be described by a single dimensionless solution ($\tilde{\bv{v}},
\tilde{\rho}$) to equations~\refeq{e:mcdimless}, \refeq{e:mmdimless}
and \refeq{e:acceldimless}.  However, rather than present dimensionless
solutions for particular combinations of ({\sc i}) through ({\sc v}),
we have chosen to present all our results in physical units for
fiducial models in order to make them more immediately accessible.
Below, we describe the manner in which the models which we have
calculated can be scaled to a wide range of parameter values, and
discuss how these scalings may relate to the relationships which have
been inferred observationally for AGN.

We have grouped $\bar\rho|\star(\tilde{r})$ and
$\tilde{\dot{q}}(\tilde{r})$ in ({\sc i}) above, since in this paper
we make the further restriction that these spatial distribution
functions are equal, which is equivalent to the assumption that
$\dot{q}|c=Q\rho|{\star,c}$.  [For this limit to be strictly valid, we
must consider young stellar clusters with core densities $\rho|{\star,c} \la
10^9 \Msun$ pc$^{-3}$, above which collisions and tidal disruption at
small radii make a significant contribution to the mass loss
\cite{mcd}.  Models A, B, C are all below this limit, but may breach
it when scaled to Seyfert parameters.]  In this case
equation~\refeq{e:mcdimless} becomes
\begin{equation}
\label{e:mc2dimless} 
\frac{\partial \tilde\rho}{\partial \tilde t} +
	\tilde\nabla . \left( \tilde\rho \tilde{\bld{v}} \right) =
	\bar\rho|\star.  
\end{equation} 
The densities and pressures in the nISM flow are then $\rho = (Q
\rho|{\star,c} r|c /c|s) \tilde\rho$ and $p=(Q \rho|{\star,c} r|c c|s)
\tilde\rho$.

Clearly in such isothermal flows, a consequence of the scaling is that
changes in the source spectrum, which result in a change in the Compton
temperature and thus in $c|s$, imply a consequent change in the time
scales.  Other things being equal, the same $\tilde{t}$ corresponds to
a shorter real time in hotter flows (with harder spectra). Note, again,
that no value dependent on $Q$ appears in the list.

Taken together, if ratios ({\sc ii}), ({\sc iii}), and ({\sc iv}) are
constant, then each of $L|{Bol}$, ${M|h}$ and $M|{cl}$ is proportional
to $r|c c|s^2$ with a known factor, so $L|{Bol} \propto {M|h} \propto
M|{cl}$.

\subsection{From High Luminosities to Low: How AGN Scale in Nature}
\label{s:nature}

One of the most striking features observed in the spectra of AGN is
the remarkable consistency of their dynamics over orders of magnitude
in total bolometric luminosity.  Line widths -- direct measures of gas
velocities -- vary almost imperceptibly as $L|{Bol}$ changes from
$\leq 10^{43}$ to $\geq 10^{47} \ergs$.  The size of the emitting
region, however, appears to track the luminosity, $r|{BLR} \propto
L|{Bol}^{1/2}$ (from reverberation and photoionization studies).
Controversy still surrounds the question of whether or not black hole
masses and/or the efficiency of the radiation process ($\langle
f|{Edd}\rangle$) vary as $L|{Bol}$ varies -- or, more accurately put,
whether $L|{Bol}$ varies because of variations in $M|h$, $f|{Edd}$, or
because of the availability of fuel.  The fuel supply is controlled by
stellar and galactic processes.  Unfortunately, to date observations
can measure neither the properties ($M|{cl}$, $\rho(r)$, age or IMF)
of the stellar cluster \cite[because it is outshone by the central
radiation by orders of magnitude]{WP94}, nor the rate of any possible
inflow of matter from the galaxy.  The relationship between the
properties of the cluster and the fuelling process is, at least in
part, a theoretical prediction of this paper.

We now consider the implications for AGN of the scale invariance of
the equations and discuss how the physical properties of AGN are
expected to vary as a function of luminosity, $L|{Bol}$.  Our models
make clear predictions about how unobserved properties (\eg\ $M|{cl}$
and $r|c$) are expected to behave as a consequence of those which are
observed.  They also provide simple rules for scaling the solutions we
present here to objects with other luminosities or cluster/BH
properties so that the reader can choose amongst the solutions we
present those which are appropriate to her or his favourite AGN.

Consider a family of {\it structurally similar} AGN (\ie\ the same
({\sc i})) all of which share the same characteristic values of ({\sc
ii}), ({\sc iii}) and ({\sc v}), but which differ in $L|{Bol}$, and
possibly in $\langle f|{Edd}\rangle$ ({\sc iv}).  The members of the
family will have different characteristic sizes, velocities, time
scales and masses.  If their spectral hardness differs, so will their
Compton temperatures, and thus their sound speeds.  We denote the ratio
of their sound speeds by $\nu$, the ratio of their typical length scales
by $\lambda$, and the ratio of their timescales by $\tau$.  For
consistency, $\tau \propto \lambda/\nu$.  Further, we denote the ratio
of their mass loss rates by $\phi$ (which, for given $\lambda$ and
$\tau$or $\nu$, sets the gas density scale).  The constancy of ({\sc
ii}) means that cluster masses scale as $M|{cl} \propto \lambda \nu^2$,
the stellar mass densities as $\rho|{\star,c} \propto (\nu/\lambda)^2$
and the nISM gas densities as $\rho \propto \phi\nu/\lambda \propto
\phi/\tau$.  The ionization parameter in the flow, $\Xi = L|i/(4\pi r^2
n|HkTc)$ (see Section~\ref{ss:pressures}), scales as $\Xi \propto
L/\phi\lambda\nu^3$.

\subsubsection{AGN with constant Eddington ratios}
\label{ss:Eddconst}

We first consider a class of AGN of different total luminosities which
are all able to accrete with equal efficiencies and, in consequence, to
radiate with the same constant Eddington ratio.  In such a population,
with ({\sc i}) through ({\sc v}) constant, the masses of the black hole
and stellar cluster scale as the luminosity ($M|{cl} \propto M|h
\propto L/\langle f|{Edd}\rangle \propto L$) and $\lambda\nu^2 \propto
\lambda^3/\tau^2 \propto L$; the ionization parameter then scales as
$\Xi \propto 1/\phi\nu$.

\noindent
(a) If, in such an AGN population, the velocities were constant
(roughly as observed), the characteristic sizes of the regions would
scale as $\lambda\propto L$ (rather than $\lambda^2 \propto L$ as is
often discussed), and typical timescales would scale as $\tau \propto
\lambda \propto L$. The stellar clusters would be very much denser and
smaller as the luminosity decreased: $\lambda \propto L$, $\rho|{\star,c}
\propto 1/L^2$.  Furthermore, the ionization parameter would be a
function only of the mass loss rate of the cluster stellar populations
through $\phi$: $\Xi\propto1/\phi$.

\noindent
(b) Alternatively, if $L/r^2$ were constant within a population, as is
often inferred from broad line studies, \cite[\eg{}]{netp97}, then
$\lambda^2 \propto L \propto M|h$.  In that case, our models predict
that the velocities would scale as $\nu \propto (L/\lambda)^{1/2}
\propto L^{1/4}$, implying also that the spectra of AGN must harden
slightly ($\langle h\nu \rangle \propto T \propto L^{1/8}$) as they
become brighter.  The stellar clusters would be more compact and
denser as the luminosity decreased, but the changes would be
significantly weaker than in (a), $\lambda \propto L^{1/2}$,
$\rho|{\star,c} \propto 1/L^{1/2}$, which seems more probable.
Furthermore, the typical timescales and the ionization parameter would
be weakly dependent on $L$, $\tau \propto L^{1/4}$ and $\Xi \propto
1/\phi L^{1/4}$.

There is some observational evidence for this kind of relation between
line width and luminosity.  Baldwin, Wampler \&
Gaskell~\shortcite{bwg} find a correlation between the width of
C{\sc\,iii}] 1909 and source luminosity, but not for the other lines
they studied.  The variation in Compton temperature also compares well
with observed change in continuum shape with source luminosity
\cite[\eg{}]{zm93}.  Mushotzky \& Ferland~\shortcite{mf84} find an
anticorrelation between ionization parameter and luminosity of this
form can also explain the Baldwin effect, an anticorrelation between
the equivalent width of lines and the source luminosity.  All such
observations are, however, plagued by selection effects and the wide
dispersion between the emission line properties of individual galaxies
with a certain luminosity.

\subsubsection{Eddington ratio varying with $L|{Bol}$.}
\label{s:scalevedd}

It is possible that the Eddington ratio varies systematically with
source luminosity (a point to which we return below).  This would seem
to exclude scaling the solution to equations~\refeq{e:mmdimless} and
\refeq{e:mc2dimless} found for one AGN to other AGNs with different
$\langle f|{Edd}\rangle$ since, as we have just seen, $\langle
f|{Edd}\rangle$ is one of the dimensionless ratios defining the
topology of the solution which we wish to apply to AGN of varying
$L|{Bol}$.  However, as we shall see from our simulations, the most
important value of the Eddington ratio for the flow is that on the
axis of the accretion disc, $f|{Edd,0}$.  Thus, for the moment, assume
that we can ignore the angular variation of the luminosity.  In that
case, we can relax the condition that $\bar{f}(\theta)$ must be
invariant (this assumption is supported by our numerical simulations,
see Section~\ref{s:results}), which is equivalent to allowing the
degree of beaming to vary with $L$ or $f|{Edd}$.  Since theoretical
studies of the structure of accretion discs find structural changes
with $f|{Edd}$, this seems a reasonable assumption.  When
$\bar{f}(\theta)$ is constant $\langle f|{Edd} \rangle \propto
f|{Edd,0}$, whereas when $\bar{f}(\theta)$ varies with $L$, $\langle
f|{Edd} \rangle$ and $f|{Edd,0}$ are independent.

Given the primacy of $f|{Edd,0}$, we find that the forces which act
upon the flow are controlled by an effective black hole mass, $(1-
f|{Edd,0})M|h$, rather than directly by $\bar{f}(\theta)$ and the BH
mass separately.  Parameters ({\sc iii}) and ({\sc iv}) then combine
into a single parameter, $(1-f|{Edd,0})(M|h/M|{cl})$ which defines the
solution and must be scaled for different AGN.  Note that in the
trivial case that the radiation is isotropic and $f|{Edd}(\theta)=1$,
the BH has no net effect on the flow and its mass is arbitrary.

Consider now such a class of AGN whose Eddington ratios, $\langle
f|{Edd} \rangle$, vary systematically with luminosity.  Requiring
$(f|{Edd,0}-1) M|h/M|{cl}$ to be constant rather than ({\sc iii}) and
({\sc iv}) introduces an additional degree of freedom, and relaxes the
strict scaling between the masses and the luminosity.  The constancy of
({\sc ii}) means that, as before, $M|{cl} \propto \lambda \nu^2$.
Since $M|h \propto L/\langle f|{Edd} \rangle$, the new condition,
$(f|{Edd,0}-1) M|h/M|{cl} =$ const.,  becomes $(f|{Edd,0}-1)/\langle
f|{Edd} \rangle \propto \lambda \nu^2/L$.  

\noindent
(c) Consider first a population for which the velocities vary as weakly
as $\nu \propto L^{1/4}$, for which $(f|{Edd,0}-1)/\langle f|{Edd}
\rangle \propto \lambda/L^{1/2}$. If, in addition, $L \propto
\lambda^2$, $(f|{Edd,0}-1)/\langle f|{Edd} \rangle = {\rm
const.}$; as we found in (b) above, $v \propto L^{1/4}$ combined with
$L \propto r^2$ implies that $f|{Edd}$ is constant; all other properties
then vary as they did in (b).   

\noindent
(d) If the velocities have constant characteristic values ($\nu = {\rm
constant}$), then the masses of the clusters must scale as $M|{cl}
\propto \lambda$ whereas the masses of the black holes scale as $M|h
\propto L/\langle f|{Edd}\rangle \propto \lambda/(f|{Edd,0}-1)$.  If,
in addition, $L \propto \lambda^2$, we find that the Eddington ratios
must scale as $(f|{Edd,0}-1)/\langle f|{Edd}\rangle \propto L^{-1/2}$.
Typical timescales, $\tau \propto \lambda \propto L^{1/2}$, in accord
with the observation that typical timescales are shorter in low
luminosity objects.  We find also that $\Xi \propto L^{1/2}/\phi$,
\ie\ for clusters with the same IMF and age, ($\phi\approx {\rm
const}$), $\Xi\propto L^{1/2}$.

If the relation $(f|{Edd,0}-1)/\langle f|{Edd}\rangle \propto L^{-1/2}$
found in (d) is imposed, and we attempt to impose $\bar{f}(\theta)$
constant, we find that both $f|{Edd,0}$ and $\langle f|{Edd}\rangle$
increase as $L$ decreases.  This correlation is in the opposite
direction from that inferred from observation, but will occur only for
a narrow range of luminosity before the luminosity becomes super
Eddington at all angles.  The flow structures change dramatically when
this occurs, suggesting that they will not scale between Seyfert
galaxies and QSOs.  Alternatively, the flows {\it can}\/ scale when the
Eddington ratio varies with $L$ and the characteristic velocities are
constant, but only if the continuum becomes closely beamed as the
luminosity decreases.  If the on-axis Eddington ratio is roughly
constant, we find that $\langle f|{Edd}\rangle \propto L^{1/2}$ which
accords well with some estimates of the variation of the variation of
$f|{Edd}$ with $L$. Furthermore, this then means that $M|h \propto
M|{cl} \propto L^{1/2}$; the stellar cluster will be more compact and
denser as $L$ decreases, $\lambda \propto L^{1/2}$, $\rho|{\star,c}
\propto 1/L$.  This situation seems in broad accord with observations.
We conclude that in the case of constant $f|{Edd,0}$ but varying
$\langle f|{Edd}\rangle$ we can scale our solutions for QSOs to low-$L$
AGN.


The variation of some physical timescales with these scaling
parameters is included in Table~\ref{t:times}.  If dynamical
velocities are constant, most of the timescales will vary in
proportion to the length scale; QSO models can then be scaled to other
situations.  Notable exceptions are the stellar collision time, and
the frequency and lifetime of supernovae.  In smaller nuclei, stellar
collisions will become an important mass source while supernovae will
become infrequent events: when they do occur, both processes will have
a catastrophic effect on the nuclear ISM.                        

\subsection{An HR Diagram for AGN?}
\label{s:parms}

\begin{table}
\caption{Derived velocities, for the models defined in
Table~\protect\ref{t:models}.  The values of $v|K$, $v|z$ and $c|s$
are set by the parameters of the model.  The ejection velocity,
$v|{ej}$, is the peak outflow speed in a region close to
$z=0.2\parsec$ on axis (except \dag\ \modG\ close to $z=0.04\parsec$,
\modK\ close to $z=2\parsec$), where the mass ejected from the region
close to the black hole has reached close to its final value.}
\label{t:derived}
\centering
\begin{tabular}{cccc}
\hline
Label & $v|K/c|s$ & $v|z/c|s$ & $v|{ej}/c|s$ \\\hline
\modC & 5.4 & 10 & 12.4 \\
\modA & 5.4 & 7 & 10 \\
\modB & 5.4 & 3.8 & 4.6 \\
\modC$_{0.005}$ & 5.4 & 10 & 12.4 \\
\modC$_{0.25}$ & 5.4 & 10 & 11.7 \\
\modC$_{0.25}'$ & 5.4 & 10 & 12.4 \\
\modE & 1.7 & 7 & 9.3 \\
\modG & 5.4 & 7 & 4.2\dag \\
\modK & 1.8 & 20 & 14.6\dag \\
\modF & 0.8 & 8 & 8.7 \\
\modD & 5.4 & 7 & 9.1 \\
\hline
\end{tabular}
\end{table}
\begin{figure*}
\epsfxsize = 8cm
\begin{centering}
\epsfbox{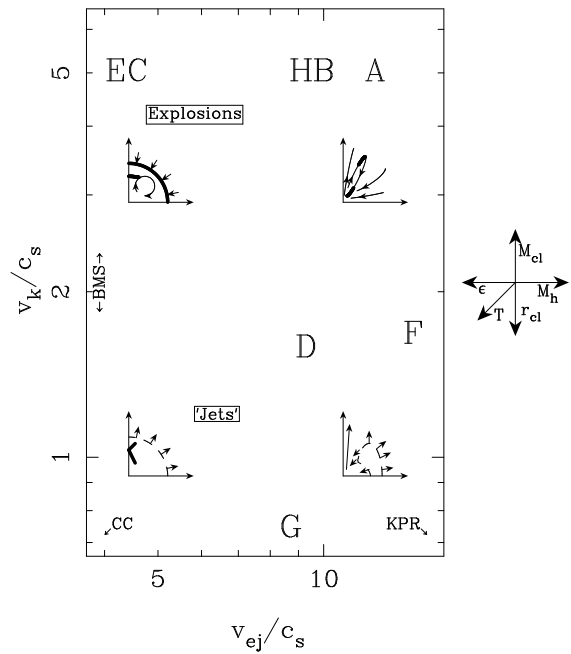}
\end{centering}
\caption[Models]{This graph compares the values of $v|k/c|s$ and
$v|{ej}/c|s$, also presented in Table~\protect\ref{t:derived}.
Smaller labels show the regimes corresponding to various models: CC --
the uniformly mass-loaded, zero-gravity models of Chevalier \& Clegg
\protect\cite[1985; \cf\ also]{wild94}; BMS -- the disc wind models of
Begelman \etal~\protect\shortcite{bms}; KMP -- the nuclear wind models
of Kippenhahn \etal\ and Beltrametti \& Perry approximate the on-axis
flows here.  The small panels show the outline structures expected
for each set of parameters: arrows show the directions of {\it
supersonic}\/ flows, solid lines show the most important shocks and
dashed lines the most significant sub- to super-sonic transitions.
The regions of parameter space in which we find explosions and
collimated jets are labelled.  The set of arrows to the right shows
the directions in which models change if the variable labelled is
changed.}
\label{f:xxx}
\end{figure*}
\begin{figure}
\epsfxsize = 8cm
\begin{centering}
\epsfbox[46 172 503 546]{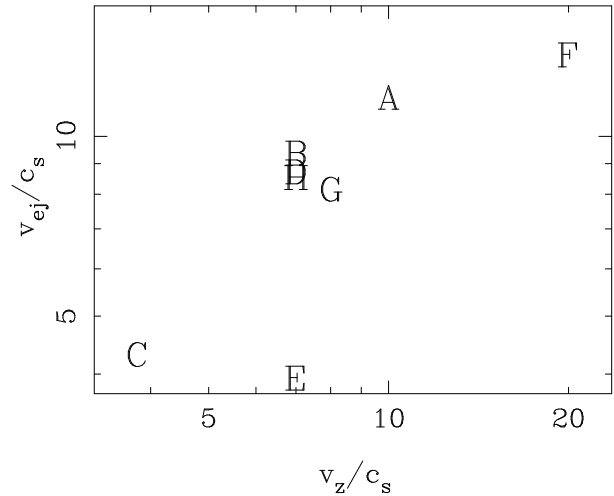}
\end{centering}
\caption[Models]{In this plot, we compare the outflow velocity, $v_z$,
estimated from eq.~\protect\refeq{e:vz} with that measured in the
simulations, $v|{ej}$.  Values are given in
Table~\protect\ref{t:derived}.  A good relationship is seen, apart
from in Models \modG\ and \modK\ where the assumptions underlying
eq.~\protect\refeq{e:vz} break down.  In Model \modG\ the smoothing
region fills much of the volume of the cluster so the cluster mass
acts to suppress the outflow.  In Model \modK, the smoothing region is
unresolved by the grid, so the effective $\epsilon$ will be larger
than that imposed.  Models \modA, \modE\ and \modD\ are coincident.}
\label{f:xxx1}
\end{figure}

As we have discussed, our models are determined by four dimensionless
ratios and a number of `form factors'.  However, extensive numerical
modelling has suggested that the number of parameters which dominate
the overall form of the flow is far smaller than is yielded by such a
formal reduction.  We find that the solutions we calculate can be well
described by just two dimensionless ratios, describing how strongly the
gas is bound to the nucleus as a whole and how strongly it is driven
away by the black hole at its centre.  In the rest of this paper, we
present models to illustrate the reasonableness of this simplified
parameterisation, and its usefulness as a classification scheme for the
numerical models.  The details of this classification scheme are
presented in Sections~\ref{s:classify} and ~\ref{s:disclasify}. While
the derivation of detailed observational diagnostics is beyond the
scope of the present paper, we intend to work to bridge the gap between
empirical classifications and this theoretical framework.  Before
describing the classification scheme we preview the physical properties
of the flows we find.

\subsubsection{A preview of the nISM flows}

Here, we preview our results, and move on, in Section~\ref{s:classify} to 
the classification scheme which we have found.  The details of 
the models are presented in Section~\ref{s:results}.  By anticipating
the conclusions here, we hope the reader will be better placed to 
understand not only the details of the individual models, but also
their relative place in the wider scheme.

In spherical models with no accreting central black hole the form of
the flow is principally determined by the ratio $v|K/c|s$ of the
Kepler velocity and the sound speed in the ISM \cite[see
also]{durb81}, where $v|K^2 \equiv GM|c/r|c$ and $M|c$ is the core
mass of the cluster.  For small $v|K\la c|s$, the cluster core fills
with gas on a dynamical timescale and the sonic surface coincides with
the edge of the cluster.  By contrast, for large $v|K/c|s$ (as in
Fig.~\ref{f:modelz2d}), most of the gas injected falls supersonically
through the cluster onto a hydrostatic core at early times.
Eventually this hydrostatic core fills up to the edge of the cluster,
causing the inflow to cease.  After this the core grows further,
quasi-statically, until it reaches the sonic surface at $r|s \sim
(v|K/c|s)^2 r|c$, when a steady isothermal wind is established.  This
scheme is illustrated by the analysis of Appendix~\ref{a:apspherical}
and the numerical models in Section~\ref{s:spherical}.

The addition of an accreting black hole to the cluster fundamentally
alters the structure of the nISM created by the stellar mass-loss.  The
non-spherical radiation, by exerting a non-spherical force on the gas,
gives rise to the in- and out-flow structures which we discuss.  We have
considered a wide range of models spanning the expected range of the
independent parameters listed above.  In this paper we present a
selected set of characteristic models which illustrate the full variety
of expected flow topologies.  Our basic model, designated \modC, is the
most likely physical scenario. We consider the systematic effects of
changing the physical structure of the region: in models \modA and
\modB\ we decrease the relative mass of the cluster; models A$_{x}$
explore accretion (where $x$ indicates the region of accretion); \modE~
and \modG~ have a dynamically less significant cluster; \modK~ has a
diffuse cluster and a heavy black hole; \modF~ has a small and diffuse
cluster.  The properties of our models are given in
Table~\ref{t:models}.

In these two-dimensional models, gas which falls close to the black hole
is driven outwards along the axis in a dynamical timescale.  Rather
than a hydrostatic core forming in the gravitational well of the
central black hole, a circulatory flow forms close to the centre.  In
addition, the amount of mass within the nuclei can continue to vary,
either because no gas escapes, as in Model \modB, or because it escapes
in an episodic fashion, as in Model \modA.

The properties of the flow may also depend on the age of the nucleus.
Where the central outflow drives directly out into the ambient ISM, the
nuclear flow rapidly reaches an equilibrium situation: it is unlikely
that the transient evolution, over a dynamical timescale, will be
observed (or even occur, depending as it does on the artificial initial
conditions).  Age will certainly be important where most of the
mass input is retained by the nucleus, since the mass content of the
ISM will then increase linearly with time.  Eventually, this will lead
the flow to break our assumptions of low optical depth and
isothermality.  In the mean time, as the mass content increases, the
dynamical importance of the mass loading decreases, so the density
scaling works for us once more.  The structure of the flows becomes a
turbulent-looking plume, with an extent determined by the balance
between driving and gravitational velocities, in which the gas density
increases linearly with time, surrounded by a constant accretion flow
which is no longer dynamically important (or observationally, as a
result of its low gas pressure).

\subsubsection{The classification scheme: An HR diagram for AGN?}
\label{s:classify}

Many authors have attempted to find out how many effective parameters
determine the structure of active nuclei, using both observational data
and theoretical analysis \cite[\eg{}]{bg92,francea92,by96,donb96}.
This may be seen as a search for diagnostics analogous to the
Hertzsprung-Russell diagram, which has led to so many insights into the
most stars very close to spherical symmetry: it is no surprise that the
situation in AGN is substantially more difficult.

To recap, we take as the starting point for our model a galactic
nucleus with an accreting black hole, surrounded by a starburst
stellar cluster. We assume that the cluster can form by prior
processes such as galaxy interactions. These two simple, relatively
well-understood input components turn out to provide a remarkable and
fascinating variety in the behaviour of the nISM.  Broadly speaking,
the results of our simulations can be understood in a two-dimensional
theoretical classification scheme, analogous to the HR diagram.

We find that we can classify the models computed in terms of two
dimensionless quantities: $v|K/c|s$, the ratio between stellar
dynamical velocity and ISM sound speed; and $v|{ej}/c|s$, the Mach
number of the gas driven away from the black hole along the axis of
the disc.  These parameters are the axes of in Fig.~\ref{f:xxx}.  Also
shown on this figure are the codes of the various models which will be
presented in Section~\ref{s:results} and the form of the flows which
characterise each region of the plot.  The flows take a variety of
forms: inflow with a small core, jets, explosions and bipolar winds,
as will be seen from the models themselves.  The application of this
plot to classifying AGN models is discussed in detail in
Section~\ref{s:disclasify}.

The interaction of the axial outflow with the gravity field and the
mass input of the stellar cluster determines the form of the global
flow: the axial wind can be confined within the core, can escape
episodically or can drive outwards as a steady flow.  The overall
effect is determined by a combination of whether the central outflow
has more than escape velocity from the nucleus as a whole (with due
account taken for mass loading), and, if so, whether it has sufficient
momentum to overcome the ram pressure of the accretion flow.

The most important effect not included in the models presented here is
the thermal balance of the flow.  As we assume that the flow is
isothermal, the two-dimensional classification of disc winds
introduced by Begelman \etal~\cite[1983; and related to global flows
by]{will97} collapses to the axis labelled BMS on Fig.~\ref{f:xxx} --
the one-dimensional classification of Section~\ref{s:spherical}.
While isothermality seems to be a reasonable first approximation for
flows in AGN (Appendix~\ref{a:nISM}), we intend to investigate the
structure of radiative flows in more detail in future papers.  This
classification must, of course, be supplemented by the inclination
angle to the observer when applied to observations.

\section{Numerical Methods}
\label{s:code}

These simulations were computed with a Godunov-type hydrodynamic
code \cite{Falle}.  The method is second order in space and time in
smooth regions, representing the flow with piecewise-linear variations
across the computational cells.  It calculates the inter-cell fluxes
by solving a Riemann shock-tube problem across the cell interfaces.
In the models presented here, the flow is treated as isothermal.  The
computational grid is two dimensional, with interfaces along the
ordinates of a cylindrical polar grid.

As discussed in Section~\ref{s:interf}, we assume that no mass passes
inwards from the galaxy to the nucleus at the outer boundaries of the
grid, at large $r$ and large $z$.  If the nuclear flow at the grid
boundary is outwards we use zero transverse gradient boundary
conditions, and keep a record of the total mass and energy flows
through these boundaries.  This allows us both to verify mass
conservation and to investigate the strength of the winds interacting
with the external medium.

In some models, the mass escaping the region has too little kinetic
energy to escape to infinity.  It is likely that some fraction of this
mass ought eventually fall back into the nucleus, probably in the
accretion disc plane, after some interval.  We comment elsewhere
(Appendix~\ref{a:loaddist}) on the insensitivity of our results to the
detailed form of the mass-loading (whether internal or extra-nuclear)
if this input mass has low angular momentum.  In the models presented
here, we have chosen to keep an independent total of the `bound
ejecta' to compare with other mass fluxes rather than impose an
arbitrary reinjection model.  Experiments with larger grids and a
variety of reinjection models have shown that while this artificial
lowering of the barrier to escape from the nucleus changes the
boundaries between the flow modes by a small amount, it does not
fundamentally alter the nature of the flows.

We do not include the effects of angular velocities about the grid
axis.  We consider only the input of gas with low net angular momentum
from stellar mass loss.  Models including disc winds, although not the
gravitational effect of a stellar cluster, have been calculated by
Woods \etal~\shortcite{woodea96}.  They require treatment of both
angular momentum and of wind acceleration very close to the accretion
disc.  The $z$ component of angular momentum is included in those
models.  While essential in that case as the mass source had high
angular momentum, this results in nearly empty vortices forming along
the poles of the simulation, which may not be physically realistic
when account is taken of non-axisymmetric perturbations to the flow.

We concentrate here on two limits, one in which there is negligible
net mass transfer to the accretion disc (so that the surface $z=0$ is
just treated as a mirror plane), and another in which mass is accreted
over some fraction of its surface.  We determine the flux onto the
accretion disc in the latter case by making the disc surface a free
flow boundary.  This mimics the effect of the strong cooling which may
well be driven by the high pressures found in the disc plane (\eg\
Figs. \ref{f:modela2d}, \ref{f:modelb2d}).

\section{Spherical models}
\label{s:spherical}

\begin{figure*}
\epsfxsize = 8cm
\begin{centering}
\begin{tabular}{ll}
a) & b) \\
\epsfbox{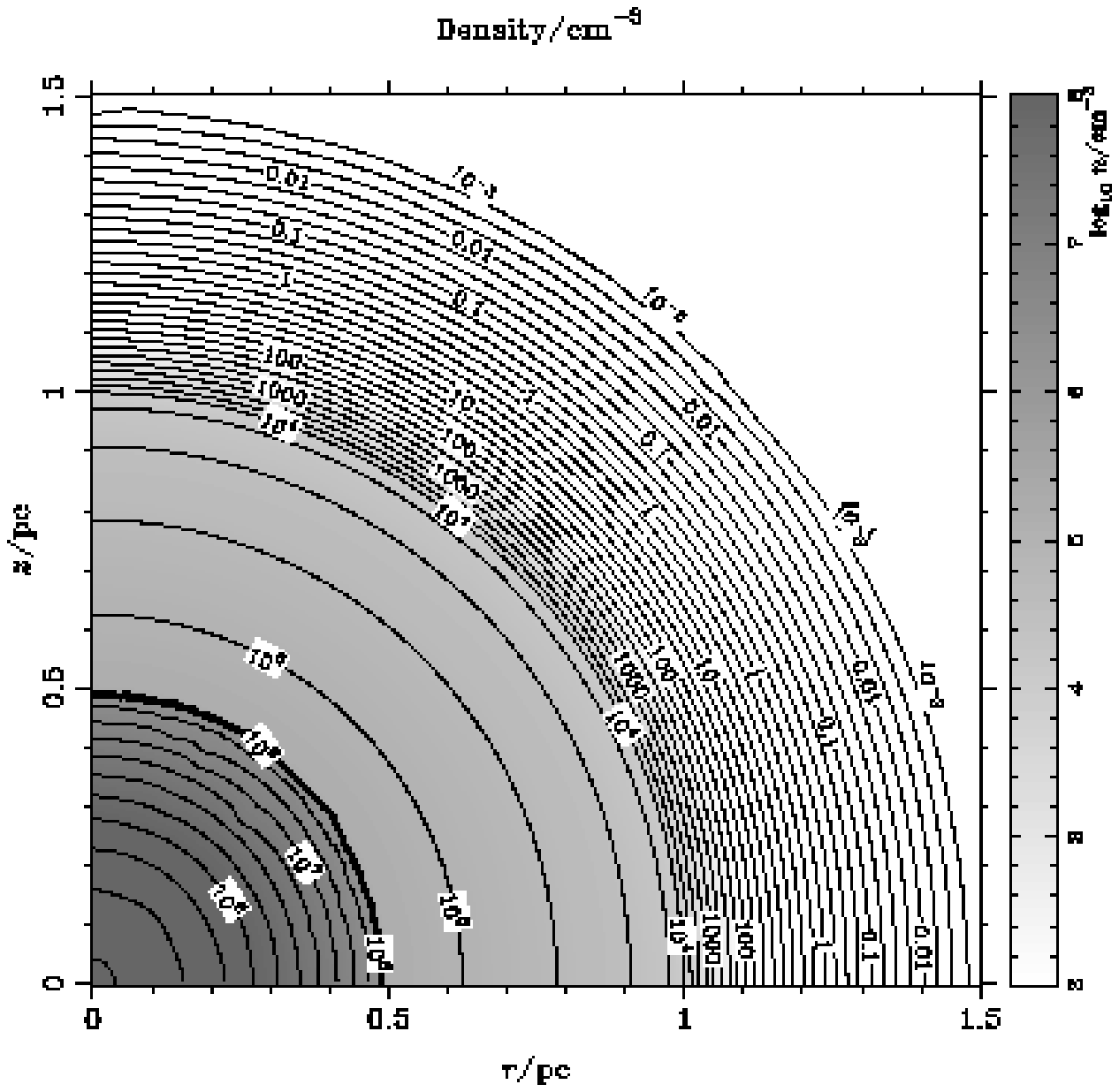} &
\epsfbox{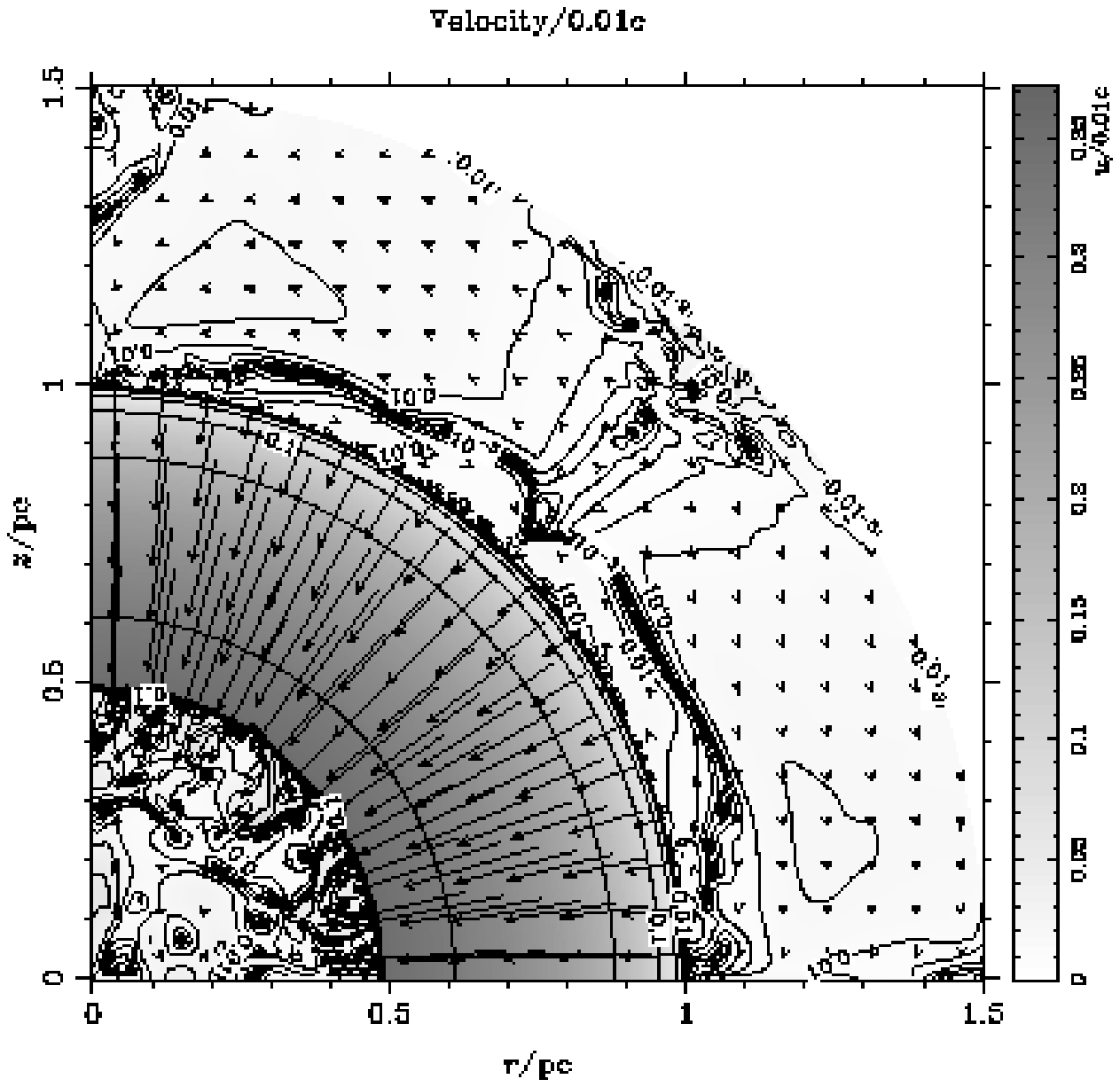} \\
c) & d) \\
\epsfbox{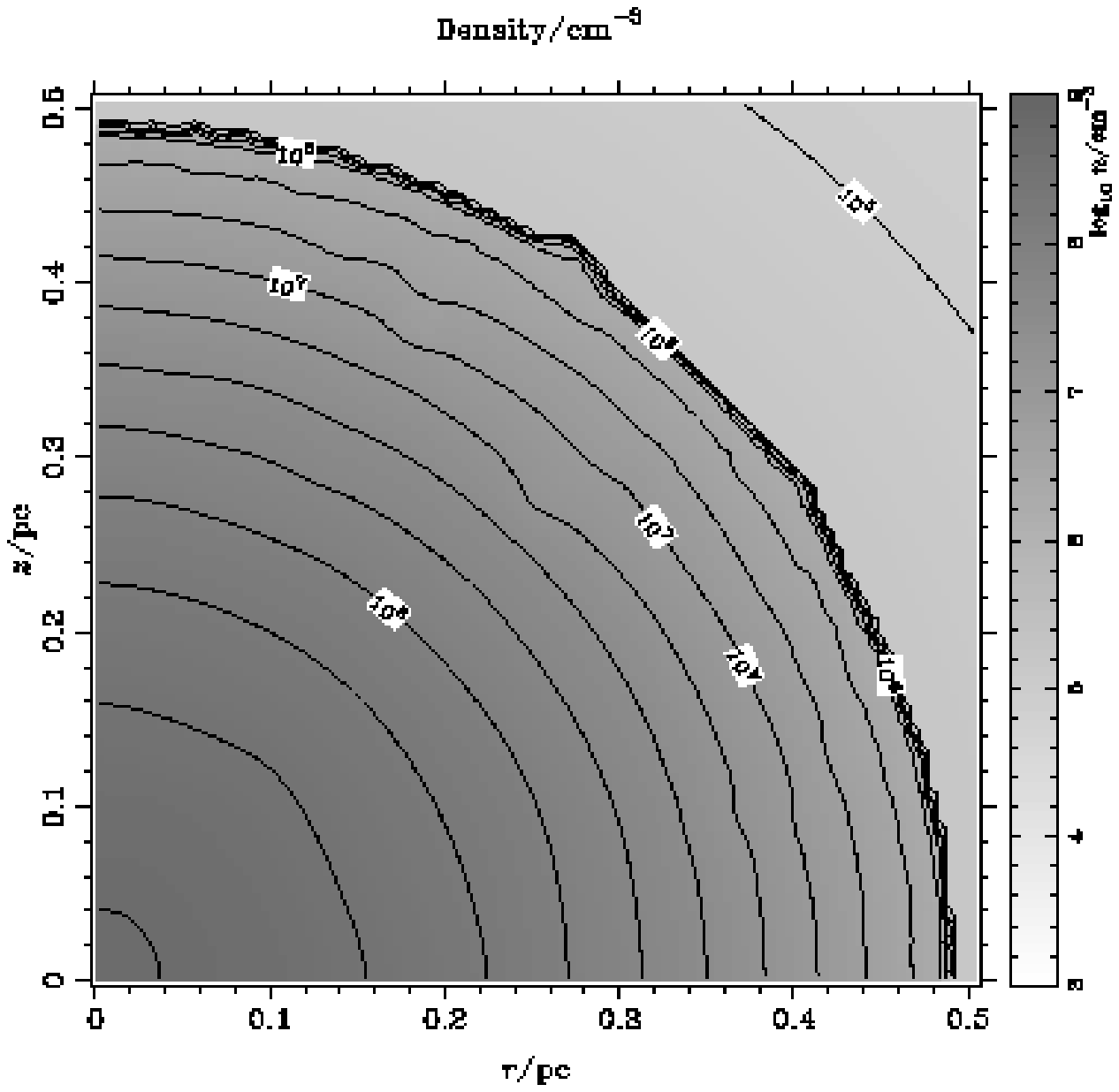} &
\epsfbox{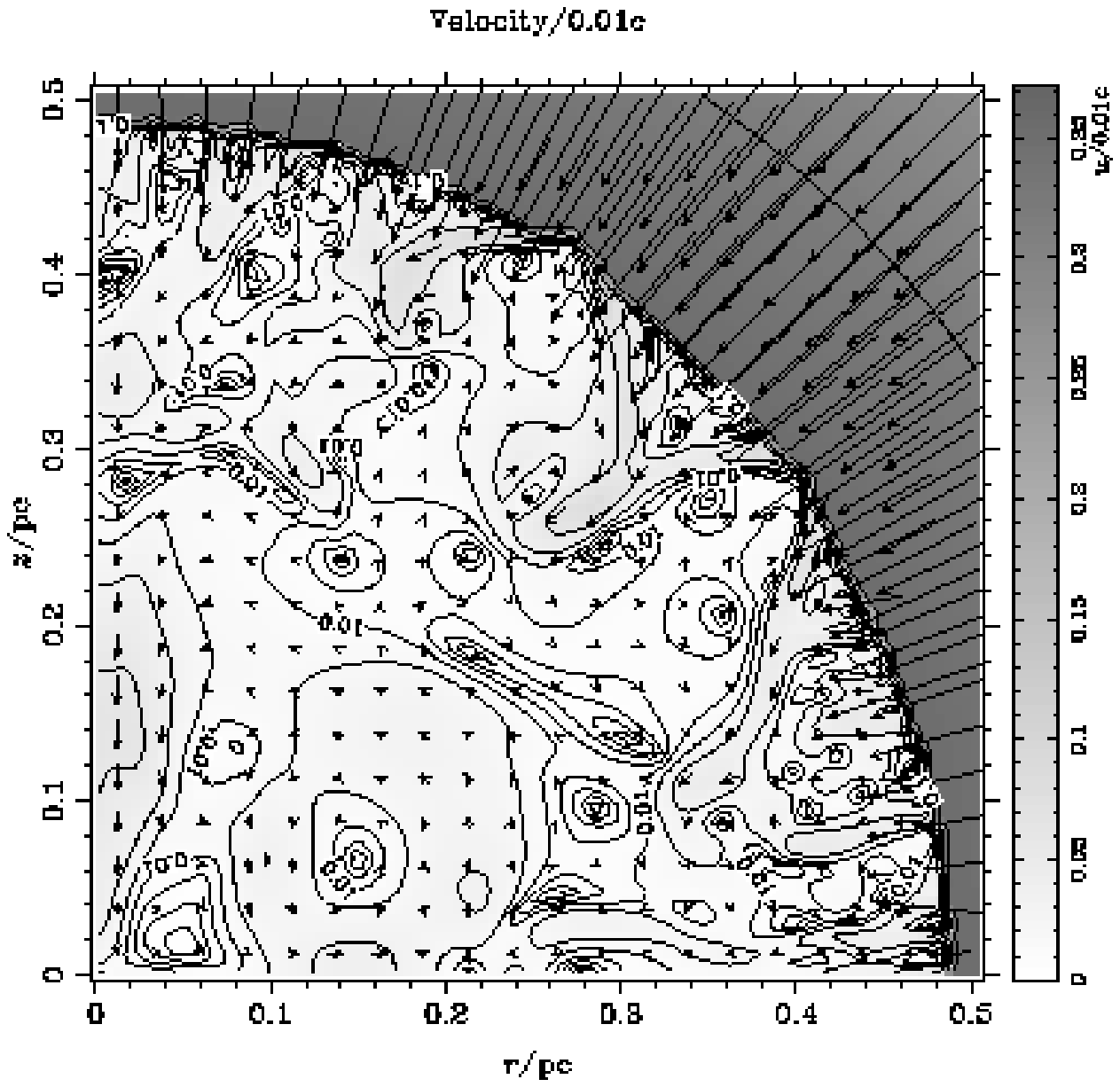} \\
\end{tabular}
\end{centering}
\caption[Model \O\ 2D plot]{A spherical flow model (Model \O\
discussed in in Section~\ref{s:spherical}, with parameters given in
Table~\protect\ref{t:models}) at $t=4.4\ee4\yr$.  The panels show a)
number density $n$ (up to $5.6\ee{9}\cm^{-3}$) b) total velocity $u$
(grey-scale and contours, up to $3.7\ee{-3}c$).  The vectors are in the
flow direction, and have lengths proportional to the velocity, sampled
at one vector in every $15\times15$ cells.  The flow is sonic at the
$u = 0.1$ contour.  c) and d) are a magnification of the central
$0.5\times0.5\parsec$ region (here, the vectors are every $5\times 5$
cells).  The flow structure is of spherical accretion onto a
near-hydrostatic core (Appendix~\protect\ref{a:apspherical}).}
\label{f:modelz2d}
\end{figure*}
One limiting case of the flow models presented in this paper is the
flow in a purely spherical system, such as an active nucleus with no
anisotropic radiation, or a starburst with no black hole, or a
globular cluster.  Prior to presenting our main results, we present
one example of a spherical flow as a test case.  This can be compared
with many treatments in the literature, with a variety of
distributions of gravitating matter and radiative forces
\cite[\eg{}]{KMP,BP,ddc87a}.  We focus on a case with no black hole
and no radiation forces to illustrate the accuracy of the numerical
method.  Our numerical results agree well with an analytic treatment
of the flow structure.

In Fig.~\ref{f:modelz2d}, we show the structure of the flow after
$4.4\ee4\yr$.  An accretion shock ($0.5\parsec$ from the centre of the
grid) then surrounds a central hydrostatic core, and is still moving
outwards.  No physical instabilities operate to drive the flow far
from spherical symmetry.  The structure is that of a central,
near-hydrostatic core surrounded by a supersonic accretion flow.  The
velocities in the core (Fig.~\ref{f:modelz2d}b) reach levels as high
as Mach $0.7$ in one isolated region, and some striated structures
appear close to the bounding shock.  Some turbulence in the core would
be expected in a real cluster, as a result of stirring by orbiting
stars and of inhomogeneities in the infalling flow.  Here, however,
turbulence is driven by the jagged form of the well-resolved shock at
the cluster edge and by the well-known numerical instability of
slow-moving shocks \cite{quirk94}, and is maintained in part by the
reflective boundaries at $r=0$ and $z=0$ (notice the vortex at $0.15,
0.6$; \cf\ the similar effects seen by Mellema \& Frank 1995, for
outgoing flows).

Beyond the sharp edge of the cluster, a near-hydrostatic atmosphere
reaches outwards.  The flow would only become transonic at a radius of
about $25\parsec$ (\ie\ $2.5M_8 T_7^{-1}\parsec$), far beyond our
grid.  The steep density gradient in the atmosphere allows us to
ignore these outer regions as they will have little dynamic effect,
and only a negligible influence on the overall mass budget.  This
density gradient amplifies small waves at the edge of the core to
generate sound waves with gas velocities up to Mach $0.2$ further out
in the halo.  Nowhere, however, are the density contours significantly
perturbed from spherical symmetry.

In Appendix~\ref{a:apspherical}, we give an analytic derivation of the
time evolution of the mass-loaded isothermal flow in an initially
evacuated spherical cluster with no central black hole.  In
Fig.~\ref{f:modelzsect} we compare these results to the numerical
model shown here, and find excellent agreement.  The smooth curve on
Fig.~\ref{f:modelzsect}a shows the good fit of the density to the
isothermal, hydrostatic solution obtained below
(equation~\ref{e:hydrostaticmodel}).  The low level features of
Fig.~\ref{f:modelzsect}b within the hydrostatic core are caused by
the turbulence in the core.

As the flow evolves, the hydrostatic core will become larger until
eventually its outer edge, having merged with the inner accretion
sonic point, reaches the outer sonic point of the flow and forms a
classic Parker-type wind.  Even after more than $4\ee4\yr$ the flow is
far from equilibrium (\cf\ Figs.~\ref{f:modelz2d} and~\ref{f:modelzsect}).
It is likely that by the time the equilibrium solution is reached the
stellar cluster will have evolved significantly.  As we discuss below,
the high densities which are reached in the hydrostatic core are also
likely to allow bremsstrahlung cooling to become important before the
core has filled, breaking our assumption of isothermality.

Adding a black hole with no aspherical radiation component or
accretion simply changes the properties of the central hydrostatic
core.  Small accretion rates can be modelled as Bondi accretion within
this core \cite{durb81,ddc87a}: as the accretion sonic point moves
further out into the core, the mass removed will weaken the wind which
eventually drives outwards from the cluster (\cf\ Model Sa below).  In
the next Section, we illustrate the dramatic effects which can occur
when the radiation field of the central black hole and accretion
structure is aspherical.

\section{Aspherical Models}
\label{s:results}

\makeatletter
\def\rightbrackfill{$\m@th \mathord- \mkern-7mu
\cleaders\hbox{$\mkern-2mu \mathord- \mkern-2mu$}\hfill
\mkern-4mu \mathord\mapstochar
$}
\def\leftbrackfill{$\m@th \mathord\mapstochar \mkern1mu
\cleaders\hbox{$\mkern-2mu \mathord- \mkern-2mu$}\hfill
\mkern-7mu \mathord-$}
\makeatother

\newcommand{\twthd}{$2/3$}
\begin{table}
\caption{Computational Grids}
\label{t:grid}
\centering
\begin{tabular}{cccc}
\hline
Models & Cells& Size/$\!\parsec$ & Smoothing radius/$\!\parsec$\\\hline
{\O}\modC\modA\modB\modC$_i$\modE & $300\times300$ & $1.5\times1.5$ & 0.05 \\
\modG SaSn & $300\times300$ & $0.15\times0.15$     & 0.05  \\
\modK & $300\times300$ & $15\times15$ & 0.05 \\
\modF & $1000\times1000$ & $4\times4$ & 0.004 \\
\modD & $600\times600$ & $1.5\times1.5$     & 0.05  \\\hline
\end{tabular}
\end{table}
\begin{table*}
\caption{Parameters of models computed.  The labels in column 1 are
used to refer to these models elsewhere in the text.  See
Table~\ref{t:glossary} for definitions of symbols.  In all the models
computed, half the luminosity is in the anisotropic accretion disc
component, $f|{disc}=0.5$, and the stellar cluster core has uniform
density, $s=0$.  The computational grids for each model are listed in
Table~\protect\ref{t:grid}.}
\label{t:models}
\centering
\begin{tabular}{cccccc|cccccccl}
\hline
Label & 
BH mass& 
\multicolumn{3}{c}{\leftbrackfill Cluster\rightbrackfill}&
Mass ratio &
ISM temp. &
& 
Accretion &
Fig. &
Comments\\
& 
$M|{h,8}$ & 
mass, $M|{cl,8}$ & 
radius, $r|{c,pc}$ & 
halo, $h$ &
$M|{h} / M|{cl}$&  
$T|{C,7}$ &
$\langle f|{Edd}\rangle$ & 
disc pc 
\\

\hline
\O     & 0   & 10  & 1      &$\infty$ 	& - 	& 1   & n/a & 0 &\ref{f:modelz2d} 			& Pure spherical cluster\\
\\
\modC  & 2   & 10  & \twthd & 5 	& 0.2  	& 1    & 1   & 0    &\ref{f:modelc2d} 			& Fiducial model \\
\modA  & 1   & 10  & \twthd & 5 	& 0.1  	& 1   & 1   & 0    &\ref{f:modela2d}			& | \\
\modB  & 0.3 & 10  & \twthd & 5 	& 0.03 	& 1    & 1   & 0    &\ref{f:modelb2d}			& | Fiducial cluster;\\
\modC$_{0.005}$& 2 & 10 & \twthd & 5 	&  |  	& 1 & 1 & $0\rightarrow0.005$ &\ref{f:modelcc2d}	& | Changes in hole\\	
\modC$_{0.25}$ & 2 & 10 & \twthd & 5 	& 0.2 	& 1 & 1 & $0\rightarrow0.25$  &\ref{f:modelcpq2d}	& | and accretion\\
\modC$_{0.25}'$& 2 & 10 & \twthd & 5 	&  | 	& 1  & 1 & $0.25\rightarrow1.5$ &\ref{f:modelcout2d}	& | \\
\\
\modE  & 1   & 1   & \twthd & 5 	& 1   	& 1    & 1  & 0   &\ref{f:modele2d}			& \\
\modG  & 1   & 1   & $2/30$ & 5 	& 1   	& 1    & 1  & 0   &\ref{f:modelg2d}			& Relative\\
\modK  & 4   & 10  & $20/3$ & 5 	& 0.4 	& 1    & 1  & 0    &\ref{f:modelk2d}			& importance of\\
\modF  & 1   & 4   & 3      &$\infty$ 	& 0.25 	& 10 & 1 & 0   &\ref{f:modelf2d}			& the cluster\\
\modD  & 10  & 10  & \twthd & 5  	& 1   & 1   &0.7 & 0   &\ref{f:modeld2d}			& \\
\\
Sa & 1   &0.01 & $2/30$     & 5 	& 100 & 1 &0.01  & $0\rightarrow5\ee{-4}$
&\ref{f:modelsa2d}\\
Sn & 1   &0.01 & $2/30$     & 5 	& 100 & 1 &0.01  & 0    &\ref{f:modelsn2d}\\\hline
\end{tabular}
\end{table*}
\begin{table*}
\caption[Results]{Computational results, for the models defined in
Table~\protect\ref{t:models}.  Values are long-term averages; mass
transfer rates are in $\!\Msun\yr^{-1}$, mass content in $\!\Msun$.
Where values they are marked `:', the simulations which had not fully
relaxed (the values given are either rough averages for intrinsically
variable simulations, or final ones when the flow could not be
integrated to convergence).}
\label{t:results}
\begin{minipage}{\textwidth}
\centering
\begin{tabular}{cccccccl}
\hline Label
& $\dot{M}|{inj}$ & $\dot{M}|{acc}$ & $\etaflow$ &
$\dot{M}|{wind}$ & $ \log ( L|{wind}/\ergs ) $ &
$M|{grid}$
& Comments\\\hline
\O & 9.5 & 0 &0 & 0   & 0 & $\dot{M}t$ & No hole\\
\modC  & 8.6 & 0 &0& 8.6 & 
43.3 & $1.1\ee4$ &
Steady outflow forms quickly\\
\modA  & 8.6 & 0 &0& 8.6 & 
39.0: & $10^5$: & Many explosions\\
\modB  & 8.6 & 0 &0& 0 & 0 & $\dot{M}t$ & Trapped circulation\\
\modC$_{0.005}$ & 8.6 & 3.2&0.37 & 5.4 &
43.1 & $9.4\ee3$ & 
Inner disc accretion\\
\modC$_{0.25}$  & 8.6 & 8.6 &1& 0 & 0 & $8.6\ee3$ &
Large accretion\\
\modC$_{0.25}'$  & 8.6 & 3.7 &0.43& 4.8 & 
43.1 & $8.9\ee3$ &
Outer disc accretion\\
\modE  & 0.86 & 0 &0& 0.86 & 
42.1 & $3.0\ee3$ & Small cluster\\
\modG  & 0.86 & 0 &0& $2\ee{-3}$ & 
38.7 & $\dot{M}t$ &Small cluster, const $v|K$ \\
\modK  & 8.6 & 0 &0& 8.6 & 
43.2 & 5.2\ee{5} & \\
\modF  & 3.8  & 0 &0& 3.7: & 
43.4: & $1.2\ee4$: & Hot flow -- jet\\
\modD  & 8.6 & 0 &0& 8.6 & 0 & 2.2\ee4
& Narrow cone -- axis force as \modA\\
Sa & 8.6\ee{-3} & 8.6\ee{-3} &1& 0 & 0 & $0.30$ &`Seyfert' parameters \\
Sn & 8.6\ee{-3} & 0 &0& 0 & 0 & $\dot{M}t$ &
`Seyfert' parameters, no accn \\\hline
\end{tabular}
\end{minipage}
\end{table*}

We present in this section the results for a variety of models chosen
to illustrate the relative importance of the black hole mass and
luminosity, the nuclear cluster and the structure of the accretion
disc.  If the outward driving by the accretion disc luminosity can
counteract the gravitational acceleration of the black hole, its
anisotropy inevitably results in a bipolar outflow.  This outflow
coexists with inflow from the outer regions of the cluster. The
outflow funnels the inflow close to the accretion disc plane at small
radii.

We explore a range of models between those in which the central flow
is strong enough to drive a steady outflow, with a classic `bipolar'
shape, and those where it is confined (smothered) within a very small
central region.  At intermediate values, the flow either drives jets or
sequences of explosions, depending on the ratio of the Keplerian velocity
to hot phase sound speed.  The balance between inflow and outflow is
determined primarily by the ratio between the dynamical velocity in the
cluster core and that at the smoothing radius around the central black
hole.

The generalizations we draw throughout this paper about the pattern of
structures which occur are based on numerous simulations -- of these,
those which we have chosen to present here are those with parameters
close to those expected from the observations (as discussed in
Appendix~\ref{a:detail}), and those where the resulting nISM structures
are non-trivial.  If, for instance, the stellar cluster has a mass far
lower than the central black hole, then its influence on the flow will
be small, a situation already treated thoroughly in the literature.

In Tables~\ref{t:grid} and \ref {t:models}, we list the parameters of
the models presented, and in Table~\ref{t:results} various numerical
results.  As can be seen, we first explore (in Models \modC, \modA,
\modB) the balance between the dynamical velocities in the cluster and
at the smoothing radius by varying the black hole parameters, leaving
the cluster parameters unchanged.  Next, we present models in which
the accretion disc is assumed to remove gas from the flow over some
fraction of its surface (\modC$_{0.005}$, \modC$_{0.25}$,
\modC$_{0.25}'$).  We then investigate the effect of varying the
parameters of the stellar cluster in Models \modE, \modG\ and \modK.
We look at the effects of changing other flow parameters, such as the
opening angle of the central outflow (\modD), and of increasing the
temperature of the ISM (\modF).

We also discuss two cases (Sa, Sn) with parameters which may be more
appropriate for Seyfert galaxies.  In Section~\ref{s:scaling} we
consider in general how our results scale with the input parameters of
the system.

The range of models presented are chosen to illustrate the
classification which we have derived for the flow morphologies, and
which we discuss in Section~\ref{s:parms}.

In the figures of this section, we present plots of the flow structures
of our simulations.  The plots are shown at the end of the
computational run, or sooner if the flow reaches a steady equilibrium
within this time.  In all cases, the structures shown are broadly
characteristic of those found throughout the evolution of the flow.
The physical time which this represents varies, as the length of the
run was specified in most cases by the total number of computational
steps taken.

\subsection{Varying the black hole mass}

\begin{figure*}
\epsfxsize = 8cm
\begin{centering}
\begin{tabular}{ll}
a) & b) \\
\epsfbox{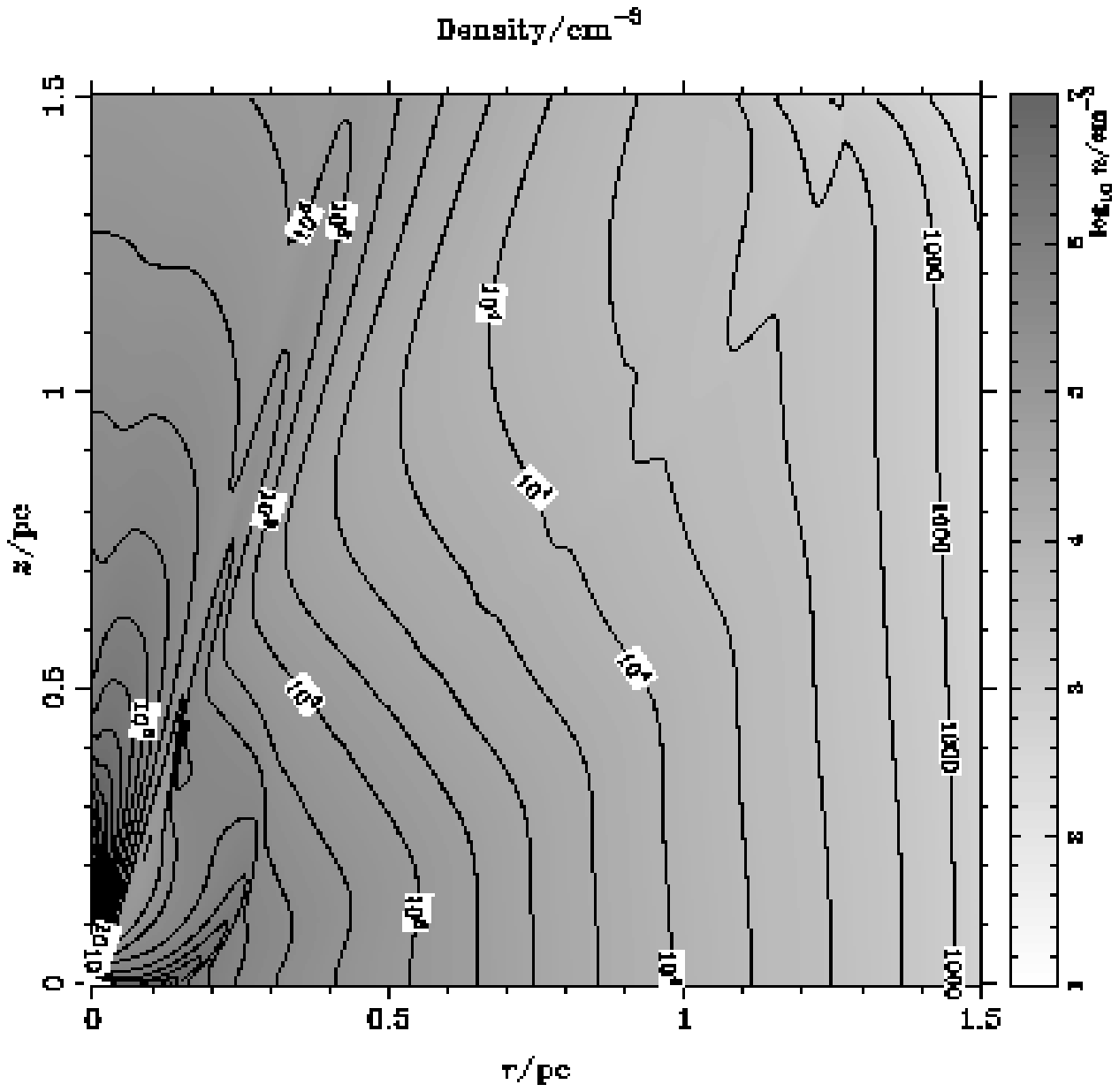} &
\epsfbox{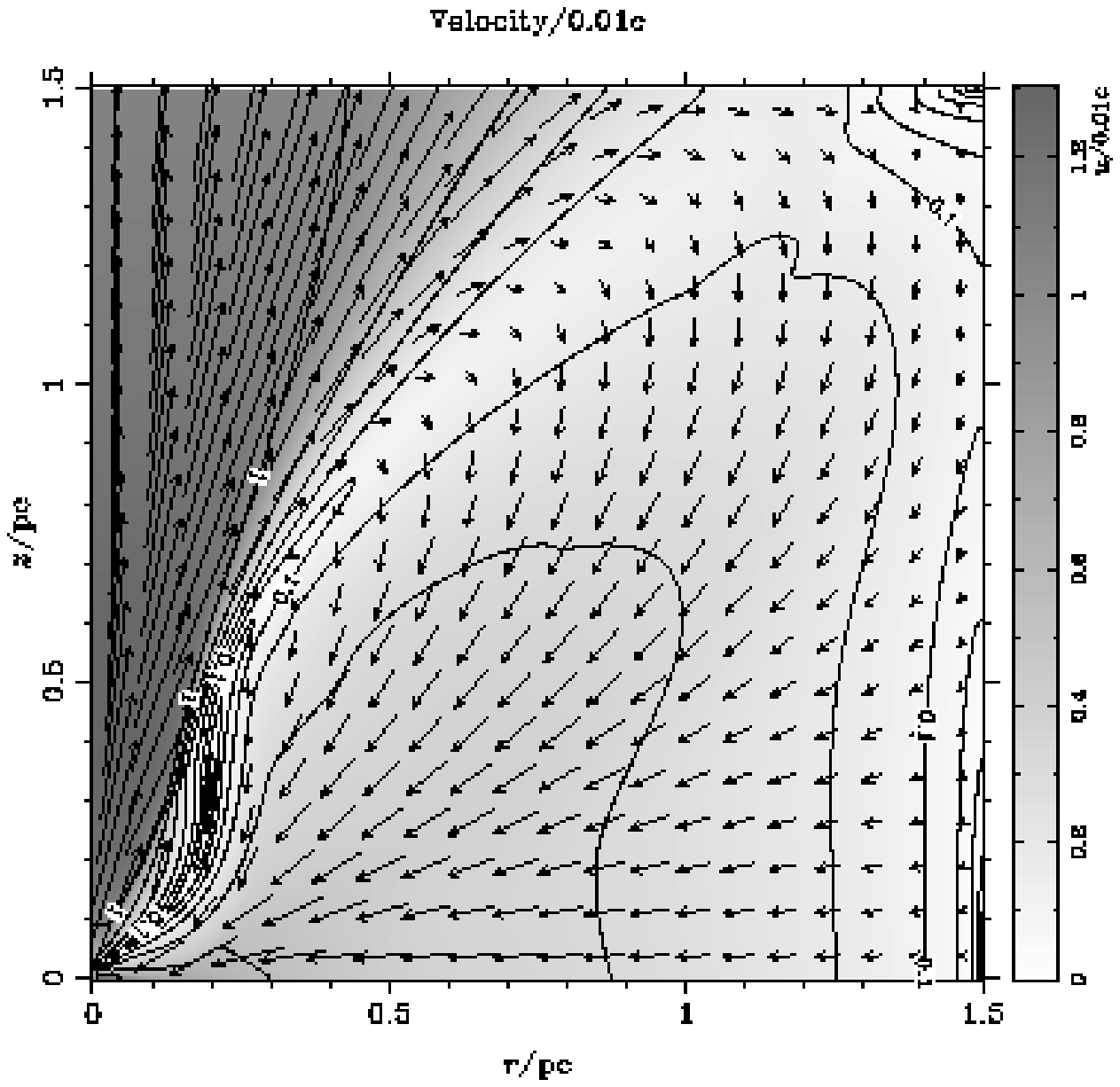} \\
c) & d) \\
\epsfbox{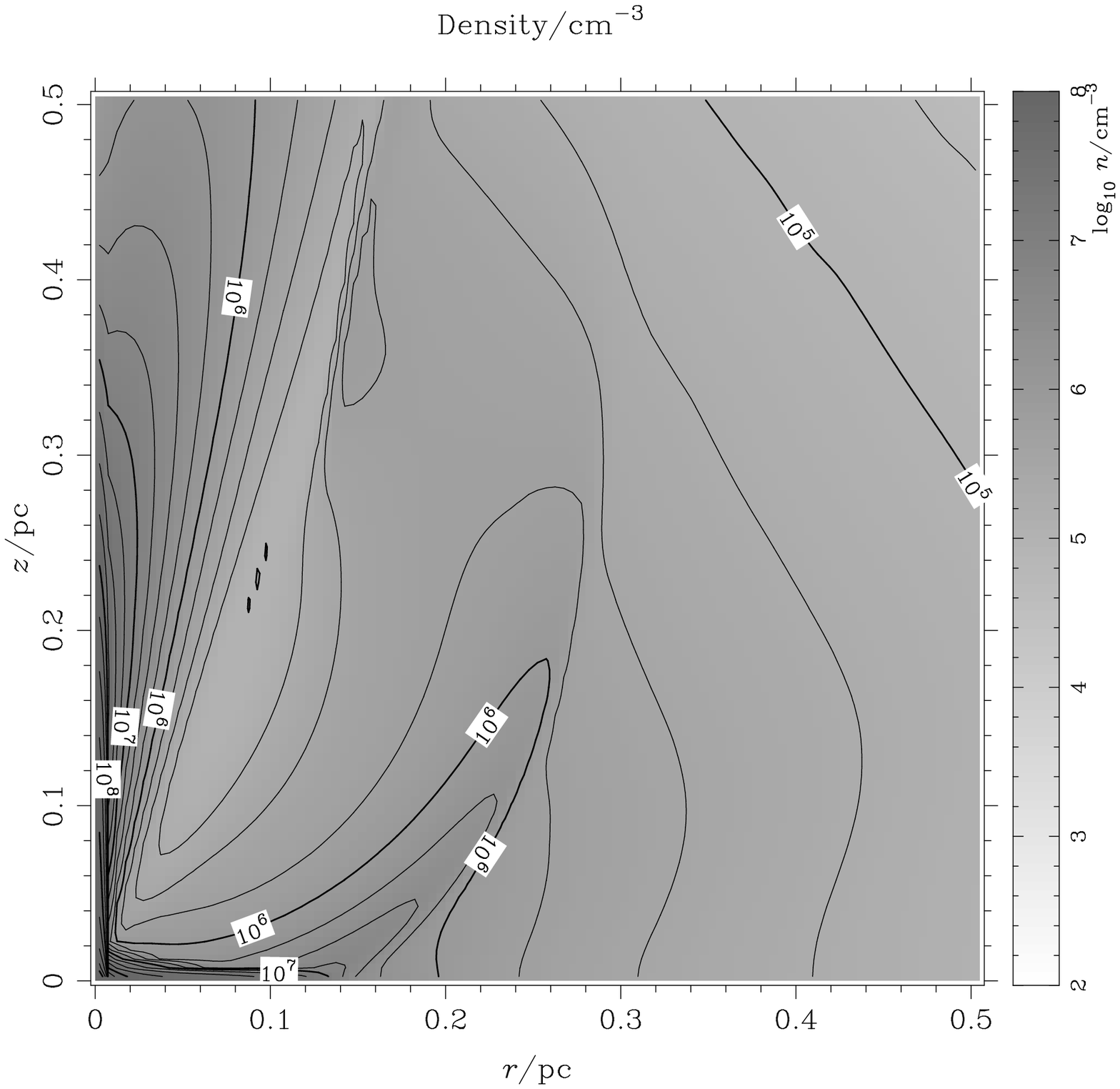} &
\epsfbox{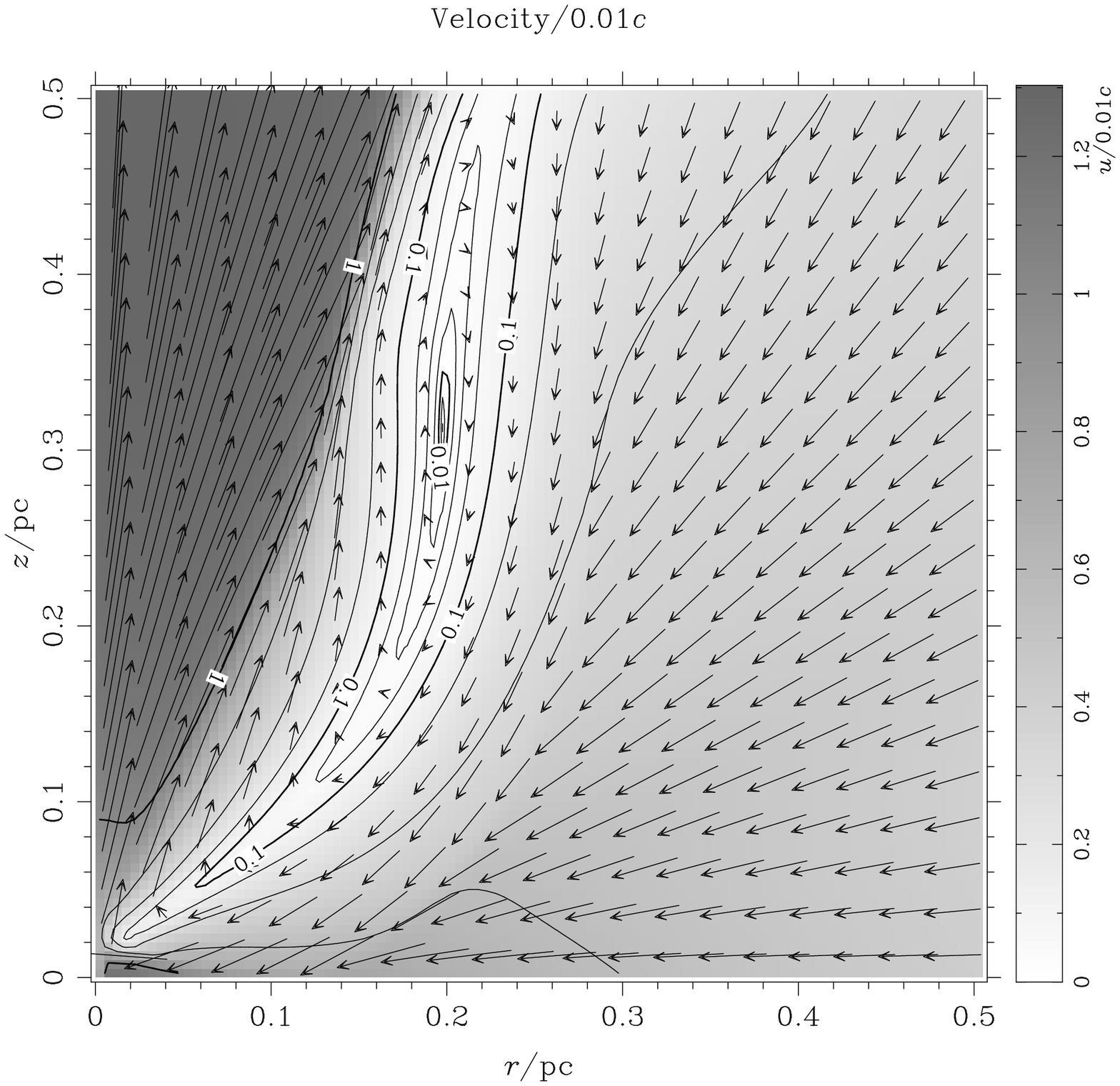} \\
\end{tabular}
\end{centering}
\caption[Model C 2D plot]{Model \modC\ at $t=2.6\ee4\yr$.  The panels show
a) density (from 323 to $1.5\ee{10}\cm^{-3}$) b) total velocity
(greyscale and contours, up to $1.3$), c), d) Magnifications of the
central region.  The vectors are in the flow direction, and have
lengths proportional to the velocity, sampled at one vector in every
$15\times15$ cells in b), every $5\times5$ cells in d).  The flow is
sonic at the $u = 0.1$ contour.}
\label{f:modelc2d}
\end{figure*}

\begin{figure*}
\epsfxsize = 8cm
\begin{centering}
\begin{tabular}{ll}
a) & b) \\
\epsfbox{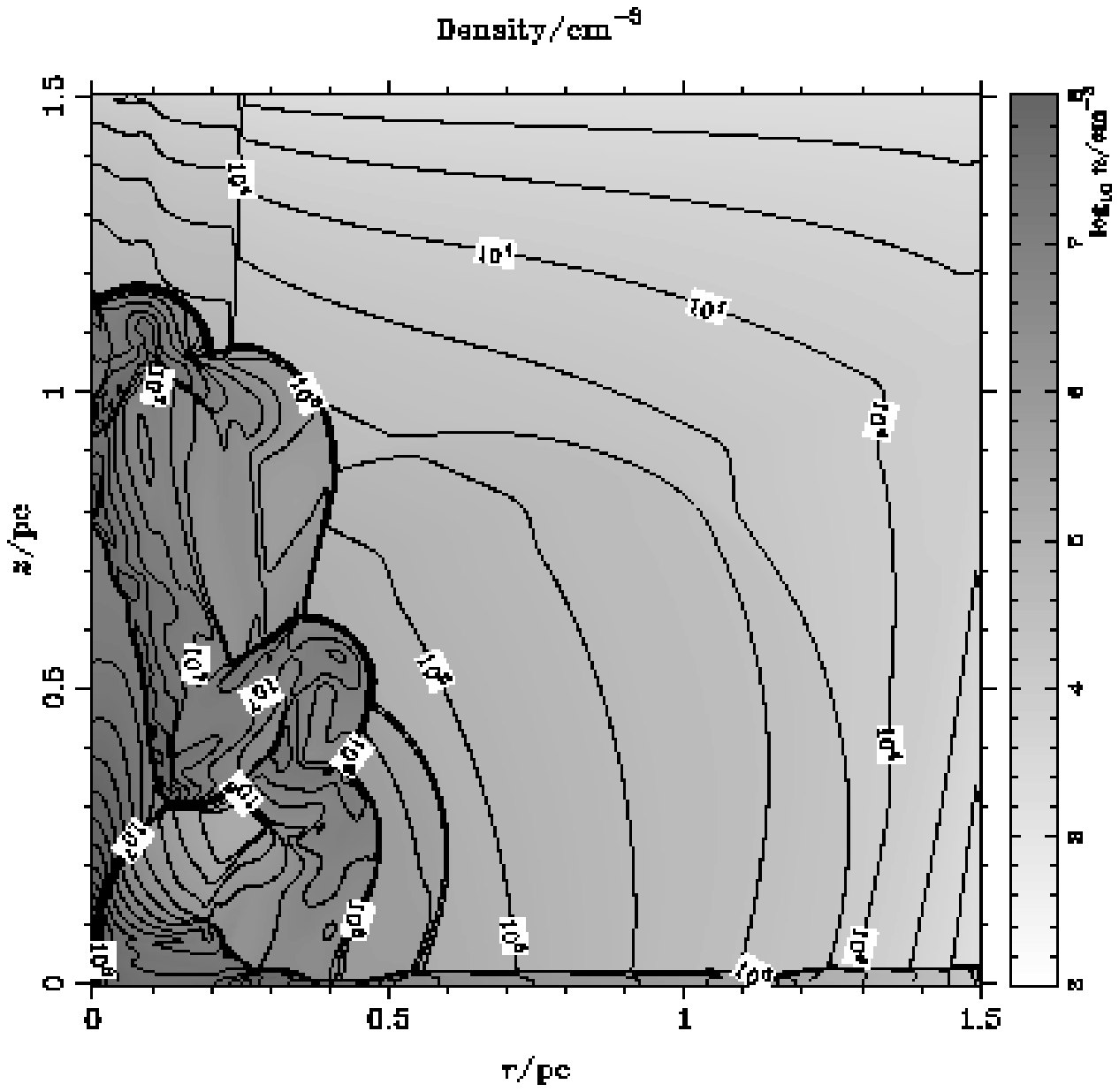} &
\epsfbox{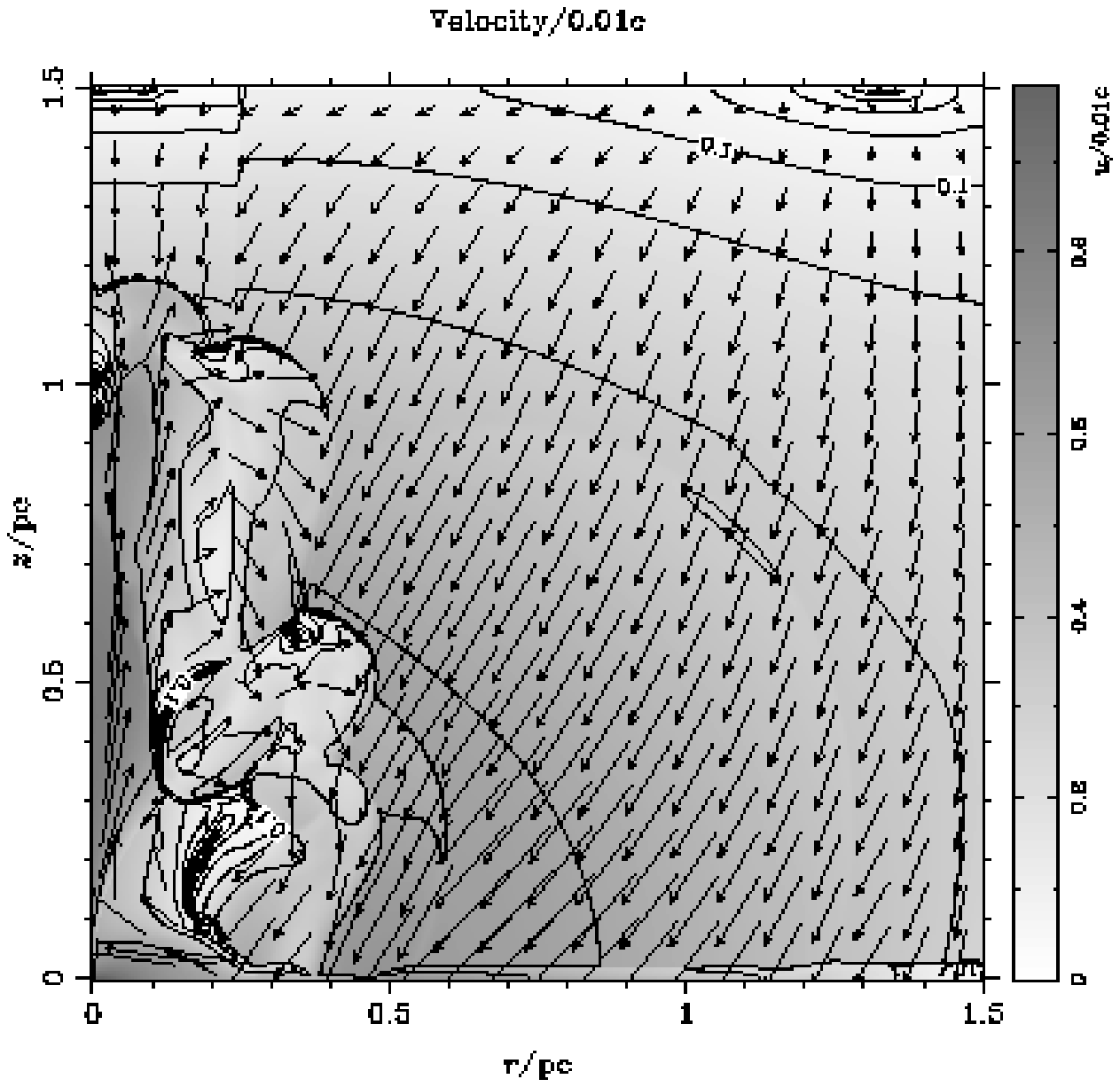} \\
c) & d) \\
\epsfbox{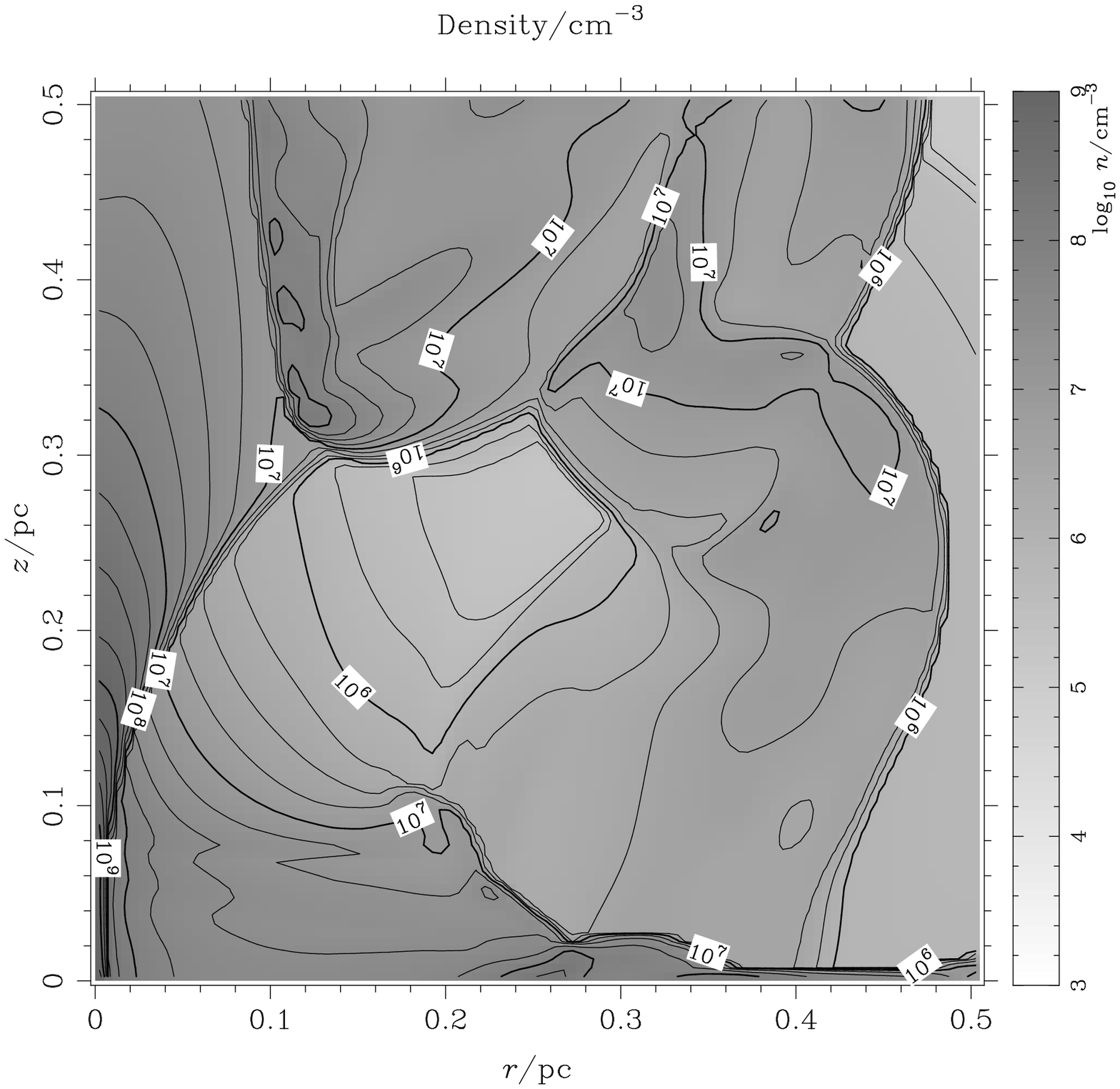} &
\epsfbox{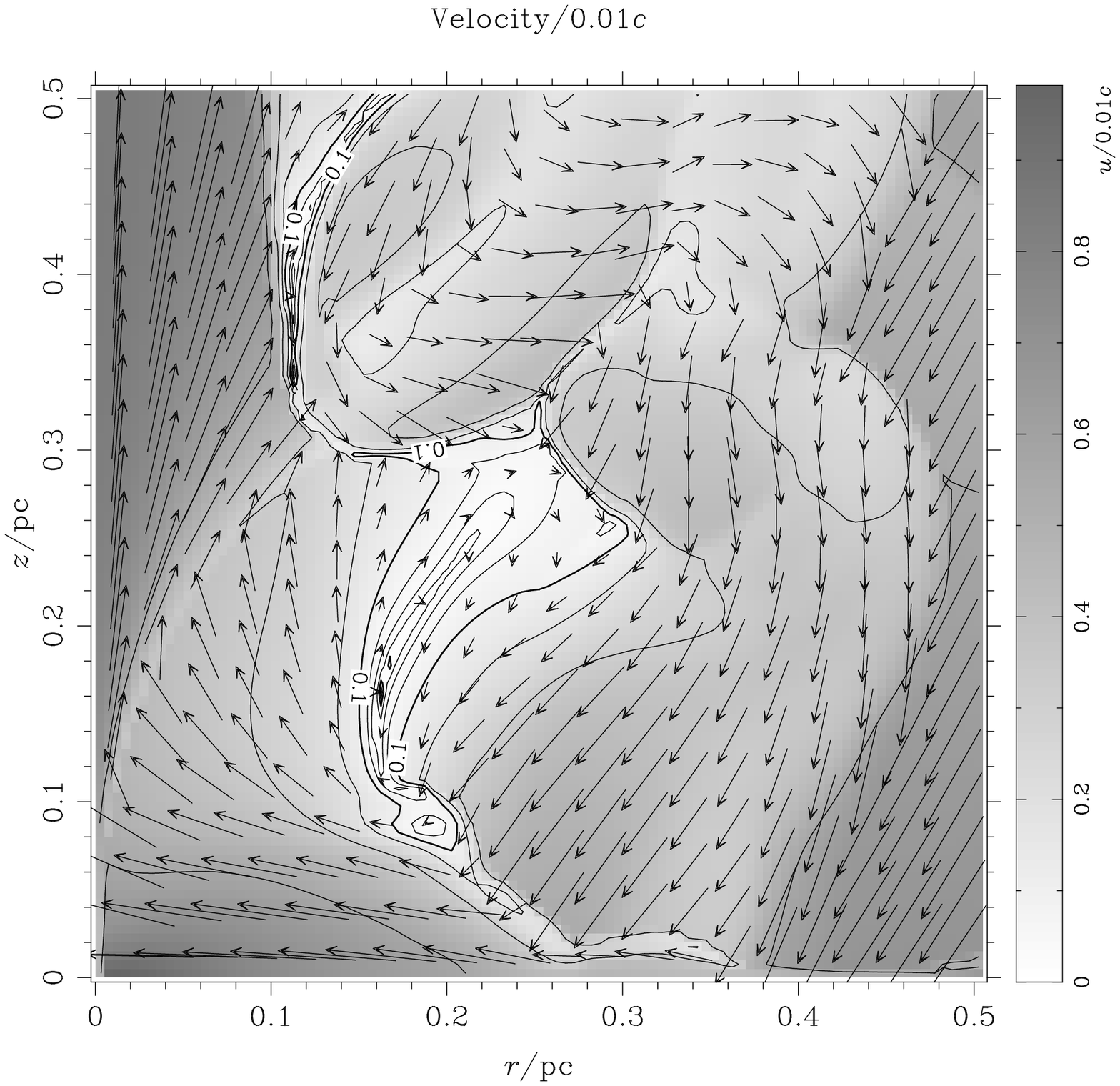} \\
\end{tabular}
\end{centering}
\caption[Model A 2D]{Model \modA\ at $t=6\ee4\yr$.  The panels are as in
Figure~\protect\ref{f:modelc2d}, except that the density range is from
$1800$ to $2\ee{10}\cm^{-3}$ and velocities reach $0.01c$.}
\label{f:modela2d}
\end{figure*}

\begin{figure*}
\epsfxsize = 8cm
\begin{centering}
\begin{tabular}{ll}
a) & b) \\
\epsfbox{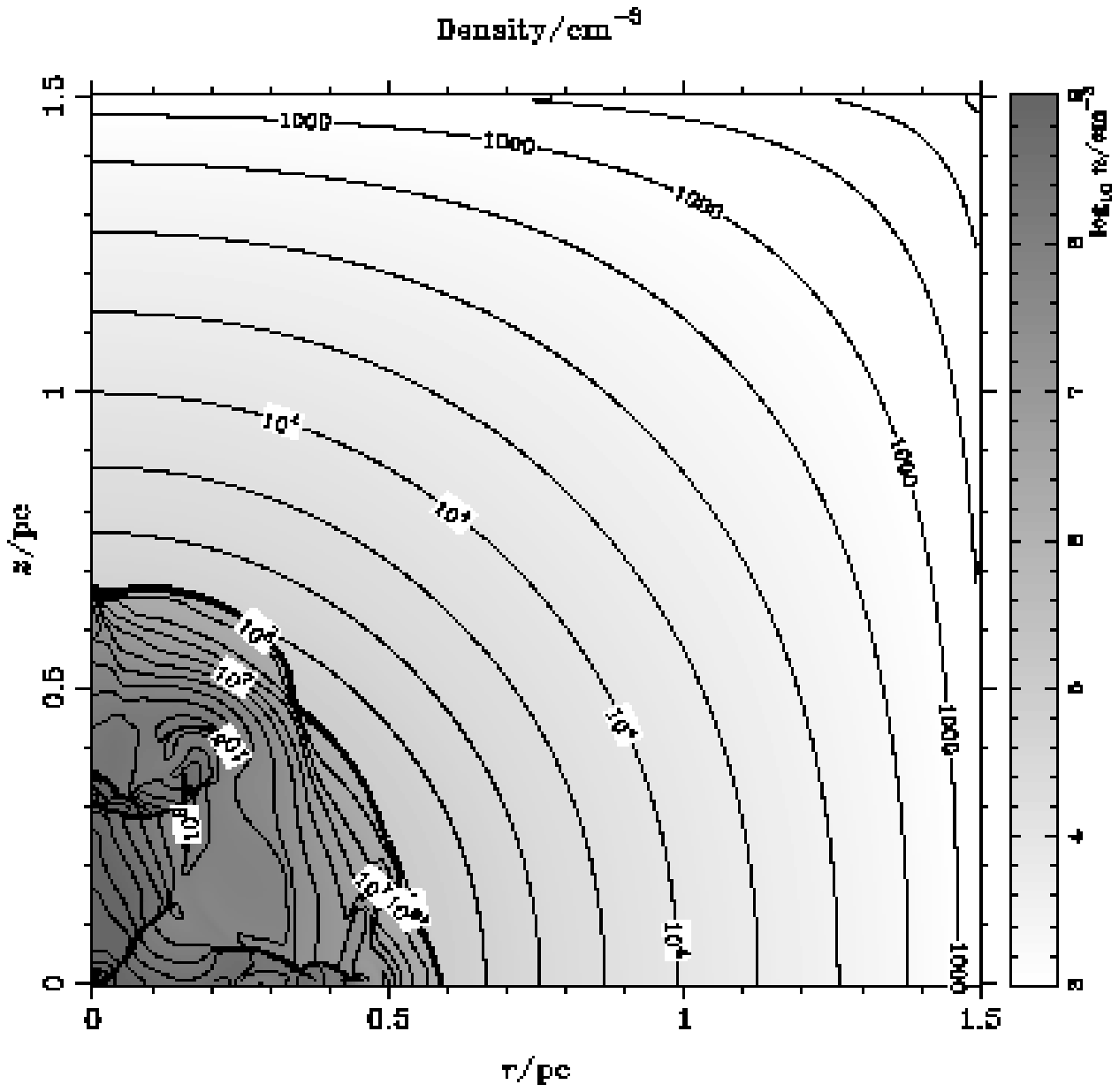} &
\epsfbox{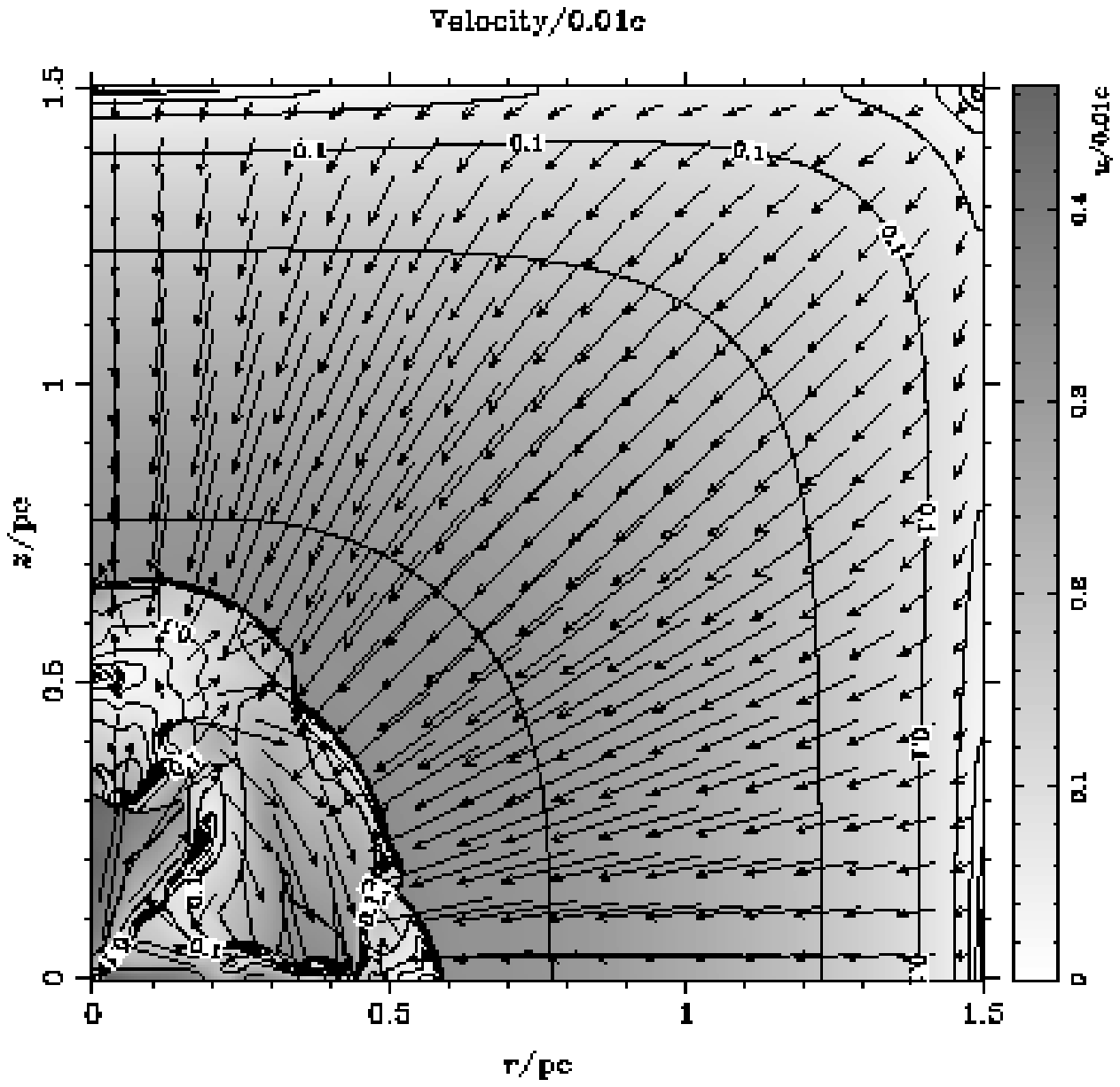} \\
c) & d) \\
\epsfbox{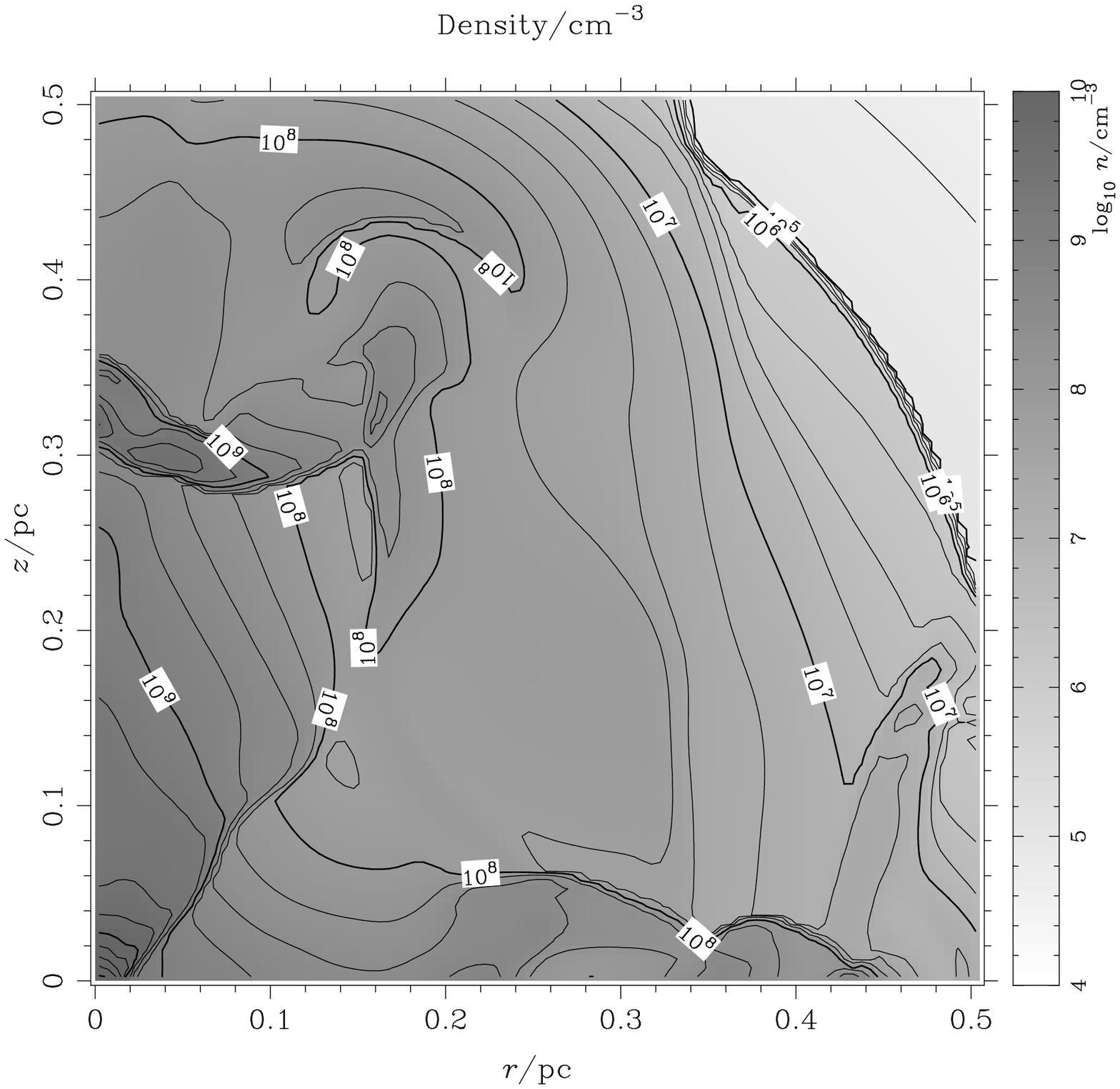} &
\epsfbox{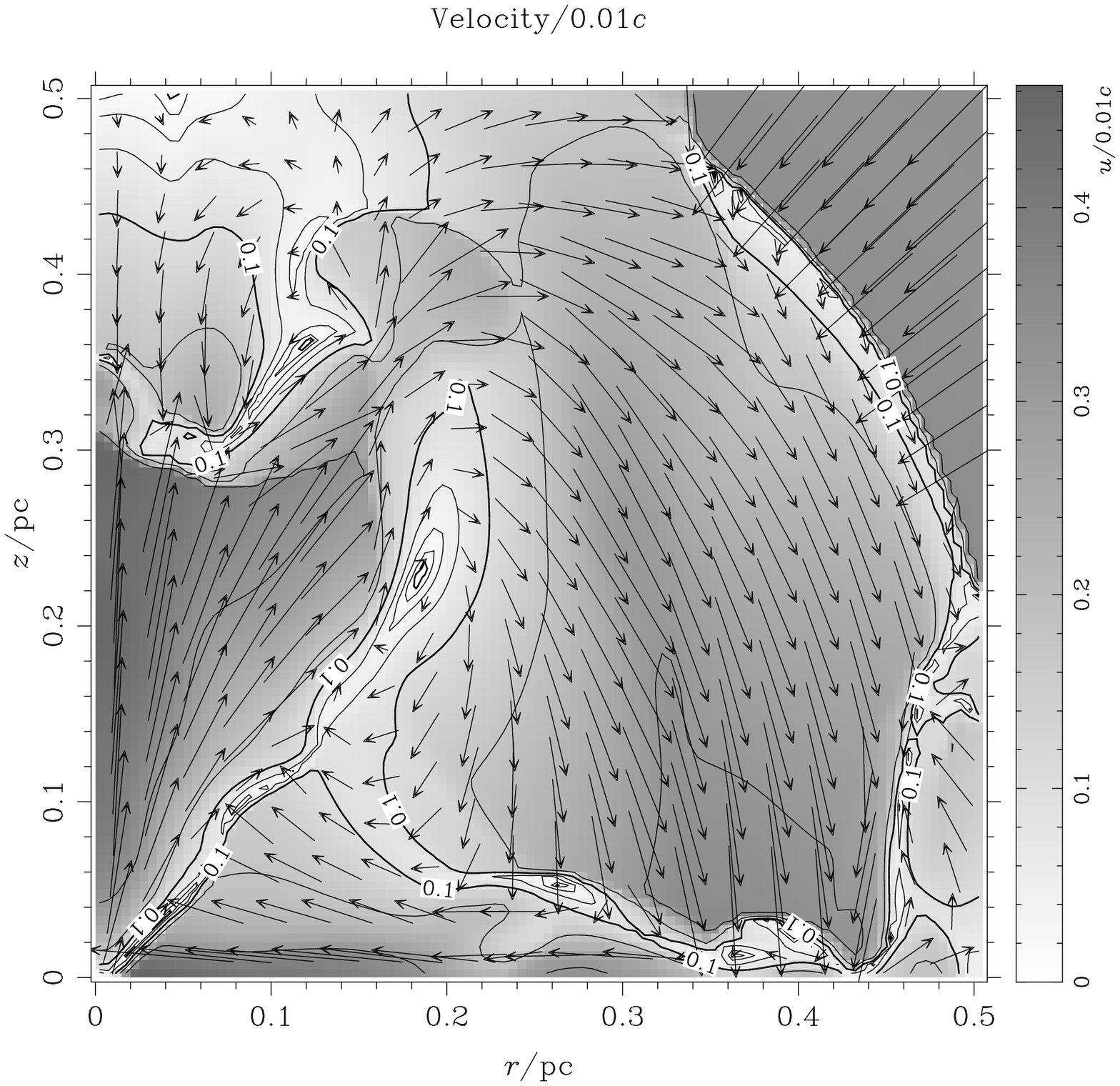} \\
\end{tabular}
\end{centering}
\caption[Model B 2D plot]{Model \modB\ at $t=10^5\yr$.  The panels are as
in Fig.~\protect\ref{f:modelc2d}, except that the density range is
from $2330$ to $1\ee{11}\cm^{-3}$ and velocities reach $4.6\ee{-3}c$.}
\label{f:modelb2d}
\end{figure*}

In the first three models (\modC, \modA, \modB;
Figs.~\ref{f:modelc2d}, \ref{f:modela2d} and \ref{f:modelb2d},
respectively), we vary the mass of the central black hole from 20 per
cent down to 3 per cent of the mass of the cluster, keeping the cluster
properties and Eddington ratios constant (see Table~\ref{t:models}).
Later, we present the structure of flows in systems where the black
hole is relatively more massive; however, we focus first on this range
of parameters since this is where mass budget requirements are best
met.  Also, the flows are distinct from the spherical accretion in a
massive cluster illustrated in the previous section and from flows
around a black hole in the absence of a cluster.  As the mass of the
hole decreases, its region of influence on the global flow contracts,
and the region in which the forces due to the stellar cluster dominate
consequently expands.

\paragraph*{Model \modC}
(Fig.~\ref{f:modelc2d}) has the highest mass black hole of the first
three models, $2\ee8\Msun$.  The flow relaxes, on a sound crossing
timescale, to a steady bipolar outflow (\cf\ Fig.~\ref{f:modelaflux}a,
which illustrates the relaxation process revealed by the mass loss rate
from the grid).  The inflow in equatorial regions interacts with the
outflow, and is funnelled by a shock into a dense region close to the
plane.  This shock is roughly coincident with the outermost section of
the $n=10^6\cm^{-3}$ contour (which hits the $r$ axis at $0.2\parsec$)
in Fig.~\ref{f:modelc2d}c, and, in Fig.~\ref{f:modelc2d}d, is most
easily seen by the sudden change in the direction of velocity vectors.
A narrow, tulip shaped region of circulating, subsonic flow separates
the inflow and outflow regions.

The sonic surface separates flow above and below the sound speed.  In
general the streamlines are almost in the plane of the sonic surface,
although both inflow and outflow from the region are seen close to its
ends.  Even at the ends, the flow is reasonably well-resolved, giving
us confidence that the transition from shock to sub-sonic to
super-sonic transition through the parallel sonic flow observed here
is a real feature.  Note, however, that this sonic transition
coincides with the outer radius of the region in which the forces are
smoothed, and will move inwards if the physical `smoothing region' is
smaller than that assumed here.

At the centre of the grid, much of the outflow is concentrated in the
grid cells adjacent to the axis (the individual cells are
$5\ee{-3}\parsec$ square).  The flow is confined by the inward radial
velocities found in these cells (not visible in the selected cells
shown in Fig.~\ref{f:modelc2d}b or d), a well-known difficulty for
axisymmetric hydrodynamical codes.  However, since the flow here is
more-or-less ballistic, and the gravitational and radiation forces
acting on the gas are chosen to vary smoothly with angle, this should
not have a strong influence on the results.  This is confirmed by the
smoothness of the velocity field shown by the contours in
Fig.~\ref{f:modelc2d}d.

\paragraph*{Model \modA} (Fig.~\ref{f:modela2d}) has a central black hole of
$10^8\Msun$, half the mass assumed in Model \modC.  The gravity of the
cluster and the friction of the mass loading are now sufficient to
stagnate the flow within the cluster.  Unsteady structures develop,
and a sequence of explosions drive out from the region.  Strong shocks
are distributed throughout the region where the inflow from the outer
cluster interacts with the central core; in the plane of the disc
(over the surface of which a thin plate of very dense, infalling gas
forms at radii between 0.5 and $1.5\parsec$), and (as bowshocks)
around clumps which form in the flow [on axis at $z=1\parsec$, and at
$(r,z) = (0.12,0.32)$].  Where shocks cross each other, regions of
particularly enhanced density (and pressure) are formed, although this
pressurization will be transient and may not be important for
generating cool gas.

In some cases, these clumps develop an independent identity, and fall
from the strong outflows close to the axis, where they are confined by
ram-pressure, towards the plane of the accretion disc.  However, as
they leave the central outflow region the ram pressure of the flow
decreases rapidly, so the clumps explode and drive shocks into the
accreting flow.

We discuss the mass budget of this model further in
Section~\ref{ss:massbudge}.

\paragraph*{Model \modB}
(Fig.~\ref{f:modelb2d}) has a yet lower mass hole, $3\ee7\Msun$.  The
accreted gas collects in a mildly unsteady, convective structure at
the centre of the grid.  Essentially no mass is lost from the central
parsec-scale region over the $10^5\yr$ for which the simulation was
run, although the density and size of the central structure gradually
increases with time.  At the end of the simulation, the mass contained
in the grid was $9\ee5\Msun$, and the mean mass loss rate was roughly
$10^{-3}\Msun\yr^{-1}$.  This total is entirely dominated by gas with
less than escape velocity flowing out from the edges of the accreting
flow.  Later, once outflows begin to be driven by the central
convective structure this mass loss rate will begin to increase.  Even
if this were to occur as early as the time illustrated in
Fig.~\ref{f:modelb2d}, and if thereafter the mass loss rate were to
increase in proportion to the flow density (hence roughly linearly in
time), it would still require 5000 times the length of the current
simulation before a balanced mass budget could be reached.  However,
it seems likely that the flow will remain unsteady in this eventual
state, as in Model \modA, with episodic loss of a tiny fraction of the
ISM mass.

The structure of the flow is broadly similar to that of Model \modA\@.
Features to note are the dense plate of gas on axis at $z=0.3\parsec$,
which forms where the inflowing halo gas and outflowing wind meet.
Oscillations of this plate drive weak shocks into the halo, and
modulate the rate at which gas falls back towards the plane of the
accretion disc.  The three-limbed structure of shocks and sonic
surfaces seen here is worth noting.  It also appears in many other
cases (but compare the single `tulip' in Model \modC).  Gas injected
close to the stagnation point does not accelerate monotonically:
rather, it spirals around in increasing spirals, passing through a
sequence of shocks until finally it can escape.

\subsection{Varying the stellar cluster properties}

We now discuss a selection of models which investigate variations in
the parameters of the stellar cluster.  In Models \modE\ and \modG,
the cluster mass is reduced. In Model \modE, we choose the radius of
the cluster to be the same as Model \modA. In Model \modG, we choose
the dynamical velocity of the stars to be roughly constant.  In Models
\modE\ and \modK, we investigate the behaviour of the flow as the
Keplerian velocity decreases relative to the sound speed, where \modK\
has a rather higher velocity of ejection from the central region than
\modE\@.

Models \modE\ and \modK\ can be paired with models \modC\ and \modA,
respectively, except that they have relatively small stellar velocity
dispersions (scaled to the sound speed of the flow, \cf\
Fig.~\ref{f:xxx}).

Model \modF\ has an cluster with a Keplerian velocity rather smaller
than the sound speed in the nISM\@.  As a result, the flow inside the
cluster core is quasi-hydrostatic, rather than the supersonic
accretion flow found in most of the other simulations presented
here.  Much of the mass loss from the cluster occurs in a spherical
wind.  A strong outflow is, however, still driven from the centre of
the grid, and a recollimating jet is formed as this interacts with the
core.

\begin{figure*}
\epsfxsize = 8cm
\begin{centering}
\begin{tabular}{ll}
a) & b) \\
\epsfbox{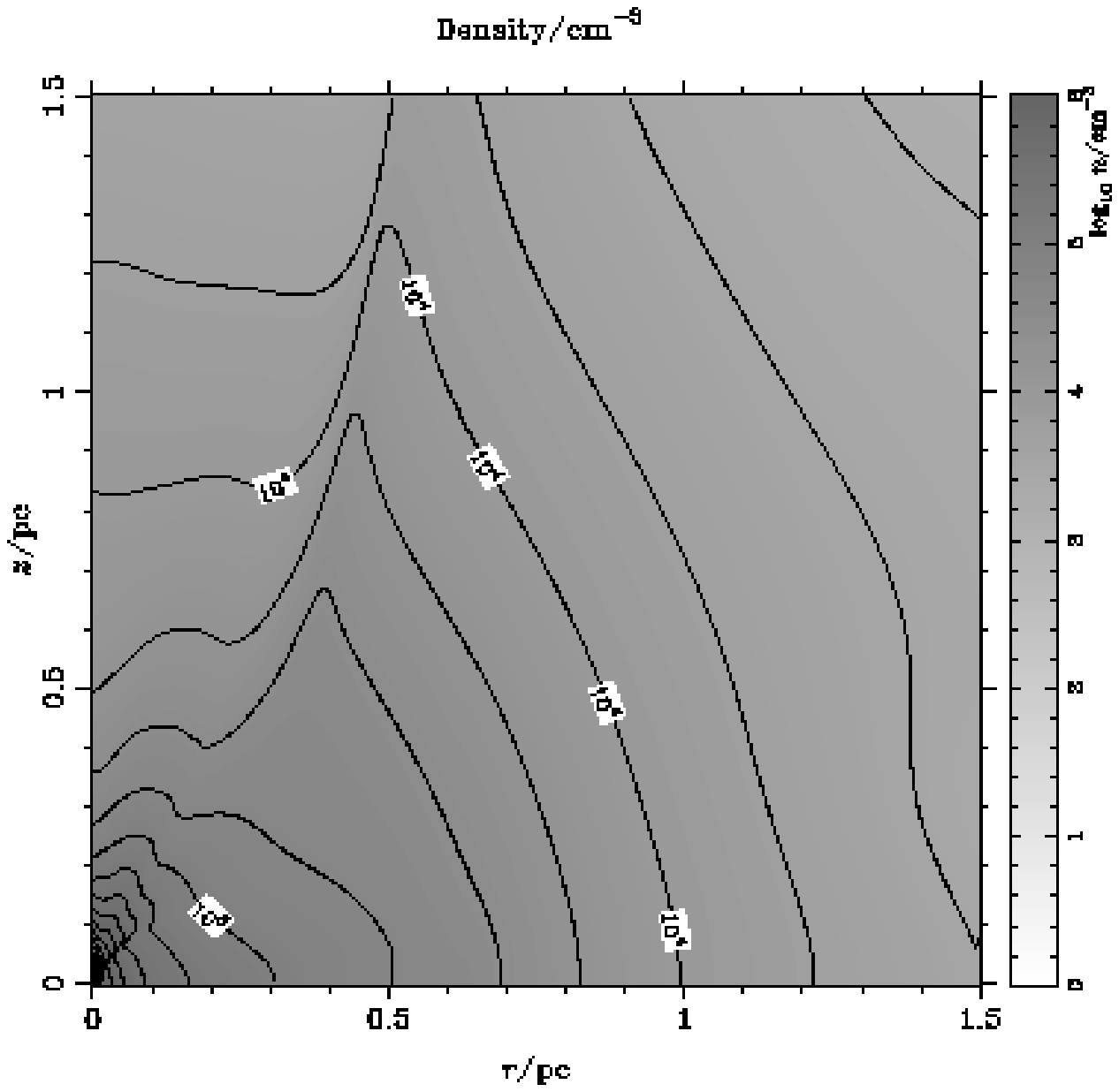} &
\epsfbox{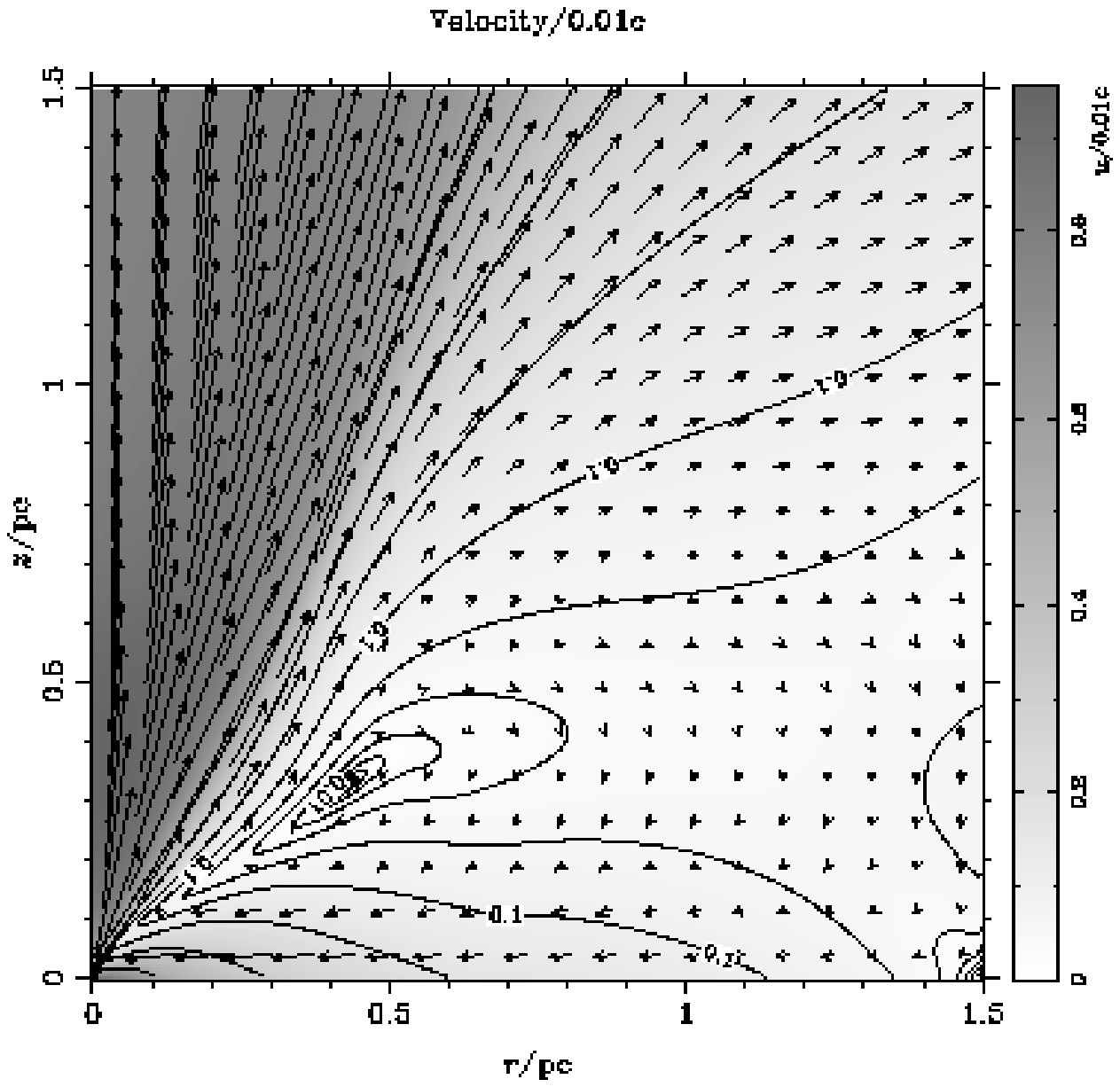} \\
c) & d) \\
\epsfbox{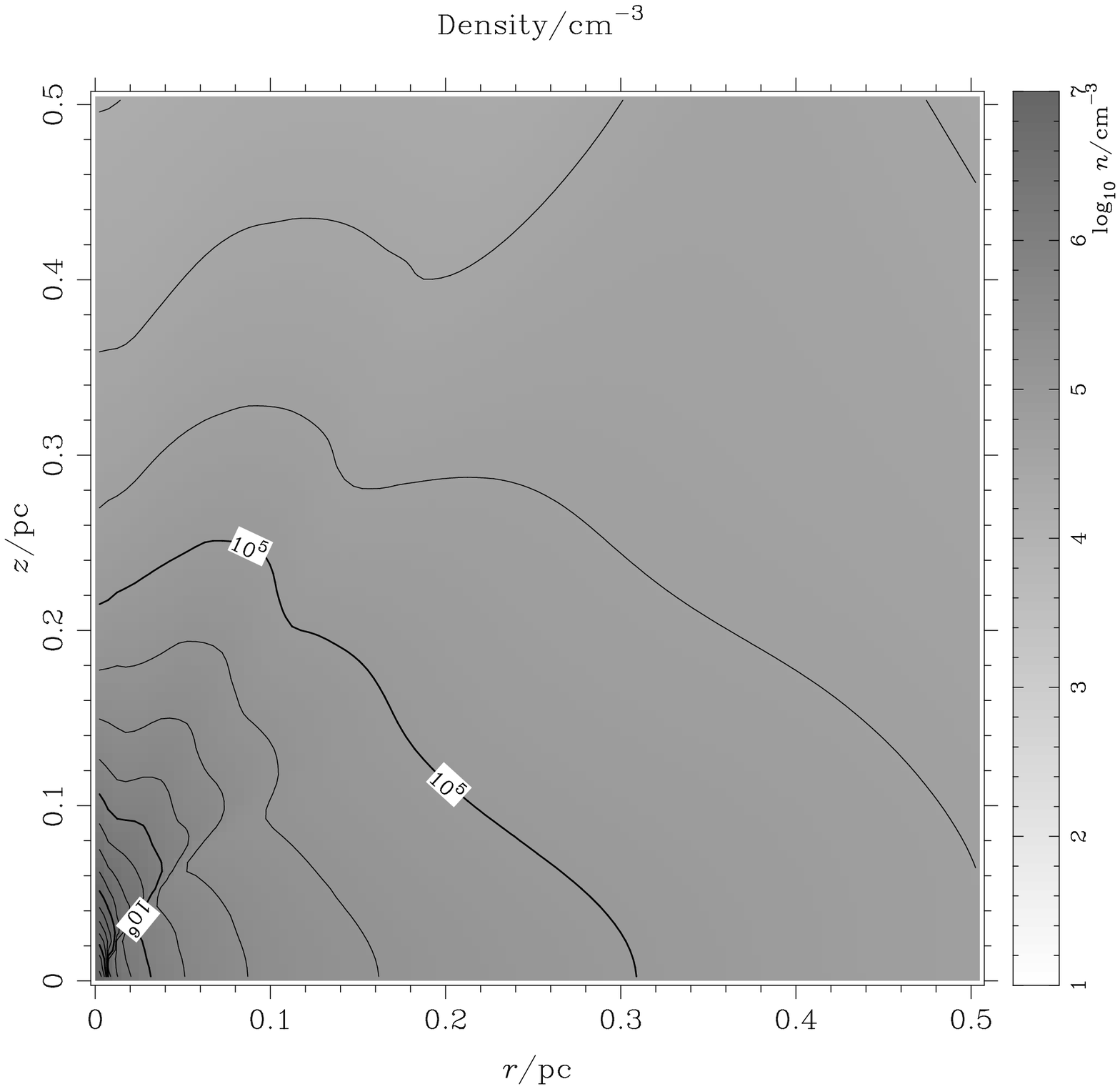} &
\epsfbox{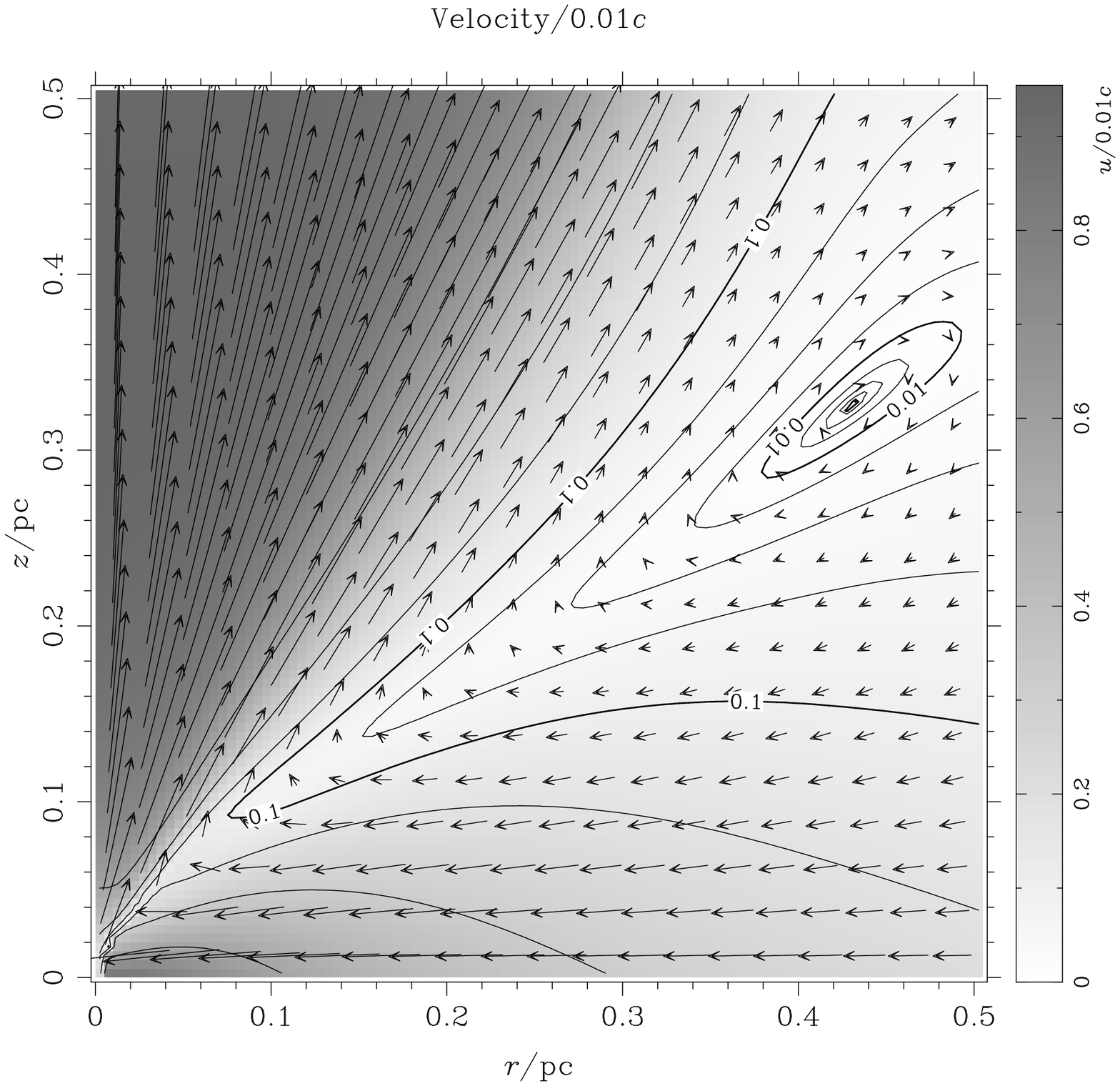} \\
\end{tabular}
\end{centering}
\caption[Model E 2D plot]{Model \modE\ at $6\ee4\yr$ (steady
equilibrium solution).  The panels show a) density (from 1513 to
$4\ee{8}\cm^{-3}$) b) total velocity (greyscale and contours, up to
$0.95$), c), d) Magnifications of the central region.  The vectors are
in the flow direction, and have lengths proportional to the velocity,
sampled at one vector in every $15\times15$ cells.  The flow is sonic
at the $u = 0.1$ contour.}
\label{f:modele2d}
\end{figure*}
\paragraph*{Model \modE}  
(Fig.~\ref{f:modele2d}) has a lower cluster mass than Model \modA, but
the same cluster core radius.  Deceleration by gravity and mass loading
have both decreased, and the flow can now drive freely from the centre
along the axis of the grid.

Compared to Model \modC, the nISM has been made dynamically `hotter'
-- \ie\ the ratio $v|K/c|s$ has been decreased.  The opening angle of
the flow which escapes the cluster core has been substantially
increased.  As expected, the sonic region substantially expanded; it
provides significant support against gravity for the gas falling in
the accretion disc plane, allowing the central outflow to widen.
However, it should also be noted that with $ M|h=M|{cl} $, the radiation
forces retain a significant influence out to the edge of the
cluster. This encourages the free outflow of gas along the axis.
\begin{figure*}
\epsfxsize = 8cm
\begin{centering}
\begin{tabular}{ll}
a) & b) \\
\epsfbox{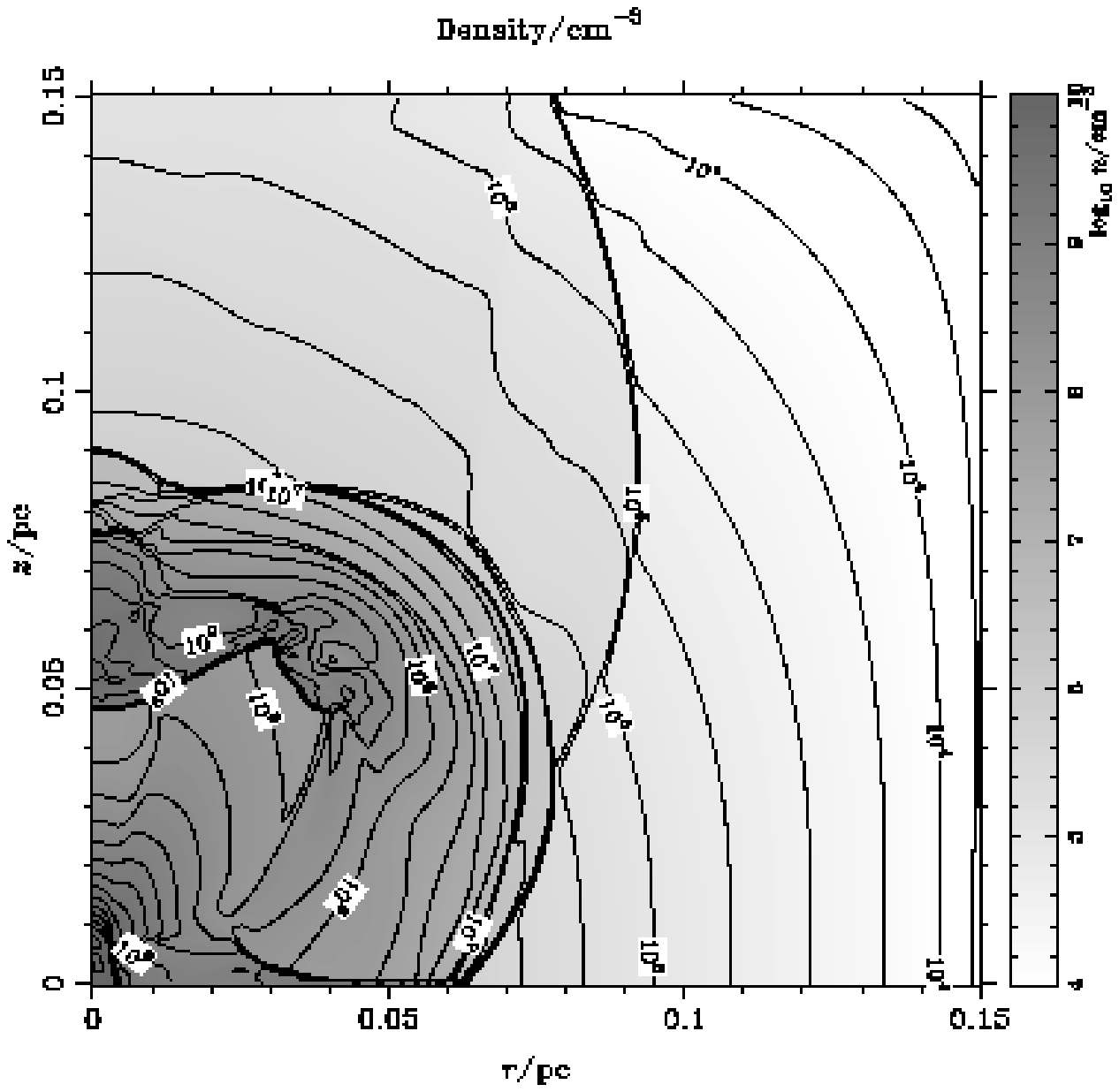} &
\epsfbox{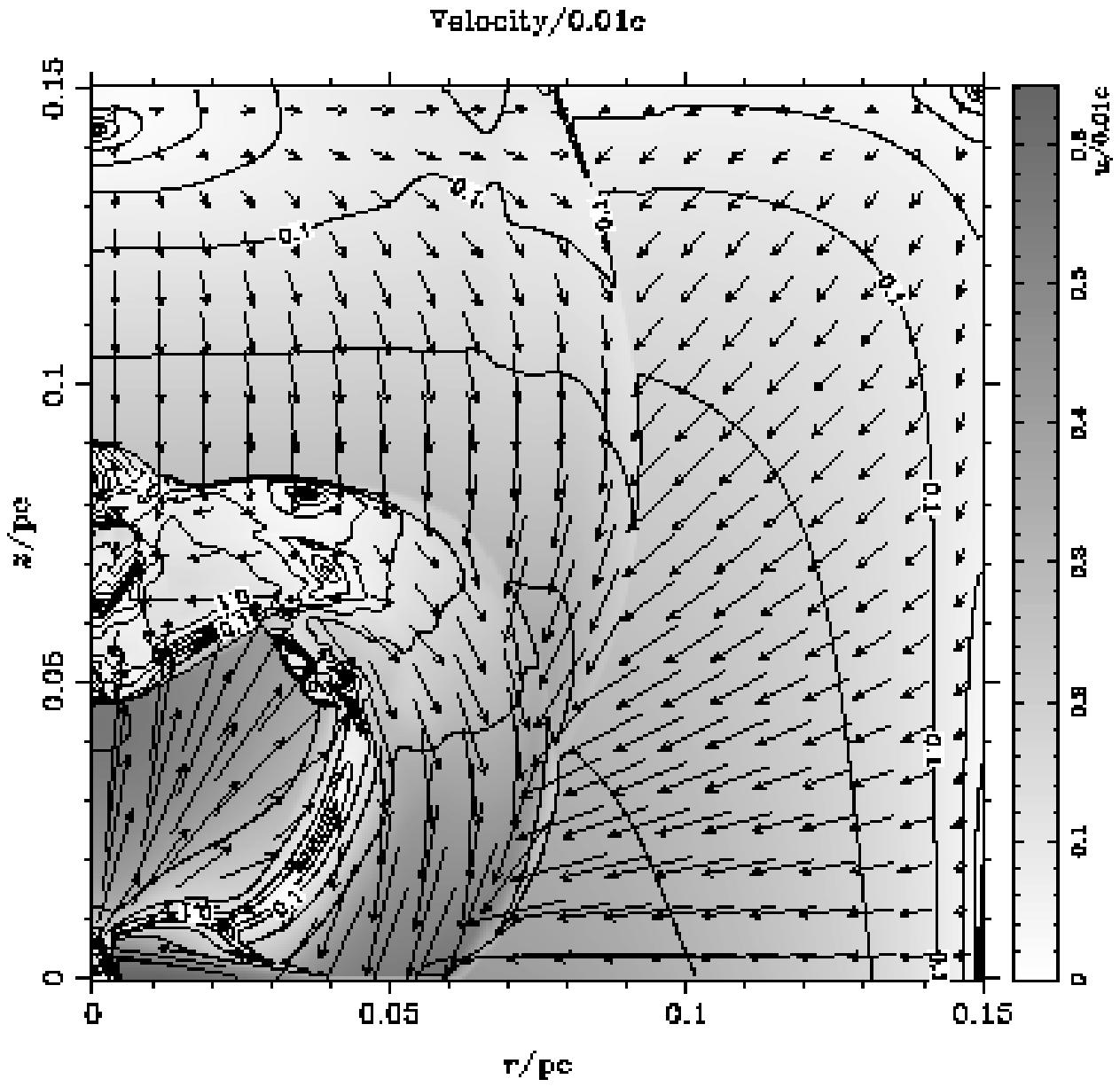} \\
c) & d) \\
\epsfbox{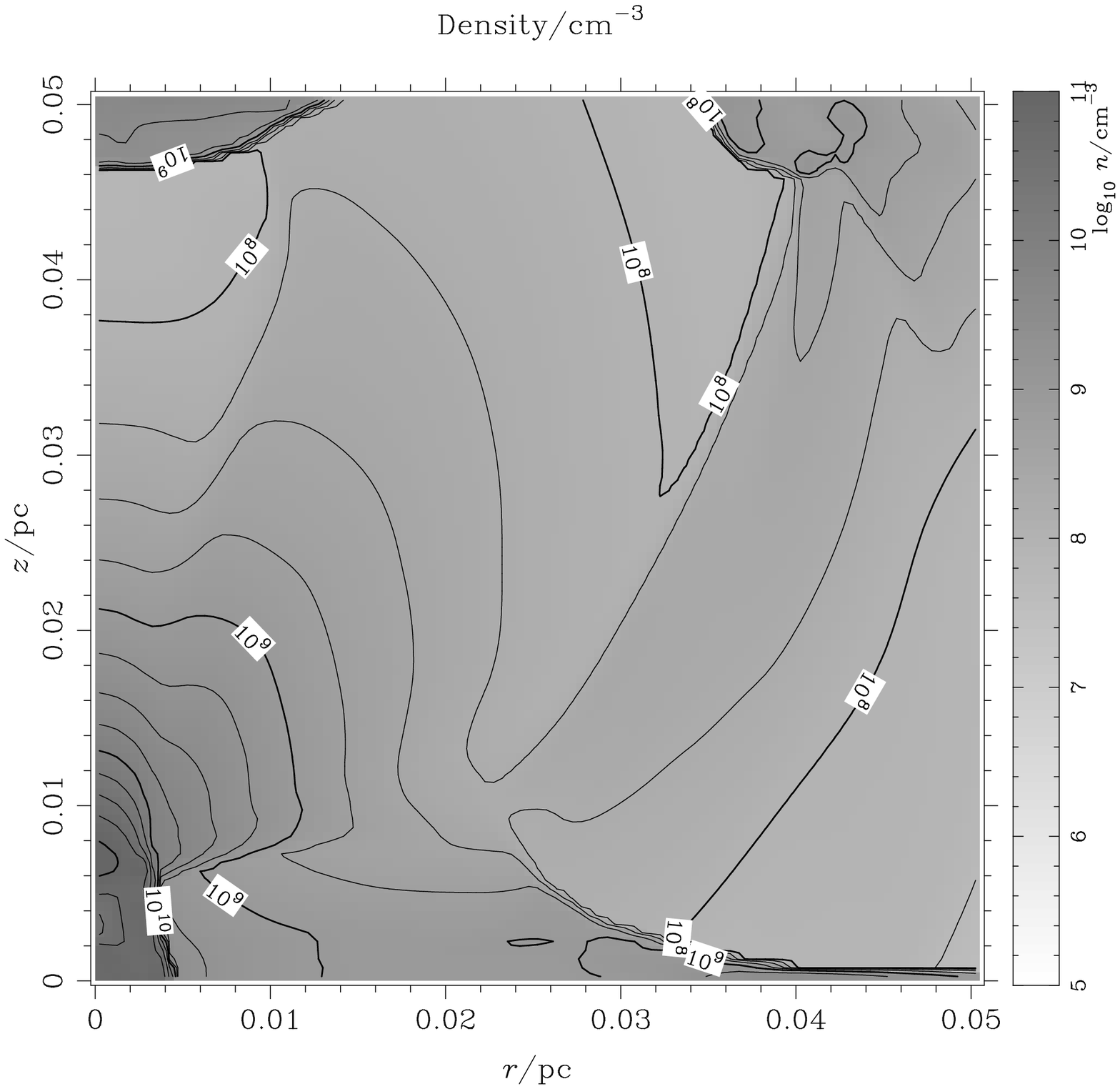} &
\epsfbox{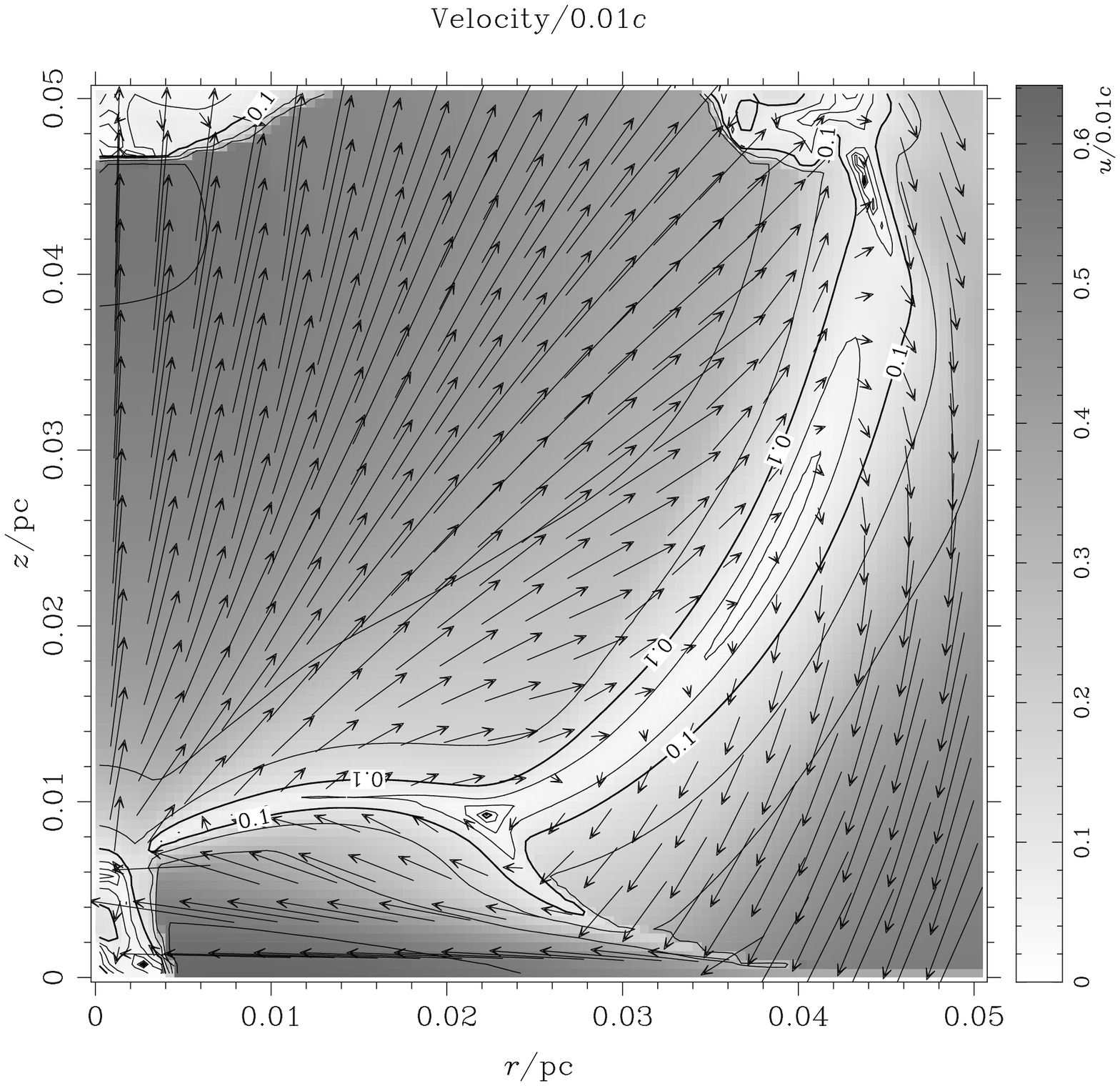} \\
\end{tabular}
\end{centering}
\caption[Model G 2D plot]{Model \modG\ at $ 9\ee3\yr $ (note that this
simulation is on a small scale, so the dynamical timescale is short).
The smoothing scale here is $ 0.05\parsec $, the same as in models C, A
and B, and takes up a large fraction of the grid.  The panels show a)
density (from 2100 to $ 10^{11}\cm^{-3} $) b) total velocity (greyscale
and contours, up to $ 0.64 $), c), d) Magnifications of the central
region.  The vectors are in the flow direction, and have lengths
proportional to the velocity, sampled at one vector in every
$ 15\times15 $ cells.  The flow is sonic at the $ u = 0.1 $ contour.}
\label{f:modelg2d}
\end{figure*}
\paragraph*{Model \modG}
(Fig.~\ref{f:modelg2d}) also has a lower cluster mass than \modC, and
has a far smaller cluster core. Weak shocks are sent into the
circumnuclear ISM.  At the end of the simulation, the mass contained
in the grid was $ 7.6\ee3\Msun $, and the mean mass loss rate was
roughly $ 2\ee{-3}\Msun\yr^{-1} $.  This mass loss was dominated by
gas with sufficient energy to escape the nucleus.  In this regime, the
mass loss rate roughly scales with the mass within the grid: a
balanced mass budget would be expected after $ 4\ee6\yr $, 500 times
later than the picture presented in the figure.

Here again we see a three-limbed sonic surface spreading out through
the cluster (note the $ v/0.01c=0.1 $ contour in Fig.~\ref{f:modelg2d}d,
which is in this case well within the extended smoothing region).  At
the very centre of the grid, the disc inflow shocks into a compact,
dense and largely subsonic structure.  The end of one of the three
limbs seems to act as a de Laval nozzle, at $ (0.004,0.008) $, where the
flow becomes supersonic.  The flow in and around this nozzle is very
variable.  This results, for the most part, from the variation of the
inflow on the disc plane: flow instabilities are expected to be
suppressed in squat nozzles such as this one \cite{ssnw83}. The small
mass of the cluster in this simulation means that the flow continues
to be accelerated beyond the nozzle, until it reaches the termination
shock which oscillates about $ 0.06\parsec $.

Note that for the parameters assumed in this case, the stellar cluster
will be highly collisional.  As we have discussed, the solution can
simply be scaled to different parameters to avoid this constraint.
However, we have chosen here to keep the parameters of the BH and
accretion disc the same as in Model~\modA, and as a result sacrificed
this freedom.

\begin{figure*}
\epsfxsize = 8cm
\begin{centering}
\begin{tabular}{ll}
a) & b) \\
\epsfbox{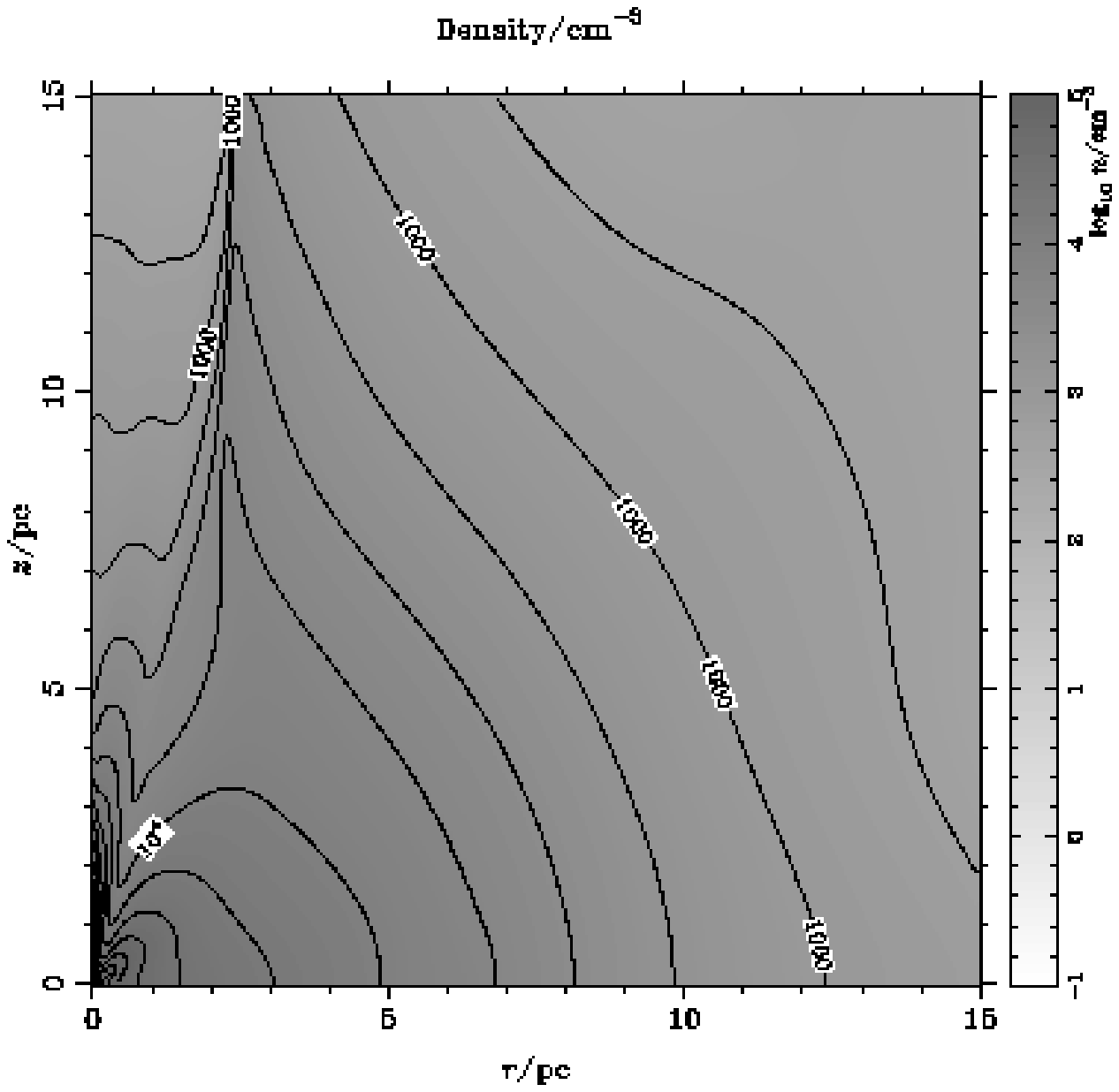} &
\epsfbox{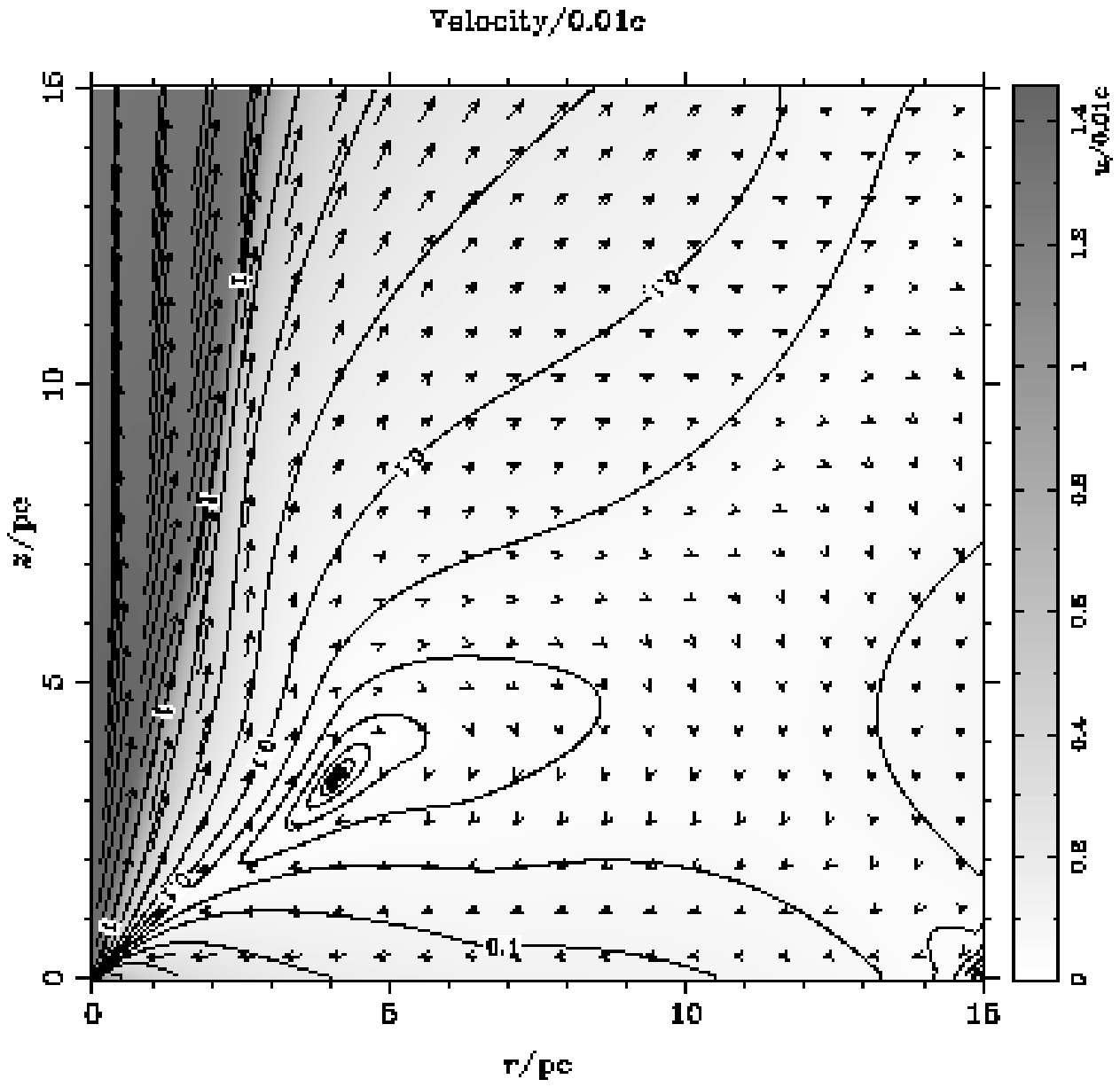} \\
c) & d) \\
\epsfbox{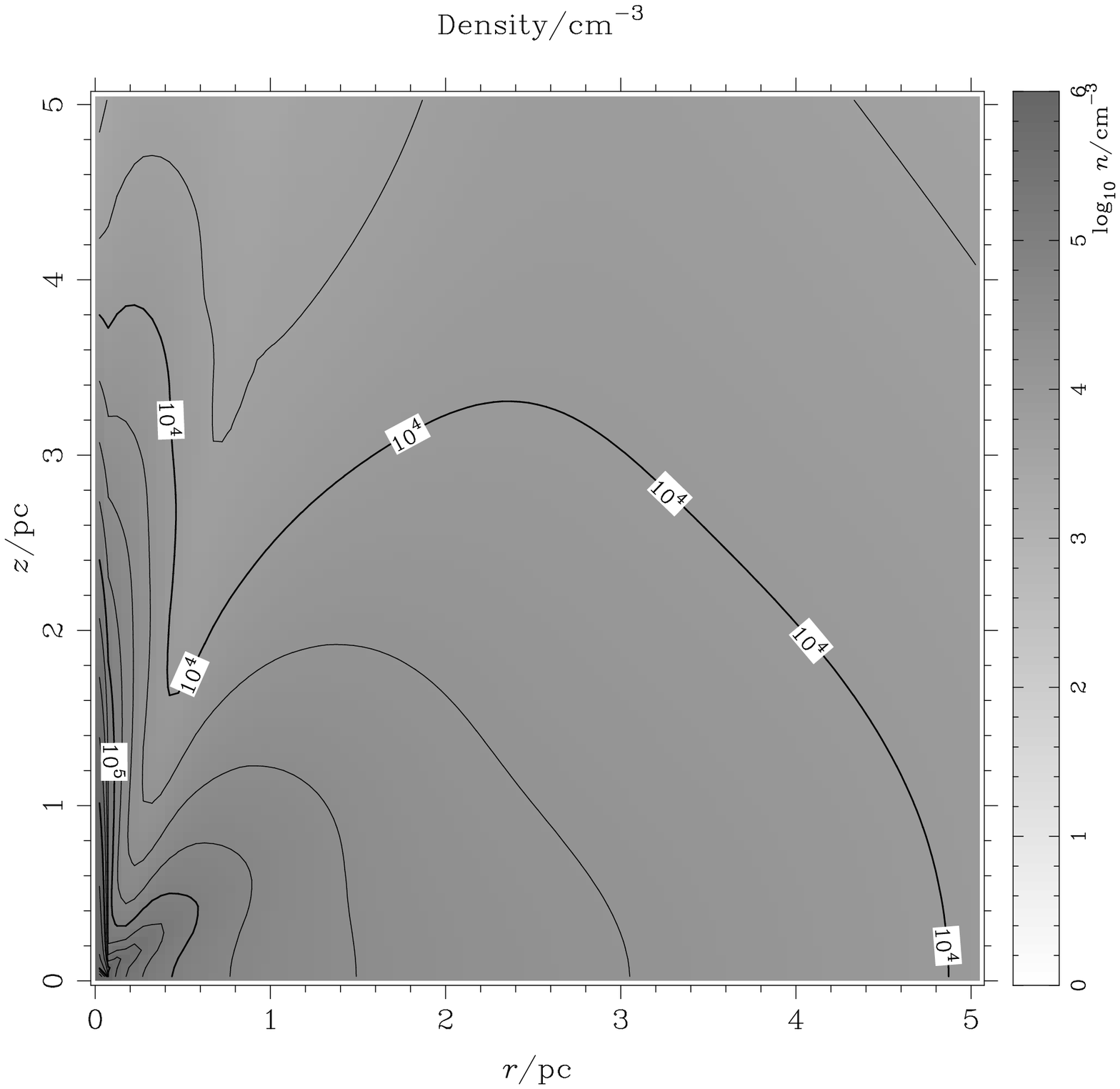} &
\epsfbox{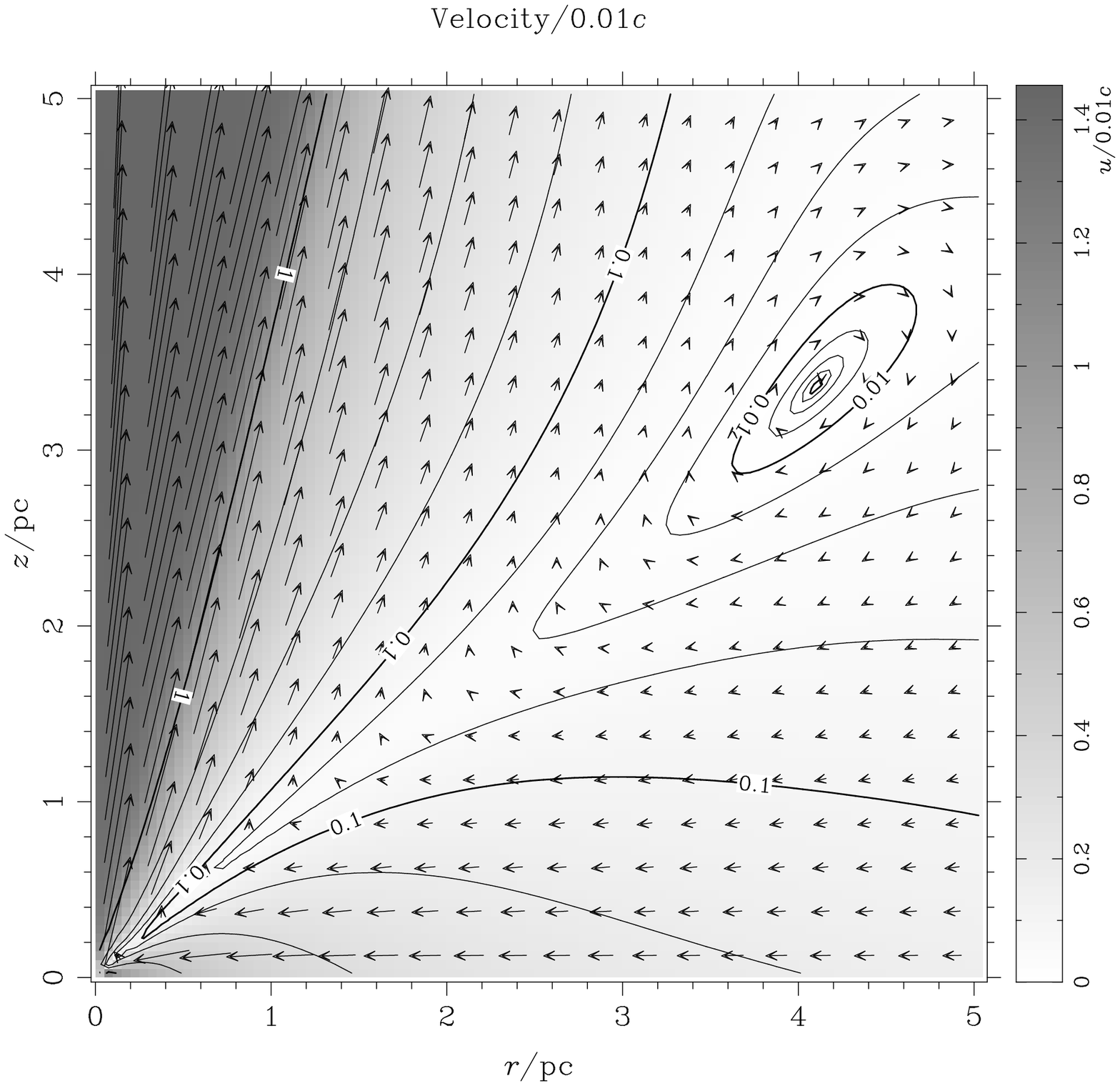} \\
\end{tabular}
\end{centering}
\caption[Model K 2D plot]{Model \modK\ at $4.2\ee5\yr$.  The panels
show a) density (from 350 to $7.5\ee7\cm^{-3}$) b) total velocity
(greyscale and contours, up to $1.46$), c), d) Magnifications of the
central region.  The vectors are in the flow direction, and have
lengths proportional to the velocity, sampled at one vector in every
$15\times15$ cells.  The flow is sonic at the $u = 0.1$ contour.}
\label{f:modelk2d}
\end{figure*}
\paragraph*{Model \modK} (Fig.~\ref{f:modelk2d}) has a cluster with a
far larger core radius than Model \modC, but with the same cluster
mass.  It extends our coverage of parameter space to higher ejection
velocities than Model \modE, but with a lower cluster dispersion than
\modC.  Hence, the cluster takes up a far larger fraction of the Parker
radius of $25\parsec$.  Comparing Figs.~\ref{f:modelk2d} and
Fig.~\ref{f:modelc2d}, the subsonic region has inflated from a narrow
tulip to take up a broad region of the flow.  The decreased flow
velocities in this region have removed the weak shocks close to the
accretion disc.  Funnelling of the flow by the broadened subsonic
region has acted to offset the decrease in mass flux in the inflow
region, keeping the density here roughly constant.  The velocity and
opening angle of the outflow are fairly similar between the models, so
the densities (at a given fraction of the cluster core radius) are
about 100 times smaller in Model \modK\ than in Model \modC\ -- in
Model \modK, the axial outflow is less dense than the surrounding
cluster ISM, in contrast to most of the other models presented here.

\begin{figure*}
\epsfxsize = 8cm
\begin{centering}
\begin{tabular}{ll}
a) & b) \\
\epsfbox{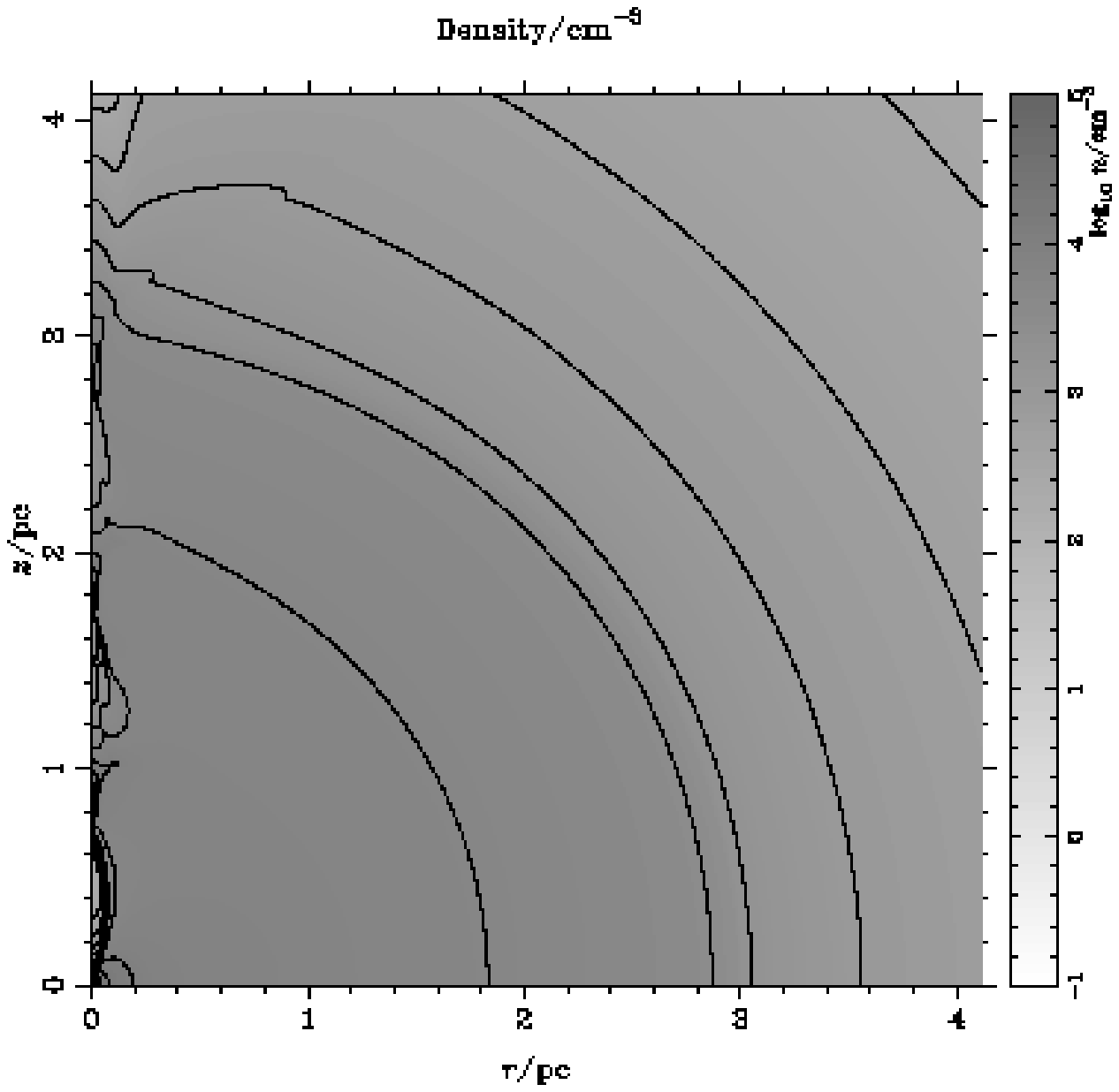} &
\epsfbox{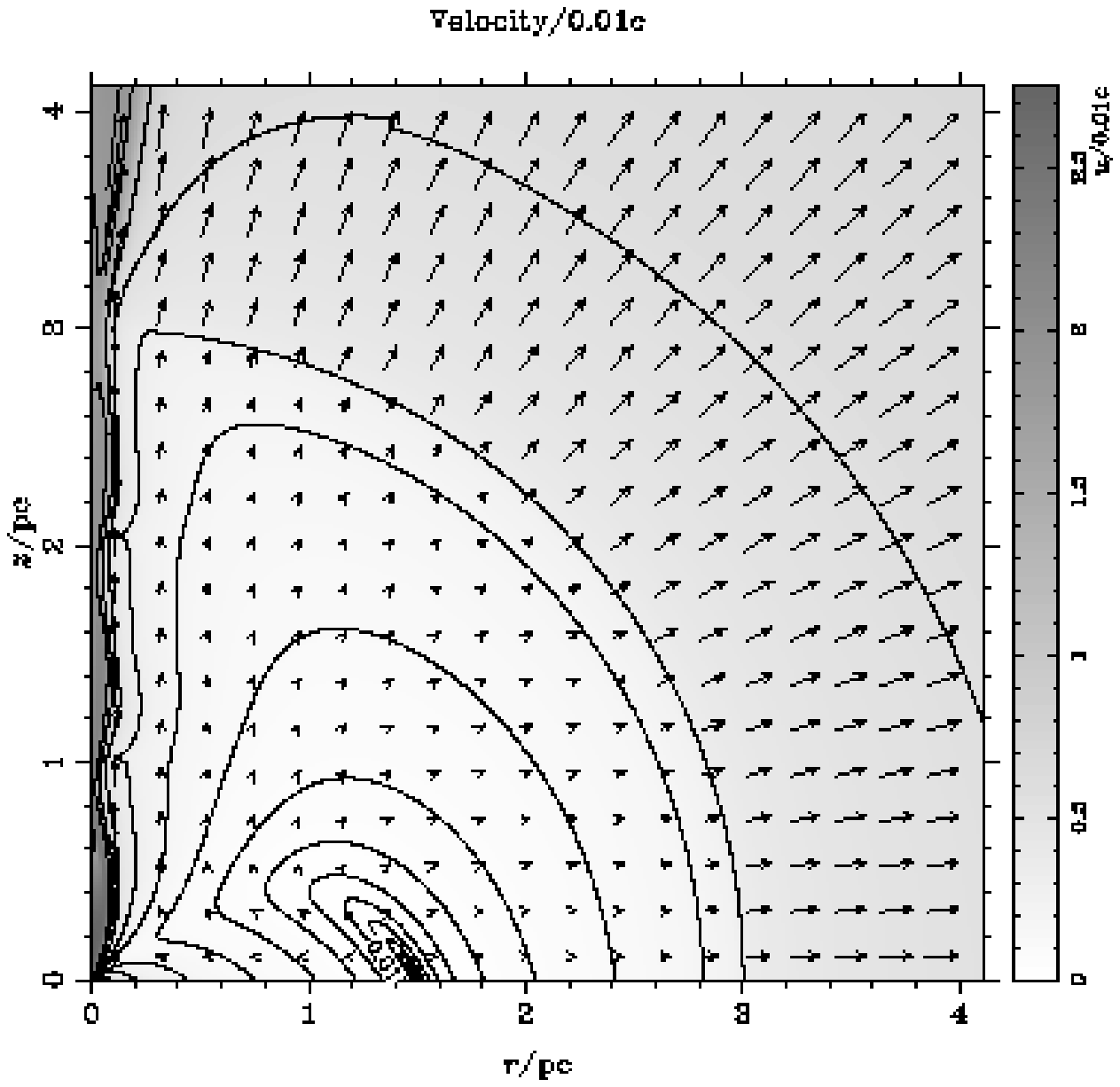} \\
c) & d) \\
\epsfbox{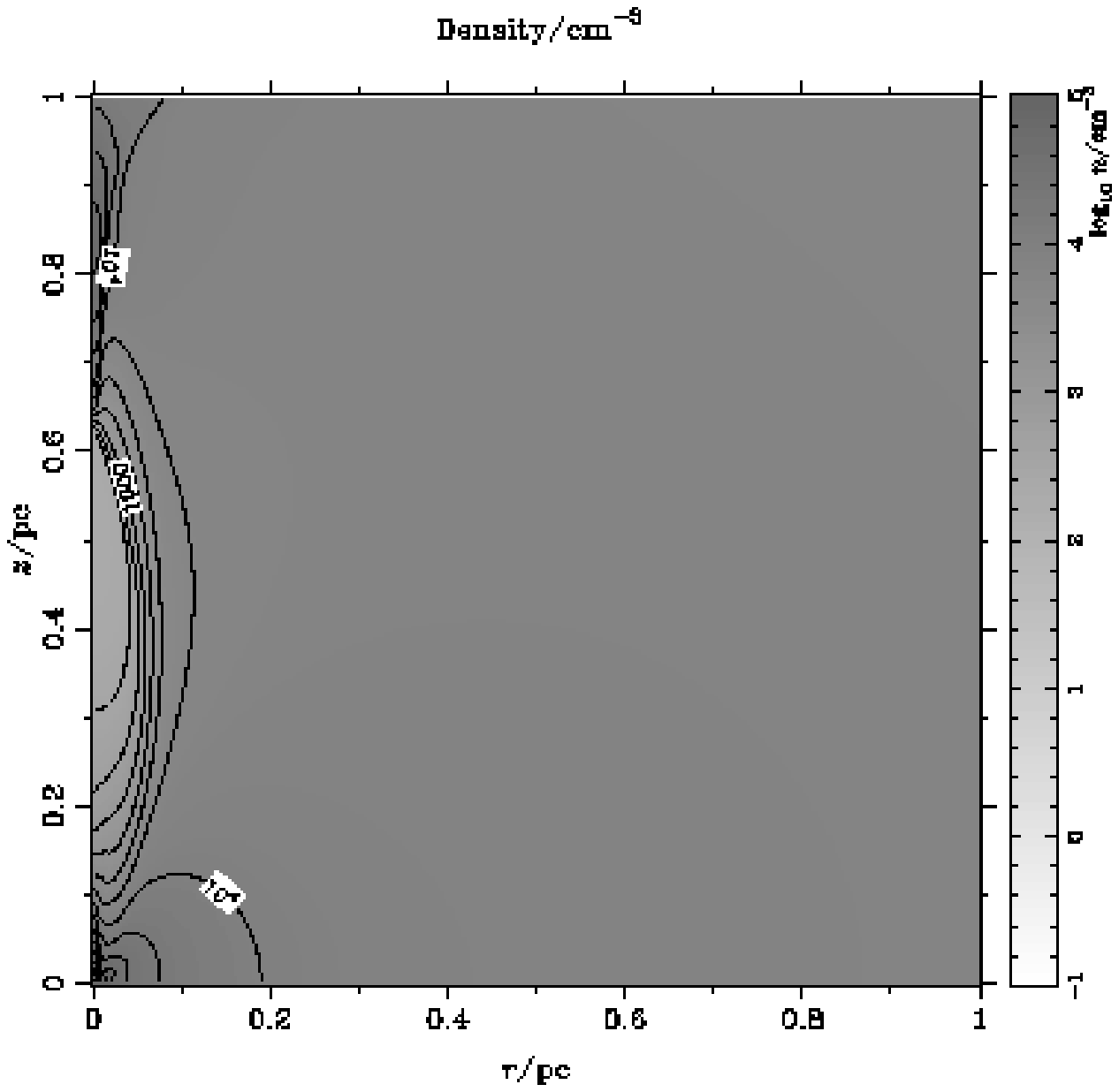} &
\epsfbox{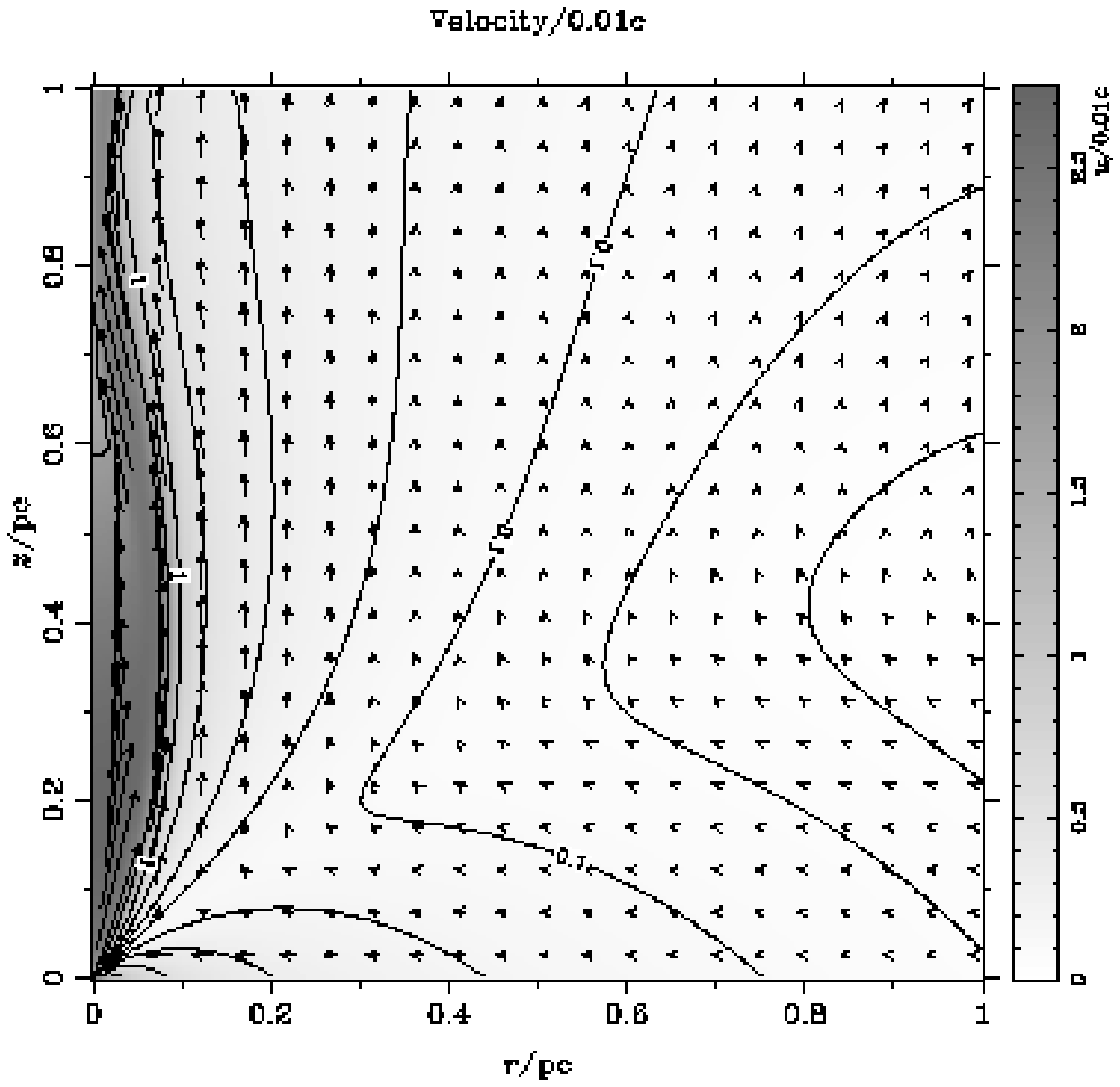} \\
\end{tabular}
\end{centering}
\caption[Model F 2D plot]{Model \modF\ at $9\ee3\yr$.  As this
simulation was very time-consuming, it was halted at this stage,
before the solution had reached equilibrium.  However, since the wind
mass loss has reached 97 per cent of the injection rate, and little
movement was seen in the solution, these plots should be fairly
representative of the eventual equilibrium.  The panels show a)
density (from 211 to $10^7\cm^{-3}$) b) total velocity (greyscale and
contours, up to $2.7$, note that for the gas temperature of
$10^8\Kelv$ used here, the sound speed is $0.32$ so this corresponds
to a peak Mach number of 8.7), c), d) Magnifications of the central
region.  The vectors are in the flow direction, and have lengths
proportional to the velocity, sampled at one vector in every
$32\times32$ cells in b), every $7\times7$ in d).  A jet close to the
axis passes through recollimation shocks which reach the axis at 1, 2
and $3\parsec$ before it escapes the cluster core.}
\label{f:modelf2d}
\end{figure*}

\paragraph*{Model \modF}  

(Fig\@.~\ref{f:modelf2d}) has an extremely low cluster core density
and a very hot ISM\@.  Looking at Fig.~\ref{f:xxx}, we see that Model
\modF\ is an extreme case in our modelling.  The Parker radius is
$1\parsec$, within the cluster core.  If there were no black hole, the
flow would relax on a dynamical timescale to a subsonic core with a
sonic transition on its surface, hardly altered by the effects of
gravity \cite[\cf{}]{cc85,wild94}.  Even when a black hole is present,
the sonic transition coincides with the edge of the cluster over much
of its surface, and a substantial fraction of the mass loss flows out
in this global wind.  In Model \modF, the central outflow is highly
supersonic.  Interestingly, this weakly collimated central flow is
soon recollimated by the hydrostatic pressure of the core: it
recollimates three times within the core, to generate a supersonic
outflow region with a high aspect ratio -- a genuine jet.

Structures similar to these have previously been studied in work
following on from that of Blandford \& Rees \cite[1974; see
also]{wiita78a}.  These authors modelled the outflow driven by a
central injection of energy, accelerated to supersonic velocities by de
Laval nozzles confined by the pressure of the surrounding nISM\@.  The
original models treated the funnel as very long and thin, but
these long funnels were found to be unstable to Kelvin-Helmholtz
instabilities, resulting in the loss of gas in a series of bubbles
rather than a continuous jet \cite{gulln73,wiita78b,nsws81}.  For
rather higher injection luminosities, continuous outflows do occur
\cite{ssnw81,ssnw83}.  However, in the latter cases the neck of the
funnel was close to the termination shock of the central wind.  In
consequence, the `jets' so formed were only slightly supersonic, and
poorly collimated.  Here we find that a dense hydrostatic envelope can
indeed lead to the formation of a well-collimated jet with a
substantial Mach number, if a highly supersonic but ill-collimated flow
is driven from the centre.

\subsection{Varying the distribution of radiation}

\begin{figure*}
\epsfxsize = 8cm
\begin{centering}
\begin{tabular}{ll}
a) & b) \\
\epsfbox{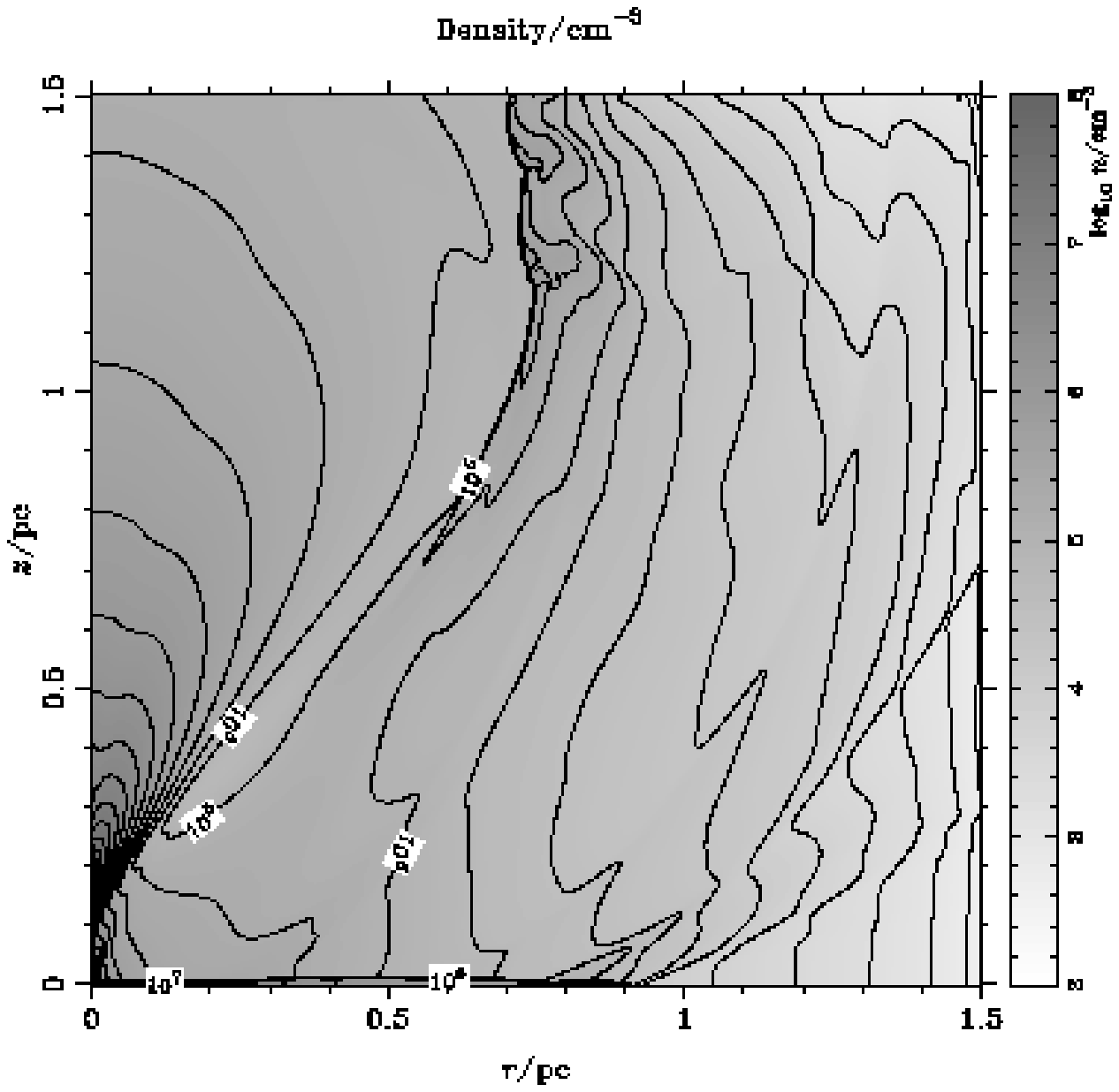} &
\epsfbox{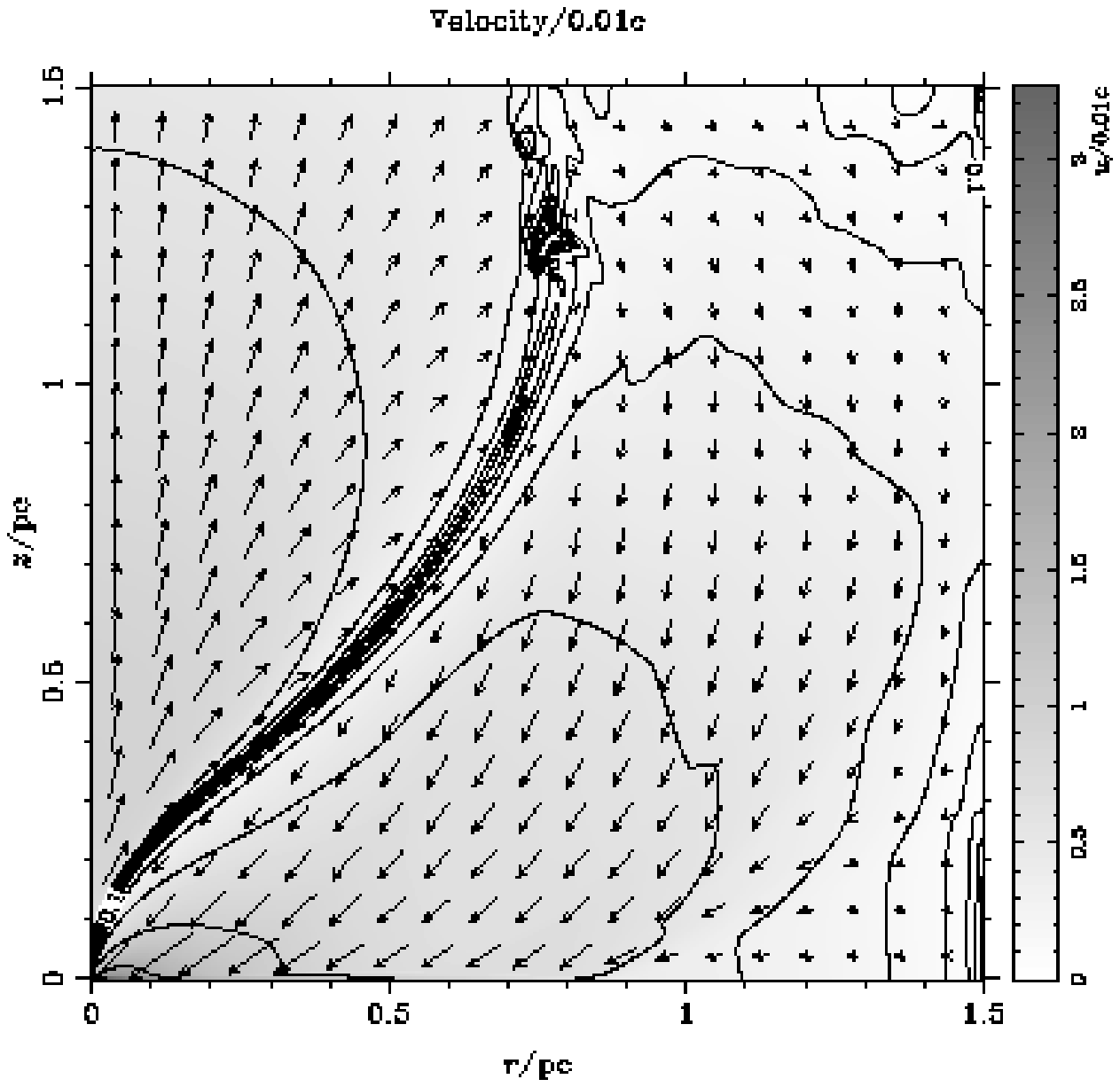} \\
c) & d) \\
\epsfbox{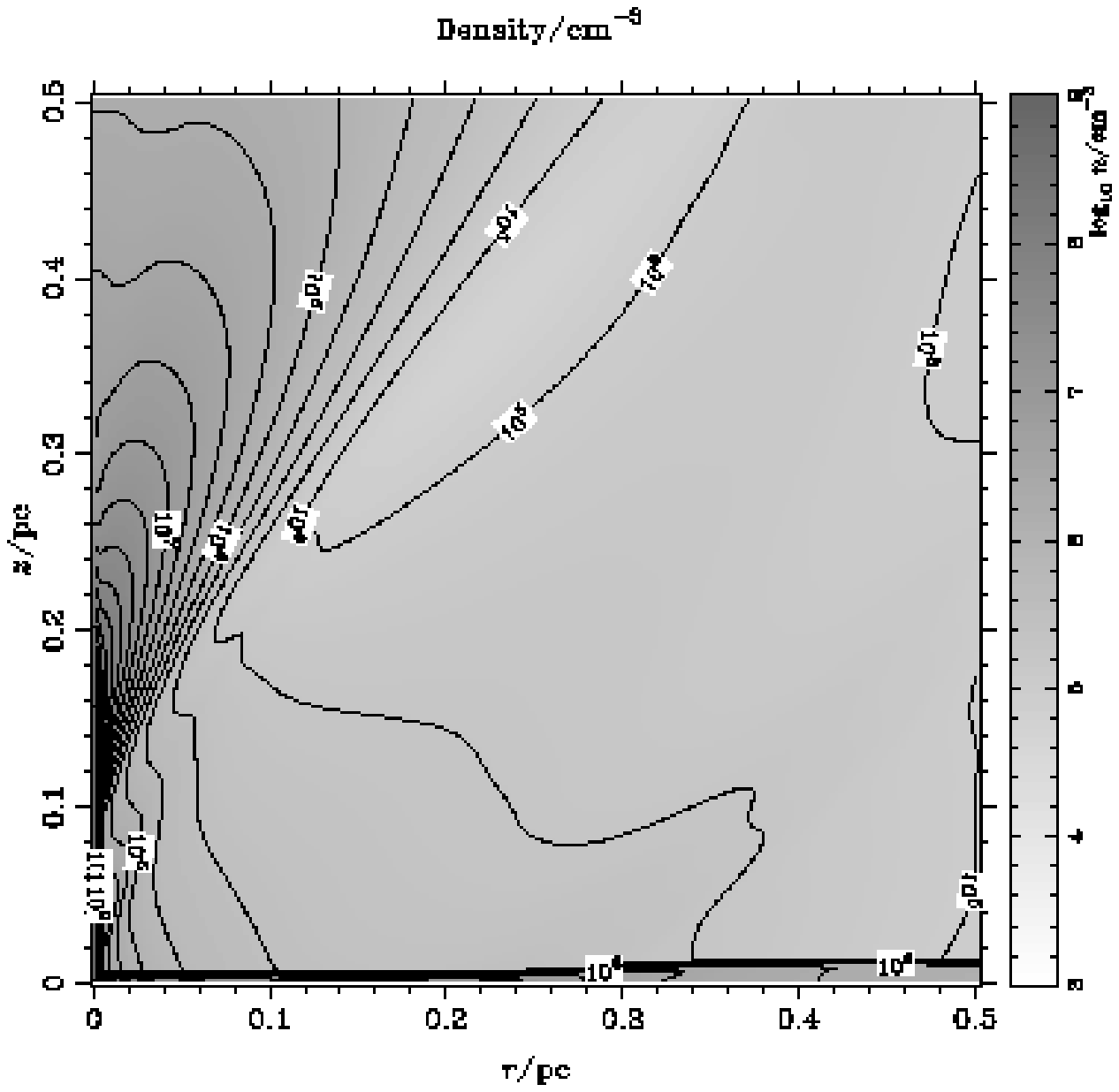} &
\epsfbox{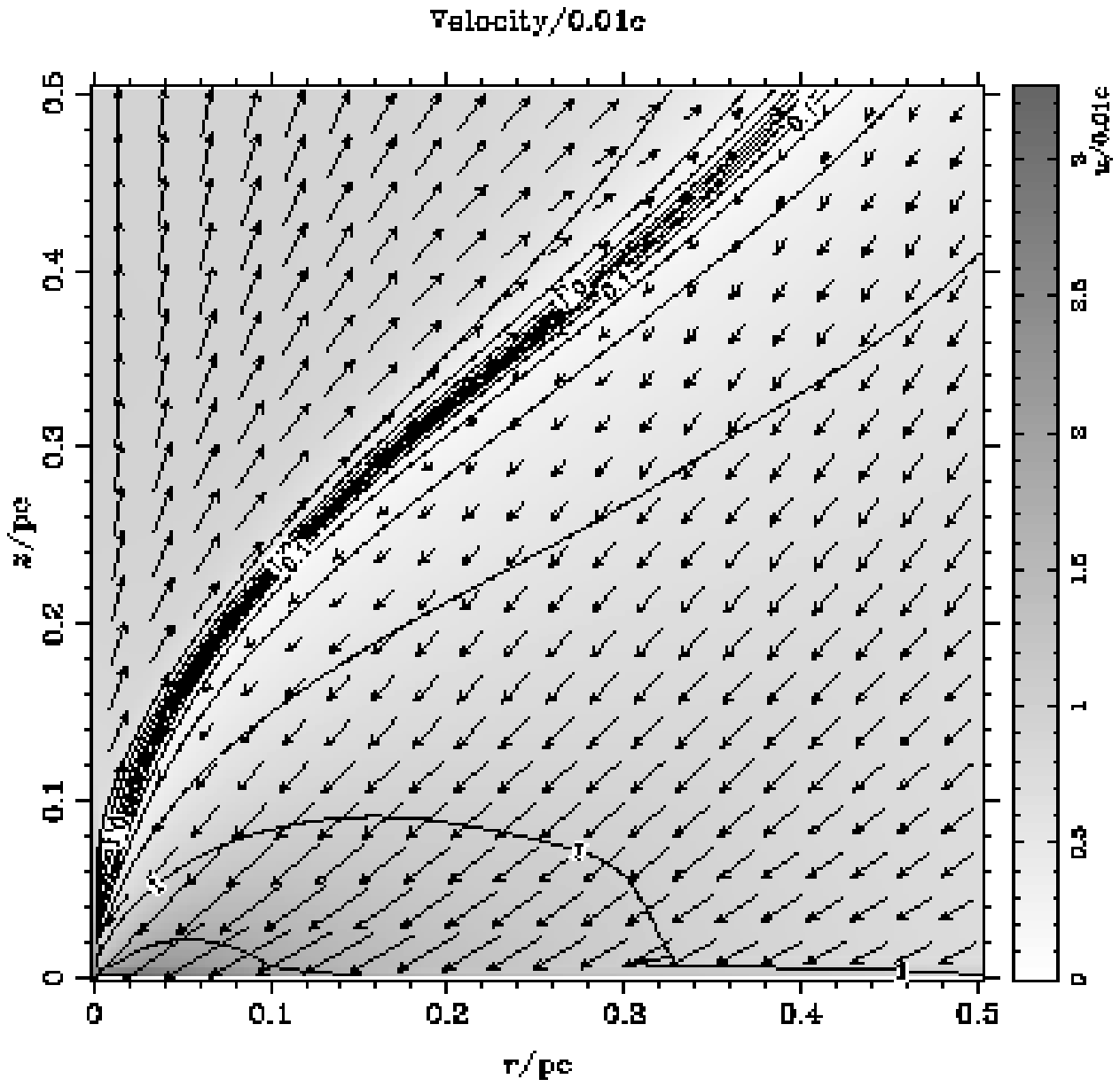} \\
\end{tabular}
\end{centering}
\caption[Model D 2D plot]{Model \modD\ at $2\ee4\yr$.  The panels show
a) density (from 400 to $2\ee{11}\cm^{-3}$) b) total velocity
(greyscale and contours, up to $3.3$), c), d) Magnifications of the
central region.  The vectors are in the flow direction, and have
lengths proportional to the velocity, sampled at one vector in every
$31\times31$ cells.  The flow is sonic at the $u = 0.1$ contour.}
\label{f:modeld2d}
\end{figure*}

\paragraph*{Model \modD}
(Fig.~\ref{f:modeld2d}) has parameters identical to those of Model
\modA, except for the Eddington ratio.  The net force is the same along
the axis of the disc as in Model \modA, but is outwards in a cone with far
smaller opening angle, and the inward force in the plane of the
accretion disc is far greater.  Model \modD\ quickly relaxes to a steady
outflow, albeit with no mass leaving the cluster with escape
velocity, whereas Model \modA\ underwent a series of explosions.
However, the flow structure of Model \modA\ close to its long-lived
mass minimum of $3\ee4\Msun$ at $2\ee5\yr$ (see
Fig.~\ref{f:modelaflux}a) is almost indistinguishable from that of the
equilibrium state of Model \modD.  Altogether, comparison of models
\modA\ and \modD\ indicates that the opening angle of the central
outflow has little influence on the overall flow structure, if the
cluster gravity and mass input are significant factors.

\subsection{Accretion over part of the disc}

We next investigate the effect of accretion on the global flows.  We
assume that the accretion from the flow onto the thin disc occurs
either in its innermost region (Model \modC$_{0.005}$), over a large
fraction of its inner surface (Model \modC$_{0.25}$) or over a large
fraction of its outer disc (Model \modC$_{0.25}'$).  These models
cover a range of ways in which mass might be lost from the global flow
to the accretion disc.  As accretion tends to weaken the net outflow
from the central regions, we use parameters which are perturbations of
those used in Model \modC\ above, which has the strongest initial
outflow.

\begin{figure*}
\epsfxsize = 8cm
\begin{centering}
\begin{tabular}{ll}
a) & b) \\
\epsfbox{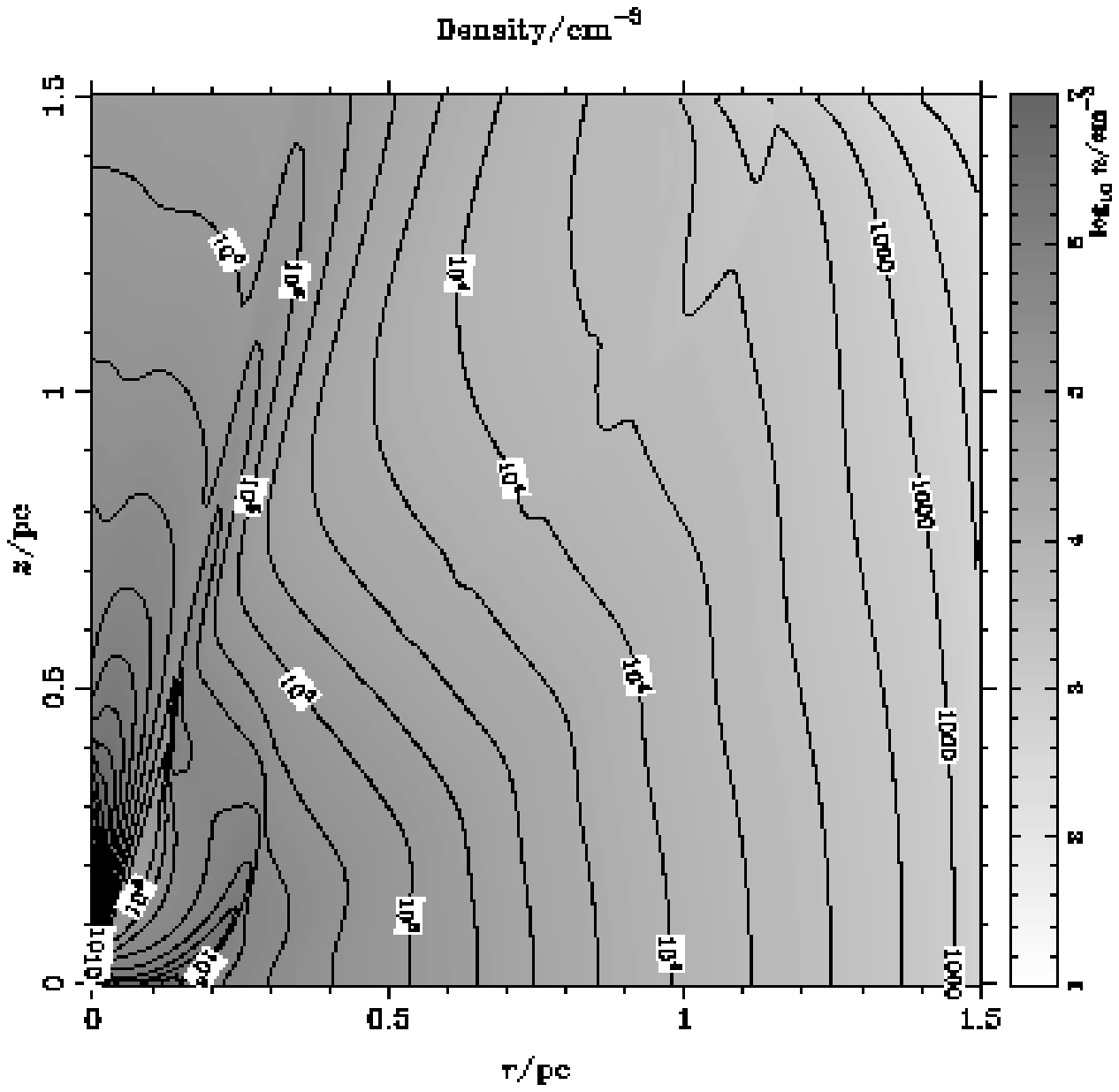} &
\epsfbox{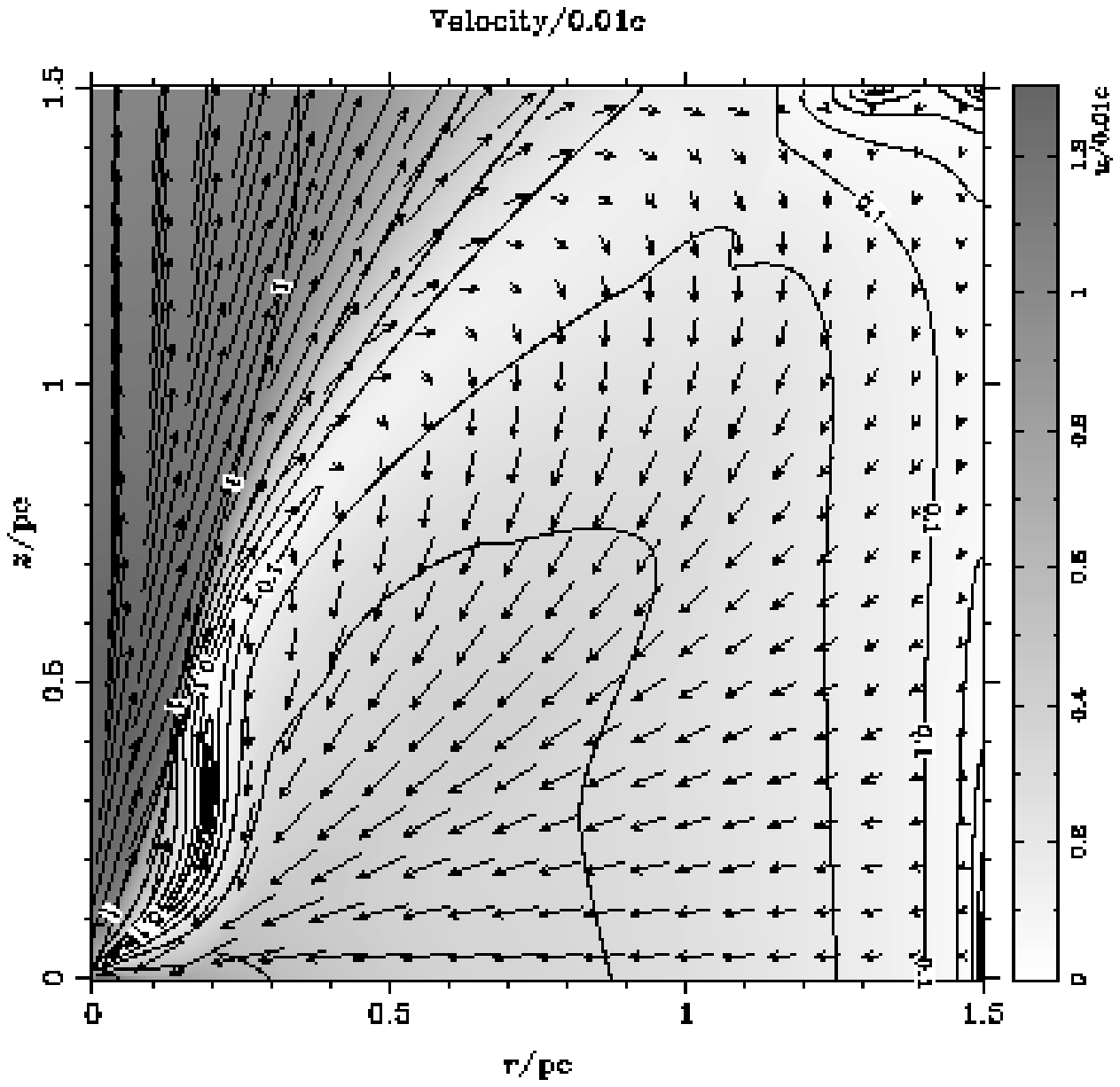} \\
c) & d) \\
\epsfbox{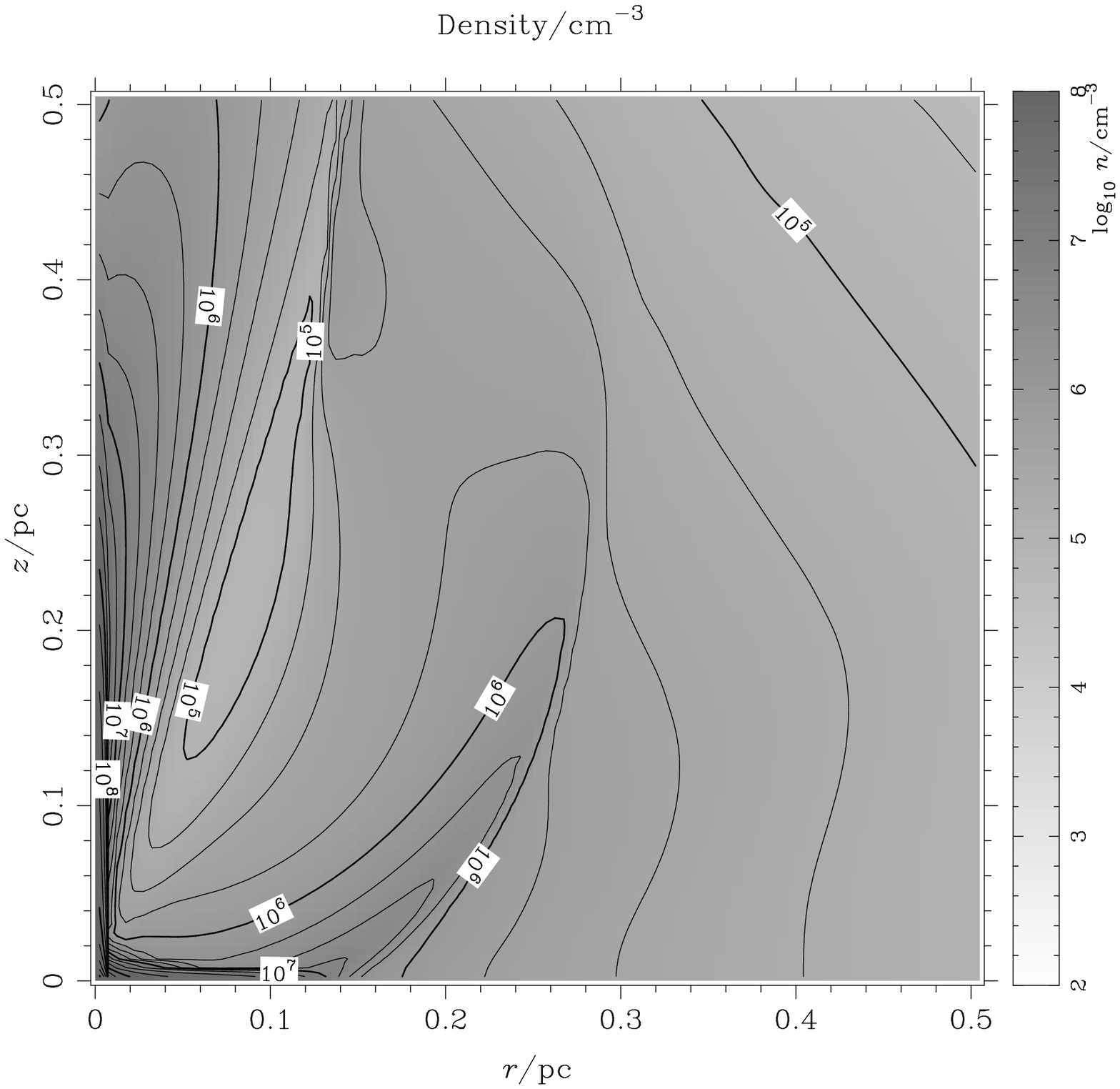} &
\epsfbox{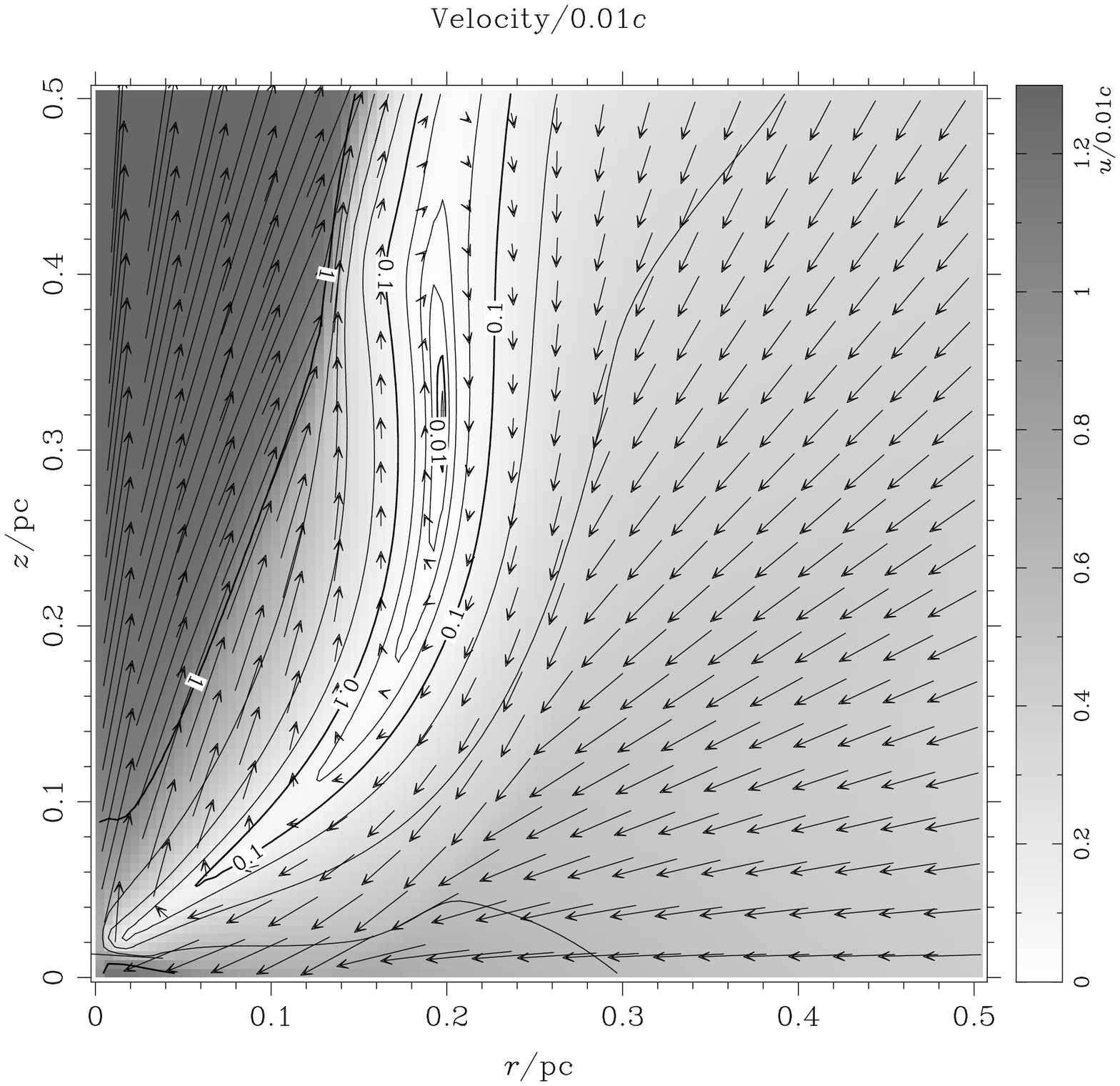} \\
\end{tabular}
\end{centering}
\caption[Model CC 2D plot]{Model \modC$_{0.005}$ at $1.8\ee4\yr$ (steady
equilibrium solution).  The panels show a) density (from 191 to
$1.1\ee{10}\cm^{-3}$) b) total velocity (greyscale and contours, up to
$1.3$), c), d) Magnifications of the central region.  The vectors are
in the flow direction, and have lengths proportional to the velocity,
sampled at one vector in every $15\times15$ cells.  The flow is sonic at
the $u = 0.1$ contour.}
\label{f:modelcc2d}
\end{figure*}
Models \modC$_{0.005} $ and \modC$_{0.25}' $, Figs.~\ref{f:modelcc2d}
and \ref{f:modelcout2d}, evolve in a manner nearly identical to model
\modC, even though around 40 per cent of the input mass flux is
diverted into the accretion disc rather than the wind.

\begin{figure*}
\epsfxsize = 8cm
\begin{centering}
\begin{tabular}{ll}
a) & b) \\
\epsfbox{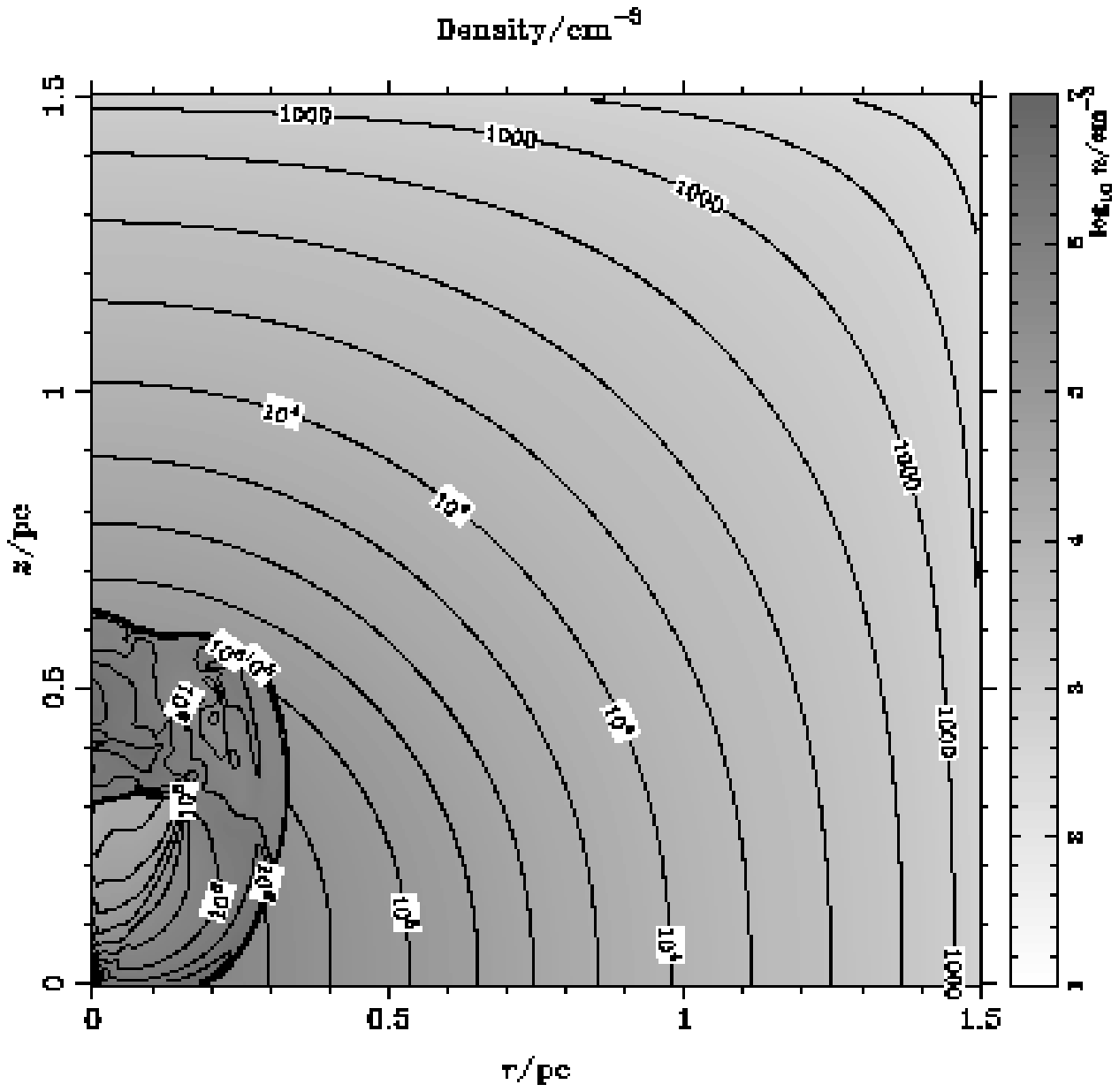} &
\epsfbox{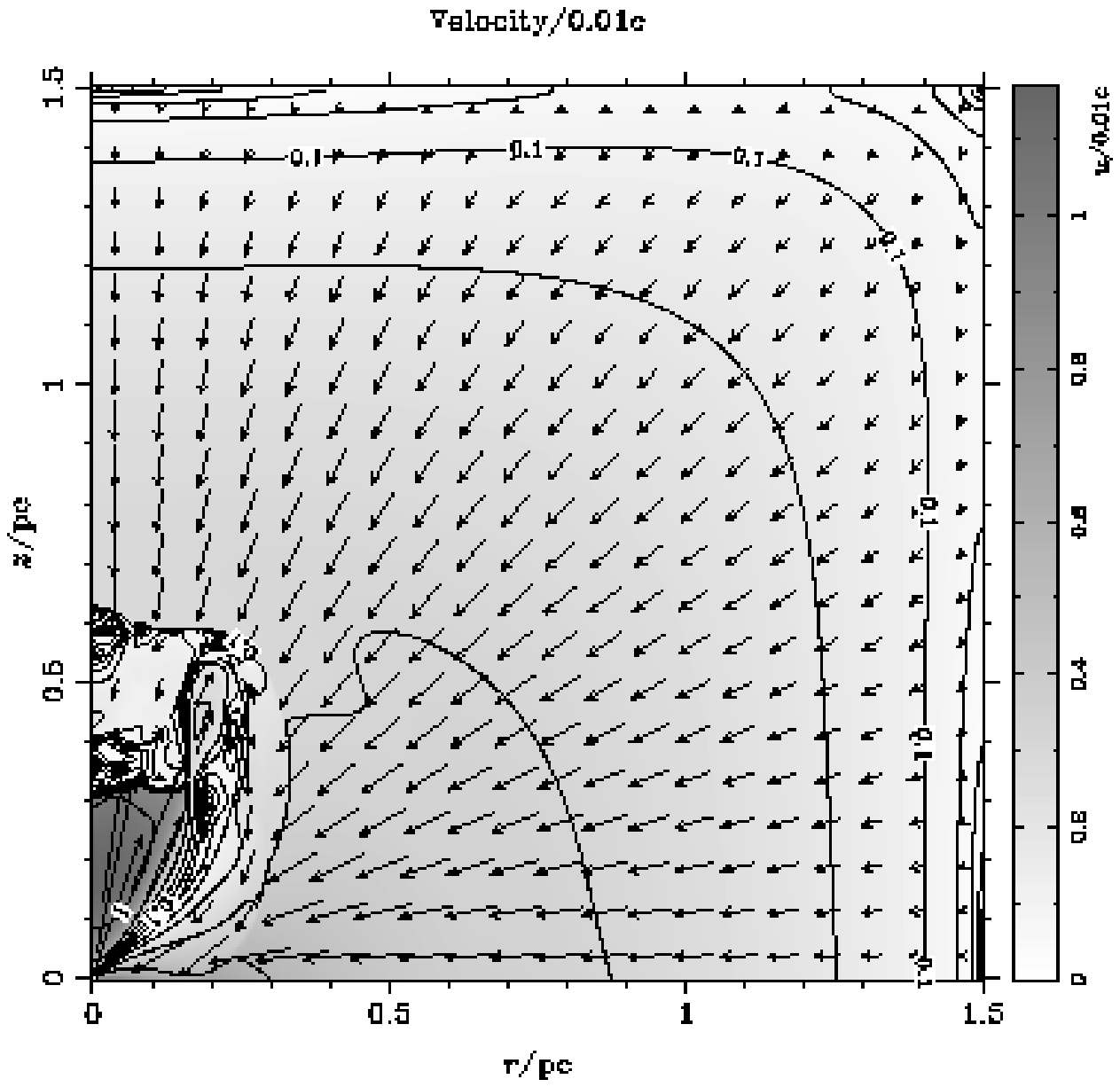} \\
c) & d) \\
\epsfbox{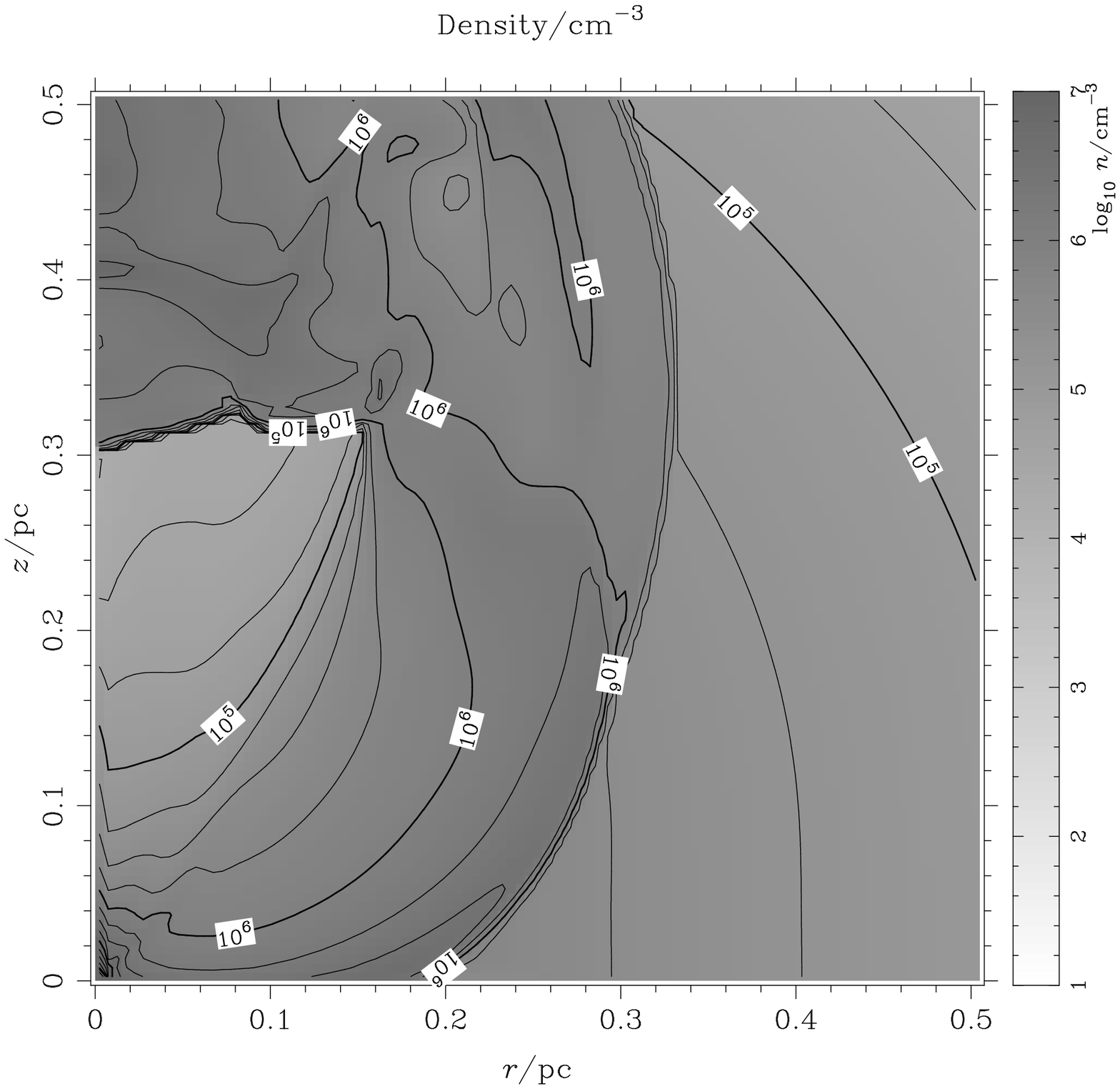} &
\epsfbox{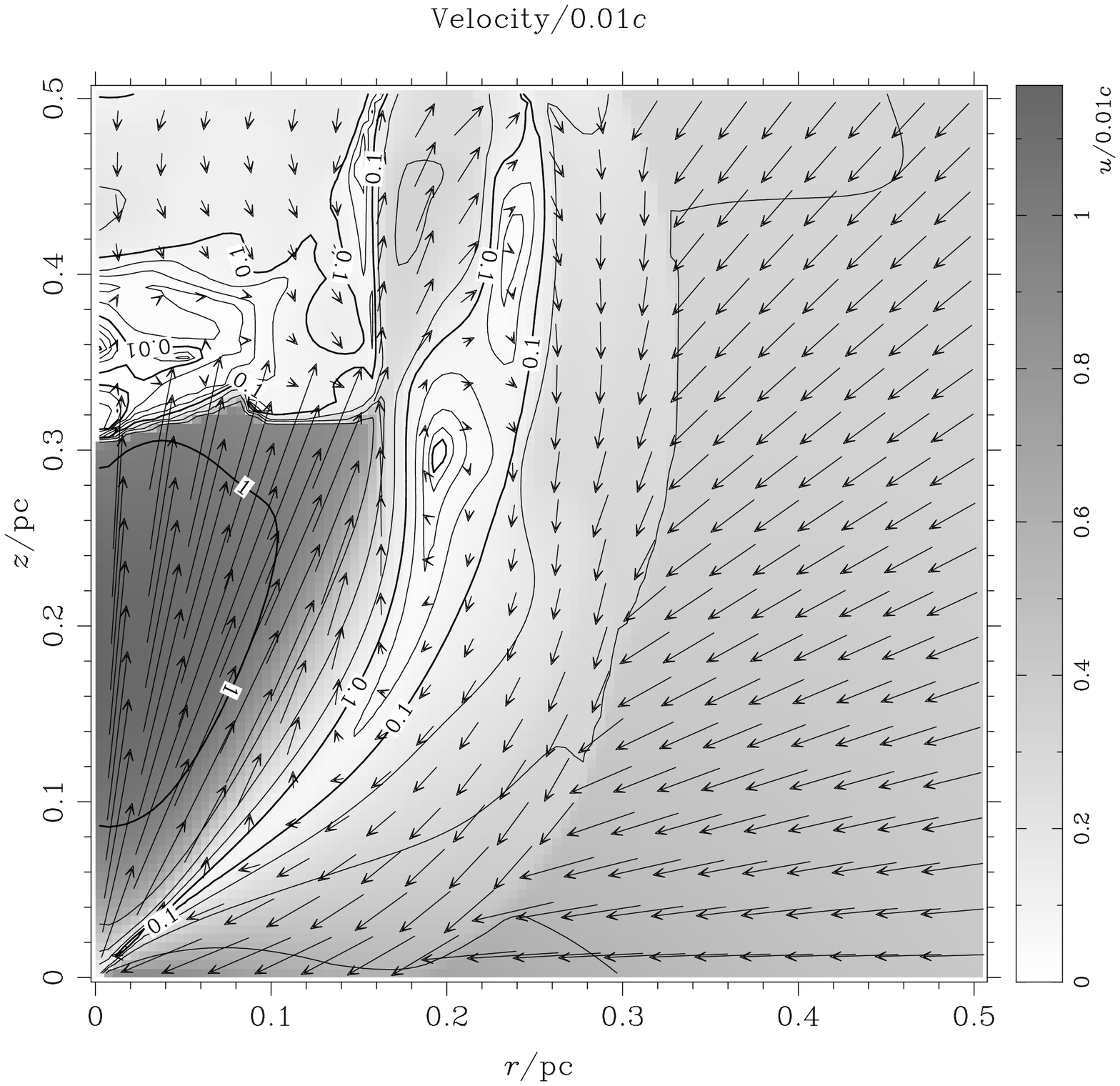} \\
\end{tabular}
\end{centering}
\caption[Model CPQ 2D plot]{Model \modC $_{0.25} $ at $ 5.1\ee4\yr $.  The
flow here is oscillating, \cf\ Fig.\ \protect\ref{f:modelcpflux}.  The
panels show a) density (from 171 to $ 5.1\ee{8}\cm^{-3} $) b) total
velocity (greyscale and contours, up to $ 1.2 $), c), d) Magnifications
of the central region.  The vectors are in the flow direction, and
have lengths proportional to the velocity, sampled at one vector in
every $ 15\times15 $ cells.  The flow is sonic at the $ u = 0.1 $
contour.}
\label{f:modelcpq2d}
\end{figure*}
There is a striking similarity between the flows illustrated in
Figs.~\ref{f:modelcc2d} and \ref{f:modelcout2d}.  This extends even to
the evolution of detailed time-dependent features, such as mass clumps
on axis (Fig.~\ref{f:modelccmp2d}).  In one, mass is removed from
the simulation at the very centre of the grid, in the other across a
wide area of the outer disc.  Some signature of this can be see in the
regions closest to the $ r $-axis in the figures, beyond
$ r=0.25\parsec $.  Overall, the comparison illustrates convincingly the
independence of the form of the solutions from the detailed
distribution of mass input, or in this case mass removal, so long as
the total mass budget is similar.

With its substantially larger area of accretion, the outflow in model
\modC$_{0.25}$ has been weakened so much that the flow structure is
more reminiscent of Model \modB, except with dramatically smaller
densities in the flow ($10^6\cm^{-3}$ is characteristic, compared to
$10^9\cm^{-3}$ in Model \modB), and with a far narrower collar of
shocked gas at the base of the central outflow.

\begin{figure*}
\epsfxsize = 8cm
\begin{centering}
\begin{tabular}{ll}
a) & b) \\
\epsfbox{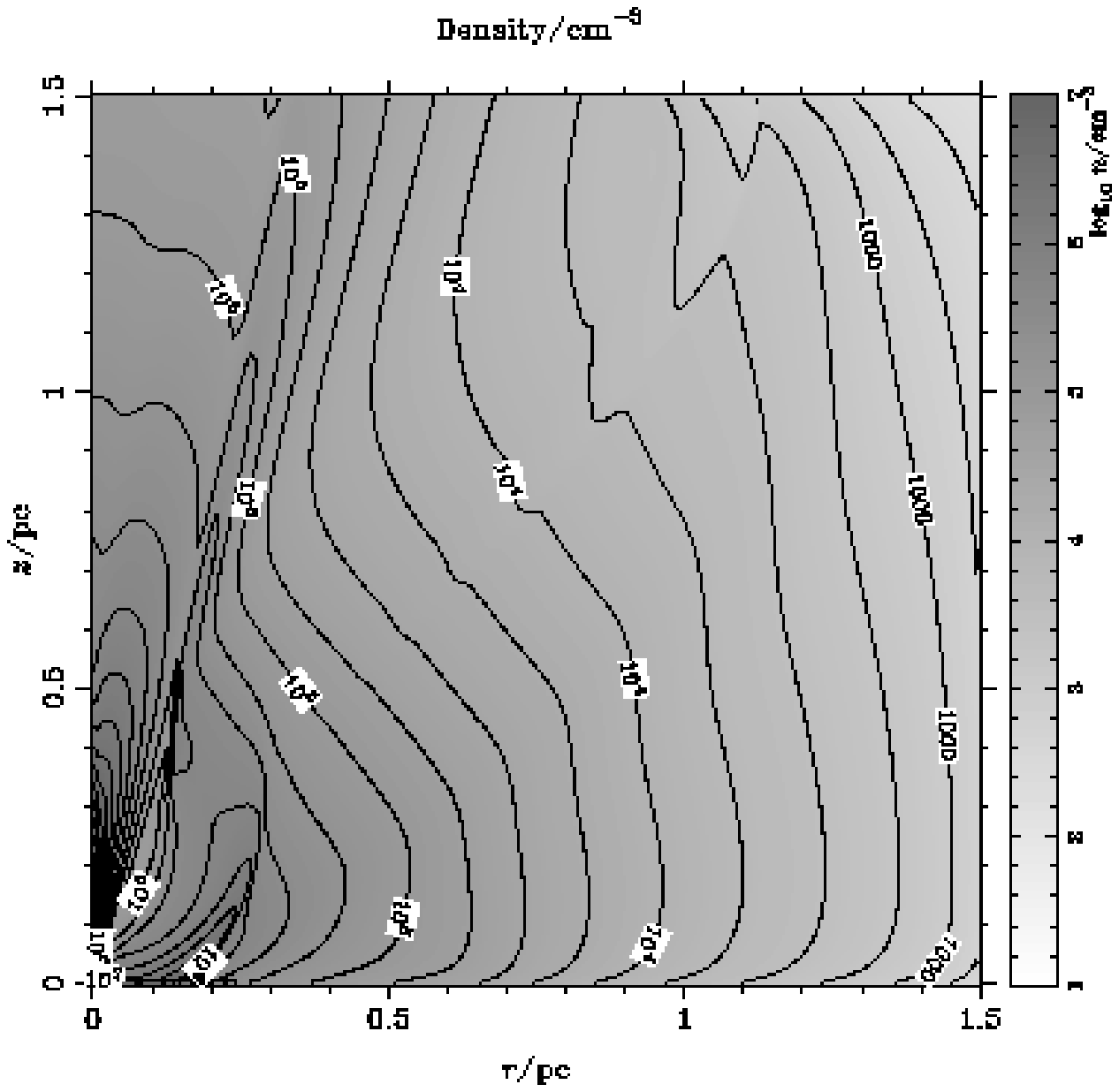} &
\epsfbox{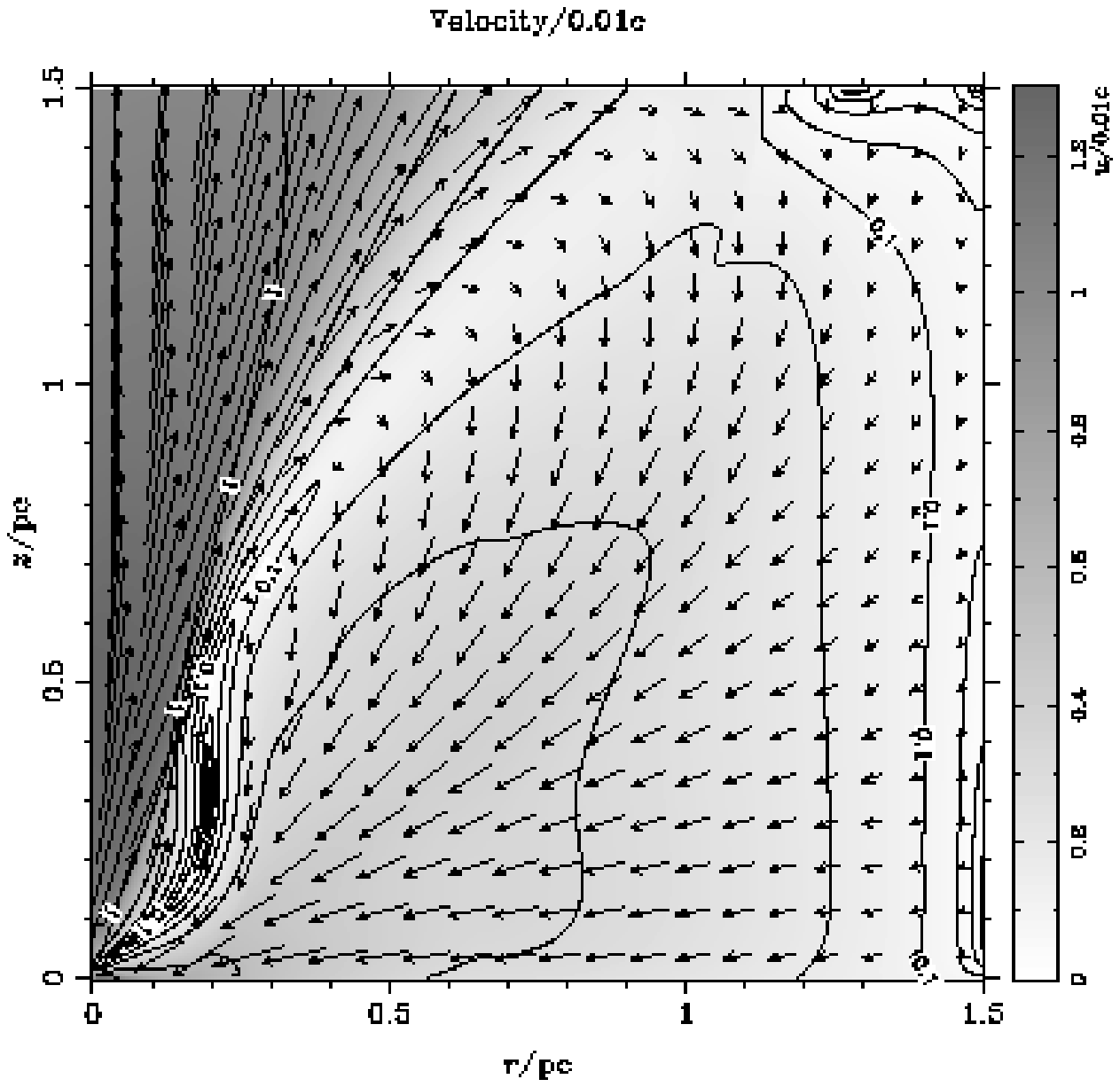} \\
c) & d) \\
\epsfbox{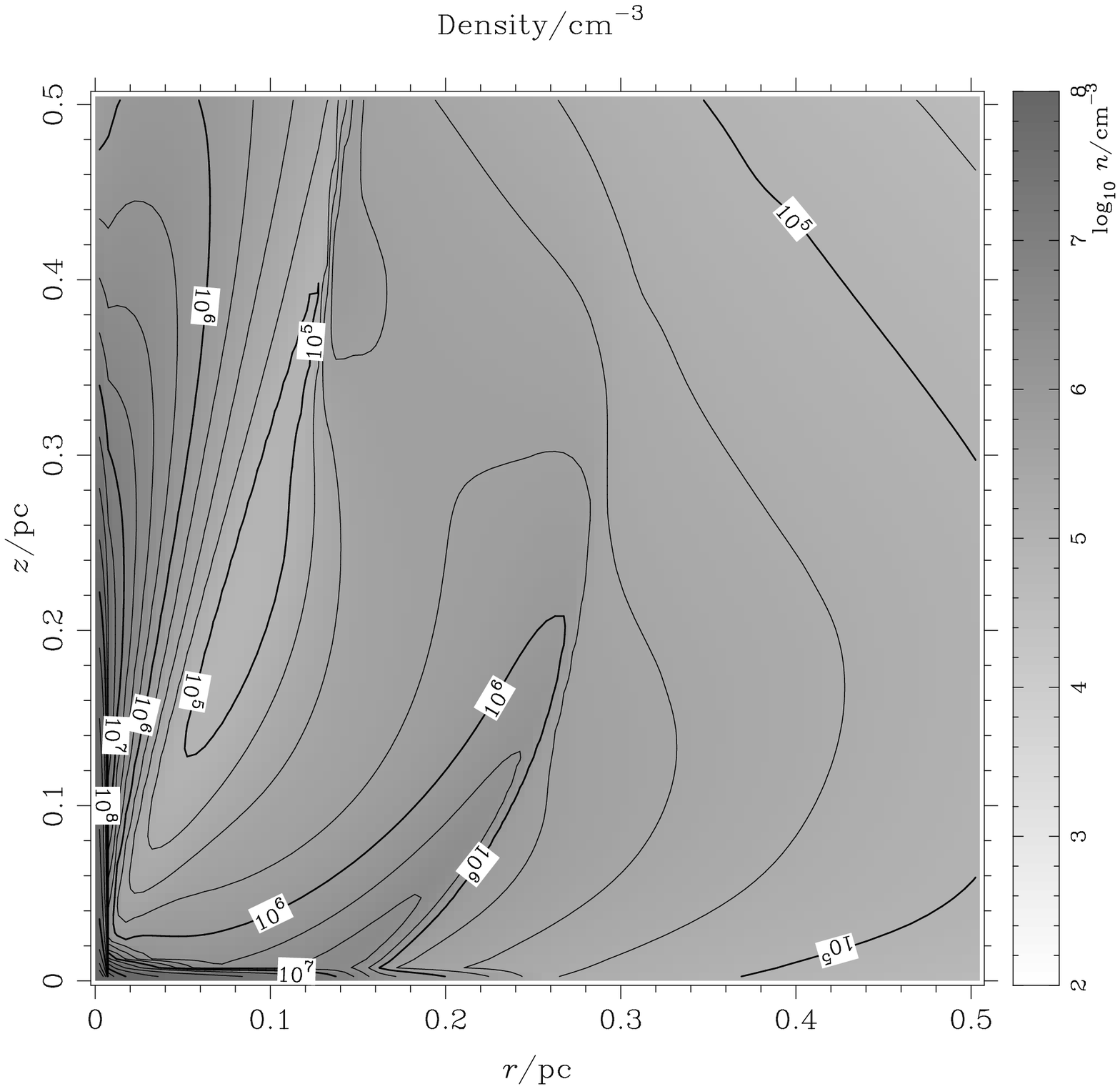} &
\epsfbox{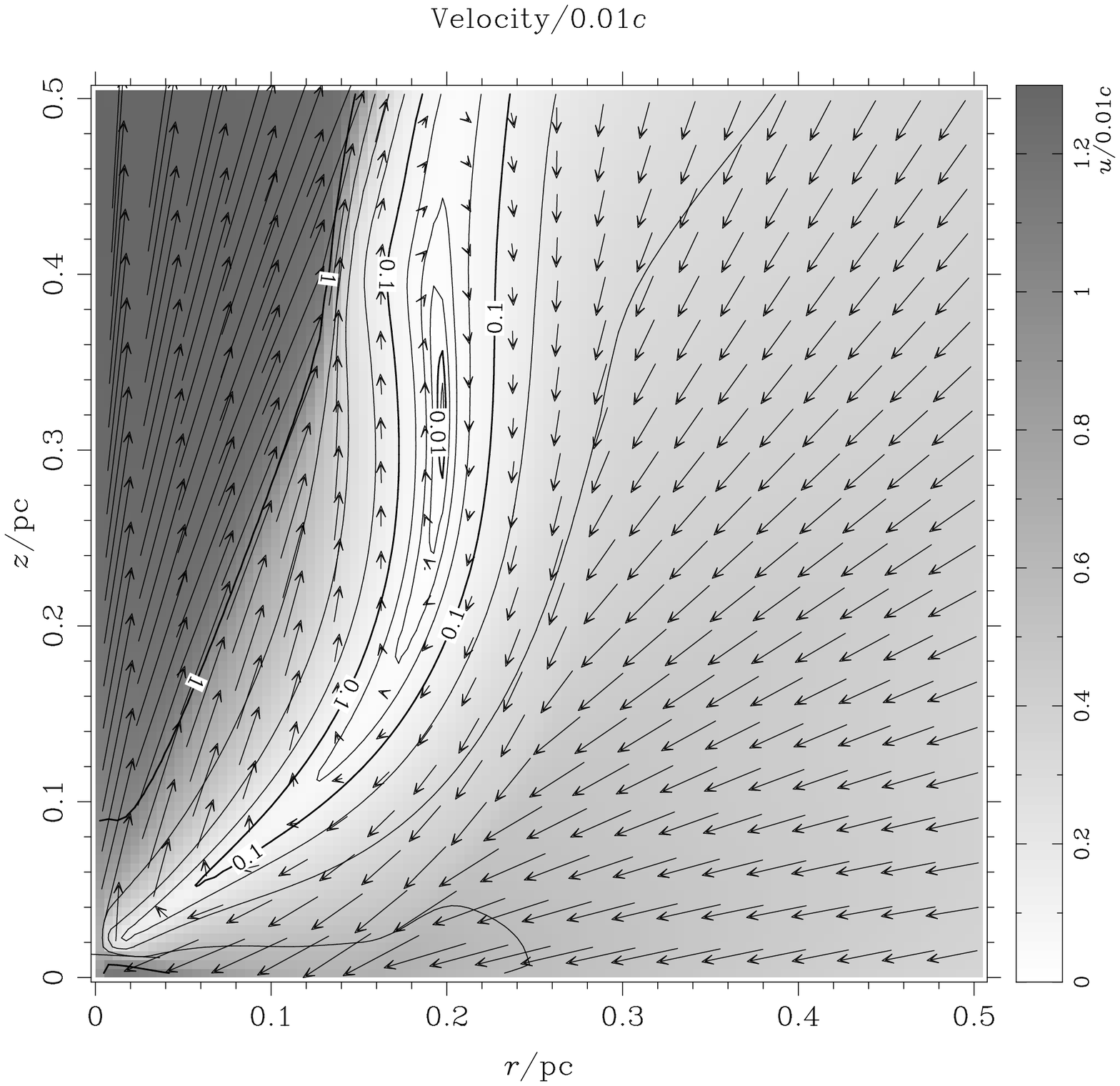} \\
\end{tabular}
\end{centering}
\caption[Model COUT 2D plot]{Model \modC$_{0.25}'$ at $4.6\ee4\yr$.
The flow here is oscillating, \cf\ Fig.\ \protect\ref{f:modelcpflux}.
The panels show a) density (from 182 to $9.6\ee{9}\cm^{-3}$) b) total
velocity (greyscale and contours, up to $1.3$), c), d) Magnifications
of the central region.  The vectors are in the flow direction, and
have lengths proportional to the velocity, sampled at one vector in
every $15\times15$ cells.  The flow is sonic at the $u = 0.1$
contour.}
\label{f:modelcout2d}
\end{figure*}

\begin{figure*}
\epsfxsize = 8cm
\begin{centering}
\begin{tabular}{ll}
a) & b) \\
\epsfbox{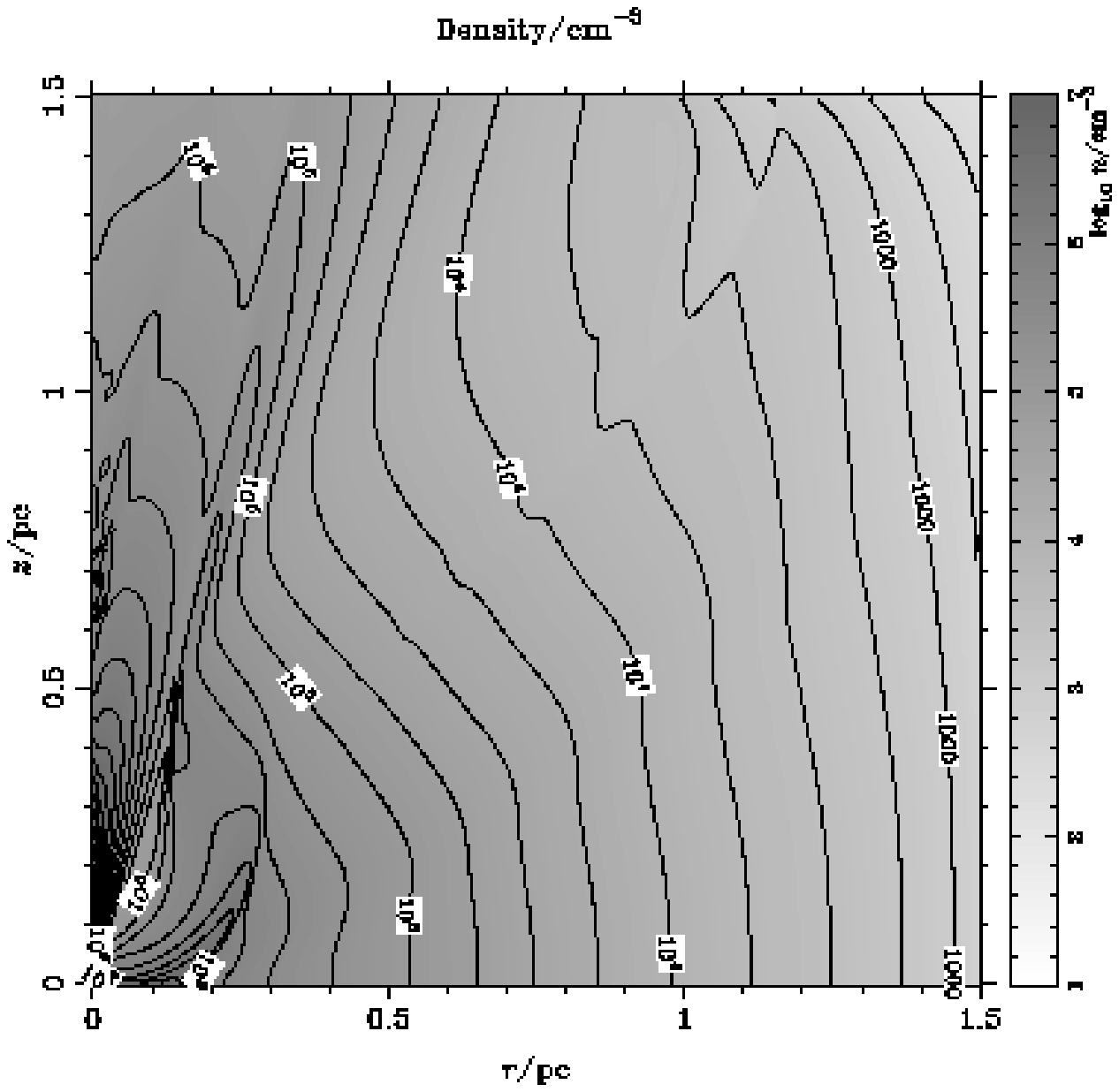} &
\epsfbox{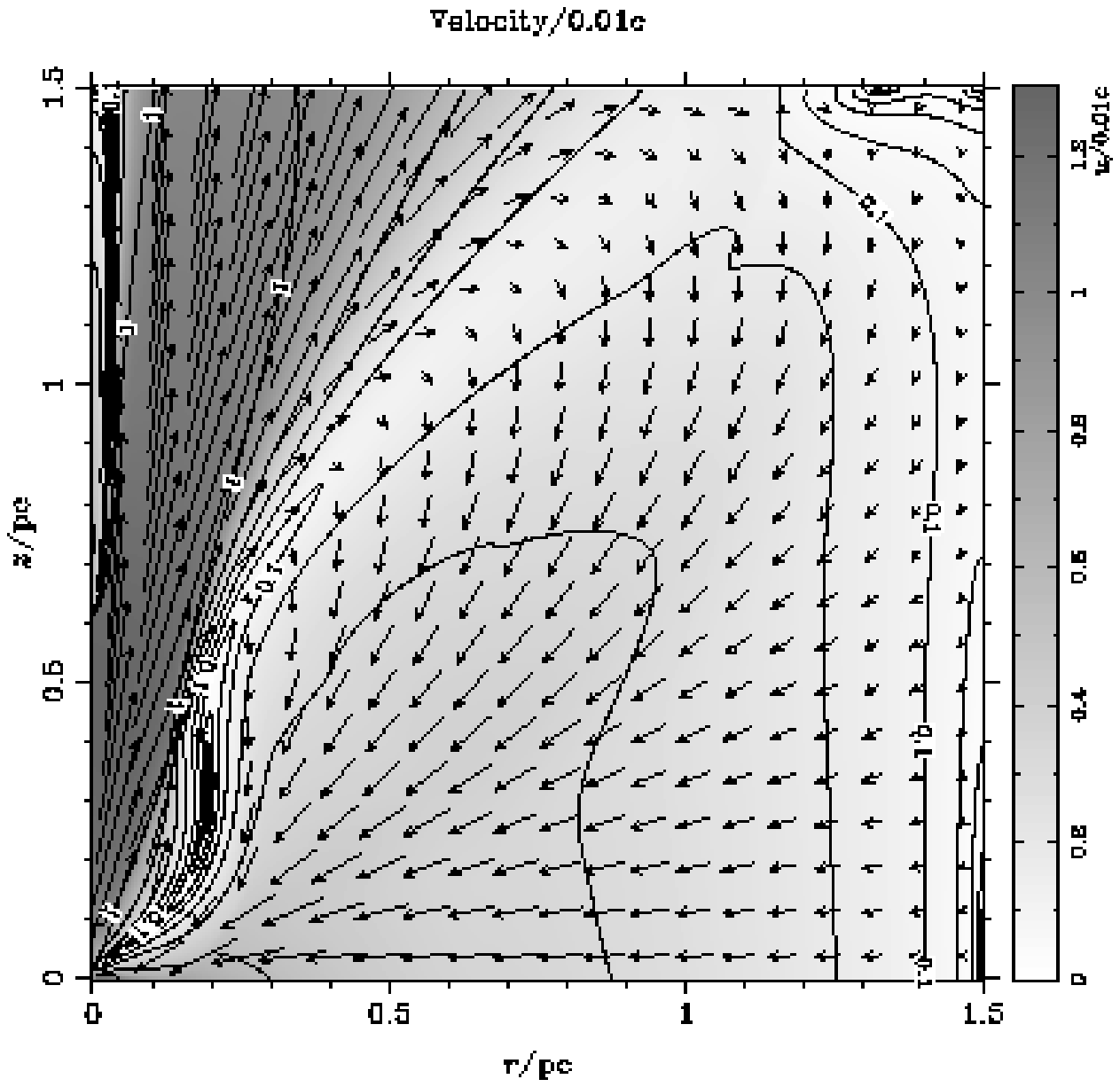} \\
c) & d) \\
\epsfbox{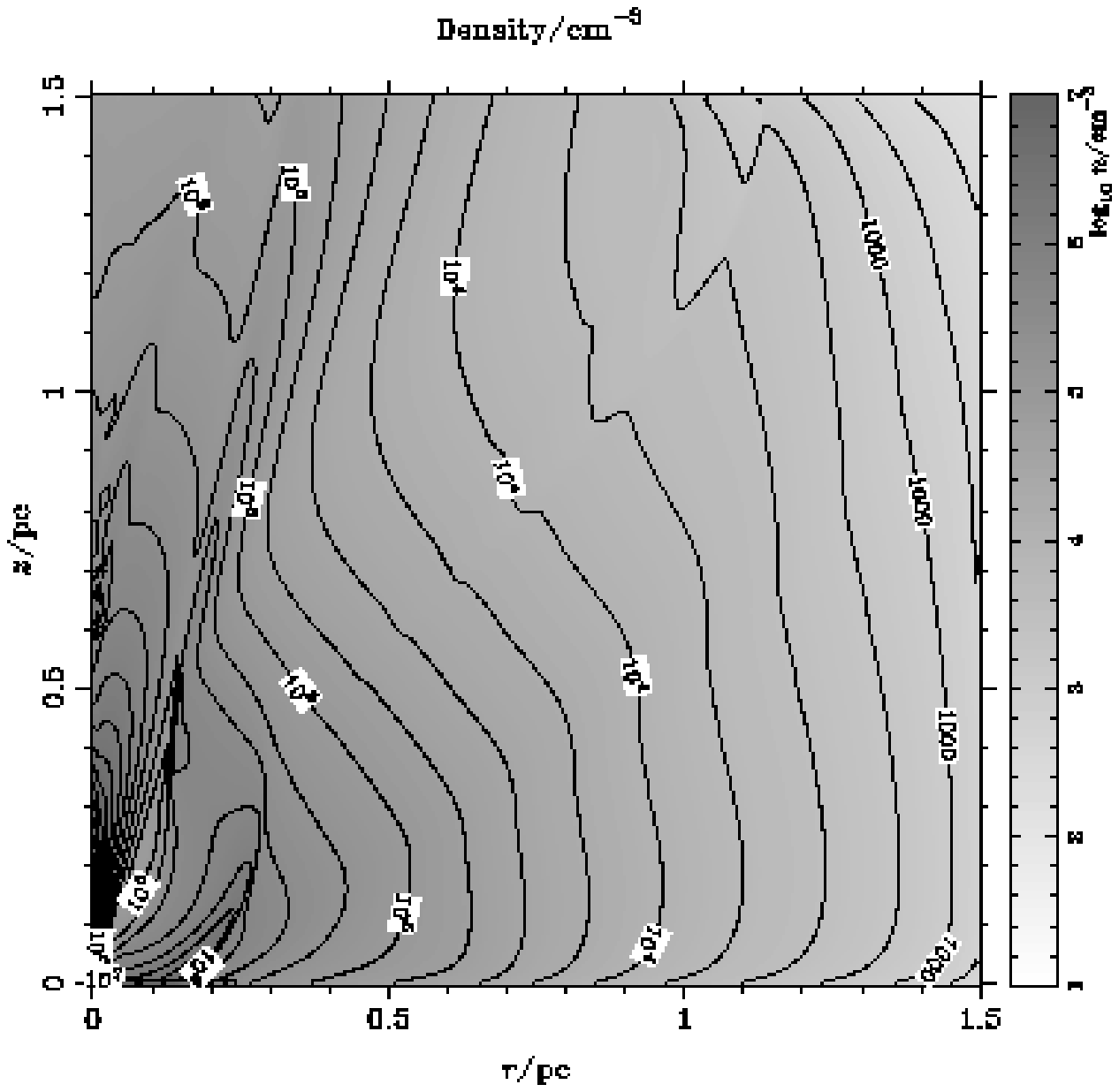} &
\epsfbox{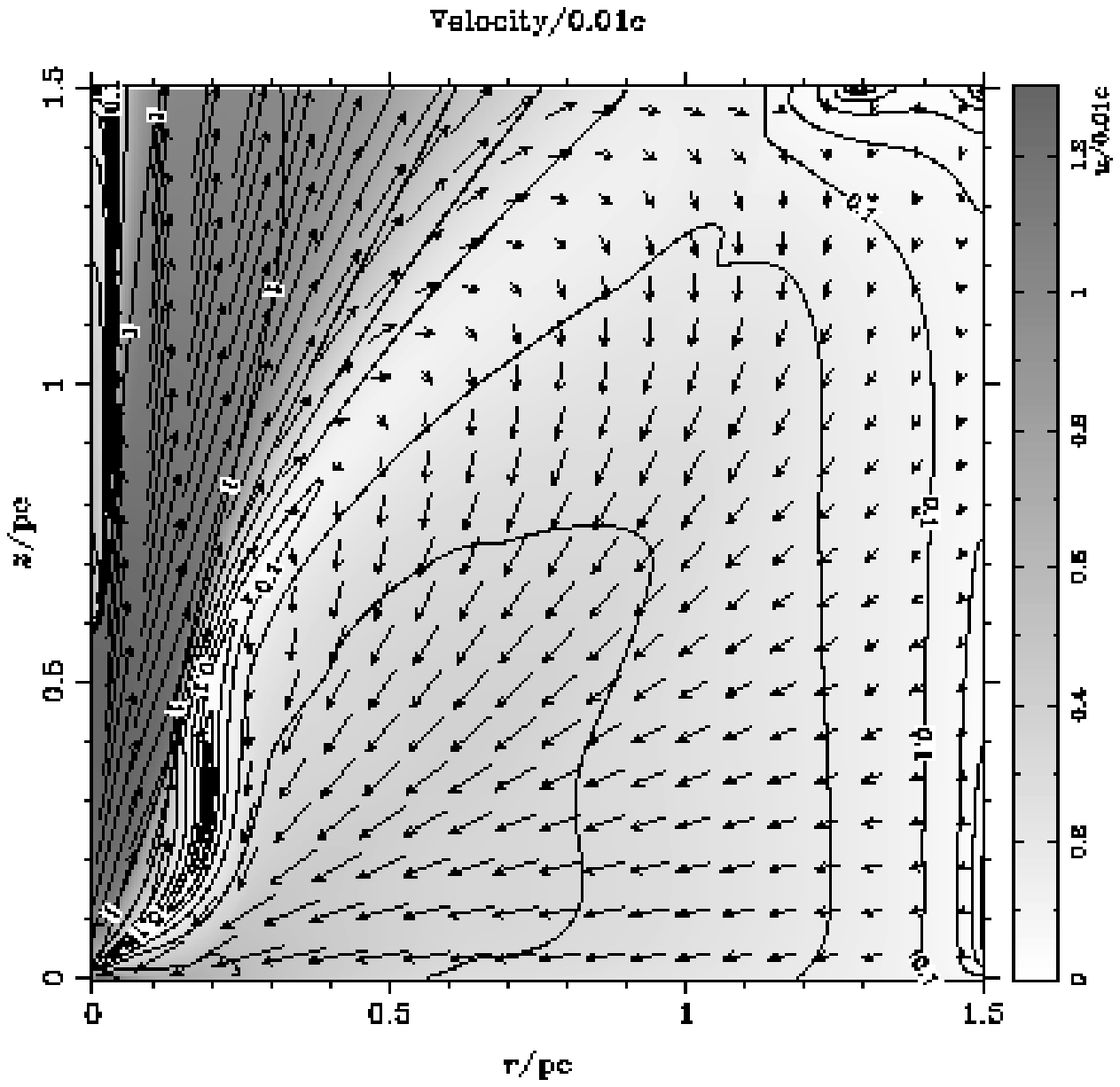} \\
\end{tabular}
\end{centering}
\caption[Model C comparison]{Comparison of (a, b) model
\modC$_{0.005}$ and (c,d) \modC$_{0.25}'$, both at $6.5\ee3\yr$.
Notice the similarity of detail in these simulations (away from the
disc plane), in particular the dense clump of gas at $z\simeq
0.65\parsec$ on axis.}
\label{f:modelccmp2d}
\end{figure*}

\subsection{Low Eddington ratio}

\begin{figure*}
\epsfxsize = 8cm
\begin{centering}
\begin{tabular}{ll}
a) & b) \\
\epsfbox{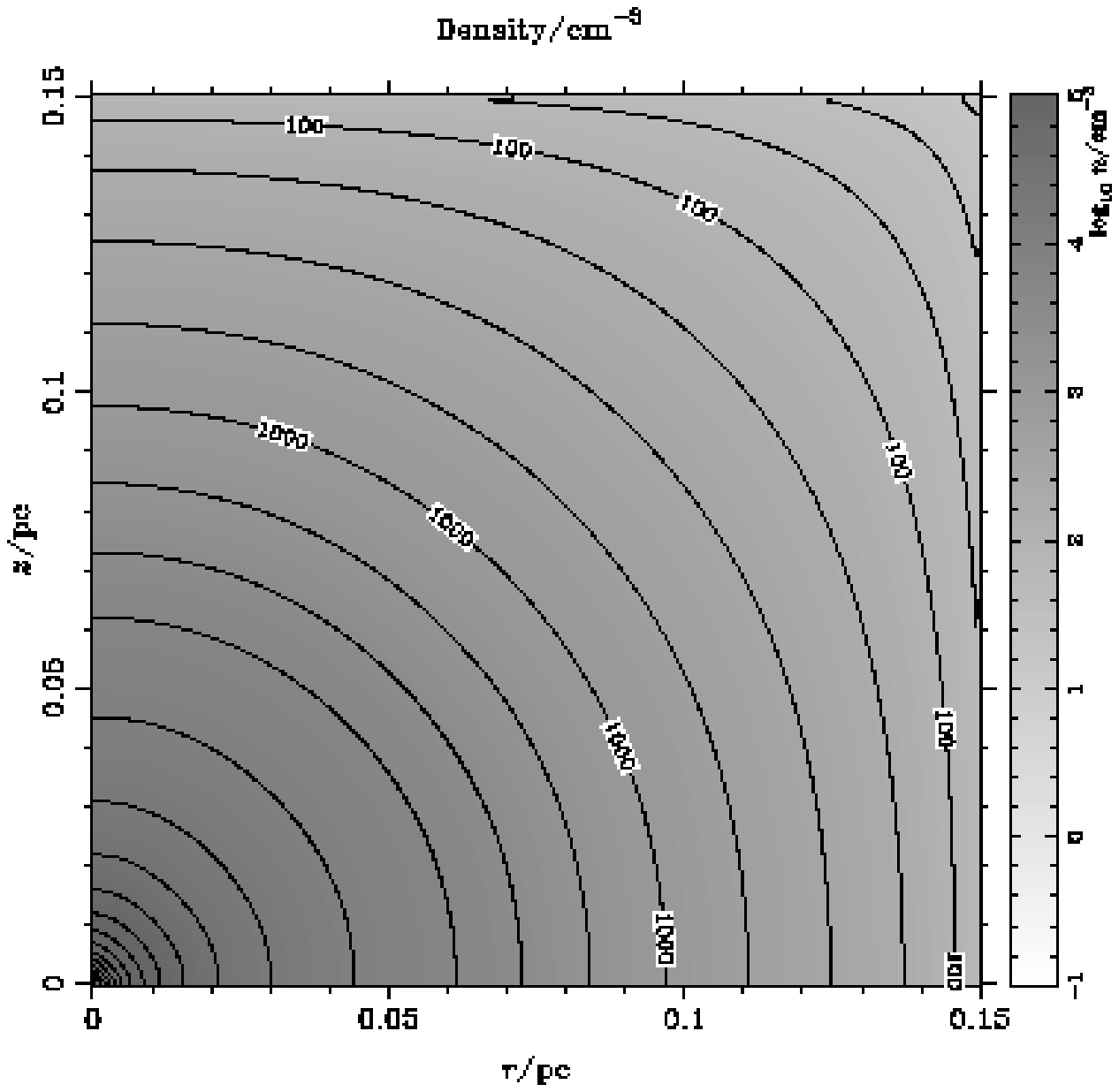} &
\epsfbox{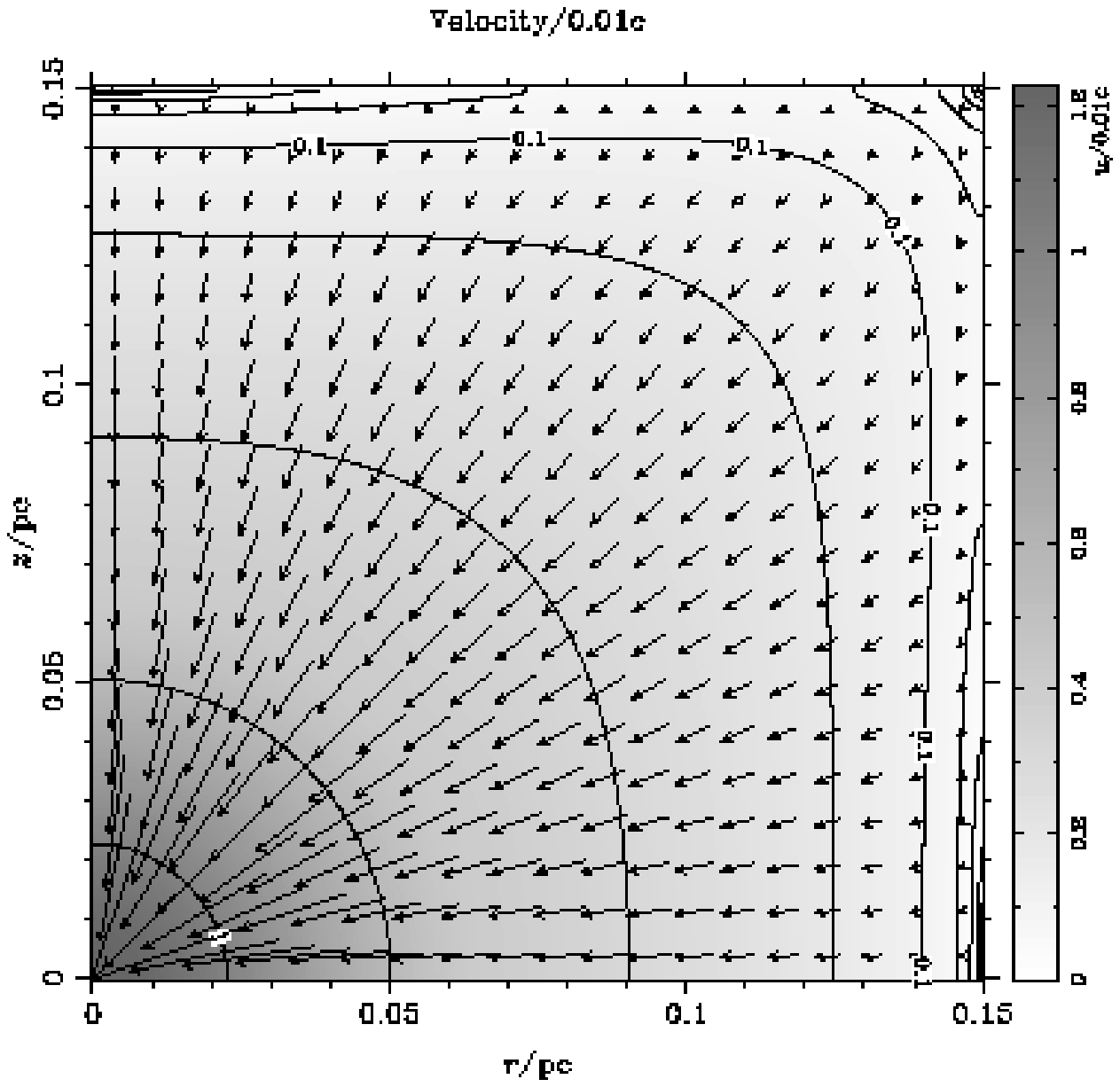} \\
c) & d) \\
\epsfbox{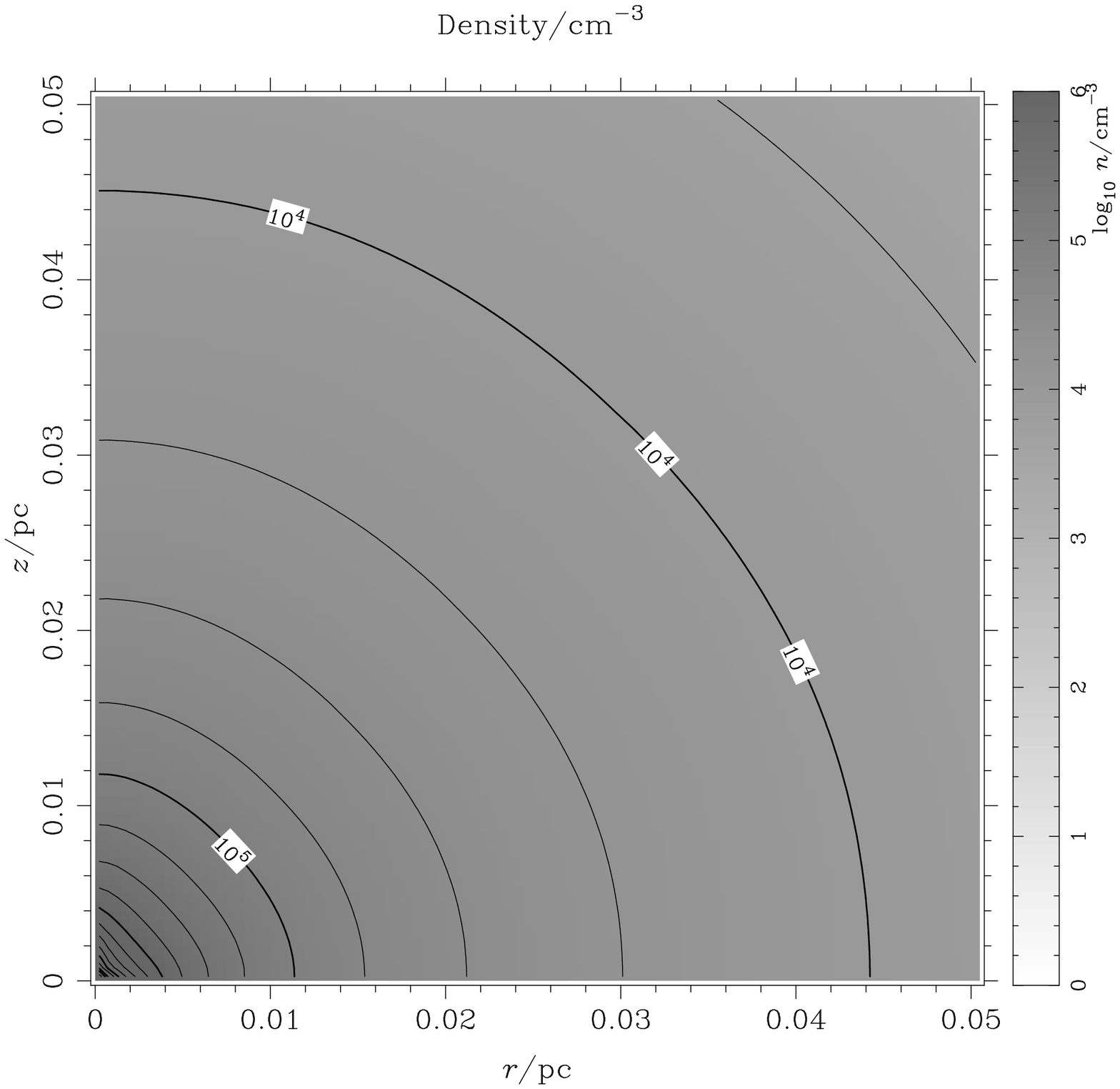} &
\epsfbox{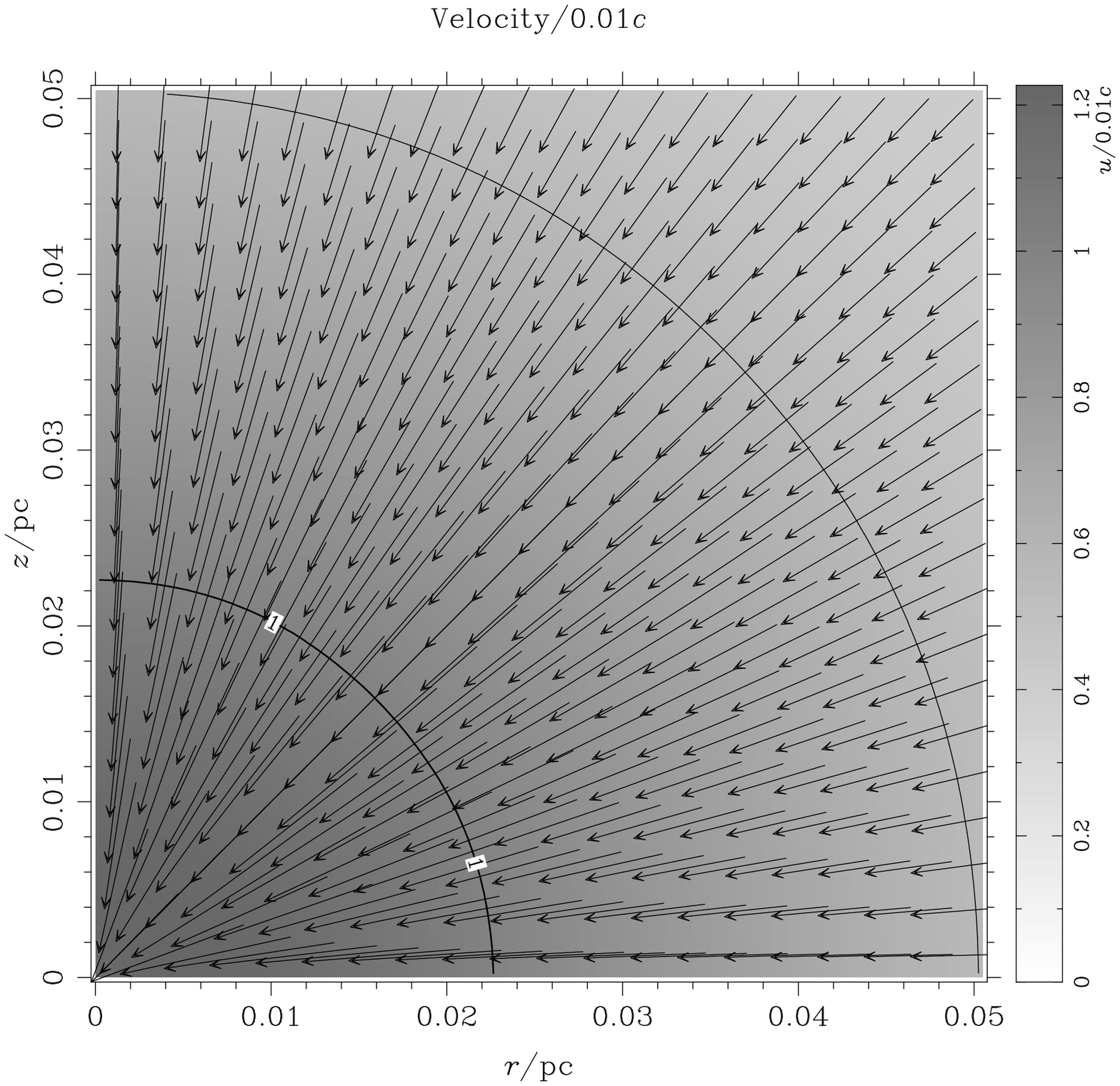} \\
\end{tabular}
\end{centering}
\caption[Model SA 2D plot]{Model Sa at $2.5\ee3\yr$.  Notice the
smaller scale of the grid in this simulation: the flow reaches
equilibrium as early as $200\yr$ as a result of the very short
dynamical timescale in this model.  The panels show a) density (from
15 to $2\ee8\cm^{-3}$) b) total velocity (greyscale and contours, up
to $1.2$), c), d) Magnifications of the central region.  The vectors
are in the flow direction, and have lengths proportional to the
velocity, sampled at one vector in every $15\times15$ cells in b),
every $5\times5$ in d).  Gas flows freely in from the cluster and is
accreted at the centre of the grid.  The anisotropic radiation, with a
mean Eddington ratio of only $0.01$, produces no noticeable
perturbations on this flow.}
\label{f:modelsa2d}
\end{figure*}

\begin{figure*}
\epsfxsize = 8cm
\begin{centering}
\begin{tabular}{ll}
a) & b) \\
\epsfbox{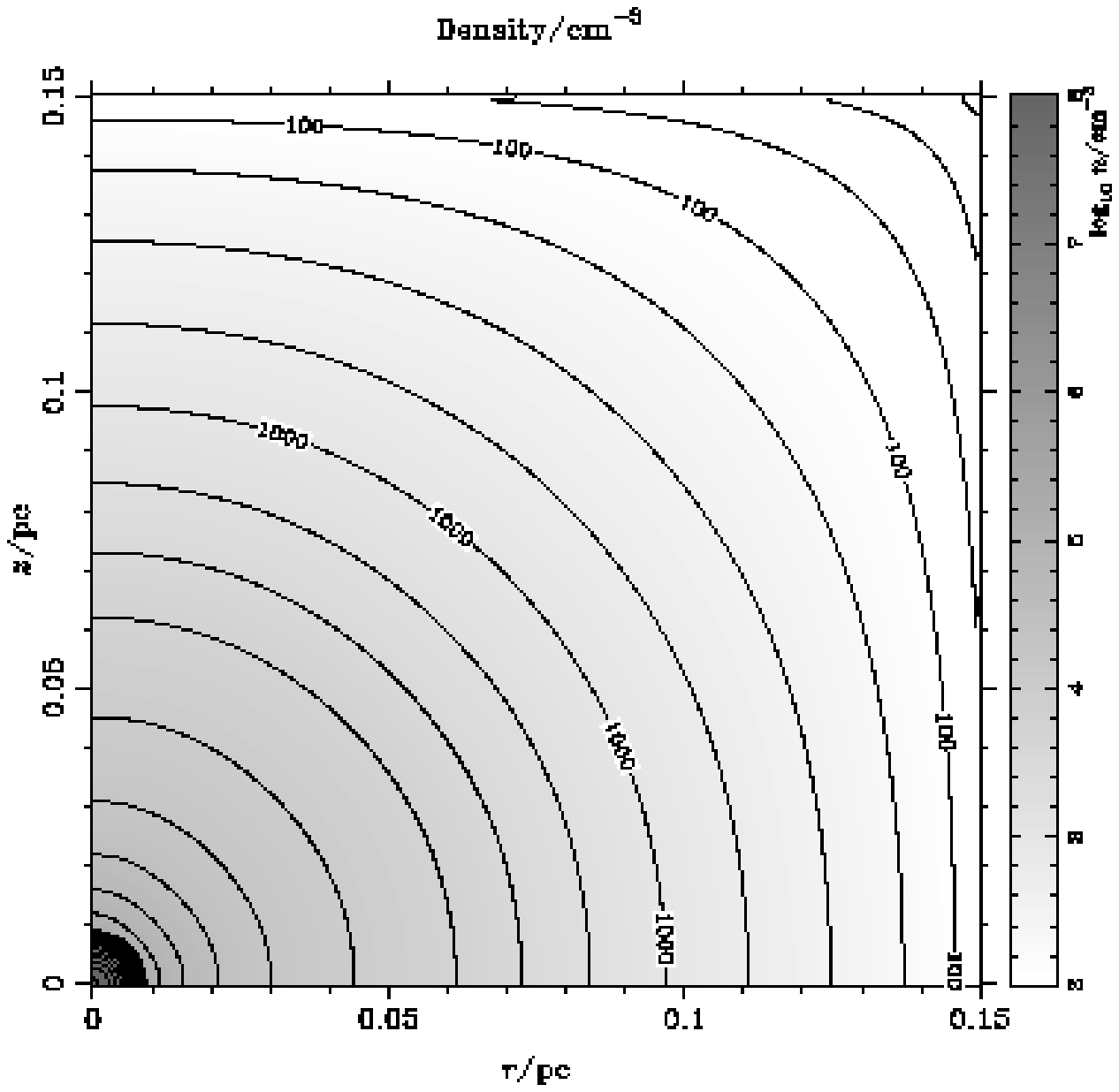} &
\epsfbox{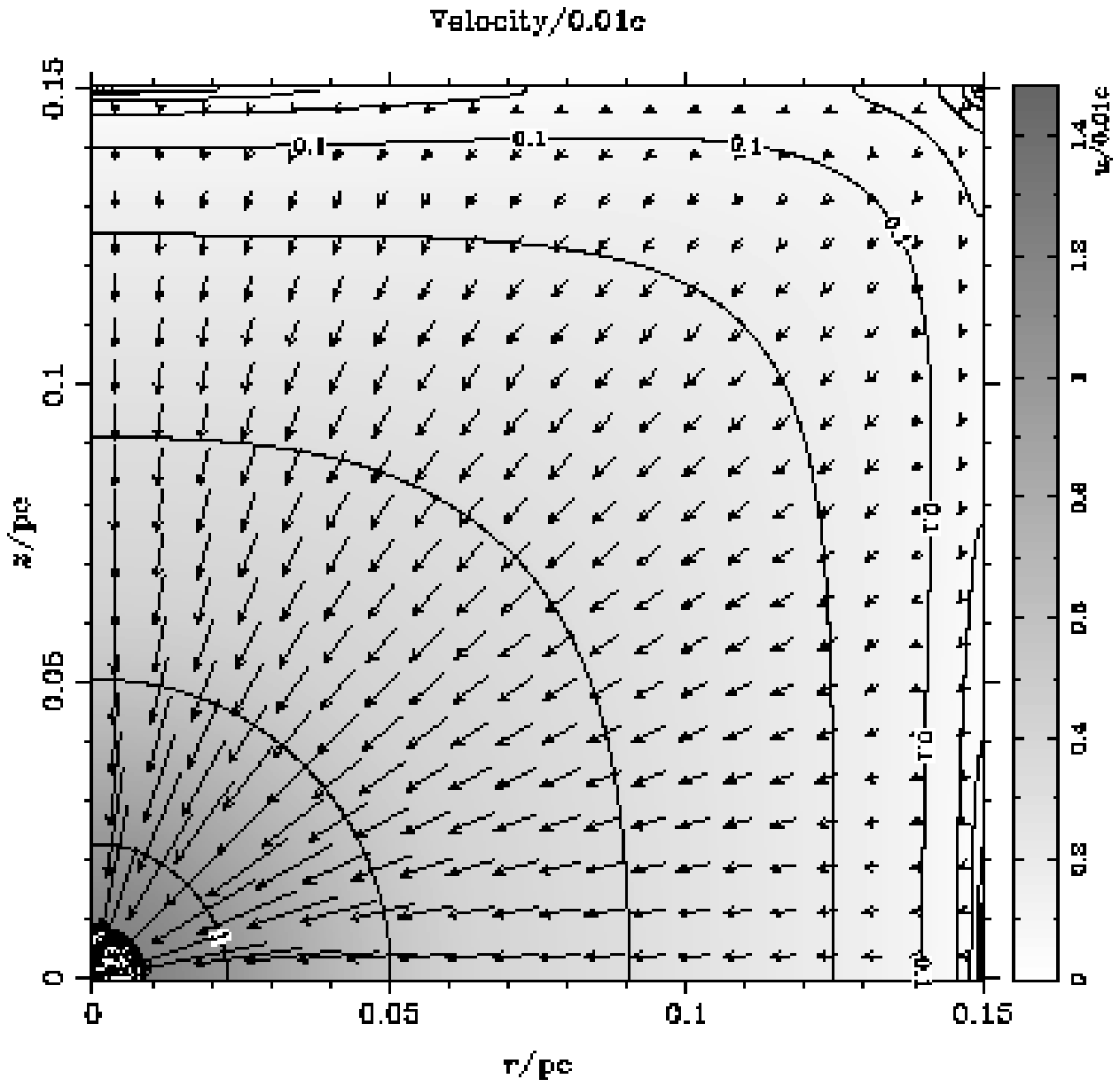} \\
c) & d) \\
\epsfbox{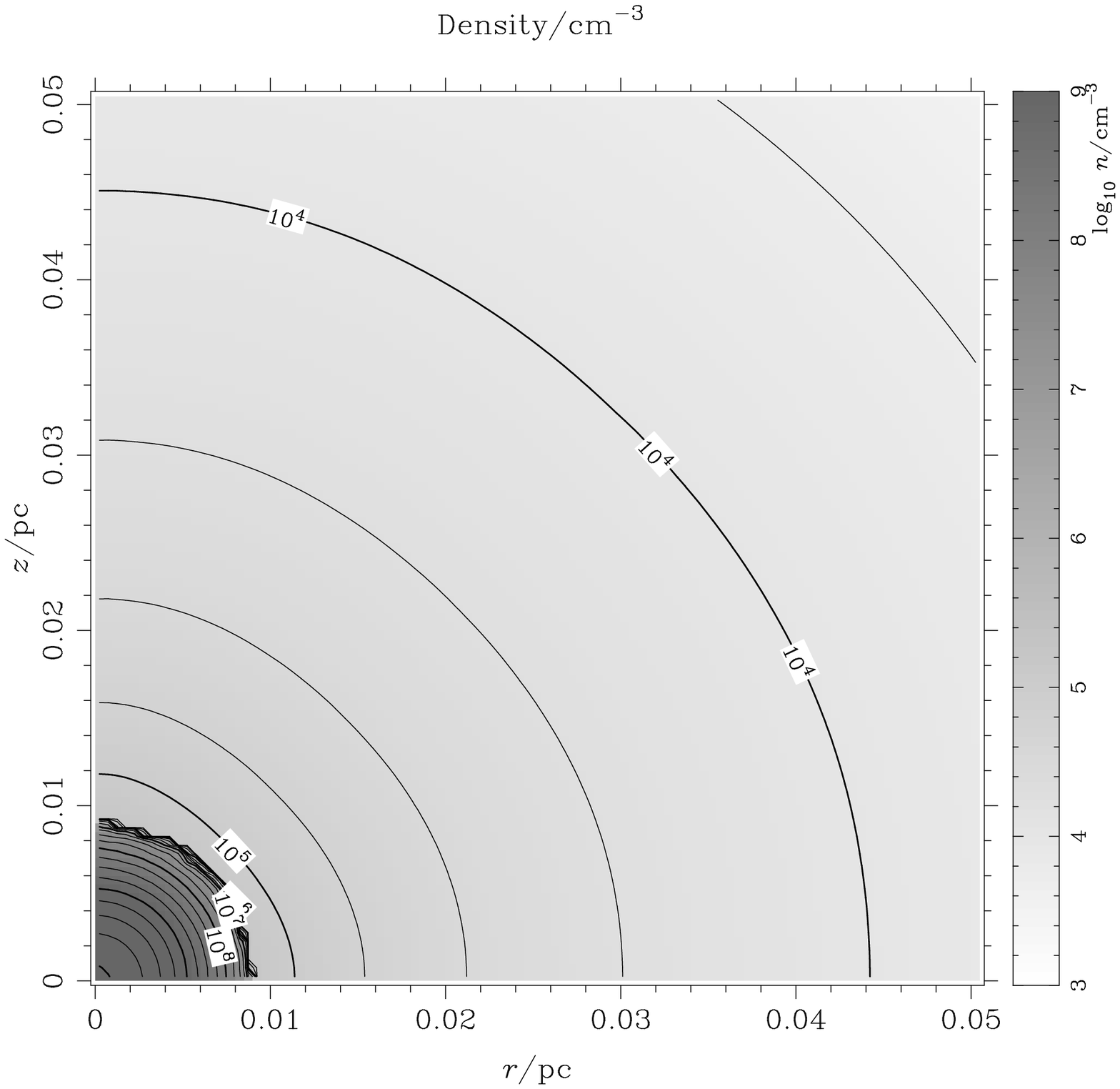} &
\epsfbox{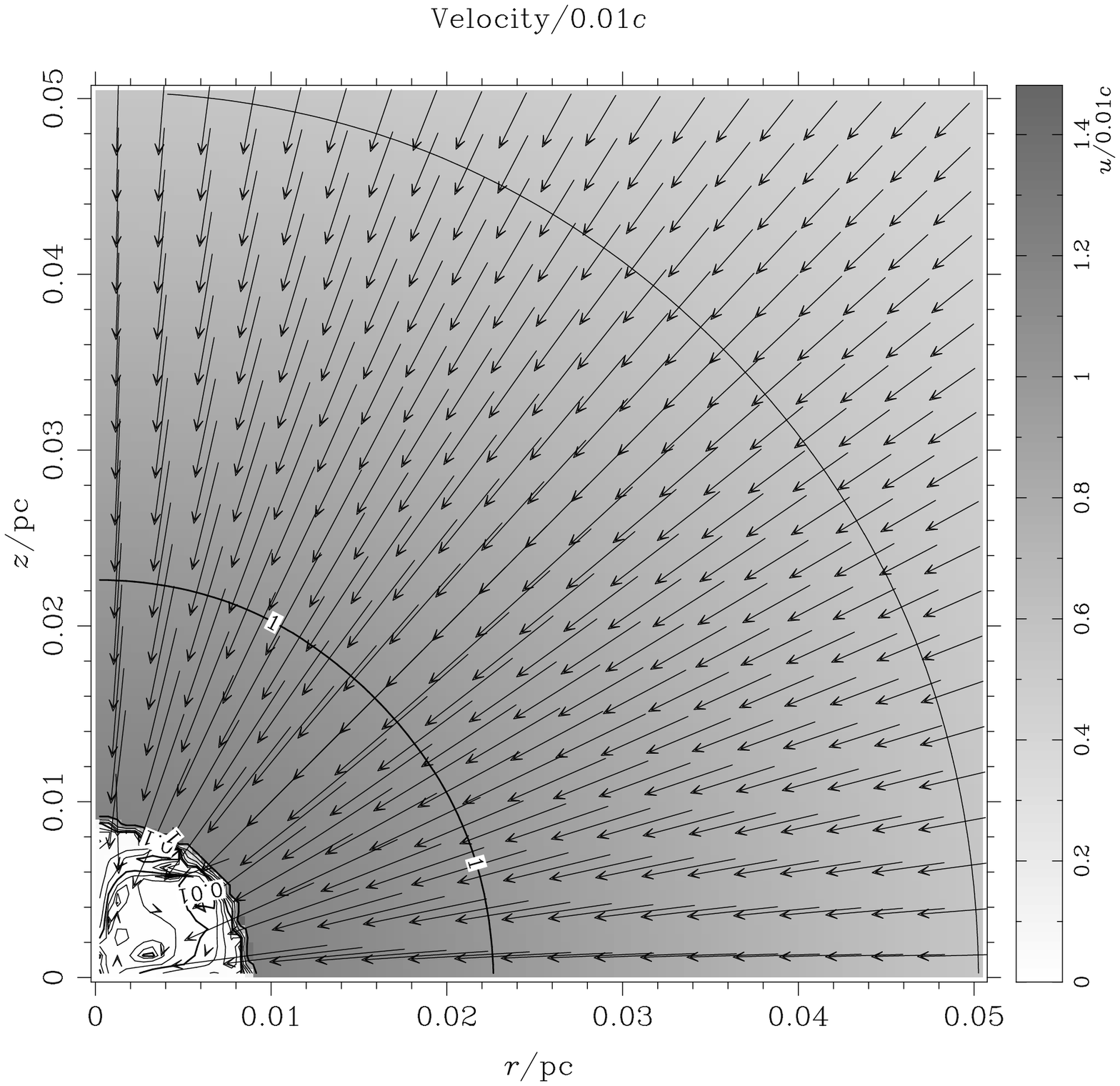} \\
\end{tabular}
\end{centering}
\caption[Model SN 2D plot]{Model Sn at $4.2\ee3\yr$.  Notice the
smaller scale of the grid in this simulation: the flow reaches
equilibrium as early as $200\yr$ as a result of the very short
dynamical timescale in this model.  The panels show a) density (from
15 to $1\ee10\cm^{-3}$) b) total velocity (greyscale and contours, up
to $1.5$), c), d) Magnifications of the central region.  The vectors
are in the flow direction, and have lengths proportional to the
velocity, sampled at one vector in every $15\times15$ cells in b),
every $5\times5$ in d).  Gas flows freely in from the cluster and is
accreted at the centre of the grid.  The anisotropic radiation, with a
mean Eddington ratio of only $0.01$, produces no noticeable
perturbations on this flow.}
\label{f:modelsn2d}
\end{figure*}

\label{s:seyfert}
The models described so far in this section were for bright QSOs.  We
now consider the effects of changes in parameters to values
characteristics of Seyfert nuclei.

The similarity in line widths and profiles between AGN of widely
differing luminosities can be interpreted as implying that the
kinematics are similar.  If this is so, in Seyfert nuclei the
characteristic length and timescales must be $\lambda \simeq \tau \simeq
0.01$, relative to the QSO case.  If the Eddington ratio, etc., remain
constant then such a self-consistent Seyfert model would have masses
smaller than those of QSOs by $\lambda$, and densities increased by
$1/\lambda$.  The ionization parameter remains constant so long as $Q$
is determined to within a small factor by stellar physics.  In this
case, the ratio between cooling and flow times is constant, and the
models scale directly from those presented above (see also the next
subsection).

However, many authors find black holes masses in Seyferts $\sim
10^8\Msun$ (as might be expected in an old QSO brought back to life),
requiring that accretion is sub-Eddington \cite{pbe,lasea96}.
Structures similar to those shown in the preceding subsections would
persist if the efficiency of outward radiation driving were increased
(\cf\ Section~\ref{s:eddeff}).  

However, if there is no such enhanced driving the topology of the
flows changes radically, as we have discussed from scaling arguments
in Section~\ref{s:scalevedd}.  In this subsection, we present
numerical results for the case in which there is no increased driving.
We consider a black hole mass of $10^8\Msun$, but a cluster mass of
only $10^6\Msun$, sufficient to power accretion at luminosity
$10^{44}\ergs$, \ie\ about $0.01L|{Edd}$.  (We shall see that the form
of the flow is similar to that for a low mass-loss cluster of
$10^8\Msun$, as might be expected in a `fossil QSO' nucleus.) In order
that stellar collisions do not dominate the mass input, the core
radius of the cluster must be larger than $0.07\parsec$ (note that in
this case the stellar velocities will be dominated by the mass of the
black hole).  This core radius is an order of magnitude greater than
variability-based estimates of the size of the BELR (\eg\ Peterson
1994), which suggests that the BELR lies entirely within the cluster
core.

We calculated two models, with and without central accretion
(Figs.~\ref{f:modelsa2d} and \ref{f:modelsn2d}, respectively).  In
both, we found near-spherical inflows of the nISM throughout the
nucleus, except for a small hydrostatic core in model Sn.

These low Eddington ratio models do not show the bipolar structures
characteristic of the QSO models, and as observed in radio maps of
Seyfert galaxies.  This is because the accretion disc luminosity is
only a fraction of the (insignificant) radiation force.  Spherical
inflow will characterise Seyferts with $f|{Edd}\ll 1$ if the opacity
is dominated by Thomson scattering. Observations imply that other
sources of opacity will produce bipolar flows in Seyferts. The
signature of this radial inflow should be detectable in the
variability of line profiles, with the red wings of the lines leading
the blue.  For the smaller nuclei observed in Seyfert galaxies,
individual supernovae will often evacuate the entire nuclear ISM
\cite{pwd98} returning the flow to initial conditions similar to those
we have assumed for our simulations.

\section{Discussion: The Properties of Flows in Active Nuclei}
\label{s:discuss}

We have modelled the central region of an AGN, assuming only that a
black hole and accretion disc with an aspherical radiation field is
embedded in a young starburst stellar cluster.  These two simple
components give rise, symbiotically, to a nuclear interstellar medium
with a remarkable and fascinating variety of structures.  Despite the
absence, in our models, of any effects due to radio-related phenomena,
the nISM shows a range of behaviour normally associated with
relativistic effects near the black hole: episodic, explosive,
percolating structures, and/or well collimated, hypersonic hydrodynamic
jets -- given only that the radiation field drives an outward flow along
the axis of the disc.  When the Keplerian velocities of the stars,
$v|K$, are $\gg c|s$, the black hole drives a meridional flow within
the previous hydrostatic core.  The mass-loaded outflow from the core
is either confined by an (apparently very unstable) internal
termination shock or can drive gas beyond the cluster core, in episodic
explosions or as a continuous wind.  For smaller $v|K/c|s$, confined
jets can form on-axis.  In all cases where a wind drives
out through the nucleus, it has a half-opening angle which is rather
smaller than that of the (conical) region close to the centre within
which the net force is outwards (typically 30 degrees compared to 60
degrees).

The models we present here are distinguished from previous studies by
the effects of the stellar cluster and by the nature of the flow at
small radii.  The poorly collimated central wind from the very centre
of the grid drives outwards supersonically into the cluster.  When
this outflow has sufficiently great momentum, it blows straight
through the cluster and onwards into the galaxy ISM: such flows soon
settle to steady equilibria.  For lower momentum flows, the outflow
shocks against the inflowing nISM of the cluster, and falls back to
the core in a circulation.  In some cases, a series of explosions are
found; rather than buoyancy-driven bubbles, these are pushed outwards
by the recycling of matter through the black hole core.  The timescale
for these explosions is the dynamical timescale of the nuclei, which
will vary between roughly $300\yr$ in QSOs and $3\yr$ in Seyfert
galaxies.  These times, too long to explain continuum variability, may
relate to variations in the underlying structure of line-emitting gas
\cite{pvgw94,wanp96}.

Gas which rains onto the disc in many of our models may also suppress
the wind from some part of its surface.  If the gas in this region can
cool, it will irradiate the outer disc and may have important
ramifications for angular momentum transport in the disc.

The structures we predict for the flows are very sensitive to the
central luminosity, if it is close to the (on-axis) Eddington limit.
The boundary between stagnated inflow and free, steady outflow is
fine, depending on whether transient explosions can drive significant
mass from the nucleus.  Indeed, there seem to be bimodal solutions in
some cases, where the flow can take either a low-mass, free-outflow or
high-mass, recirculating form.  Quite small changes in luminosity may
lead to ejection of large masses of nISM gas within the dynamical
timescale of the nuclear flow.  In a fully self-consistent model,
variations may be even more catastrophic than those presented here
(although these variations may be damped by lags due to viscous
transport within the accretion disc).

In many cases the time-steady system does not lead to structures in
the nISM which are time steady, unlike the models of the wind from the
surface of a molecular torus by Balsara \& Krolik \shortcite{bals93}.
Our results indicate that the flows may remain chaotic for all time,
with shocks forming an important part of the structure.  In the models
of Balsara \& Krolik, outflows are driven from scales $\ga 1\parsec$
-- the flow within the BELR is purely accretion -- and have
significantly lower velocities than those we find in our models.

This is also true of the Compton-heated disc winds modelled by Woods
\etal~\shortcite{woodea96}.  In their very detailed treatment, they
found low-level variability in the flows -- it is our eventual aim to
model the thermal structure of the flow in similar detail, but cooling
of gas behind global shocks will make this problem even more
computationally taxing than the cases calculated by Woods \etal{} 

We now consider the physical properties of the nISM flows found
above. We first discuss the free parameters which are most important
in determining the flow structures, and the physical limits of the
models.  Next, we derived the dynamic pressures and ionization state
of the flowing gas, especially in Section~\ref{ss:pressures}, where we
also consider the interaction of the nISM with the shocks driven by
stars and supernovae.  We calculate the overall mass budget of the
nucleus in Section~\ref{ss:massbudge} and the mechanical energy
carried in the winds and their effects in
Section~\ref{s:windpow}. Throughout this section, we highlight the
observational consequences of the model.

\subsection{Classification Scheme}
\label{s:disclasify}

Comparison of the flow structures shown in the previous section with
the sketched structures in Fig.~\ref{f:xxx} illustrates our suggested
classification in terms of the first two parameters.  In terms of the
physical parameters of the nuclei, the Keplerian velocity is $v|K^2
\equiv GM|c/r|c$ and the sound speed is determined by the shape of the
AGN continuum.  We can estimate the ejection velocity, $v|{ej}$, by
assuming that the central outflow is dominated by gas accelerated in
the $z$ direction within the central smoothing region, so
\begin{equation}
{v|{ej}\over c|s} \simeq
{v_z\over c|s} \equiv \left\{\left[1-f|{Edd,0}\right] {G
M|h\over\epsilon c|s^2}\right\}^{1/2}.\label{e:vz}
\end{equation}
The comparison between the calculated $v_z$ and the measured $v|{ej}$ is
shown in Fig.~\ref{f:xxx1}.  The comparison is good except for the
cases (described in the figure caption) where the approximations made
in applying eq.~\refeq{e:vz} break down.

Where the outward driving from the black hole is large, the flow
relaxes to a steady bipolar form: gas falls inwards in the plane of
the disc and is then driven away when reaches the super-Eddington
region on-axis, close to the central black hole.  When the retarding
effects of cluster gravity and mass loading become great enough, the
outflow shocks within the stellar cluster.  If the sound speed in the
hot gas is sufficiently great, a steady recollimating jet structure
can be produced.  More characteristic, however, is a structure where
episodic explosions drive gas from the nucleus in bursts.  As the
retarding forces increase further, the outflow region becomes more
compact and can less often drive mass out from the nucleus, eventually
becoming a turbulent core in the flow onto which gas injected in the
outer regions of the stellar cluster accretes.

For our approximations to hold, we require that the forces due to the
black hole dominate close to it, but that the gravity of the cluster
dominates at large radii (this assumption is strained in models \modE\
and \modG). Various other parameters come into play; although they are
less important, they limit the strict application of our
classification scheme.  Changes in the opening angle of the outflow
(Model \modD) and the distribution of stars within the cluster will
have only a small effect.  Accretion of gas from the flow on a
dynamical timescale can also serve to weaken the central outflow
(models \modC$_i$).  In all these cases, the morphology of the flows
are still broadly within our classification, and analytic estimates of
flow velocities and densities are still readily calculable.The net
effect of accretion on the flow structures is to weaken the winds and
dampen the variability, although this depends on where the accreted
mass is removed from the global flow.  Changing this region does not
fundamentally alter the structure of the flows until the fraction of
the mass accreted exceeds $\sim 0.5$.  The overall fraction is more
important in determining the flow structure than the location of the
sink.  The structures are basically similar to those found in
non-accreting cases, albeit with different black hole parameters
(cases in which there is a central wind blowing through a mass-loading
cluster will again be similar, so long as this wind does not have a
high angular momentum).

This classification should be supplemented by the number of dynamical
times for which mass is retained by the nucleus (which will often
relax to a constant value).  Most importantly,
the gas may not remain in Compton equilibrium, in which case the
scalings we have derived will have to be recalibrated in terms of an
`effective' Eddington ratio.

\subsection{Densities and velocities}

The models we have calculated imply a range of local properties
(densities, pressures, velocities) for the flows.  We now compare
these with simple estimates, based on spherically symmetric flows,
similar to those often given in the literature.  Conservation of mass
in spherical symmetry gives a density for the global flow (on scales
of $r = r|{pc}\parsec$) of
\begin{equation}
n \simeq 4\ee3 
\left(f|{Edd}\over\etaflow\eta|{acc,-1}\right) 
{M|{h,8}\over r|{pc}^2(v|{flow}/0.01c)} \cm^{-3}
\label{e:pdn}
\end{equation}
where $v$ is a characteristic flow velocity within the nucleus (by
comparison, for the sound speed, $v=10^{-3} T_7^{1/2} c$ and for the
Keplerian velocity, $v\simeq 7\ee{-3}(M|{cl,8}/10 r|{pc})^{1/2}c$).

This estimate of density is in line with the values seen in the
accretion flows in, \eg, model~\modC.  The flow velocity in the inflow
region is about $3\ee{-3}c$, similar to the Keplerian velocity, and
the density in this region corresponds to equation~\refeq{e:pdn} to
reasonable accuracy.
In the axial outflow region, the velocity is about 3 times higher
(corresponding to non-Keplerian velocities that are obviously required
by the widths of the observed optical emission lines).  Here, the
density is {\it higher}\/ than in the inflow region, with a value more
than ten times larger than suggested by equation~\refeq{e:pdmcl}.
This is because the 60 per cent of the mass input not accreted by the
black hole flows out in a cone of half opening angle of only $\sim
15^\circ$, (3.5 per cent of the solid angle) rather than the large
fraction assumed in deriving the approximate result.

As the outflows are driven from very close to the central black hole,
their velocities are characteristic of these radii.  While the highest
speeds in the models presented here are $\sim 8700\kms$, this value
can be rescaled by choosing different flow parameters, and might be
significantly increased if the effective smoothing length was lower
than that assumed.  The physical processes which determine the
smoothing length are discussed in Section~\ref{s:smooth}.  It is
possible that values as small as $\sim 3\ee{15}\cm$ may be appropriate
in some cases, rather than the values of $\sim 1.5\ee{17}\cm$ we used
principally because of computational necessity.  Nuclei with very high
outflow velocities are, however, likely to have structures similar to
our model~\modC, with little retained gas and few global shocks.

If mass is retained by the nucleus for $N$ dynamical times, as for
instance in models~\modA\ and~\modB, the density and pressure
[equations~\refeq{e:pdn} and \refeq{e:pdp}] will be boosted by a
factor $N$, and the ionization parameter [equation~\refeq{e:pdx}]
decreased by a similar factor, within the central convective
structure.  Where this occurs, this will increase the emission from
gas in the central plume beyond that, retention occurs.

Therefore, we find that the simple arguments of PD and Perry (1993a)
are borne out throughout much of the nucleus.  However, they did not
anticipate several important effects which we have found boost the gas
density in some regions.  These are the very regions which may
dominate the line emission from these nuclei (as we have seen in the
discussion of Fig.~\ref{f:modela2dv} above).

\subsection{Stagnation pressures, ionization parameters and the
multi-component BELR}
\label{ss:pressures}

\begin{figure*}
\epsfxsize = 8cm
\begin{centering}
\begin{tabular}{ll}
a) & b) \\
\epsfbox{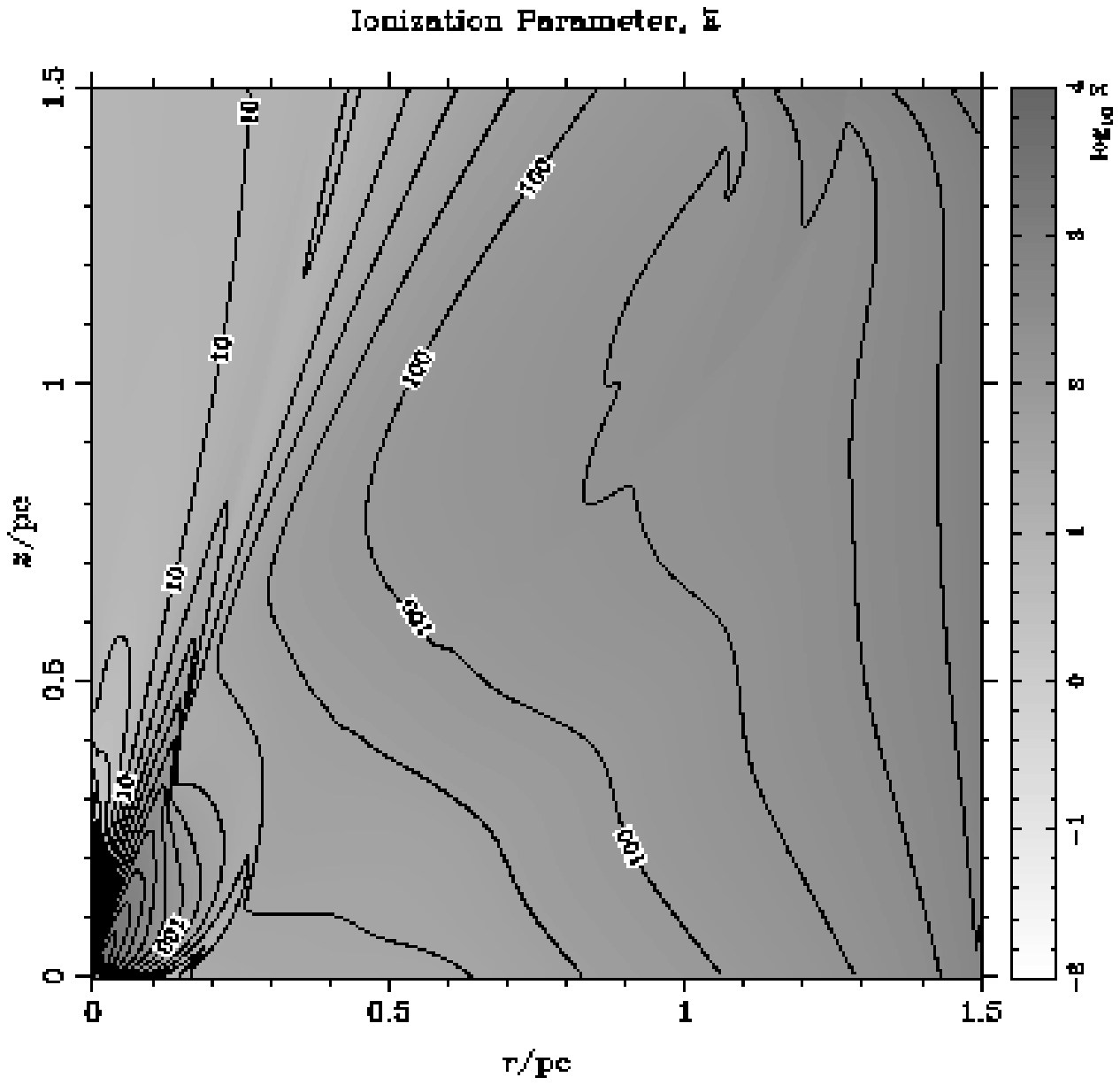} &
\epsfbox{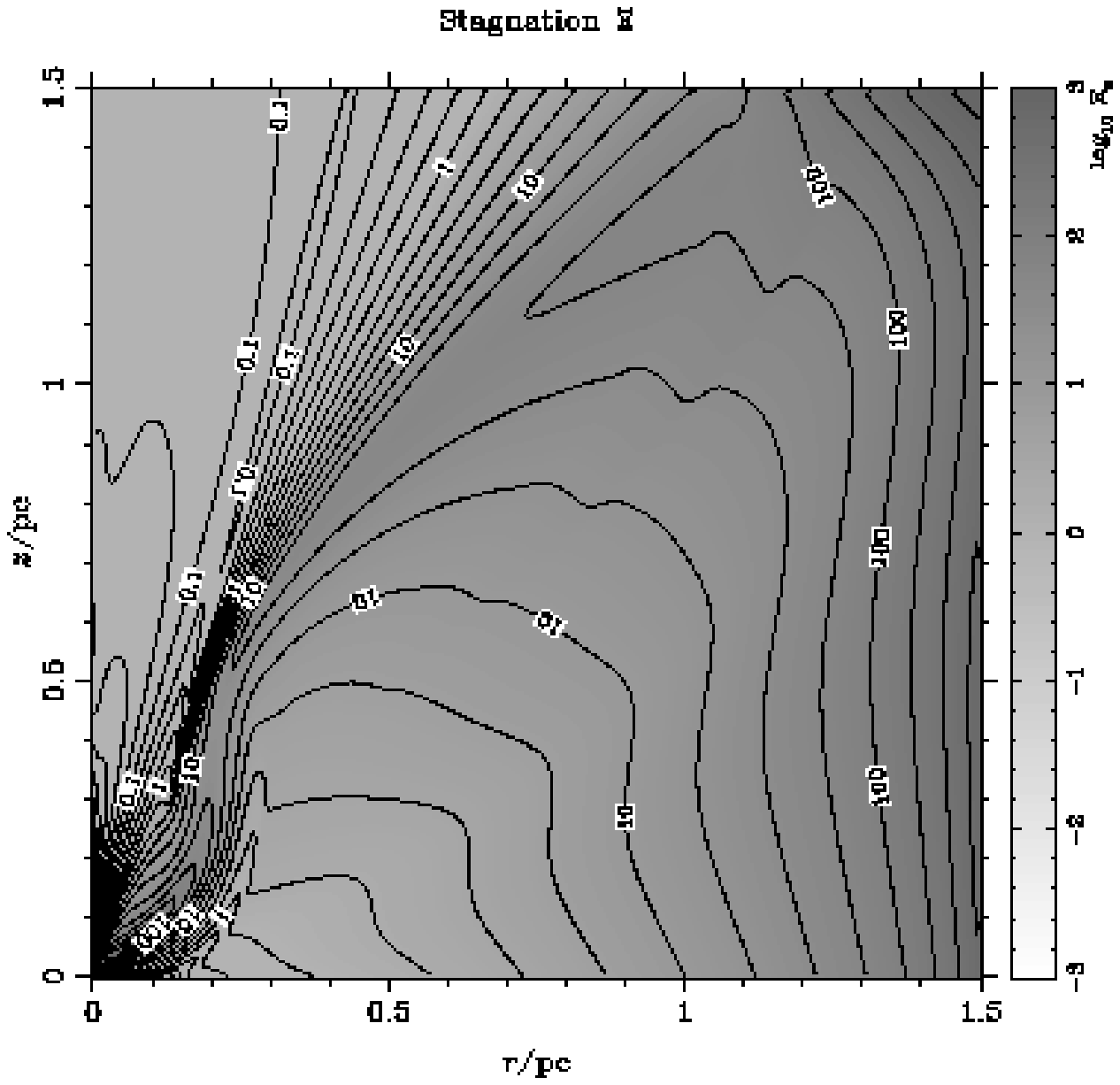} \\
c) & d) \\
\epsfbox{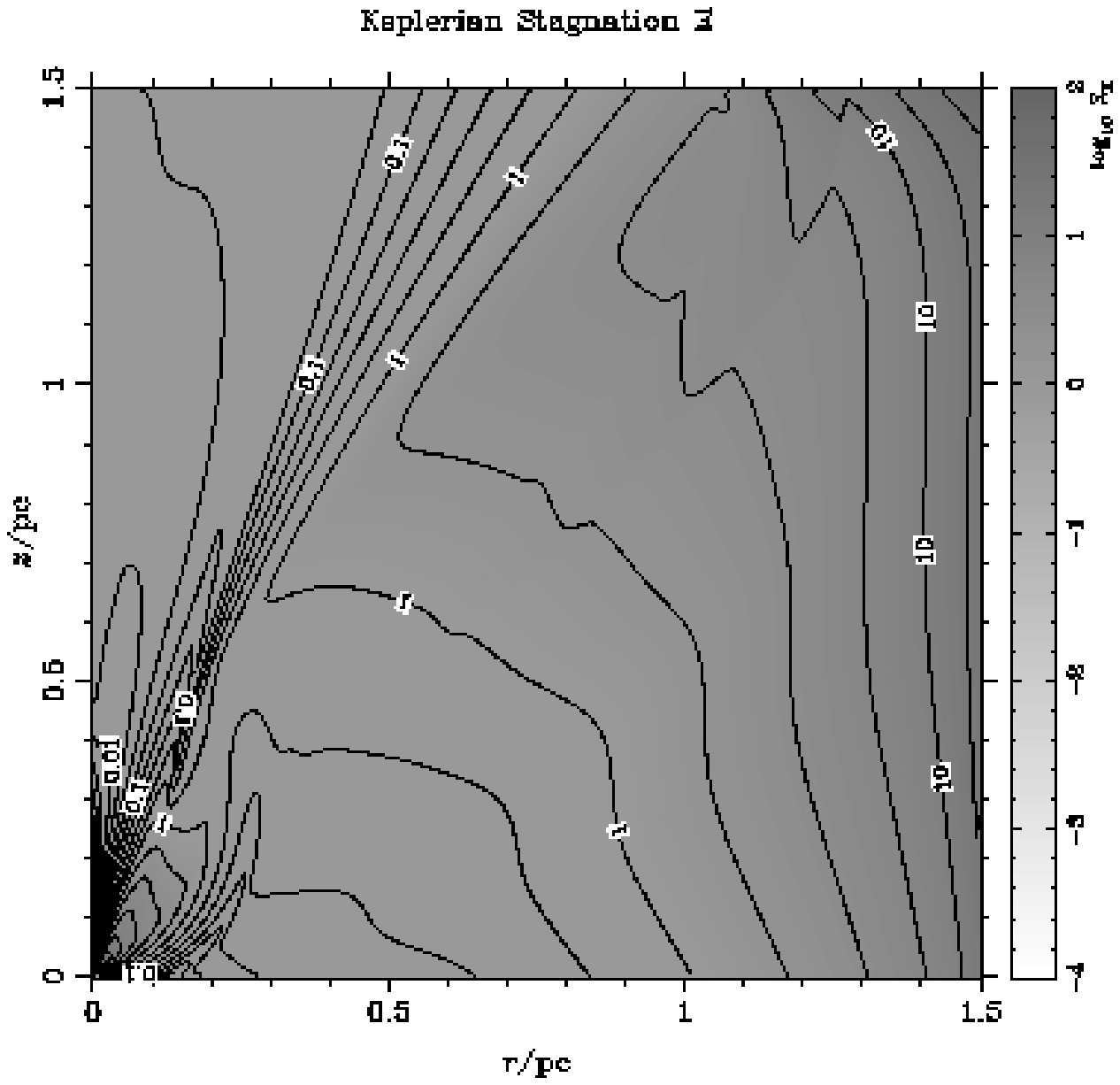} &
\epsfbox{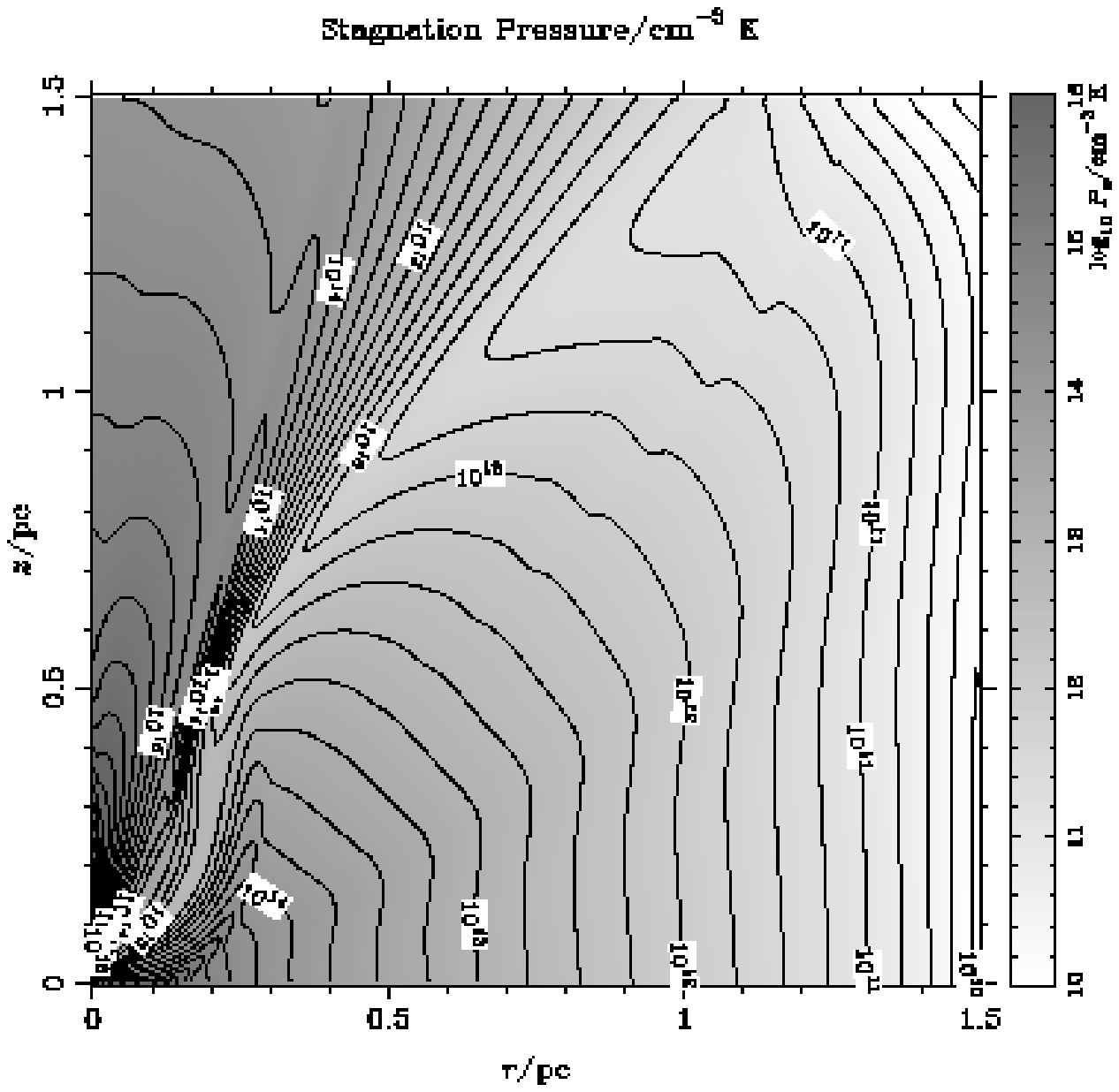} \\
\end{tabular}
\end{centering}
\caption[Model C 2D various plot]{Model \modC\ at $2.6\ee4\yr$.  The
panels show a) ionization parameter, $\Xi$, in the free flow, b)
ionization parameter, $\Xi|s$, at stagnation in the rest frame of the
nucleus, c) `minimum' stagnation ionization parameter, $\Xi|k$, at
stagnation against a counter-moving Keplerian obstacle, d) gas
pressure, $p|s$, at stagnation in the rest frame of the
nucleus (in $\!\cm^{-3}\Kelv$).  }
\label{f:modelc2dv}
\end{figure*}

\begin{figure*}
\epsfxsize = 8cm
\begin{centering}
\begin{tabular}{ll}
a) & b) \\
\epsfbox{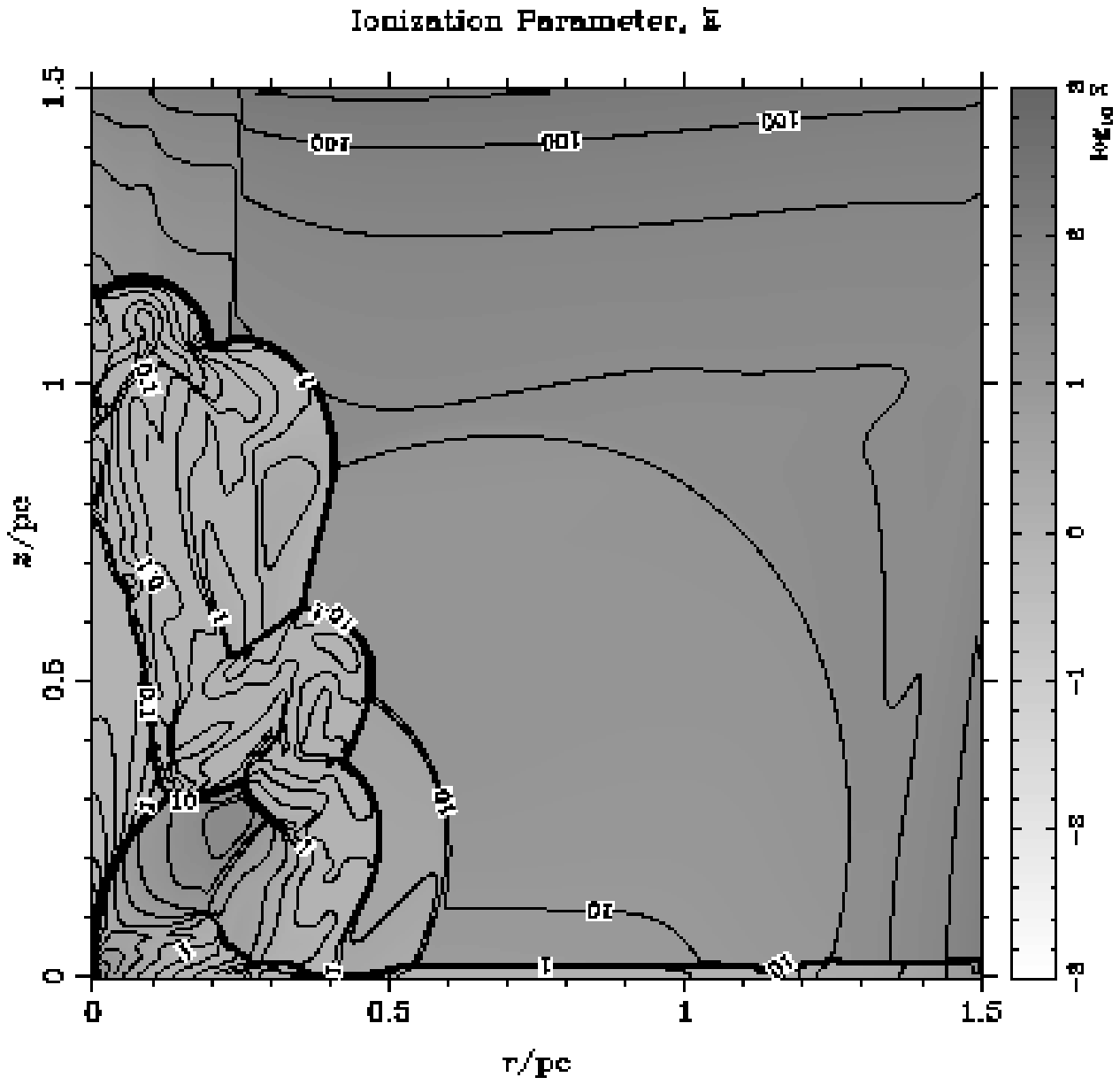} &
\epsfbox{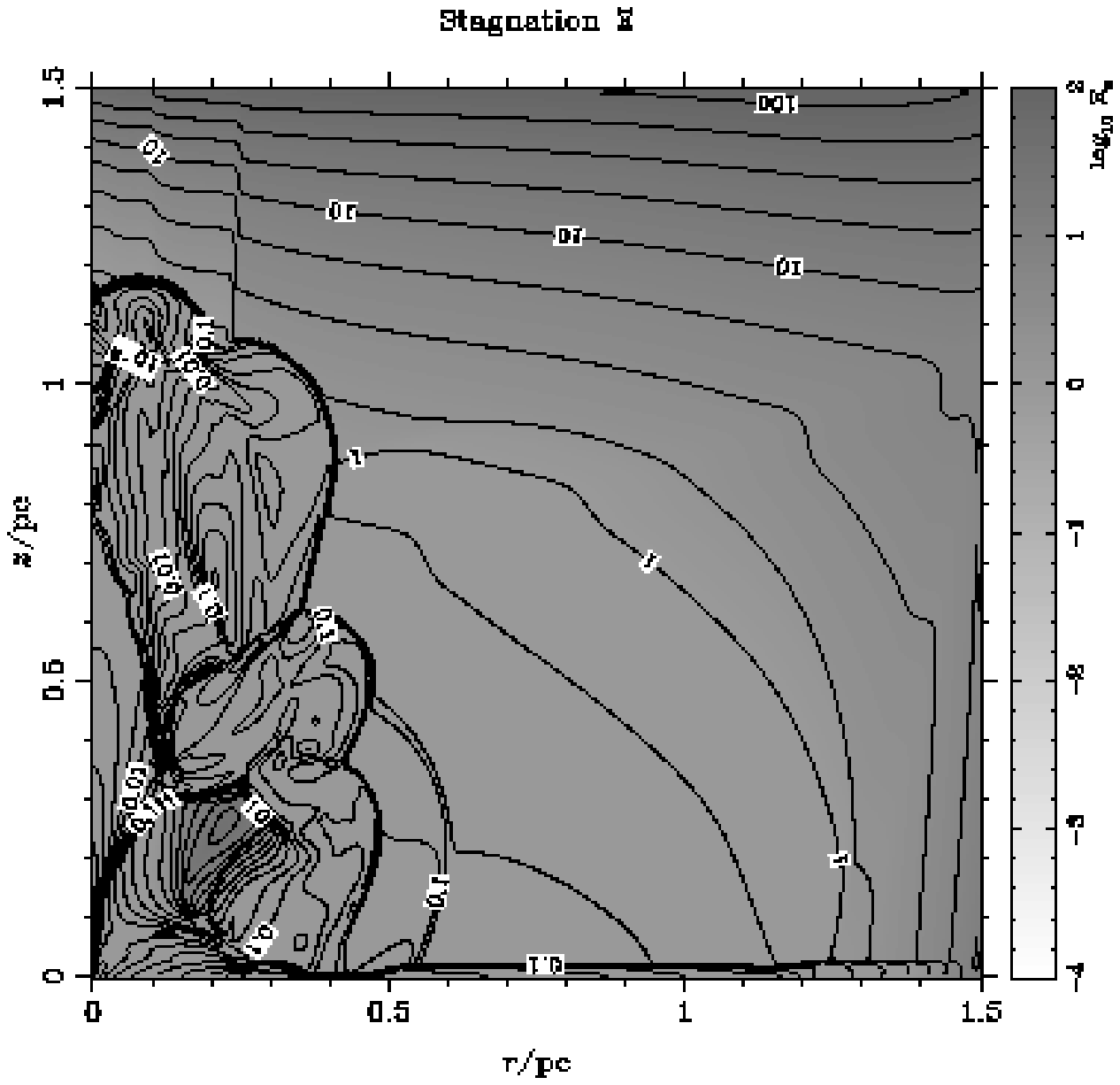} \\
c) &d) \\
\epsfbox{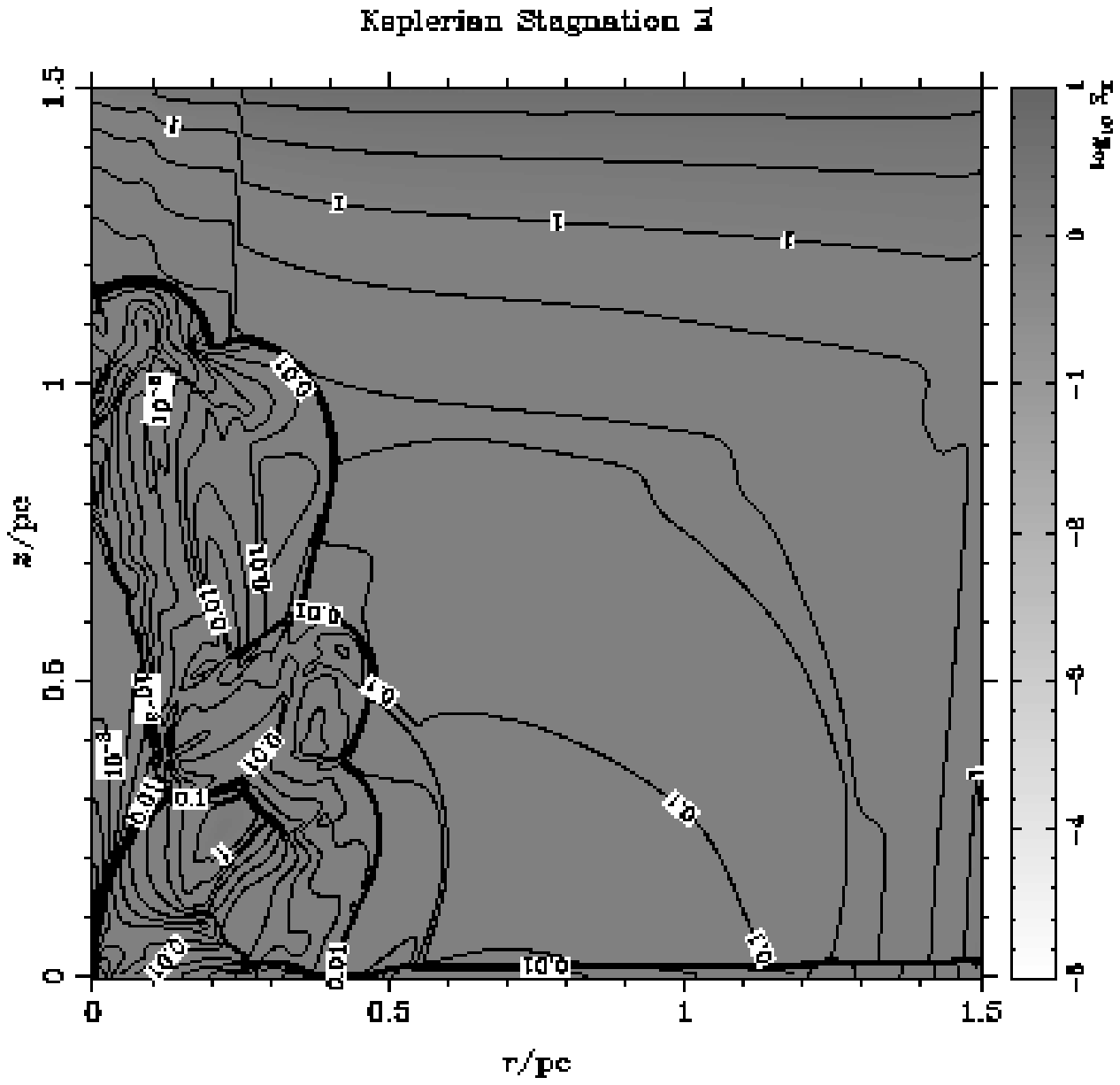} &
\epsfbox{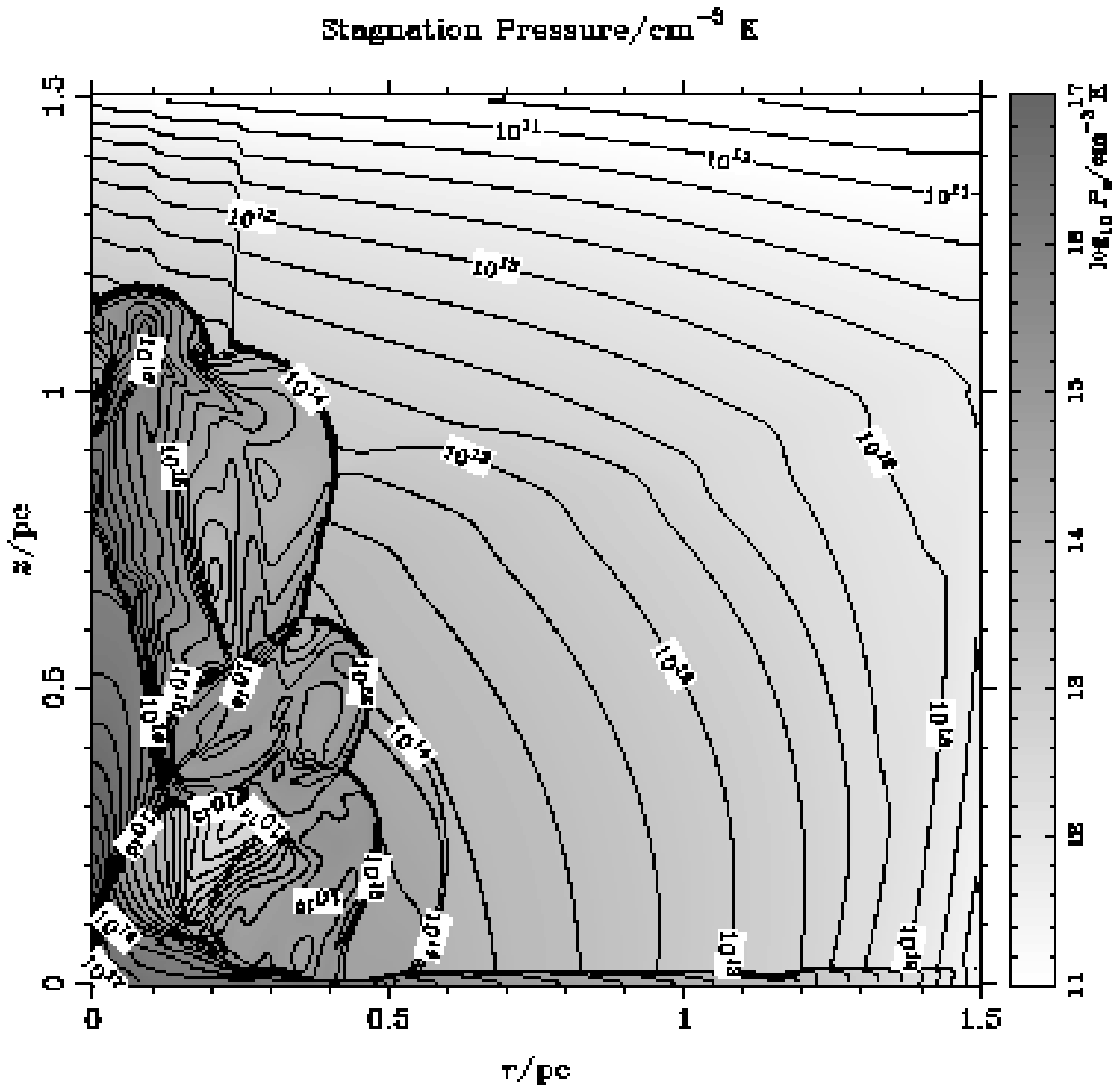} \\
\end{tabular}
\end{centering}
\caption[Model A 2D various plot]{Model \modA\ at $6\ee4\yr$.  The
panels show a) ionization parameter, $\Xi$, in the free flow, b)
ionization parameter, $\Xi|s$, at stagnation in the rest frame of the
nucleus, c) `minimum' stagnation ionization parameter, $\Xi|k$, d) gas
pressure, $p|s$, at stagnation in the rest frame of the nucleus (in
$\!\cm^{-3}\Kelv$).  }
\label{f:modela2dv}
\end{figure*}

\begin{figure*}
\epsfxsize = 8cm
\begin{centering}
\begin{tabular}{ll}
a) & b) \\
\epsfbox{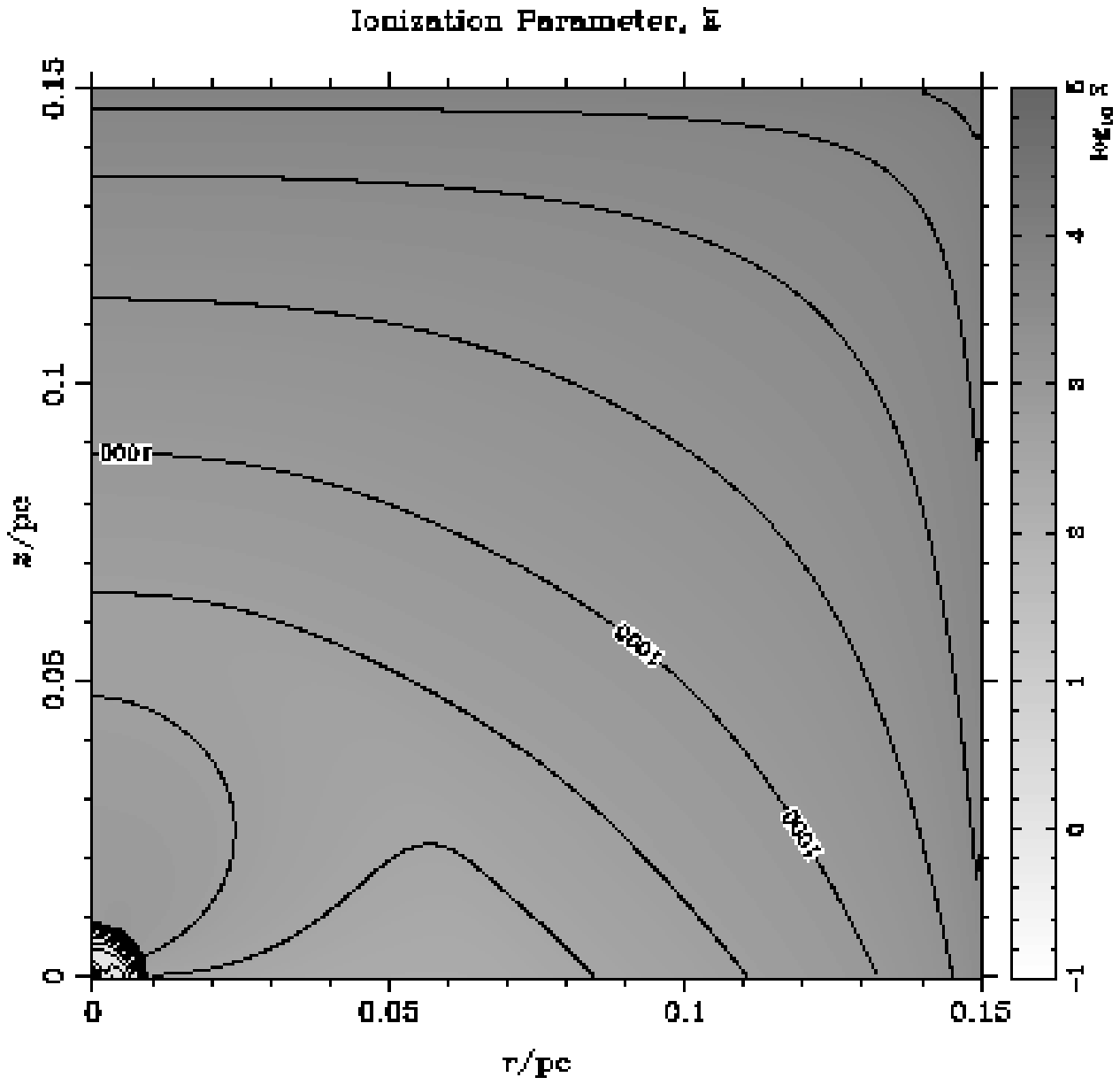} &
\epsfbox{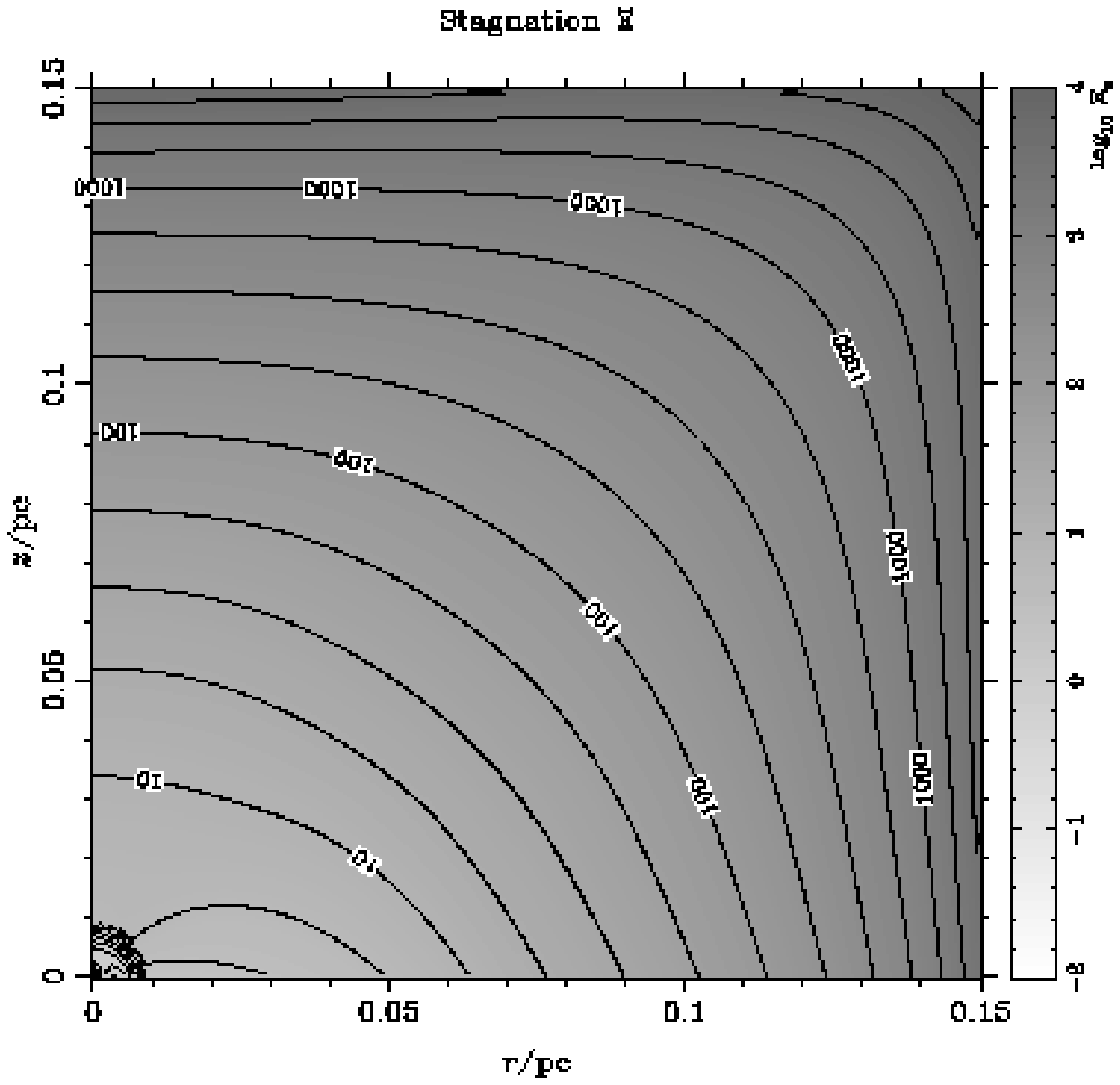} \\
c) & d) \\
\epsfbox{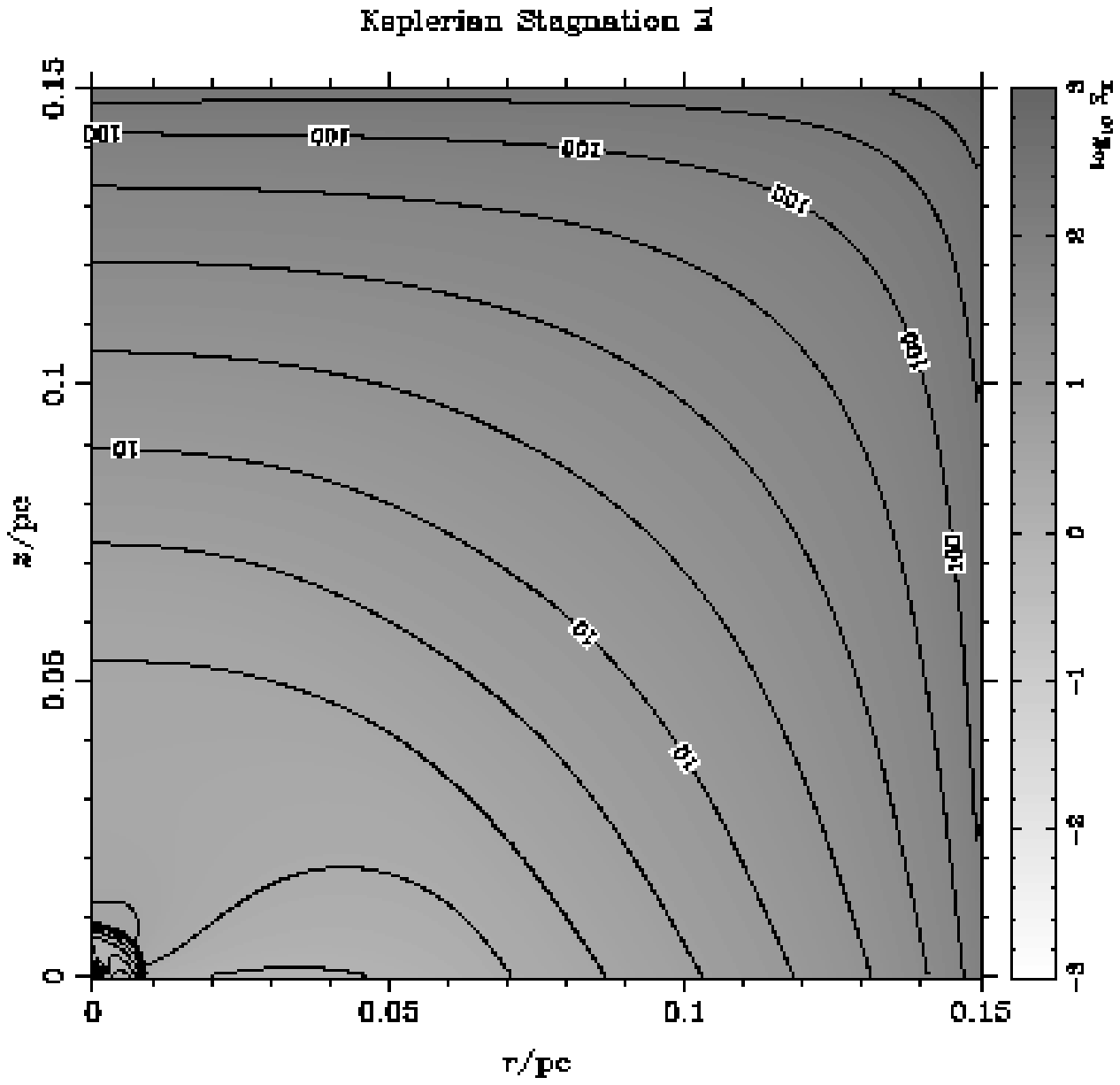} &
\epsfbox{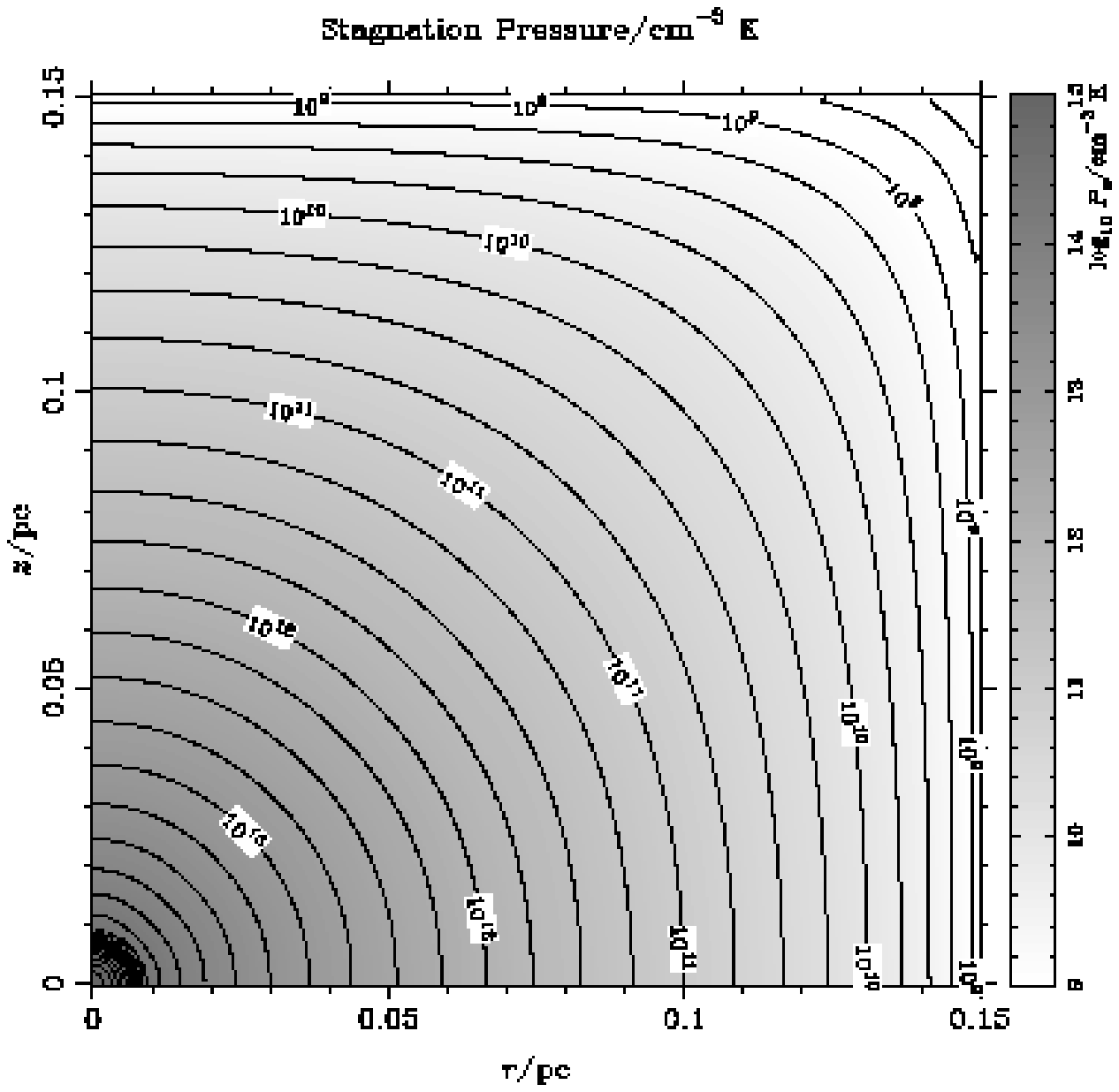} \\
\end{tabular}
\end{centering}
\caption[Model Sn 2D various plot]{Model Sn at $4.2\ee3\yr$.  The
panels of this plot show a) ionization parameter, $\Xi$, in the free
flow, b) ionization parameter, $\Xi|s$, at stagnation in the rest
frame of the nucleus, c) `maximum' stagnation ionization parameter,
$\Xi|k$, d) gas pressure, $p|s$, at stagnation in the rest frame of
the nucleus (in $\!\cm^{-3}\Kelv$).  The anisotropic form of the
ionization parameter plots results from the variation of nuclear
luminosity and spectrum with angle to the nucleus. }
\label{f:modelsn2dv}
\end{figure*}

Both the ionization parameter and the stagnation pressures within the
nISM provide checks on the consistency of our models with our
assumptions, and are first steps towards assessing the observational
consequences of the symbiotic starburst-black hole model for AGN\@.
In this section, we present results for the ionization parameter in
the freely flowing nISM, and the ionization parameter and gas pressure
at stagnation in the nISM, for two characteristic shock velocities.

The ionization parameter we use is that as defined by \shortcite[Krolik
\etal{} ]{kmt} 
\begin{equation}
\Xi = {F|i\over n|H k|B T c},
\label{e:xi}
\end{equation}
where $F|i$ is the local radiation flux between $1$ and
$1000\unit{Ryd}$.  This choice of ionization parameter is that which is
most appropriate for discussions of the interplay between hydrodynamics
and photoionization equilibrium. Collin-Souffrin (1993) discusses the
relationship between the various ionization parameters in common use.
We plot $\Xi$ based on the local intensity of radiation, assuming that
the ionizing flux is a constant fraction of the total flux.

PD and Perry (1993a) emphasized the importance of shocks in the flows
in AGN.  A simple estimate of the pressure on cool gas behind a
stationary shock, \ie\ the stagnation pressure of the flow at $v|s$,
is
\begin{equation}
p|s \simeq 4\ee{11} 
\left(f|{Edd}\over\etaflow\eta|{acc,-1}\right) 
{M|{h,8}\over r|{pc}^2} {v|s^2\over0.01cv|{flow}}
\cm^{-3}\Kelv
\label{e:pdp}
\end{equation}
and the ionization parameter of shocked gas is
\begin{equation}
\Xi|s \simeq 0.1 \eta|{acc,-1}\etaflow{c v|{flow}\over v|s^2}.
\label{e:pdx}
\end{equation}

Hot phase gas will begin to cool when its ionization parameter falls below
that at the knee of the equilibrium curve, 
\begin{equation} 
\Xi|{knee} \simeq 100 {L|i\over L|{Bol}} T|{C,7}^{-3/2}, \label{e:knee}
\end{equation}
where bremsstrahlung cooling begins to dominate over Comptonization.
For a recent composite spectrum, \cite{hazy} $L|i/L|{Bol} = 0.45$ and
$T|{C,7} = 1.3$.

While this limit is consistent with our choice of constant Compton
temperature in the flows we analysed, the anisotropic component of the
radiation field may well have a different spectrum to the isotropic
one.  For example, for the outer regions of the disc plane, it might
be more appropriate to consider only the luminosity under the blue
bump. In the case of Ferland's continuum, the luminosity in the
$1\mbox{--}10^3\unit{Ryd}$ wave-band then decreases by a factor more
than 5 with a concomitant decrease in $\Xi$ (and, as discussed below,
the Compton temperature increases by a factor 2).  These gross changes
in spectral shape may have considerably more far-reaching effects than
simply a change in the ionization parameter.  The effects of
variations in spectrum and Compton temperature with angle will be
discussed in detail in future papers.

If $\Xi|f \la 10$, our assumption of Compton equilibrium for the hot gas
will be strained by the increasing influence of bremsstrahlung and
atomic cooling.   Conservatively, we take $\Xi|f = 10$ as a limit below
which cooling is likely to become important.

The pressure in any gas which manages to cool will rarely be less than
that in the surrounding free flow, and so the free-flow ionization
parameter $\Xi|f$ is an upper limit to that which acts on the cool gas.
This free-flow $\Xi|f$ is shown in panel (a) of the graphs in this section.  

In panel (b) of the graphs in this section, we show the ionization
parameter $\Xi|s$ in gas shocked to stagnation in the rest frame of
the nucleus.  This ionization parameter characterises the `average'
ionization parameter in shocks which form around local obstacles to
the flow, as well as in steady features within the flow.

In reality, most of the obstacles in the flow will be driven by stars
and supernovae.  The speed of shocks driven by these stars will depend
on their orbital velocities, mass ejection velocities and the amount of
gas swept up by the shell, as discussed extensively by Perry (1992,
1993a,b, 1999).  As an indication of the likely importance of these
random velocities, we also plot ionization parameters behind shocks of
velocity $v|{flow}+v|{K}$, where $v|{K}$ is the local velocity of a
circular orbit in the potential of black hole and cluster.  This is the
maximum shock velocity for a rigid obstacle.  Note that in the inner
regions of the flow, where the gravity is dominated by the black hole,
the Keplerian velocities depend on the absolute mass of the central
black hole (and thus the degeneracy between assumed values of $\langle
f|{Edd}\rangle$ and $M|h$ is lifted).

In Fig.~\ref{f:modelc2dv}a we show the ionization parameter, $\Xi|f$,
for Model \modC.  As $\Xi|f>10$ through most of this region, our
assumption that the flow is isothermal at the Compton temperature is
self-consistent, except for a very small region close to the origin.
There are no direct observational tracers of gas with this high
temperature and ionization parameter.  Even iron K-shell absorption,
one component of the so-called `warm absorbers', which can be
significant in gas with temperatures as high as $10^7\Kelv$, will only
be important in gas with typical densities $\ga 10^7\cm^{-3}$ for the
radiation flux assumed here \cite{matfer87}. 

While a wide range of densities and pressures are predicted by these
models \cite[\cf{} empirical evidence for this,]{loc}, broad
systematic trends are also expected.  For instance, in Model \modC,
there are two regions where the gas shocked to typical stagnation
pressures can form line-emitting clouds. In the rest of the flow,
post-shock gas has high ionization parameters, and so does not cool
sufficiently to radiate emission lines (Fig.~\ref{f:modelc2dv}b).  The
two line-emitting regions are near the outflow cone, and just above
the disc. The stagnation ionization parameter is $\la 0.1$ through
most of the conical outflow around the axis.  The pressures here are
high ($\ga 10^{14}\cm^{-3}\Kelv$, Fig.~\ref{f:modelc2dv}d).  Above the
accretion disc, in the other low ionization parameter region,
inflowing gas shocks to a rather smaller pressure, $\sim
10^{13}\cm^{-3}\Kelv$.  The relative contributions of these two
regions to the observed broad emission lines will be weighted by their
relative volumes of rotation about the $z$-axis.  Many physical
effects other than the overall pressure and ionization parameter must
be included in determining the volume emissivity of the cool gas from
hot phase properties (\cf\ PD, and below).  However, the obvious
distinction between the properties of gas in the two regions means
that, broadly-defined, two `components' will describe these two,
dynamically separate flows.

The fastest shocks, where Keplerian velocities add to the flow
velocity, can cool gas throughout the entire cluster core as well as
in a broad cone outwards along the flow axis, {\it if}\/ the gas can
be maintained at the high pressure at the head of the shock for at
least a cooling time (Fig.~\ref{f:modelc2dv}c).  In future papers
\cite{ap99}, we will predict the detailed line profiles and
variability signatures expected as a result of the combination of all
these effects.

The flow in Model \modA\ (Fig.~\ref{f:modela2dv}a) is too dense in
most of the central `plume' to be maintained in isothermal equilibrium
by Compton heating and cooling.  Except for a small conical area about
the central black hole, all the gas with $r\la 0.5\parsec$ and $z\la
1\parsec$ has an ionization parameter less than unity.  This low
free-flow ionization parameter puts considerable strains on our
assumption of Compton equilibrium.  Net heating of the gas in  
smaller mass-loading shocks may be able to offset some of the
bremsstrahlung cooling.  In addition, the additional cooling will have
the side-effect of increasing the opacity of the gas, and hence the
radiative forces on the flow (\cf\ Section~\ref{s:eddeff}).  As these
radiative forces increase, the topology will gradually become more like
Model \modC.  This self-regulation will enhance the frequency with
which flow structures intermediate between Models \modC\ and
\modA\ are found.  The particularly high pressures found in the
accretion disc plane will encourage cooling, perhaps leading to the
addition of mass from the free flow to the outer accretion disc, as in
Model \modC$_{0.25}'$.

As might be expected, the distribution of stagnation ionization
parameters, in Model \modA\ (Fig.~\ref{f:modela2dv}b) is rather
smoother than the distribution of free-flow ionization parameters
discussed above (since gas at rest has the stagnation pressure by
definition, slow moving shocks have a less marked effect on the
ionization parameter).  The stagnation ionization parameter is,
however, less than 1 throughout the central $1\parsec$ sphere of the
nucleus.  In the near-radial inflow region outside the central plume,
gas pressures are $\ga10^{13}\cm^{-3}\Kelv$.

The stagnation ionization parameters are far lower than this in the
central plume itself, with pressures reaching more than
$10^{16}\cm^{-3}\Kelv$.  Shocks in the plume will cool fast, and the
cool gas will radiate strongly, since this is the region in which the
radiation from the central accretion disc is at its strongest.

In Fig.~\ref{f:modelsn2dv}, we show the ionization parameters derived
from the low Eddington-ratio model with no accretion, Sn (\cf\
Fig.~\ref{f:modelsn2d}).  The free-flow ionization parameter, $\Xi|f$
(Fig.~\ref{f:modelsn2dv}a), shows that the nISM is always hot and
diffuse.  The stagnation ionization parameter, $\Xi|s$
(Fig.~\ref{f:modelsn2dv}b), is sufficiently small to allow cooling in
only a very small region in the plane of the accretion disc where the
radiation field is weak.  This region is expanded to fill most of the
cluster core when the influence of Keplerian velocities is accounted
for (Fig.~\ref{f:modelsn2dv}c) -- however, the cooling length behind
such shocks would be a substantial fraction of the size of these
nuclei.

Behind shocks in this region the cooling will be very rapid
\cite{ip98}, and consequently the size of the BELR will be close to
that derived from variability studies \cite{p94}.

We conclude that the BELR in many AGN must be highly complex,
anisotropic, unsteady and irregular.  Physical effects not included in
the current treatment will modify our results, but the general form of
the flows will remain similar to those presented here.  Nevertheless,
Figs.~\ref{f:modelc2dv} and \ref{f:modela2dv} give a general
impression that the structures should be able to be broken down into a
few components, with varying mean properties.  We will undertake the
detailed modelling required to confirm this in future papers
\cite{ap99}.

\subsection{Column density and opacity}
\label{s:eddeff}

In some of our models, in which mass is retained for several dynamical
times, the column depth of the nISM reaches values around
$10^{24}\cm^{-2}$.  This will have potentially observable
consequences.  At these column densities, the nucleus will be getting
close to a Thomson optical depth of unity.  Electron scattering will
weaken the continuum observed along the densest lines of sight to the
nucleus, smooth out continuum variability and lead to polarisation of
the scattered component. The emission lines will also develop a
thermally broadened component, although the optical depth to
line-emitting clouds distributed in the nISM will be smaller than that
to the central continuum source.

X-ray absorption by highly ionized oxygen and iron, at energies of
$0.3\mbox{--}10\kilo\eV$, is seen in many low luminosity AGN
\cite{np94,rf95,kk95}.  The location of the absorbing gas is uncertain:
some of the absorption may be from the gas which is observed in Seyfert
2 galaxies to scatter the obscured central radiation source, while some
may also be far closer to the nucleus.  This `warm absorption' is
currently the best diagnostic of gas which is more widely distributed
than the locally confined broad emission line clouds.

The observational results are, in general, given in terms of the
column density and ionization parameter required for a uniform
intervening slab of solar-abundance gas in thermal equilibrium to
generate the absorption features seen.  Typical column densities are
$10^{22}\mbox{--}10^{24}\cm^{-2}$ and temperatures $10^5 \mbox{--}
10^6\Kelv$.  Reynolds \& Fabian~\shortcite{rf95}, for instance, base
their discussion of the observations of a particular Seyfert on a
column density of $10^{22}\cm^{-3}$ and $L|i/nr^2 =
30\erg\cm\secnd^{-1}$, where $L|i$ is the total ionizing luminosity
and $n$ is the electron density.  These values imply that the gas
temperature is $10^5\Kelv$ -- they find the gas resides in a toe-hold
of thermal stability between the usual cool and hot equilibria.  The
poor spectral resolution of (all but the most recent) X-ray
observations is a valid reason for using this approach to get the most
out of the available data. Krolik \& Kriss~\shortcite{kk95} emphasise
that the material observed may well be out of thermal equilibrium,
cooled by adiabatic expansion (although it is likely to be in
ionization equilibrium).  This is an even more important constraint
where the gas is being evaporated from a local evaporation centre
rather than expanding as part of a global flow.

Until our models are extended to explicitly include the thermal
balance of the gas, we cannot model the warm absorber in detail.  It
is, however, clear that a number of potential locations for the warm
absorber arise naturally in these models.  The global flow may dense
enough that it has begun to cool in some regions, or may be cooled at
large distances from the nucleus by adiabatic expansion \cite{kk95}.
In addition, in the shocks which create the broad-line emitting gas
must cool through this temperature range; subsequently the gas will
evaporate from the cooled clouds and rejoin the flow.  For a plane
radiative shock, the column between $10^5\Kelv$ and $10^6\Kelv$ is
roughly equal to the mass flux multiplied by the cooling time, which
is $\la 10^{19}\cm^{-2}$ for post-shock pressure of
$10^{14}\cm^{-3}\Kelv$.  The Compton heating time is $\sim 10^5$ times
longer than this cooling time, and may be lengthened further by
adiabatic cooling and residual atomic cooling.  However, the
evaporation will only occur once the gas has reached a region with
substantially lower pressure than behind the radiative shock and
further divergence of streamlines will occur while the gas is heating
\cite{PD}, both of which effects may act to decrease the column of gas
available to generate absorption.  The heating gas may be a
significant component of the absorption \cite{rf95}, but this needs to
be studied in greater detail.

\subsubsection{Cooling and the effective Eddington ratio}

Cooling leads to significantly enhanced opacity and hence radiative
forces on nISM gas: in consequence, the densities will be reduced, so
an equilibrium may be found with a substantial mass loss rate even
where the Eddington ratio is less than unity.  The overall radiation
force can be expressed as an angle-averaged {\it effective}\/
Eddington ratio,
\begin{equation}
\langle f|{Edd}^{\rm Effective}\rangle = 
{\bar\sigma\over\sigma|T}\langle f|{Edd}\rangle
\simeq 0.4{\eta|{acc}\etaflow\over 0.1}
{\bar\sigma\over\sigma|T}
Q_{-8}{M|{cl}\over M|h},
\end{equation}
where $\eta|{acc}$ is the efficiency with which the central accretion
disc radiates the rest-mass energy delivered to it by the global flow;
$\etaflow$ is the fraction of the mass input from the cluster
delivered to this central region, and $\bar\sigma$ is the mean
effective cross-section for gas close to the black hole.  Of all the
various uncertainties implicit in the terms of the above equation, the
determination of the effective cross-section is perhaps the greatest
\cite[\cf{}]{turnea93}.

In most of the models presented here, we assume that the overall
Eddington ratio is close to one.  Various independent observations
suggest that typically $ 10^{-3} < f|{Edd} < 1$, given specific model
assumptions (see Appendix \ref{a:lum}). As $f|{Edd}$ approaches unity,
the structure of the accretion disc will become bloated by radiation
pressure, although this bloating is reduced if the luminosity of the
accretion disc is dominated by reprocessing.

We have assumed $\bar\sigma\equiv \sigma|T$ (and thus
$f|{Edd}^\star\equiv f|{Edd}$, the classical Eddington ratio) in the
models presented here.  However, if the density of the gas is high
enough, some fraction of the gas may be able to cool, enhancing the
radiative driving.  For optically thin gas at equilibrium in the BELR
\cite[modelled by Cloudy]{hazy} the force multiplier
$\bar\sigma/\sigma|T$ increases to roughly 2 at $10^6\Kelv$, 40 at
$10^5\Kelv$ and to $10^4$ and beyond when the gas has cooled fully
\cite[see also]{alb94}.  It is reasonable to take $g = 7\ee{14}\rho$
when the gas is cooler than $\sim 5\ee4\Kelv$
\cite[\cf{}]{roeser,DFP}, although this force is dominated by
resonance line scattering and will weaken by a factor of a few if the
lines are optically thick.  Hence even in Seyfert galaxies, where
$f|{Edd}$ (determined by electron scattering) may be far smaller than
unity, a dynamical feedback may lead to flows of very similar forms to
those presented in here.

\subsection{Mass budgets: winds and accretion efficiency}
\label{ss:massbudge}

Here, we discuss the mass budgets found for the models we have
calculated:  how much of the mass lost by the stellar cluster -- and
its associated heavy elements -- remains in the nISM, how much is
delivered to the accretion disc, and how much is transferred to the
wider galaxy. Numerical values for the equilibrium mass processing
rates have been given in Table~\ref{t:results}.

\begin{figure*}
\epsfxsize = 8cm
\begin{centering}
\begin{tabular}{ll}
a) \hfil Model \modC \hfil & b) \hfil Model \modA\hfil\\
\epsfbox{cflux.ps} &
\epsfbox{aflux.ps}
\\
\end{tabular}
\end{centering}
\caption[Mass fluxes]{Mass processing rates of models a) \modC, b)
\modA.  Model \modC\ soon reaches a steady equilibrium between mass
input and mass outflow.  In Model \modA, mass is lost episodically
from the nucleus.  While as much as 60 per cent of this loss does not
have escape velocity, we argue in the text that the effect of assuming
this mass is lost to the grid will be small.  Model \modB\ has no
appreciable mass loss in $10^6\yr$.}
\label{f:modelaflux}
\end{figure*}

The mass processing rates of models \modC\ and \modA\ are shown in
Fig.~\ref{f:modelaflux} (we do not show Model \modB, since almost no
mass is lost from this nucleus over our integration period).  For
Model \modC, the mass loss rate reaches an equilibrium after roughly
the sound crossing time of the nucleus.  Model \modA\ produces a
series of explosive mass loss episodes, continuing for times as long
as we have simulated.  The characteristic timescale of these
explosions, typically $10^3\yr$, is comparable to the gravitational
timescale, $1.6\ee3r|{c,pc}^{3/2} M_8^{-1/2} \yr$ .

\begin{figure*}
\epsfxsize = 8cm
\begin{centering}
\begin{tabular}{ll}
a) \hfill Model \modC$_{0.005}$ \hfill &
b) \hfill Model \modC$_{0.25}$ \hfill \\
\epsfbox{ccflux.ps} &
\epsfbox{cpqflux.ps}
\\
c) \hfill Model \modC$_{0.25}'$\hfill \\
\epsfbox{coutflux.ps}
\end{tabular}
\end{centering}
\caption[Mass fluxes]{Mass processing rates of models a)
\modC$_{0.005}$, b) \modC$_{0.25}$, c) \modC$_{0.25}'$. Models
\modC$_{0.005}$ and \modC$_{0.25}'$ reach a steady equilibrium
solution, very similar to Model \modC\ but with a fraction of the mass
input diverted into accretion.  Model \modC$_{0.25}$, on the other
hand, has a global structure rather similar to Model \modB.  Only a
small fraction of the gas escapes accretion, and the central wind is
consequently too weak to drive out through the cluster core.  The mass
input and accretion rates soon reach equilibrium, modulated by
near-periodic instability of the accretion/wind interface.}
\label{f:modelcpflux}
\end{figure*}

\begin{figure}
\epsfxsize = 8cm
\begin{centering}
\epsfbox{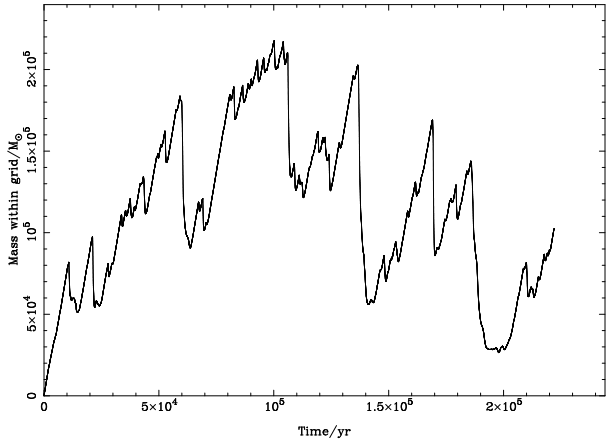}
\end{centering}
\caption[Model A flux]{Mass within the grid for Model \modA, \cf\
Fig.~\ref{f:modela2d}.}
\label{f:modelamass}
\end{figure}

We plot the variation of total mass within the grid of Model \modA\ in
Fig.~\ref{f:modelamass}. This is the integral of the variation shown
in Fig.~\ref{f:modelaflux}b.  As we have already described, the flow
is highly variable, alternating between times when the input mass is
almost entirely retained (the steepest upward slopes in
Fig.~\ref{f:modelamass}), and epochs of rapid mass loss.  In
particular, this figure emphasises the long-term chaotic behaviour of
the flow structure we have simulated.  It also shows how close Model
\modA\ is to the border between steady-outflow and chaotic flow.
During the extended minimum around $2\ee{5}\yr$, the structure and
density of the flow is very similar to that shown, for Model \modD, in
Fig.~\ref{f:modeld2d}.  Indeed, we would hope this were so, since the
fundamental parameters for Models \modA\ and \modD\ are identical
(\cf\ Fig.~\ref{f:xxx}) -- they differ only in the opening angle of
the central outflow cone.

The fraction of stellar mass loss delivered by the flow to the
accretion disc depends critically both on the structure of the flow
and on details of the processes by which mass is added to the
accretion disc.  A primary aim of this paper is to address the first
of these issues.  Therefore, we treat the second by three extreme
assumptions -- no mass addition to the disc, or addition only in an
inner or outer region of its surface.  

The fractional rate of accretion, $\etaflow$, was {\it assumed} to be
zero in models \modC, \modA\ and \modB.  In models \modC$_{0.005}$ and
\modC$_{0.25}'$ it was found to be around 40 per cent, while for model
\modC$_{0.25}$ almost all the mass input is accreted.  The area of the
disc over which accretion took place in models \modC$_{0.005}$ and
\modC$_{0.25}'$ was chosen so that the accretion rate and wind flux
were nearly in balance.  In general, $\etaflow$ tends to be close
either to zero or to unity: we leave the precise elaboration of this
dependence as a function of physical parameters for a future paper.

We show the mass fluxes for models \modC$_{0.005}$, \modC$_{0.25}$ and
\modC$_{0.25}'$ in Fig.~\ref{f:modelcpflux}.
Fig.~\ref{f:modelcpflux}a shows how the accretion region in model
\modC$_{0.005}$, which is relatively small, diverts a fraction of the
mass which would otherwise have been driven out in the wind, perhaps
slightly extending the length of the variable period compared to model
\modC.  In contrast, if accretion occurs over a large area (Model
\modC$_{0.25}$, Fig.~\ref{f:modelcpflux}b), the outgoing wind is
suppressed.  The solution soon relaxes to a near-equilibrium between
mass input and accretion, modulated at a 10 per cent level by the
instability of the termination shock and accretion.  The evolution of
model \modC$_{0.25}'$ (Fig.~\ref{f:modelcpflux}c), in which accretion
takes place in the outermost regions of the disc, is strikingly
similar to that of Model \modC$_{0.005}$. This can be understood, as
already discussed, on the basis of morphological arguments.

To supply the luminosity of a QSO, $L \simeq 1.3\ee{46}
f|{Edd}M|{h,8}\ergs$ requires accretion at a rate
\begin{equation}
\dot{M} \simeq 2.3 \left(f|{Edd}\over
\eta|{acc,-1}\right)M|{h,8}\Msun\yr^{-1},
\end{equation}
where $\eta|{acc} = 0.1\eta|{acc,-1}$ is the accretion efficiency of
the black hole.  The accretion rate is given, over the long term, as a
fraction of the mass-loading rate from the stellar cluster by
$\etaflow$, which is determined by the global hydrodynamics of the
nucleus.  The fraction of input mass which must be accreted is given
by
\begin{equation}
\etaflow \simeq 0.23 \left(f|{Edd}\over
\eta|{acc,-1}Q_{-8}\right){10M|h\over M|{cl}}.
\label{e:pdmcl}
\end{equation}
Perry~\shortcite{dublin} argues that $\etaflow\simeq 0.5$.  
The accurate calculation of this fraction is one eventual aim of this
work: some values of $\etaflow$ are given in Table~\ref{t:results}.
To the accuracy of the present treatment, models \modC$_{0.005}$ and
\modC$_{0.25}'$ are self-consistent, and closest to the case
considered by PD and Perry (1993a).

In a Seyfert nucleus residing in a rejuvenated QSO, the ratio of $M|h$
to $M|{cl}$ may be substantially larger than 1:10.  From
equation~\refeq{e:pdmcl}, a self-consistent model (with $\eta|{acc} <
1$) must be significantly below the Eddington limit unless gas is
provided to the black hole from non-stellar sources or the stellar
lifetime is very short.

{}From Table~\ref{t:results}, we see that rates for these models range
between 3.2 and $8.6\Msun\yr^{-1}$.  Comparing these with the
accretion rate required to fuel the central luminosity,
$4.6\eta|{acc,-1}\Msun\yr^{-1}$, we see that these models are
reasonably self-consistent, \ie\ the accretion rate is sufficient to
power the central luminosity for a reasonable assumed radiative
efficiency.  If we impose the condition that the flow is
self-consistent over the short-term, then the flow may to relax to one
of these solutions, {\em however sensitive} the dependence of
$\etaflow$ on flow parameters.  Note, however, that delays in the
conversion between mass accretion onto the disc and the output of
radiation are likely to imply that the equilibrium is, in fact,
unstable.

If all relevant dimensionless quantities, including the Eddington
ratio, are kept constant (\cf\ Section~\ref{s:scaling}), the ratio
$L/\dot{M}$ varies as $1/\phi$ and is independent of the length and
timescales chosen (because $L/\dot{M} \propto M|h/\phi M|{cl}$; by
contrast, if $L\propto\lambda^2$ and $\lambda=\tau$, then
$L/\dot{M}\propto\tau/\phi$, and any solution with finite accretion can
be made self-consistent).  Note, however, that delays and instabilities
in the processing of gas through the accretion disc may mean that, on
the short timescales,  not all AGN are strictly self-consistent, in the
formal sense defined above.

\begin{figure}
\epsfxsize = 8cm
\begin{centering}
\epsfbox{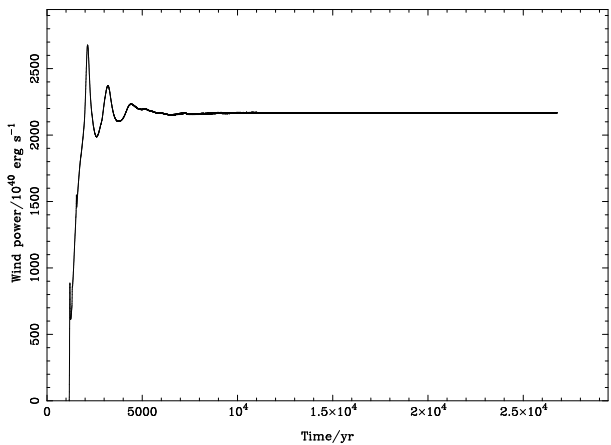} 
\end{centering}
\caption[Model C wind]{Wind power in Model \modC.}
\label{f:modelcwind}
\end{figure}
\subsection{Wind powers}

\label{s:windpow}

The average wind powers found for `mature' flows in the various models
presented here are included in Table~\ref{t:results}.  As can be seen
by comparing Fig.~\ref{f:modelcwind} to Fig.~\ref{f:modelaflux}a, the
time variation in the power of the wind out from the nucleus was very
similar to the variation in mass loss rate.  By contrast, in model
\modA, only a very little of the mass lost to the grid has escape
velocity, so the value quoted in Table~\ref{t:results} is very
uncertain.

The highest efficiency with which the central luminosity was converted
into outflow power in the models presented here was in Model \modC,
where the efficiency was $L|{wind}/L|{Bol} \simeq 8\ee{-4}$.  Mass
loss rates from the nucleus to the surrounding galaxy vary from
effectively nil to $8.6\Msun\yr^{-1}$ (or, rather, that the entire of
the mass input to the cluster is lost).  Rather lower values are
typical in most flows.  By scaling the models, as discussed in
Section~\ref{s:scaling}, we see that the wind efficiency varies as
$\phi\lambda^2/t^2$: it would seem to be able to be increased to an
arbitrary degree.  However, the total efficiency obtainable is limited
in reality by the Thomson optical depth of the nISM (which limits the
column density, $\propto \phi\lambda/t$) and by the onset of
relativistic effects as dynamical velocities ($\propto\lambda/t$)
approach $c$.  The values of the parameters for Model \modC\ were also
chosen to reflect those likely in real nuclei, except that the driving
radius of the flow (\ie{} the smoothing region) is larger than is
likely to obtain in real AGN, so values rather greater than
$L|{wind}/L|{Bol}\sim 10^{-3}$ may be possible.

The effect on the galaxy of the winds from a symbiotic nucleus will be
difficult to distinguish from those of winds driven by larger-scale
galactic starbursts or of relativistic jets driven from smaller scales
\cite[\eg{}]{veilea94,colbert96}.  Heckman \etal~\shortcite{ham}
observed starburst superwinds with wind powers up to several
$10^{43}\ergs$, with overall $L|{wind}/L|{Bol} \sim 10^{-2}$ and wind
velocities from a few $100\kms$ to above $1000\kms$ (although again
the winds in some of their sample may be driven, in part, by AGN).
Whether the source is a starburst or an AGN may be distinguished by
whether the starburst is strong enough to drive the wind, and to
produce the observed gas velocities, or whether the morphology of the
wind bubble suggests an unresolved, dynamically distinct source (\eg,
if it is misaligned with galactic structure).  Ill-collimated winds
may be distinguished from jets by the length-to-breadth ratios of the
structures which result, and by evidence for relativistic motions
(\eg{} synchrotron emission or superluminal motion).

A wind with the strength we find can have a major effect on the ISM of
the galaxy and on its surroundings.  As the wind-blown bubble moves
out, it will generate dense shells of gas which may be important for
narrow line emission; eventually it may evacuate the ISM completely
\cite{DFP,FPD,tda,smithkc93,steffea97}. As it blows outwards, it will
be an important source of energy and processed material to the
intergalactic medium \cite{voit94,sp98}.

\section{Conclusions}
\label{s:concl}

The central thesis of this paper is that stars are of fundamental
importance in the nuclei of active galaxies.  The symbiosis between a
compact nuclear starburst stellar cluster and a massive black hole can
self-consistently explain the properties of active nuclei.  

These two simple ingredients -- a spherically symmetric starburst
stellar cluster and an accreting black hole -- conspire symbiotically
to produce flows of immense variety and complexity.  Where the gravity
and radiation field of black hole and accretion disc dominate at the
very centre of the nucleus, a bipolar circulation is found.  Outflows
along the axis perpendicular to the accretion disc coexist with inflow
in a broad flattened structure above the disc.  The interaction between
this central flow and that in the region where the gravity of the
stellar cluster dominates the dynamics results in highly complex
nuclear wide flows.  These resulting parsec-scaled nISM bipolar flows
range from episodic, explosive, percolating structures to
well-collimated, narrow, hypersonic hydrodynamic jets.  The jets arise
when the flow through most of the core is quasi-hydrostatic, while
episodic `bubbles' occur when the Keplerian velocity of the cluster is
substantially supersonic.  It is important to note that these dramatic
structures arise as a result of pure hydrodynamics in a relatively
simple system; they do not require the intervention of radio-related
nuclear phenomena.  Our results dramatically illustrate why broad
emission line studies have consistently failed to identify any simple,
global flow patterns. The real flow structures in AGN are {\it
inevitably} more complex.

Our simulations confirm the overall flow properties of derived by PD
and Perry (1993a).  We also find details that their treatment
overlooked.  The gravitational attraction of the stellar cluster can
lead to gas being retained within the nucleus for many dynamical
times, with the line emission likely to be dominated by gas in a
central plume which is retained by the gravity of the cluster.

We propose (see Section~\ref{s:classify}) a classification of the flow
structures in terms of dimensionless dynamical parameters: the
velocity dispersion of the cluster and the outflow velocity from close
to the black hole, both expressed as fractions of the sound speed in
the nISM\@. These relate to the gravitational and thermal energies in
the nISM\@. This classification scheme, which is analogous to the
Hertzsprung-Russell diagram for stars, will be broadened by the
various secondary effects which depend on the dominant parameters. But
the morphology of the flows are likely to remain similar and analytic
estimates of flow velocities and densities are still readily
calculable.

We have discussed other observational diagnostics of these flows.  The
mass-losing stellar cluster acts as an optically-thin fuel reservoir
which chemically enriches the nISM.  When we consider the properties we
expect in the broad emission line region (BELR) we find that the nISM
itself has a multi-component structure.  This fact, combined with the
close relationship between the accretion disc and the kinematics of the
global wind helps explain the observed multi-component structure of the
line spectra and the observed correlations between low- and
high-ionization broad emission lines.  The varied kinematics of the
flows seems to explain the contradictory evidence that has been
found for inflow, outflow, chaotic or rotating flows by observers.  The
winds driven from the nuclei will also affect the surrounding galaxy and
the IGM\@.

Our study clearly shows that active galactic nuclei will only be fully
understood once the symbiotic relationships between the black hole,
the nuclear starburst stellar cluster, and the wider galaxy are
considered.  Our conceptually simple {\it Symbiotic}\/ Model provides
a self-consistent explanation for the observed complexity of active
galaxies.

\subsection*{Acknowledgements.}

We thank Sam Falle for providing a version of his hydrodynamic code
for us to abuse, and an anonymous referee for a constructive report
which helped clarify the structure of this paper.  Computer resources
were provided by the University of Leeds and by Starlink (in
Cambridge, Cardiff, Leeds and Manchester).  ACB and RJRW acknowledge
support from PPARC; ACB acknowledges the EU TMR Network
(FMRX-CT96-0068) for Infrared Surveys; JJP thanks the Leverhulme Trust
for generous support in the early stages of this research.

\appendix
\section{Details of the physical assumptions}

\label{a:detail}

In this appendix, we discuss in more detail the assumptions of our
model.  The presentation progresses from the structures within the
nucleus -- the black hole, starburst stellar cluster and accretion
disc -- to the properties of the radiation field, thermal equilibrium
of the nISM and treatment of accretion.

\subsection{Black Hole}
\label{a:black}

Since the bolometric luminosities of most AGN can be determined to
order-of-magnitude or better, the combined issues of black hole masses
and Eddington ratios are strongly linked.

Indications of the masses of dark objects in nearby galaxies have been
obtained from the brightness distribution and stellar kinematics in
their cores.  The masses inferred to reside in dark, unresolved cores
vary from $2\ee6\Msun$ in the Galaxy and M32 to around $3\ee9\Msun$ in
M87 \cite{kormer95}.  The sample of galaxies observed is at the moment
rather sparse and heterogeneous, but the presence of these `dark
objects' suggests that there may be enough nuclear black holes of
sufficient mass to satisfy our model.

Limits on have also been derived on black hole masses in active AGN\@.
Some attempts have been made to estimate the mass function using
simple estimates of the black hole mass \cite[\eg{}]{pbe}, but has
only been very recently that radio observations of maser emission and
detailed line profile studies have begun to give well-constrained
dynamical masses for black holes, in the range
$10^7\mbox{--}2\ee8\Msun$ \cite{green96,lasea96,newmea97}.

\subsection{The Stellar Cluster}
\label{a:cluster}

\paragraph*{Dynamics}
\label{a:dynamics}

We base our assumptions about the dynamics of the nuclear stellar
cluster in the models of MCD\@.  They modelled the evolution of the
stellar cluster around a central black hole by a multi-mass
energy-space Fokker-Planck code, including the effects of tidal
disruption, physical collisions between stars and of stellar
evolution.

As our aim is to model the hydrodynamics of the nuclei, we only
include their results in a schematic fashion.  MCD find that the
dynamical evolution of the stellar clusters occurs, for the most part,
in the cluster cores.  For initial stellar densities below
$10^7\Msun\parsec^{-3}$, the dynamical evolution of the cluster is
dominated by tidal disruption, leading to a core density law $s=7/4$
(as predicted by early models of this process), while for higher
stellar densities collisions dominate the core evolution and mass
loss, giving a core power-law of $s=1/2$.

Beyond the core, the density law in the models presented by MCD
remains roughly constant, except for the effects of stellar evolution.
In their initial Plummer model, MCD take the halo power-law index to
be $h=5$.  Beyond $10\parsec$, the stellar density will be dominated
by the normal stellar distribution of the central regions of the
galaxy.  The precise form of the halo power law and the surrounding
galactic bulge are not well determined.  However, they will have only
a small effect on the hydrodynamic models of the flows inside the BELR
that we present here.

The relationship between the enclosed mass and radius in the Galactic
centre between 1 and $100\parsec$ is roughly $M \simeq 3\ee6
r|{pc}^{1.2} \Msun$ \cite{gt}, implying stellar densities $n_\star
\simeq 10^5 (M_\star/1\Msun)^{-1} r|{pc}^{-1.8}\parsec^{-3}$.  Within
the Galactic centre, the cluster of $10^8$ or $10^9\Msun$ which we
discuss here would be a dynamically independent system.  Indeed, any
such cluster would dominate the dynamics of at least the central
$100\parsec$ of the Galaxy.  In NGC~3115, a dense stellar cluster of
mass $\sim4\ee7\Msun$, distinct from the core of the galaxy, appears
to surround a dark object (perhaps black hole) of mass $\sim10^9\Msun$
\cite{kormea96}, although such distinct stellar clusters are not found
around all candidate massive dark objects \cite{kormea97}.

HST observations have provided evidence for stellar densities at least
$>10^7\Msun\parsec^{-3}$ in M32 \cite{lauea98}.  It is more difficult to
observe such dense compact clusters in more distant galaxies, but the
properties at larger radii seem similar to M32, and nuclear clusters
of $10^7\Lsun$ are certainly observed \cite{lauea95}, although there
is no strong evidence for stellar densities above
$10^6\Msun\parsec^{-3}$ \cite{faber97}.  Faber \etal{} find that the
density decreases quite strongly as the mass of the cluster increases,
while the velocity dispersion in the clusters varies only weakly; they
also suggest that the cluster masses in these objects are between 3
and 6 times those of the massive dark objects they harbour.  A cluster
with mass $\sim 6\ee7\Msun$ contained within $6\parsec$ of an
$4\ee6\Msun$ black hole is seen in the Circinus galaxy, the most
nearby known Seyfert 2 \cite{mai98}.

The clusters which we discuss here would only be resolvable in nearby
galactic nuclei (although their gravitational effects would extend
over a larger radius).  While the nucleus was active, their presence
would be masked by the active QSO and at later epochs much of their
mass may have been lost to the nucleus in a wind or accreted by a
black hole as a result of stellar evolution or stellar collisions.
Clusters with these high densities are a requirement of the PD model
if stellar mass loss envelopes are to generate the required covering
fractions of cool gas within the BELR.  However, clusters at least as
dense are required by many models of the formation of the central
massive black hole \cite[\eg{}]{br78,qs90}, so it is not unreasonable
to consider the properties of a black hole surrounded by the flotsam
of its formation.

The stellar cluster models of MCD are, as a matter of computational
necessity, spherically symmetric.  In the initial hydrodynamic models
presented here, we follow this assumption.  In general, if the young
star cluster were to form as a thin disc, non-axisymmetric
instabilities will thicken its central regions substantially within a
dynamical timescale, to generate a bulge-disc structure reminiscent of
a spiral galaxy.  If the clusters remain aspherical enough to
influence our models, they would either consist of stars on plunging
box-type orbits or have significant net angular momentum.  In the
first case, the aligned orbits would rapidly be randomised by the
influence of the central black hole \cite{gb85}.  In the second case,
for self consistency, we should then also include the angular momentum
of the flow, a topic we leave for a subsequent paper.

Even in the core of a nuclear starburst, the pressure, typically
$10^7\Kelv\cm^{-3}$ \cite{sbhb96}, is significantly less than the
pressure in the BELR of AGN.  We have chosen to neglect this
extranuclear gas in the simulations presented here, as the rate of
mass input into the central parsec-scale region will depend strongly
on the transport of angular momentum.  Detailed simulations of the gas
dynamics of galaxies with active nuclei will allow us to correct this
in future papers.

\paragraph*{Stellar mass loss}
\label{a:massloss}

The manner in which mass is lost from stars and added to the flow has
an effect on our simulations on various scales.  The overall
normalization of the stellar mass loss, within our assumptions, sets
the overall normalization of the density, but does not alter the
kinematics of the flow.  Variations in mass input across the core (for
instance, if the stellar cluster is oblate rather than spherical) will
alter the form of the flow, but we argue (in
Appendix~\ref{a:loaddist}) that this will not have a major influence
on the structure of the flow.  The graininess of the mass input, on
the scale of the individual interactions between stars and the global
ISM, is determined by the mode of stellar mass loss (through
supernovae, stellar winds, collisions, \etc).  This can have
consequences for the thermal structure of the flow.  Perhaps the most
important consequence of these interactions will be the production of
cool gas which will emit optical and UV emission lines \cite{PD}.

In all the models presented above, we take the input rate of gas into
the cluster to be proportional to the stellar density.  The scaling
chosen, $\dot{q} = Q\rho_\star$ with $Q=3\ee{-16}\secnd^{-1}\simeq
10^{-8}\yr^{-1}$, is appropriate for mass loss from a young stellar
cluster with a lower mass cut off above $4\Msun$ (Shull 1983; Williams
\& Perry 1994, in particular their Fig.~3b).  In this way, our
assumptions differ from those of MCD who used a broad IMF, extending
from stellar masses as low as $\Mstar = 0.3 \Msun$.  For IMFs
extending to such low masses, collisional mass loss exceeds stellar
mass loss at the very smallest radii \cite{mp}.

Rieke \etal~\shortcite{rieke80} suggested that IMFs with lower mass
cutoffs above a few solar masses were necessary to explain the high
supernova rates and $2\micron$ luminosities for the derived dynamical
mass in the best studied starburst galaxy, M82 \cite[see
also]{rieke93,doane93,bern93}.  Studies of the variation of the
mass-to-light ratio and metallicity with galaxy mass for ellipticals
also suggest that `bimodal' star formation occurs in some
circumstances \cite{larson86,zs96}.  High-biased IMFs may be
particularly favoured in the highly sheared environment of a galactic
nucleus, if the increase in pre-collapse density required for higher
shears is overweighed by increased internal energy resulting from
turbulence and heating of the cloud gas (\cf\ Silk 1977, 1995; Larson
1985).

It should be noted that the observational evidence for high-biased
IMFs is not clear cut \cite{scalo90}.  Detailed models of the
distribution of dust can explain the broad band spectra at the
required mass-to-light ratio with normal IMFs \cite{deve89,calz97}.
In M82, the distribution of sizes of the radio supernova remnants (and
the unchanged population over a 10 year period) suggest that the
supernova rate was initially overestimated \cite{mux94}.  While some
Galactic open clusters {\it are}\/ deficient in the $1\mbox{--}10\Msun$
range \cite{scalo90}, in the best nearby analogue of a starburst
system, the central region of 30 Doradus, imaging studies find
populations of stars down to $3\Msun$ with no sign of differences in
the global IMF \cite{hunt95,bran96}.

Note that we assume the formation of the cluster occurs on a timescale
of a few million years or less.  Many of the constraints described on
the number of quiescent low-mass stars in the starburst IMF become yet
more stringent if stars have formed over a period significantly longer
than the main sequence lifetime of the massive stars, since they
depend not on the rate of formation but on the total mass of low-mass
stars.  There is good evidence that the formation of massive stars in
OB associations is coeval to within $20\Myr$ \cite{shull95}, with the
massive stars destroying their nascent star forming region (perhaps
encouraging further generations of massive stars to form at a larger
distance).  It is not unreasonable for closely synchronized star
formation to have occurred in the compact stellar clusters we discuss,
since the dynamical timescales are short, and even at $10^4\Kelv$ the
sound-crossing time of the nucleus is only $10^5\yr$.

Less high-biased stellar IMFs will decrease the mass input rate, and
hence the density scale will be lower.  Different modes of stellar
mass loss, driven by stellar collisions, tidal disruption or the
effects of the nuclear radiation, will boost the mass input rate,
albeit with a different spatial distribution.  It does become
difficult to produce the required luminosity in such low-loading
models, unless the cluster mass is extremely large.

Stellar mass-loss through collisions is a threshold process: the
collisions lead to disruptions when the impact velocity is greater
than the surface binding energy of the stars (\cf{} MCD).  If the
velocity dispersion varies with radius, this implies that collisional
mass loading can occur from any individual class of stars wherever the
dispersion is greater than the stellar escape velocity.  From the
above discussion, we see that within the core, head-on collisions are
likely either to be disruptive through most of the core (where $v|K >
v_\star$), or not at all.  The collisional mass-loading term may then
be given by
\begin{equation}
\dot{q}|{coll} = \Mstar n_\star^2v|KA_\star
\end{equation}
when $v|K > v_\star$, and zero otherwise.  For our purposes, this only
need be applied to main sequence stars, since giants will soon lose
their extended envelopes as a result of stellar winds and supernovae.
Collisional mass loss will thus only be important when $\tau|{coll} =
1/nv|KA_\star < \tau_\star$.  Since for a uniform cluster $\tau|{coll}
\simeq 10^{10} r|{pc}^{7/2}M|{c,8}^{-3/2}\yr$, roughly independent of
stellar mass \cite{pw93}, collisions are only likely to be an
important factor in stellar mass loss only for the small fraction of
stars in the central density cusp, or in the densest of nuclear
clusters \cite[\cf{} MCD,]{mp}.

Close to the central black hole there is an exceptional region of
increased velocity dispersion.  An `extended loss cone' may form, for
which the vigour of the stellar environment in the velocity cusp
performs a similar destructive role to the high tidal field of the
black hole at yet smaller radii.  This region, together with that in
which mass is input by tidal disruption, is not resolved by our
simulations.  Central collisions could most easily be catered for by
an unresolved central mass addition term, to be traded off against the
condensation of mass onto an accretion disc.

\paragraph*{Effects of Loading Distribution}
\label{a:loaddist}

We assume spatially smooth mass loading.  The validity of this
assumption depends on the efficiency of mixing between mass input and
global flow, and on the manner in which stars lose mass.  Supernovae,
tidal disruption and head-on collisions will add mass in very
localized regions, but initially with a wide range of velocities.
Smooth mass loading may be a rather rough approximation where these
effects dominate the mass input: indeed in Seyfert nuclei, individual
supernovae will be rare, but may be able to blow away much of the nISM
(PWD).  However, sufficiently many supernovae should occur in a QSO
that smooth loading is a reasonable first approximation (PD).

Mass loss by stellar winds will result in a smoother spatial
distribution of mass-loading, as many sources are active at any time,
and each one is active for many dynamical timescales.
Maeder~\shortcite{maed92} suggests that for solar-abundance stars over
$25\Msun$, most of the stellar mass is ejected in before the supernova
explosion.  For stars of $40\Msun$ or over, the mass of the
pre-supernova star is less than $8\Msun$.  The fraction of mass lost
pre-supernova may be further increased by the higher metallicities
expected in nuclear stellar clusters.  Glancing collisions,
particularly with the envelopes of giant stars, will also load the
region in a gradual fashion.

The large-scale distribution of loading is important when $r\dot{q}
\ga \rho v$ [from comparison of terms in the equations of mass
conservation and motion, eqs~\refeq{e:masscons} and~\refeq{e:motion}].
Hence, the details of the mass input distribution are important in the
outer accretion flow and close to stagnation points (where $v\sim 0$).
If, in the absence of accretion, the central convective region has
size $r|{con}$ at time $t$ (where $r|{con}$ saturates at $r|c$ at time
$t|{fill}$ when the wind mass loss becomes equal to the mass input,
after which time we should substitute $t=t|{fill}$ in the following
expression), we find that within the convective structure
\begin{equation}
{r|{con}\dot{q}\over \rho v} \simeq \left(r|{con}\over
r|c\right)^3{t|{dyn}\over t},
\end{equation}
where $t|{dyn} = r|{con}/v$.  Unless the onset of outflow is prompt
(\ie\ within a dynamical timescale), the precise form of mass loading
distribution through the cluster core will be unimportant.  Even if
the outflow is prompt, its structure will be simple: the central
outflow will still be independent of the form of the mass loading, and
the outer structures will be ballistic (\cf\ Model \modC).  We will
investigate these assumptions in more detail in a future paper.

\subsection{The Accretion Disc}

In the limit of low external pressure, a thin accretion disc will have
a corona-wind structure similar to stellar wind models.  The rate of
mass loss in the wind would depend on the incident flux (whether
direct from the nucleus or back-scattered from the nISM) and the
structure of the disc (self-gravity may well lead to a rather clumpy
structure at these radii), and will be influenced by magnetic fields
and angular velocities.

However, the models we have calculated show that the flow pressure
tends to be high in the plane of the disc.  The radiation intensity is
low in the disc plane.  Confined, high pressure gas may cool, adding
gas to the disc.  The importance of this process depends on the ratio
of the cooling time of the shocked gas to the flow time through the
high-pressure region.  In one limit, essentially all the gas will
`collapse' onto the disc surface.  In the other, mass addition from
the wind would have a negligible effect.

To treat these processes accurately requires detailed heating/cooling
models of the flow in and around the disc.  The length- and
time-scales important for such models are difficult to include in a
global model for the nuclear flow.  Woods \etal~\shortcite{woodea96}
treat the evaporation of gas from the disc surface, using an adaptive
grid technique.  In their model, the regions of cool gas are localized
near the disc surface: in general, where the unstable {\it cooling}\/
of the gas may be important, complicated turbulent structures would
quickly overwhelm any adaptive grid model with irrelevant detail.  To
study these cases, local results on the detailed structure of the flow
must be smoothed to predict the mean properties of gas on a global
scale.  

\subsection{Gravity}
\label{a:gravity}

The gravitational forces are dominated by the central accreting black
hole (point source), and the nuclear starburst stellar cluster.  We
discuss observational constraints on the masses of these components in
Appendices~\ref{a:black} and \ref{a:cluster}, respectively.

We do not include the gravitational field of the interstellar gas.  At
any time, the integrated mass in the ISM interior to $1\parsec$ is far
smaller than that of the stellar cluster, since gas will not be
retained in the nuclear ISM for longer than the average stellar
lifetime $\ga10^6\yr$.  If the mass of the ISM were gravitationally
significant, the nucleus would be optically thick to Thomson
scattering and X-ray absorption, and the ISM would rapidly cool by
bremsstrahlung.

The accretion disc is limited by its vulnerability to local
self-gravitating instabilities, to have a density no greater than the
tidal density,
\begin{equation}
\rho|T = {M|h + M|{cl}(\mbox{$<$}r)\over (4/3)\pi r^3},
\end{equation}
and hence it can contribute no more than $H/r$ of the gravitating
mass.

For the model stellar cluster given by equation~\refeq{e:cluster}, the
cluster mass is given by
\begin{equation}
M|{cl} = {h-s\over(3-s)(h-3)} 4\pi\rho|{\star,c}r|c^3 = {h-s\over h-3}M|c,
\end{equation}
where the core mass $M|c$ is
\begin{equation}
M|c = {1\over3-s} 4\pi\rho|{\star,c}r|c^3.
\end{equation}
The gravitating mass of stars inside radius $r$ is then
$\mu_\star(r/r|c)M|{cl}$, where
\begin{equation}
\label{e:mint}
\mu_\star(r/r|c) = \left\{
\begin{array}{ll}
\displaystyle
{h-3\over h-s}\left(r \over r|c\right)^{3-s}			& r<r|c\\
%
\displaystyle 		
1-{(3-s)(r/r|c)^{3-h}\over (h-s)}	& r>r|c.
\end{array}
\right.
\end{equation}

\subsection{Radiation}
\label{a:rad}

The dynamical effects of the black hole are not limited to to its
gravitational attraction -- the high luminosity generated by accretion
will generate a significant outward radiation force. Accretion on to a
black hole at a rate $\dot{M}|a$ will produce a luminosity of
\begin{equation}
L|{Bol} = \eta|{acc} \dot{M}|a c^2
\end{equation}
where $\eta|{acc}$ is an efficiency factor, of order 0.1.  In this
paper, we have taken the driving to result from optically thin Thomson
opacity only, as it is the dominant opacity when the nISM is at the
Compton temperature. The dynamical effect of this radiation can be
parameterised in terms of the angle-averaged Eddington ratio
\begin{equation}
\langle f|{Edd}\rangle \equiv L|{Bol}/L|{Edd},
\label{e:fangle}
\end{equation}
between the central bolometric luminosity and the Eddington limit
luminosity, \cf{} equation~\ref{e:frtheta}.

The broadband continua of many AGN are dominated by a quasi-thermal
`bump' at UV wavelengths \cite{Sanders89b,zm93}, which is believed to
be emitted by the central regions of an accretion disc
\cite{shields78,CDMP}.  If this is so, the intensity of the UV
radiation field will have an angular dependence similar to $F(\theta)
\propto (\cos\theta)^\alpha$, with $\alpha=1$.
Netzer~\shortcite{netz90} suggests that using $F(\theta) \propto
\cos\theta\,(1+1.5\cos\theta)$ better accounts for the effects of
limb-brightening at optical wavelengths \cite{ln89,pensea90} find that
for NGC~4151, intrinsic continuum anisotropy (as opposed to reddening)
must be part of the reason for the decrement in ionizing luminosity by
a factor $\sim13$ between that observed and that inferred to excite
the ENLR emission in the galaxy.  While using $\alpha = 1.5$ gives an
almost identical $F(\theta)$ to this, we can also obtain an very
similar fit (at the important angles where the net force is outward)
by rescaling the hole mass and Eddington ratios, keeping $\alpha=1$.
For example, as a test case, we treated a model in which $\alpha=1.5$,
with all other parameters as in Model \modA, were almost identical to
those for Model \modC\ with $\alpha=1$ (with flow velocities about
15 per cent smaller, as expected).  Hence we will only present results the
simplest model here.

We smooth the radiation field close to the centre of the nucleus,
principally for numerical reasons.  However, the radiation field may
be physically smoothed in many sources, if the nISM is optically thick
to scattering close to the nucleus.  The anisotropic component of the
radiation will mostly be emitted within a radius roughly determined by
the black body emission law
\begin{equation}
r|{emit} \simeq (L|{disc}/2\pi\sigma|{SB} T|m^4)^{1/2} \simeq 0.017
L_{46}^{1/2} T|{m,4}^{-2}\parsec,
\label{e:remit}
\end{equation}
where $\sigma|{SB}$ is the Stefan-Boltzmann constant, and $T|{m,4}
\simeq 3$ for observed accretion discs \cite[with a broad
scatter,]{zm93}.  The temperature profile $T(r)$ of a
geometrically-thin, optically-thick disc heated either by external
irradiation or viscous dissipation is
\begin{equation}
T \simeq 500 L_{46}^{3/8} T|{m,4}^{-1/2} r|{pc}^{-3/4}\Kelv.
\label{e:trad}
\end{equation}

It should be noted that $\langle f|{Edd}\rangle$, $f|{disc}$ and $M|h$
are not actually independent parameters of the hydrodynamical model.
The net force close to the centre (taking account of radiative forces)
can be described by an equivalent gravitating mass
\begin{equation}
M|{net}(\theta) = M|h - \left[1-f|{disc}+2f|{disc}\cos\theta\right]
\langle f|{Edd}\rangle M|h,
\end{equation}
which is determined by only two parameters.  However, we retain all
three parameters, because of their physical relevance -- processes
outside the domain of our hydrodynamic models, such as the stellar
dynamics close to the black hole or the details of the inner accretion disc
do, of course, depend on the separate values of these parameters.

\subsection{The thermal state of the nISM}
\label{a:nISM} 

The thermal equilibrium of the nISM is controlled by Compton heating
and cooling.  Neglecting relativistic and occupation number effects,
the net Compton heating rate, per unit volume, is
\begin{equation}
Q|{C} = {4\pi n|e \sigma|T\over m|e c^2}\left[ \int J_\nu h\nu\id\nu -
4k|BT|e\int J_\nu\id\nu \right].\label{e:tc}
\end{equation}
The ratio between the Compton cooling time and the flow time ($\simeq
r|{pc}/{\cal M}T_7^{1/2}10^{11}\secnd$) is
\begin{equation}
\label{e:compton}
\frac{t|{C,v}}{t|f} \approx 8\ee{-2}
\frac{r|{pc} {\cal M} T_7^{1/2}}{f|{Edd}M|{h,8}},
\end{equation}
where ${\cal M}$ is the flow Mach number, and $T_7$ its temperature in
units of $10^7\Kelv$.  The Compton heating time is $T/T|C$ times
$t|{C,v}$.  Only at velocities comparable to or smaller than the sound
speed can the temperature structure of the ISM gas have a significant
dynamical effect on the gas flows. Thus, adiabatic expansion can begin
to cool the global gas flow substantially below the Compton
temperature only at large radii ($\ga 10\parsec$). The wind velocity
will already be supersonic by the time gas reaches radii large enough
that it can cool -- it will coast outwards on its residual momentum
until it hits the galaxy ISM.

We have taken the Compton temperature as uniform throughout the flow.
This approximation may break down for a variety of reasons.  The
variation in the contribution of the accretion disc compared to the
direct radiation from the central source will almost certainly change
the spectrum; indeed, the spectrum of the accretion disc itself
depends on azimuthal angle in some models \cite{sm89}, although not
all \cite{ce87}.  It may also vary with distance from the nucleus as a
result of, for instance, photoelectric absorption or oblique
scattering by moving gas (although these processes will only be
important once the Thomson optical depth though the ISM is close to
unity).  The temperature will also increase, at smaller distances than
our simulations resolve, as a result of stimulated Compton scattering
\cite[\cf{}]{reesnf89}.

The likely variation in Compton temperature of the gas with the
strength of the accretion disc component may be investigated by
varying the strength of the UV bump in the spectrum, and solving
equation~\refeq{e:tc} \cite[or, in these results, more accurately by
including relativistic corrections to the cross section, as
in]{reesnf89}.  We compared the results for the default spectrum given
in Cloudy \cite{hazy} with the same spectrum with a power-law
interpolation under the UV bump component.  Without the bump, the
Compton temperature increases from $1.3\ee7\Kelv$ to $2.6\ee7\Kelv$,
and the luminosity is cut to 0.52 of its previous value: the UV bump
has a negligible effect on Compton heating, so $T|C\propto
1/(1+f|{disc})$.  Thus, the variation in the Compton temperature with
azimuthal angle is likely to be small enough to be negligible (if the
broad-band continuum is isotropic).  This also suggests that, taking
this mean spectrum to correspond to a population average, the
luminosity of the reflected component is characteristically equal to
the broad-band component \cite[detailed studies of form of the X-ray
spectrum, as determined by Compton reflection, suggest a similar
result, \eg{}]{gf91}.

Whether Comptonization is sufficient to maintain strict isothermality
or not, it is likely that the flow will remain hot, at least until
bremsstrahlung cooling becomes comparable to Comptonization, which
occurs when the flow density increases above
\begin{equation}
n|{Br} \simeq 3\ee5 {f|{Edd} M|{h,8}\over r|{pc}^2 T_7^{1/2}}
\cm^{-3},
\end{equation}
\ie\ the hot phase ionization parameter decreases below $\Xi \simeq 50
T_7^{-3/2}$.  In some of the simulations presented here, particularly
where gas is retained for many dynamical times, the densities violate
this limit in some places.  The extension of the models presented here
to treat the effects of gas with cooling not dominated by Compton
upscattering is discussed in Section~\ref{s:eddeff}.

We can, however, maintain the physical relevance of our models either
by rescaling dimensional parameters -- loading rates, sizes, sound
speeds -- or by accepting that the gas will cool (thereby increasing
the mean opacity and the effective radiative driving force).  We
discuss these implications more fully in Section~\ref{s:eddeff} --
for the present, it sufficient to warn that scale of the densities in
the models we present must be treated with caution: relative values
and flow topologies are likely to accurate nevertheless.

\section{Evolution of isothermal flows in stellar clusters}
\label{a:apspherical}

\begin{figure*}
\epsfxsize = 8cm
\begin{centering}
\begin{tabular}{ll}
a) & b) \\
\epsfbox{z_d.ps} &
\epsfbox{z_u.ps} \\
\end{tabular}
\end{centering}
\caption[Model \O\ section]{A section through Model \O, shown in
Fig.~\ref{f:modelz2d} (in which $m|K = 7$).  The points plotted show
a) density b) total velocity, along the line $r=z$.  In a), dashed
lines show the density derived from the hydrostatic model for the core
and the ballistic model for the accreting halo.  The velocity profile
in the accreting halo model is shown in b).  The numerical results
agree well with the core model inside the shock at $r \simeq
0.5\parsec$, and with the accreting halo model beyond it.  Small
deviations in the core probably result from vibrations or some
supra-thermal support by turbulence.}
\label{f:modelzsect}
\end{figure*}

In this appendix, we derive an approximate analytic model for the
transient behaviour of an isothermal ($c_s =$ const) flow in a
uniform, spherical stellar cluster with no black hole.  This model
serves as a test case for our numerical models.  It also illustrates
that even in spherical symmetry, the flows within stellar clusters can
take very many dynamical timescales to reach equilibrium.

We consider the case where the Keplerian velocity at the edge of the
cluster is supersonic, $v|K \ga c|s$.  When this is not the case,
gravitational effects are negligible, and the flow will relax to
equilibrium on a sound-crossing timescale.  Here, we derive how this
relaxation timescale increases as the gravitational field increases.

We take the stellar cluster to have a uniform core and no halo, \ie\
$s=0$ and $h=\infty$.  The mass loading rate, $\dot{q}$, is uniform
throughout the cluster following its abrupt turn-on at time $t=0$.
The gravitational force is then
\begin{equation}
g = \left\{
\begin{array}{lc}
\displaystyle\left(v|K\over r|c\right)^2r & \mbox{for }r\le r|c \\
\displaystyle\left(v|K\over r\right)^2r|c & \mbox{for }r> r|c,
\end{array}
\right.
\end{equation}
where $v|K^2 \equiv GM|c/r|c$.

If there were free accretion at the centre of the cluster, the flow
would relax to a doubly-sonic equilibrium solution.  This solution has
an outward-going sonic point at a radius $r|s^+ \simeq (v|K^2/2c|s^2)
r|c$ (where the flow density is $\rho^+$), and an inner sonic point
within the cluster.  The ratio of mass fluxes between outer and inner
sonic radii is
\begin{equation}
{4\pi (r|s^+)^2 \rho^+ c|s \over 4\pi (r|s^-)^2 \rho^- c|s} =
\left(r|s^+\over r|s^-\right)^2 {\rho^+\over\rho^-}.
\end{equation}
While the area ratio increases between the sonic radii, the density
decreases exponentially in the near-hydrostatic region between the
sonic points and the total mass flux ratio is small, so little mass is
lost to the cluster at this stage.  The stagnation radius, $r|0$,
(where the flow velocity is zero) is only just inside the edge of the
cluster core.

To determine the flow within the cluster, we use the mass and momentum
equations, equation~\refeq{e:masscons} (integrated, assuming the
stagnation radius is coincident with the edge of the cluster)
and~\refeq{e:motion}, to obtain
\begin{eqnarray}
3 r^2 \rho v & = & -\dot{q}\left(r|c^3-r^3\right) \\
(v^2-c|s^2)\rdif{v}{r} & = & -{v|K^2\over r|c^2} rv 
- - {\dot{q}\over \rho}\left(v^2+c|s^2\right)
+ 2{v c|s^2\over r}.\label{e:inmach}
\end{eqnarray}
Below, we use dimensionless variables $m=v/c|s$ and $x=r/r|c$ (so that
$m|K=v|K/c|s$ is the Mach number at the Keplerian velocity).

We approximate the accretion flow by free-fall collapse.  This will be
strictly applicable only when there is an inner sonic point, which we
find is at $x|s$ given by
\begin{equation}
(m|K^2 x|s^2-2)(1-x|s^3) = 6 x|s^3.\label{e:inson}
\end{equation}
This equation has no solutions for $x|s$ between 0 and 1 if $m|K <
2\times2^{2/3} \simeq 3.17$.  For $m|K > 3.2$, the relevant solution,
$x|s^-$, may be approximated by expanding equation~\refeq{e:inson}
about $x|s^-=1$:
\begin{equation}
x|s^- \simeq {m|K^2-4\over m|K^2-2}.\label{e:machx}
\end{equation}

Assuming $x < x^-|s \sim 1$, we take $c|s \simeq 0$ in
equation~\refeq{e:inmach} to obtain
\begin{equation}
m\rdif{m}{x} = - m|K^2 x + {3x^2m^2\over 1-x^3}.
\end{equation}
Taking $m=0$ at $x=1$ we find the inward velocity is
\begin{equation}
v = -v|K\,\times\, (1-x^3)^{-1}\sqrt{9-20x^2+16x^5 - 5x^8\over 10}.
\label{e:freefallv}
\end{equation}

When the accretion rate at the centre is less than that at which mass
is added throughout the flow, gas will pile up in a hydrostatic core,
surrounded by an accretion shock.  The density distribution in the
core (assuming $u \simeq 0$ here) is
\begin{equation}
\rho = \rho|s \exp\left[\left(v|K^2\over 2c^2\right)
\left(x|s^2-x^2\right)\right],
\label{e:hydrostaticmodel}
\end{equation}
where $\rho|s$, $x|s$ are the values just inside at the accretion
shock.  These may be found, at any time, by equating $\rho|s c^2$ with
the ram pressure of the accretion flow, $\rho v^2$, and by setting the
mass accumulated in the core equal to the total mass input prior to
$t$:
\begin{equation}
{4\pi\over 3}r|c^3\dot{q}t \simeq r|c^3\int_0^{x|s} \rho(x) 4\pi x^2 \id{x}.
\label{e:banichi}
\end{equation}
The fit between equations~\refeq{e:freefallv}
and~\refeq{e:hydrostaticmodel} and the time-dependent numerical
model, Model \O\ above, is illustrated in Fig.~\ref{f:modelzsect}.

Together, these equations (\ref{e:hydrostaticmodel} and
\ref{e:banichi}) give the time at which the accretion shock reaches
$x|s$, $t(x|s)$, as a function of $m|K$.  The function $t(x|s)$ is
zero at $x|s=0$ and $x|s=1$, with a maximum at $t=t|{merge}$ at a
value of $x|s$ close to the edge of the stellar cluster for large
$m|K$.
At times later than $t|{merge}$, the inner
sonic point merges with the accretion shock, and supersonic inflow
ceases.  After this time, the stagnation radius begins to move inwards
within the cluster (when equilibrium is finally reached, the stagnation
radius is at the centre of the cluster).

For large $m|K$, the timescale over which the hydrostatic core fills
the central stellar cluster can be approximated by
\begin{equation}
t|{fill} \simeq 3\sqrt{2\pi} t|K e^{m|K^2/2}/m|K^3,
\end{equation}
where $t|K = r|c/v|K$ is the free-fall timescale within the stellar
cluster.  When $m|K = 3$, the filling timescale is roughly 25 times
the free-fall timescale, rising to roughly 350 times for $m|K=4$ and
very rapidly thereafter.  A yet longer time, roughly $4t|K/m|K^5
e^{m|K^2-2}$, is then taken to fill the region beyond the cluster core
out to the outer sonic radius, $r|c m|K^2/2$.  Only after this time,
far longer than any of the obvious dynamical timescales, is
equilibrium finally achieved throughout the subsonic region.  The wind
through the sonic radius, which has gradually been strengthening, then
removes a mass flux equal to the mass input in the core.

\bsp
\label{lastpage}
\end{document}